\documentclass[apj]{emulateapj}
\usepackage[colorlinks=true,linkcolor=blue,citecolor=blue]{hyperref}

\usepackage{lipsum, babel}
\usepackage[left]{lineno}

\usepackage[percent]{overpic}
\usepackage{xcolor}
\usepackage[export]{adjustbox}

\usepackage{amssymb}
\usepackage{amsmath}

\usepackage{sans}

\makeatletter\AtBeginDocument{\let\@elt\relax}\makeatother

\usepackage{graphicx}
\usepackage{graphics}

\newbox{\bigpicturebox}

\usepackage{dcolumn}
\usepackage{color}
\usepackage{apjfonts}
\usepackage{rotate}
\usepackage{fancyhdr} 
\usepackage{hyperref}
\usepackage{indentfirst}
\usepackage{booktabs} 
\usepackage{fontawesome} 
\usepackage{physics} 
\usepackage{umoline}
\usepackage{placeins}

\def\apjl{ApJL}

\def\be{\begin{eqnarray}}
\def\ee{\end{eqnarray}}
\def\ba{\begin{eqnarray}}
\def\ea{\end{eqnarray}}
\def\no{\nonumber}

\def\V{\bf \color{blue}\surd}
\def\X{\bf \color{red}X}

\definecolor{darkred}{rgb}{.743,0,0}

\begin{document}

\title{Globular cluster distributions as a dynamical probe of dark matter}
\author{Nativ Ben-Yeda}\email{nativ.ben-yeda@weizmann.ac.il}
\affiliation{Weizmann Institute of Science, Rehovot 7610001, Israel} 
\author{Kfir Blum}\email{kfir.blum@weizmann.ac.il}
\affiliation{Weizmann Institute of Science, Rehovot 7610001, Israel} 
\author{Inbar Havilio}\email{havilio@tauex.tau.ac.il}
\affiliation{Weizmann Institute of Science, Rehovot 7610001, Israel} 
\affiliation{Tel Aviv University, Tel Aviv 69978, Israel} 
\begin{abstract}
Globular clusters (GCs) act as massive probe particles traversing the dark matter halos of their host galaxies. Gravitational dynamical friction due to halo particles causes GC orbits to contract over time, providing a beyond-mean field test of the cold dark matter paradigm. We explore the information content of such systems, using N-body and semianalytic simulations and scanning over a range of initial conditions. We consider data from the ultradiffuse galaxies NGC5846-UDG1 and UDG-DF44, and from the Fornax dwarf spheroidal galaxy. The GC systems of UDG1 and Fornax indicate the presence of dark matter halos, independent of (but consistent with) stellar kinematics data. UDG-DF44 is too diffuse for dynamical friction to give strong constraints. Our analysis can be extended to many additional galaxies. 
\end{abstract}
%
\maketitle
%


\section{Introduction}

Globular clusters (GCs) act as massive probe particles traversing the halo of their host galaxy. Dynamical friction (DF) causes GC orbits to contract over time, so the distribution of GC projected radial positions in dark matter (DM)-dominated galaxies provides information on the DM halo~\citep{Hernandez:1998hf,ohlinricher2000,Lotz:2001gz,Read:2006fq,Goerdt2006,SanchezSalcedo:2006fa,angus2009resolving,Cowsik:2009uk,Petts2015,Hui:2016ltb,Contenta:2017jph,Leung_2019,Berezhiani2019,orkney2019,boldrini2020embedding,Hartman:2020fbg,Lancaster2020,Dutta_Chowdhury_2019,Dutta_Chowdhury_2020,RenaCampos2021,Bar:2021jff,Bar:2022liw,Liang:2023ryi,Modak:2022kxv}. This information is distinct from the information available from stellar kinematics: while kinematics relies on the mean field of the halo, DF is an beyond-mean field effect that arises from gravitational scattering between GCs and halo particles. 

Dwarf and ultradiffuse galaxies (UDGs)~\citep{vanDokkum:2014cea,2017MNRAS.468.4039R,Lim_2018,Saifollahi:2022yyb,Forbes:2024esb} are promising systems to explore DM-induced DF, because they tend to host large numbers of GCs relative to their stellar mass, and because  kinematics suggest that at least some dwarf and UDGs are DM-dominated. 
A good example is NGC5846-UDG1 (UDG1)~\citep{muller2020spec,muller21udg,Forbes:2020vly,2022ApJ...927L..28D,Bar:2022liw,Mahdavi:2005pr}, which 
hosts about 30 GC candidates within twice its stellar Sérsic radius $R_{\rm e}\approx2$~kpc. \cite{muller21udg} noted that UDG1's GCs show evidence of mass segregation, and \cite{Bar:2022liw} and \cite{Liang:2023ryi} demonstrated that this mass segregation can indeed be attributed naturally to DM-induced DF. A similar result was reported in~\cite{Modak:2022kxv} for the dwarf galaxy UGC7369. 
If these interpretations of GC data are correct, they amount to a rare beyond-mean field positive signature of DM, and can lead to constraints on the fundamental nature of DM (see, e.g., \cite{Bar:2021jff,2024A&A...690A.119B}; related effects are dynamical heating and relaxation~\citep{Hui:2016ltb,Bar-Or:2018pxz,Church:2018sro,DuttaChowdhury:2023qxg,Graham:2023unf}\footnote{Other beyond-mean field analyses focus on DM substructure, via, e.g., gravitational lensing magnification anomalies~\citep[][]{Dalal:2001fq,Vegetti:2014wza,Hezaveh:2016ltk,Wagner-Carena:2022mrn}, stellar streams~\citep[][]{Bovy:2015mda,Amorisco:2018dcn,Banik:2019smi}, and precision  astrometry~\citep[][]{Necib:2018iwb,Ravi:2018vqd}.}.

While the motivation is compelling, the interpretation of GC data is affected by a long list of observational and theoretical challenges. On the observational side, robust identification of GC candidates is important, accounting for background and incompleteness\footnote{GC identification can be far from trivial; see, e.g., the discussion in~\cite{2021MNRAS.502.5921S}, concerning GCs of the Coma cluster's UDG DF44, and the analysis devoted in~\cite{2022ApJ...927L..28D} to GCs in UDG1.}. In addition, the true power of the data comes into play if we have mass estimates for GC candidates. This requires spectroscopy for at least a sample of the GCs, and luminosity and color information for all GCs and for the stellar body of the galaxy. Uncertainties in mass-to-light ratio and in GC age should also be considered\footnote{For reference, Galactic GCs have an $M/L_V$ spread of about 25\%~\citep{2020PASA...37...46B}. The $M/L_V$ spread of extragalactic GCs (including both intrinsic and modeling uncertainties) is necessarily larger.}. 
On the theory side, a central unknown is the initial conditions of the GC radial distribution and the GC initial mass function (GCIMF). 
Any observed snapshot of the GC system today can be mapped back to some (in general, non unique) set of initial conditions. Thus, the problem of identifying the impact of DF and other dynamical effects boils down to the problem of 
defining a sufficiently complete parameterization of the space of plausible initial conditions, surveying the predictions arising from each model in that space, and finding either success or contradiction in the comparison with observational data.
 
Our approach towards this problem, and the layout of this paper, are as follows. 
In Sec.~\ref{s:sims} we describe our method to survey the space of initial conditions, using fast semianalytical simulations that treat GC-GC interactions in detail, but implement DF approximately via the Chandrasekhar formula~\citep{Chandra43,BinneyTremaine2}. 
Sec.~\ref{s:model} describes our modeling parameters, DM halo models, parameterization of the GCIMF, GC mass loss, and GC mergers. 
Secs.~\ref{s:udg1},~\ref{s:fornax}, and~\ref{s:df44} contain our results focusing on three case studies: UDG1, the Fornax dwarf spheroidal (dSph), and the Coma cluster UDG DF44. 
For UDG1, our results suggest that a DM halo is needed for a reasonable description of the GC system. This conclusion remains robust under various initial conditions, within what we think are reasonable limits. 
For Fornax, we show that the GC system -- despite its meager statistics -- also provides a strong indication of a DM halo. 
Naive estimates of the DF time for some of Fornax's GCs are much shorter than their age~\citep{Tremaine1976a,Hernandez:1998hf,ohlinricher2000,Lotz:2001gz,Goerdt2006,Cowsik:2009uk,angus2009resolving,Cole2012,Kaur2018,Hui:2016ltb,Leung_2019,boldrini2020embedding,Berezhiani2019,Hartman:2020fbg,Bar-Or:2018pxz,Lancaster2020,Meadows20}, leading to suggestions of exotic properties of DM (see \cite{Bar:2021jff} for review of ideas). We show that exotic models of DM are not needed and that the minimal theory of DM-induced DF can explain the GC data. 
For UDG-DF44, we show that reasonable initial conditions can roughly match the data, despite massive off-center GCs that could naively appear to challenge the DM-induced DF paradigm. 
We discuss and summarize our results in Secs.~\ref{s:disc} and~\ref{s:sum}. A number of technical issues and simulation verification tests are reserved to Appendices~\ref{a:calib},~\ref{a:chandra}, and~\ref{a:checks}.

\section{Preliminaries and calculation set up}\label{s:sims}
The Chandrasekhar formula for the DF deceleration of a GC of mass $M_{\rm GC}$ moving with instantaneous velocity $V$ in a background gas of density $\rho$ reads~\citep[][]{Chandra43,BinneyTremaine2}
\be\label{eq:Chandraa}  \dot V&=&-\frac{1}{\tau_{\rm DF}}V,\\
\label{eq:Chandratau}\tau_{\rm DF}&=&\frac{V^3}{4\pi G^2\,M_{\rm GC} \,\rho\,C_{\rm DF}\ln\Lambda},\\
\label{eq:ChandraC}C_{\rm DF}&=&4\pi\int_0^Vdv\,v^2\,f(r,v).\ee
Here, $\tau_{\bf DF}$ is the DF deceleration time. $C_{\rm DF}$ is a dimension-less constant, encoding information from the phase space distribution function $f(r,v)$. 
We define $f(r,v)$ such that $4\pi\int dv\,v^2\,f(r,v)=1$. $G$ is Newton's constant. The Coulomb log $\ln\Lambda$ encodes the dynamical range of the two-body impact parameter. It reveals the sensitivity of the semianalytic formulation of DF to the boundary conditions of the system. We set
\be \Lambda&=&\sqrt{1+\left(\frac{2V^2r}{GM_{\rm GC}}\right)^2}.\ee
For circular orbits $V= V_c(r)$ this becomes $\Lambda\approx2M_h(r)/M_{\rm GC}$, where $M_h(r)$ is the halo mass enclosed in $r$ (valid for $M_h(r)\gg M_{\rm GC}$).

Eqs.~(\ref{eq:Chandraa}-\ref{eq:ChandraC}) hold for a homogeneous distribution of background gas, but our application involves an inhomogeneous galaxy. The extension of the semianalytic result to more realistic set-ups was discussed in many studies. For early analyses, see~\cite{Tremaine1984,Weinberg1986}, and for more recent studies see e.g.~\cite{Inoue:2009em,Inoue:2009wd,Kaur2018,Kaur:2021bfs}. Numerical experiments show that Eq.~(\ref{eq:Chandraa}) succeeds quite well to describe DF in the outskirts of a galaxy halo, where the halo mass enclosed within the GC orbit is much larger than the GC mass. On the other hand, in the inner halo Eq.~(\ref{eq:Chandraa}) can fail in a way that depends on the details of the halo profile and the particular orbit of the GC.  

We address this problem using a mixed approach, performing the following sets of calculations:
\begin{itemize}
\item Semianalytic: Our workhorse is semianalytic orbit integration, that uses Eq.~(\ref{eq:Chandraa}) with the (inhomogeneous) local halo density as input and solves GC orbits by direct integration. The halo potential is treated as a mean field, but the GC system itself is treated as an N-body problem such that GC-GC interactions are included. DF is modeled using the Chandrasekhar formula~\citep[][]{Chandra43}. GC mergers are treated using an effective prescription, similar to that described in \cite{Bar:2022liw}. This method allows to simulate thousands of  GC systems using relatively little computing power. 
We compare the semianalytic computation to N-body simulations (next item) in App.~\ref{a:calib}. 
\item N-body: We complement the semianalytic calculations by N-body simulations of GC motion in a ``live" halo, using a Barnes-Hut modified tree code~\citep[][]{Barnes:1986nb}. By default, we use an opening angle parameter $\theta=0.5^o$ and Plummer softening length of 3~pc. Our convergence tests show that modifying these parameters by a factor of few does not affect our results. The halo phase space is initialized using the Eddington formalism~\citep[][]{BinneyTremaine2}. DM and stars in the simulation have the same mass per particle, with the ratio of DM and star particle numbers matching the ratio of respective halo masses. GCs are added as point particles with mergers modeled as in the semianalytic method. 

To speed up the calculation, DM and star particles in the simulation are confined inside a spherical region of radius $R_{\rm sim}$. The velocity vector of a DM or star particle that encounters the boundary is reversed in the next time step. GCs travel freely in and out of $R_{\rm sim}$. To calculate GC orbits at $r>R_{\rm sim}$ we replace the N-body gravitational field by the mean field of the halo, and implement DF via the semianalytic formula. 
The $R_{\rm sim}$ approximation allows to concentrate resources in the most interesting region. We verify convergence by comparing results for different $R_{\rm sim}$. As a rule, we find convergence when $R_{\rm sim}$ is comparable to the characteristic scale of the halo density profile.

Our tree code is written in GO \footnote{\url{https://go.dev/}. Code available on request.}, and we call it GONBY. 

We repeat some calculations using full N-body integration, implemented in Matlab \footnote{\url{https://ch.mathworks.com/products/matlab.html}. Code available on request.}. 
\end{itemize}
In the rest of this section we comment on our (standard) treatment of the phase space distribution function, used for initialization of N-body simulations, and to compute the Chandrasekhar coefficient in the semianalytic calculation.  

We construct statistically ergodic, isotropic, spherically-symmetric distribution functions as follows. 
Defining $\Psi(r)=-\Phi(r)$, where $\Phi(r)$ is the mean halo gravitational potential including all particle populations; and $\varepsilon=\Psi(r)-\frac{1}{2}v^2$; the distribution function can be obtained via~\citep[][]{BinneyTremaine2}\footnote{We consider distributions that satisfy $\lim_{r\to\infty}\left(r^2\frac{d\rho}{dr}\right)=0$. Otherwise, a boundary term needs to be added in Eq.~(\ref{eq:f}).}:
\be\label{eq:f} \rho(r)\,f(r,v)&\equiv&F(\varepsilon)=\frac{1}{\sqrt{8}\pi^2}\int_0^\varepsilon\frac{d\Psi}{\sqrt{\varepsilon-\Psi}}\frac{d^2\rho}{d\Psi^2}.\ee
Given a choice of $\rho(r)$, we compute the resulting $\Psi(r)$, assemble the function $\rho(\Psi)=\rho(\Psi(r))$, compute the derivative $\frac{d^2\rho}{d\Psi^2}(\Psi)$, 
and then perform the integral of Eq.~(\ref{eq:f}) to obtain $F(\varepsilon)$ and $f(r,v)=F(\varepsilon(r,v))/\rho(r)$. It is convenient to use $d\Psi/dr=-GM(r)/r^2$ and calculate $d^2\rho/d\Psi^2$ from the expression
\be\label{eq:d2rhodpsi2simple}\frac{d^2\rho_i}{d\Psi^2}(\Psi)&=&\frac{r^2}{GM(r)}\frac{d}{dr}\left[\frac{r^2}{GM(r)}\frac{d\rho_i}{dr}\right].\ee
Then
\be\label{eq:f2} F_i(\varepsilon)
&=&\frac{1}{\sqrt{8}\pi^2}\int_{r_\varepsilon}^\infty\frac{dr}{\sqrt{\varepsilon-\Psi(r)}}\frac{r}{GM(r)}\times\no\\
&&\left[\left(2-\frac{4\pi r^3\rho(r)}{M(r)}\right)\frac{d\rho_i}{dr}+r\frac{d^2\rho_i}{dr^2}\right],
\ee
where $\Psi(r_\varepsilon)=\varepsilon$. This expression can be convenient if derivatives of density profiles are known analytically. 

Here (and implicitely in Eq.~(\ref{eq:f})) we allow for different distributions $f_i$ to coexist. Our fiducial models include star and DM particles, and Eq.~(\ref{eq:f}) is evaluated using each $\rho_i$ separately with the same total self-consistent $M(r)$. 

As is well known, Eq.~(\ref{eq:f}) does not guarantee a physically-acceptable distribution function. It can produce negative $f(r,v)$ if a stationary solution does not exist for the prescribed form of $\rho(r)$. 
For example, if $M(r)$ is dominated by a cusp DM density profile $\rho_\chi(r)\propto1/r$, for which the inner halo satisfies $r^2/M(r)\approx$~Const., then a stellar density component $\rho_*(r)$ must satisfy $d^2\rho_*/dr^2>0$ in that region to yield non-negative $f_*(r,v)$. This condition is not satisfied, for example, by a Plummer profile: spherical isotropic stellar cores cannot formally exist in statistical equilibrium with a dominant DM cusp~\citep[][]{Almeida:2023xmx,Almeida:2024cqa,Almeida:2025ggc}. 
In practice, no galaxy halo is exactly spherical, isotropic, or in equilibrium, and the formal inconsistency may be unimportant if it only affects a small region of the halo. We comment on this point again in Sec.~\ref{ss:dmhalo}. 

In the semianalytic calculation we evaluate $C_{\rm DF}$ directly from $f(r,v)$. 
App.~\ref{a:chandra} compares this direct implementation to analytic results based on the Maxwellian velocity distribution. When different particle distributions co-exist, the total DF deceleration rate is obtained from $\tau_{\rm DF}^{-1}=\sum_i\tau_{{\rm DF},i}^{-1}$ using the relevant $\rho_i$ and $C_{{\rm DF},i}$ in Eq.~(\ref{eq:Chandratau}).
%

\section{Theoretical models and modeling assumptions}\label{s:model}
\subsection{Stellar and dark matter halo models}\label{ss:dmhalo}
The contribution of stars to the halo total density profile is added based on the observed surface brightness $\mathcal{S}_*(r_\perp)$, which is modeled by the S\'ersic profile with parameters $n$ and $R_{\rm e}$, and total stellar mass $M_*$~\citep[][]{Sersic1963}:
\be \mathcal{S}_*(r_\perp)&=&S_{0}e^{-b\left(\frac{r_\perp}{R_{\rm e}}\right)^{\frac{1}{n}}}.\ee
$b$ is the solution of $\Gamma\left(2n\right)=2\Gamma\left(2n,b\right)$, so the projected cumulative luminosity profile $\mathcal{L}_{*}(r_\perp)=2\pi \int_0^{r_\perp} dxx\mathcal{S}_*(x)=\frac{2\pi n}{b^{2n}}S_{0}R_{\rm e}^2\left[\Gamma(2n)-\Gamma\left(2n,b(r_\perp/R_{\rm e})^{\frac{1}{n}}\right)\right]$ satisfies $\mathcal{L}_{*}(R_{\rm e})=0.5L_*$, with $L_*=\mathcal{L}_{*}(\infty)$ the total luminosity. $S_0$ is related to $M_*$ via $M_*=\left(M/L\right)_*L_*=\left(M/L\right)_*\frac{2\pi n}{b^{2n}}S_{0}R_{\rm e}^2\Gamma(2n)$, with $\left(M/L\right)_*$ the stellar mass-to-light ratio.
The 3D stellar density profile is given by the inversion formula,
\be\label{eq:sers}\rho_*(r)&=&-\left(M/L\right)_*\frac{1}{\pi}\int_r^\infty\frac{dx}{\sqrt{x^2-r^2}}\frac{d\mathcal{S}_*(x)}{dx}\no\\
%
&=&\frac{M_*}{R_{\rm e}^3}\frac{b^{2n+1}}{2\pi^2 n^2\Gamma(2n)}\int_{\frac{r}{R_{\rm e}}}^\infty\frac{dy}{\sqrt{y^2-\left(\frac{r}{R_{\rm e}}\right)^2}}y^{\frac{1}{n}-1}e^{-by^{\frac{1}{n}}},\;\;\;\;\\
&&\;\;\;\;\;\;\;\;\;\;{\rm Sersic\;\;(stellar~core)}.\no
\ee

The fiducial DM halo models we consider are the Burkert~\citep[][]{Burkert:1995yz} core and Navarro-Frenk-White (NFW)~\citep[][]{Navarro:1996gj} cusp profiles: 
\be\label{eq:burk}\rho(r)=\frac{\rho_0}{\left(1+\frac{r}{R_0}\right)\left(1+\frac{r^2}{R_0^2}\right)},\;\;\;\;\;\;\;\;\;\;{\rm Burkert\;\; (core)},\;\ee
\be\label{eq:nfw}\rho(r)&=&\frac{\rho_s}{\frac{r}{R_s}\left(1+\frac{r}{R_s}\right)^2},\;\;\;\;\;\;\;\;\;\;{\rm NFW\;\; (cusp)}.\ee
To investigate halos of different mass, we vary the density parameters $\rho_0$ and $\rho_s$ such that the DM halo mass enclosed inside 3D radius $r=2R_{\rm e}$ is equal to some multiplier of the total stellar mass of the galaxy. 
We set the radial parameters $R_s$ and $R_0$ fixed and equal to $3\,R_{\rm e}$. 
%
We do not vary $R_0$ or $R_s$ because the key difference between the Burkert and NFW models is the core vs cusp. Changing $R_0$ or $R_s$ either blurs this difference, or is degenerate with changing $\rho_0$ or $\rho_s$. 
{ 
The truth may lie in between, e.g., baryonic feedback may flatten the inner region of otherwise cusp profiles~\citep{Pontzen2012,Cole2012,Oman2016,Read:2018fxs}. Intermediate configurations would give results in between our two scenarios. 
}

As noted in Sec.~\ref{s:sims}, a spherical isotropic stationary stellar core cannot coexist in the background of a dominant dark matter cusp profile: the Eddington formalism produces phase space distribution function for the stars that goes negative in the inner core. In this work we simply set $f(r,v)\to0$ for values of $r$ or $v$ in which the Eddington procedure prescribes negative $f(r,v)$. Numerical simulations show that the profiles produced this way evolve slightly to adjust for this inconsistency, then remain stationary at a slightly modified profile, without significant effect to our results.

\subsection{GCIMF}\label{ss:gcimf}
At the start of a simulation run, we assign GC masses from the distribution
\be\label{eq:GCIMF}\frac{dN_{\rm GC}}{dM_{\rm GC}}&=&A_MM^{-\alpha_M}_{\rm GC},\;\;\;\;M_{\rm min}<M_{\rm GC}<M_{\rm max}.\ee
The power-law slope of this distribution is $\alpha_M$, and the distribution is understood to extend only between a minimal mass $M_{\rm min}$ and a maximal mass $M_{\rm max}$. We use $\alpha_M=2$, consistent with observations of YMCs in the Milky Way and near-by galaxies~\citep[][]{Kravtsov:2003sm,Gnedin:2013cda,Longmore:2014epa,Krumholz2019}\footnote{It should be noted that the identification of GCs as an older phase of YMCs is not free of debate~\citep[][]{Charbonnel:2015xgm}.}. We also set $M_{\rm min}=5\times10^4~M_\odot$. Changing the value of $M_{\rm min}$ in the range $10^3-10^5~M_\odot$ has little effect on the results, as GC mass loss (discussed further below) during $\sim10$~Gyr of evolution eliminates GCs with initial mass below $\sim10^5~M_\odot$. 
The value of $M_{\rm max}$ is important to the results, as noted in the context of Milky Way GCs~\citep[][]{Gnedin:2013cda}. We treat $M_{\rm max}$ as a free parameter. We will see that consistent models for the GC population of UDG1 require $M_{\rm max}\sim10^6~M_\odot$, a smaller value than the $M_{\rm max}\sim10^7~M_\odot$ used by~\cite{Gnedin:2013cda} for the Milky Way and M31, but consistent with observations for lower mass galaxies in the Virgo cluster~\citep[][]{Jordan:2007xc}. A smaller $M_{\rm max}$ for a lower mass halo is qualitatively consistent with cosmological simulations~\citep[][]{Kravtsov:2003sm,Kim:2017gsp,Hughes:2021got}.

The normalization constant $A_M$ in Eq.~(\ref{eq:GCIMF}) is also a free parameter, but it is constrained by observations once the rest of our procedure has been laid out. We fix this parameter such that in each of our models, the final GC distribution surviving dynamical evolution and mass loss gives a total mass in GCs that is approximately consistent with observations. 

\subsection{GC mass loss}\label{ss:gcmassloss}
The GCMF is modified by mass loss and mergers during the simulation. We follow the simplified mass loss prescription adopted in \cite{Gnedin:2013cda}, in which GCs lose mass via three main mechanisms~\citep[][]{PortegiesZwart:2010cly,Lamers:2010vv,Gnedin:2013cda,Krumholz2019}. The first is stellar evolution, which causes a newly formed star cluster to expel as much as $\sim50\%$ of its mass in a time scale $t\lesssim100$~Myr. We do not explicitly account for this early mass loss in our simulations, because the dynamical evolution we are most interested in occurs on a time scale of Gyrs. Instead, the GCIMF we use can be considered as the GCMF after a few 100~Myr of evolution. 

The second mechanism is dynamical evaporation of stars from the cluster due to two-body interactions. In our baseline calculations we adopt 
\be\label{eq:Mdotiso} \left(\frac{dM_{\rm GC}}{dt}\right)_{\rm iso}&=&\frac{10^{5}~M_\odot}{8.5~\rm Gyr},\ee
considered in \cite{Gnedin:2013cda} for isolated GCs without a background tidal field. Eq.~(\ref{eq:Mdotiso}) is independent of $M_{\rm GC}$, reflecting the assumption of a two-body evaporation rate proportional to the half-mass relaxation time, at an approximately constant half-mass  density~\citep[][]{1987degc.book.....S}. This is probably a crude estimate, a point we return to later. 

\begin{figure*}
\centering
         \includegraphics[scale=0.55]{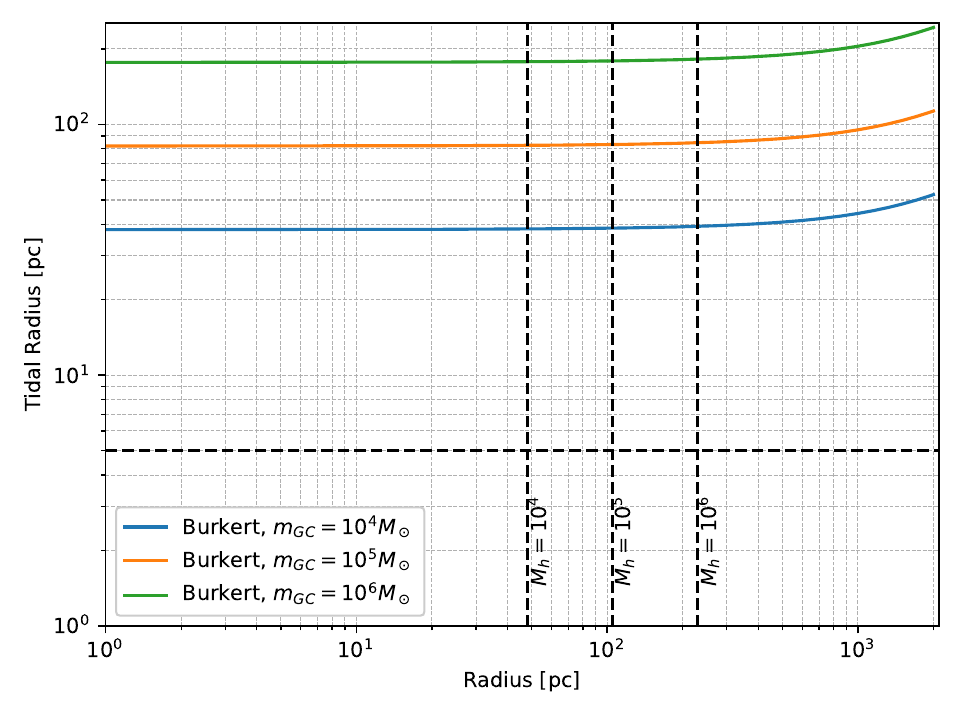}
                  \includegraphics[scale=0.55]{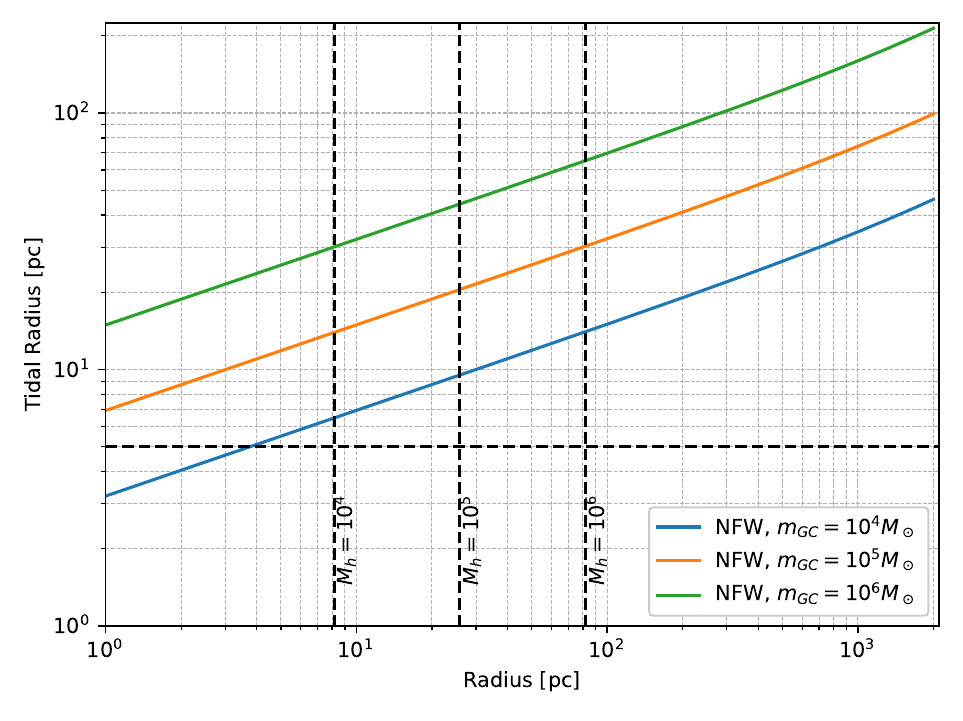}
 \caption{Green, orange, and blue lines (with green being top-most and blue bottom-most) show the tidal radius $r_t$ (Eq.~(\ref{eq:rt})) for a GC with mass $10^4,10^5,10^6~M_\odot$. Left panel: Burkert halo; right panel: NFW halo. Vertical dashed lines mark the radii where the enclosed halo mass $M_h(r)$ is equal to $10^4,10^5,10^6$~M$_\odot$. Horizontal dashed line marks a characteristic half-mass radius for a GC, taken here as 5~pc. Tidal dissolution of the GC would be naively expected when $r_t$ becomes small than the half-mass radius. This situation is never realized in the Burkert halo. For the NFW halo, this situation can appear to occur for the lightest GC of $10^4$~M$_\odot$, however at that point the enclosed NFW halo mass is smaller than the GC mass, which suggests that a naive tidal disruption estimate is not applicable. 
 }
 \label{fig:rJ}
\end{figure*}

The third mass loss mechanism is tidal mass loss due to the halo mean field. To model this, we again follow \cite{Gnedin:2013cda}, setting
\be\label{eq:Mdottid} \left(\frac{dM_{\rm GC}}{dt}\right)_{\rm tid}&=&\left(\frac{1~{\rm kpc}}{r}\right)\left(\frac{V_{\rm c}}{10~{\rm km/s}}\right)\left(\frac{M_{\rm GC}}{10^5~M_\odot}\right)^{\frac{1}{3}}\frac{10^{5}~M_\odot}{25~\rm Gyr}.\no\\
&&\ee
In principle, GCs can also experience rapid tidal destruction if the effective local halo density ($3M_h(r)/(4\pi r^3)$) becomes larger than the mean GC density, provided that the halo mass enclosed inside this region is significantly larger than the GC mass. This situation, however, is not expected in a UDG-like galaxy. To see this, Fig.~\ref{fig:rJ} shows the tidal radius $r_t$ as a function of orbital radius $r$, calculated for a GC on a circular orbit via
\be\label{eq:rt} r_t&=&\left[\frac{M_{\rm GC}}{2M_h(r)}\right]^{\frac{1}{3}}r.\ee
(The factor of $2$ in Eq.~(\ref{eq:rt}) says we take the mean GC density at $r_t$ as $3M_{\rm GC}/(4\pi r_t^3)$, a rough approximation, but sufficient for the current estimates.)
The {left (right) panel} shows the result for the Burkert (NFW) halo, for a GC mass of $10^4,10^5,10^6~M_\odot$. 
In the core profile, at $r\ll R_0$ the tidal radius saturates to $r_t\approx\left(\frac{3M_{\rm GC}}{8\pi\rho_0}\right)^{\frac{1}{3}}\approx100\left(\frac{M_{\rm GC}}{10^5~M_\odot}\right)^{\frac{1}{3}}\left(\frac{10^7~M_\odot/{\rm kpc}^3}{\rho_0}\right)^{\frac{1}{3}}$~pc. In the NFW halo, at $r\ll R_s$ the tidal radius approaches $r_t\approx\left(\frac{M_{\rm GC}r}{4\pi\rho_sR_s}\right)^{\frac{1}{3}}\approx60\left(\frac{M_{\rm GC}}{10^5~M_\odot}\frac{2\times 10^6~M_\odot/{\rm kpc}^3}{\rho_s}\frac{r/(100~{\rm pc})}{R_s/(6~{\rm kpc})}\right)^{\frac{1}{3}}$~pc. These numbers suggest that for the UDG1-like halo models considered in this work, tidal disruption by the halo mean field should be insignificant for the Burkert core model for all GCs of interest, and remain insignificant even for the NFW cusp model for $M_{\rm GC}\gtrsim10^5~M_\odot$. For lower mass GCs, the case for tidal disruption by an NFW cusp requires careful assessment: the radius $r_p(M_{\rm GC})$ at which the enclosed halo mass equals the GC mass is $r_p\approx\left(\frac{M_{\rm GC}}{2\pi\rho_sR_s}\right)^{\frac{1}{2}}\approx10\left(\frac{M_{\rm GC}}{10^4~M_\odot}\frac{2\times10^6~M_\odot/{\rm kpc^3}}{\rho_s}\right)^{\frac{1}{2}}$~pc, and one can check that $r_t\sim r_p$. The halo mean field cannot be used in this case to estimate tidal disruption. Instead, we expect the GC and the inner NFW cusp halo particles to participate in approximate two-body dynamics, smearing out the cusp alongside, perhaps, disrupting the GC. We leave a detailed investigation of this problem to a separate study. 
 
We implement mass loss in the simulations by updating the GC mass every 2~Myr. As in \cite{Gnedin:2013cda}, at each evaluation we choose
\be\label{eq:Mdot}\frac{dM_{\rm GC}}{dt}&=&-{\rm max}\left\{\left(\frac{dM_{\rm GC}}{dt}\right)_{\rm iso},\left(\frac{dM_{\rm GC}}{dt}\right)_{\rm tid}\right\}.\ee
Using at any given time just the one dominant mass loss mechanism, rather than combining both, is of course a crude approximation.

For a GC population evolving over duration $t$, assuming the GCIMF of Eq.~(\ref{eq:GCIMF}) with $\alpha_M\approx2$, the mass loss rate dictates the peak of the GCMF at $M_{\rm GC,peak}\approx\left(\frac{dM_{\rm GC}}{dt}\right)_{\rm iso}t$ (for this  estimate we neglect tidal losses). The rate assumed in Eq.~(\ref{eq:Mdotiso}) then leads to $M_{\rm GC,peak}\approx10^5~M_\odot$, consistent with observations~\citep[][]{Fall:2001ti,Gnedin:2013cda,Brodie:2006sd,Jordan:2007xc,Saifollahi:2022yyb}. The UDG1 GCMF analysis of \cite{2022ApJ...927L..28D} also seems consistent with these estimates: judging from Fig.~5 {\it there}, GCs with initial mass of $10^5~M_\odot$ lose about 85\% of their mass during the evolution, and essentially no GCs which start at $M_{\rm GC}<0.5\times10^5~M_\odot$ survive to the present day.
%
 
{
We note that the mass loss rate for an isolated GC, quoted in Eq.~(\ref{eq:Mdotiso}), is larger by a factor of $\sim$20 than the corresponding mass loss rate adopted in \cite{Liang:2023ryi} (guided by \cite{2016MNRAS.463L.103G}). We discuss this issue in more detail in Sec.~\ref{ss:comp}; and with this discrepancy in mind, we repeat several of our main calculations using a low mass loss rate compatible with that in \cite{Liang:2023ryi}. 
%
%
Low mass loss rate requires stronger DF at fixed GC radial distribution today~\citep{Bar:2022liw,Liang:2023ryi}, and numerical exploration in App.~\ref{a:massloss} shows that it also increases the statistical spread of the evolved GC radial distribution compared to its spread at larger mass loss rate. 
}
%

\subsection{GC mergers}\label{ss:mergers}
In simulations, a pair of GCs is merged if (i) the total energy of the pair is negative, (ii) the distance between GC centers of mass is smaller than 20~pc. In computing the total energy of the pair we approximate each GC by a Plummer sphere of radius 5~pc. This prescription is the same as that in~\cite{Bar:2022liw}, and very similar to that in~\cite{Modak:2022kxv}. We test for and implement mergers every 2~Myr (compared to every 20~Myr as done in~\cite{Bar:2022liw}).

Our merger prescription is a crude simplification of a complicated dynamical process. However, as long as we define our primary observables of the GC system in a way that does not depend strongly on the GC distribution in the deep inner region of the halo, we find that GC mergers are not a crucial ingredient in our main results. This point is demonstrated in App.~\ref{a:nomerge}, where we compare results using the baseline merger prescription to results in which GC mergers are disabled altogether. Note that the implementation of mergers becomes important if one tries to quantify the efficiency of NSC formation, and use it as a discriminator between models~\citep[][]{Modak:2022kxv}. We will comment on NSC formation in our simulations, but will not use it as a quantitative test of the models.

\section{Results: NGC5846-UDG1}\label{s:udg1}
%

\begin{figure*}
\begin{tabular}{@{}cc@{}}
    \raisebox{-\height}{\includegraphics[width=0.3\textwidth]{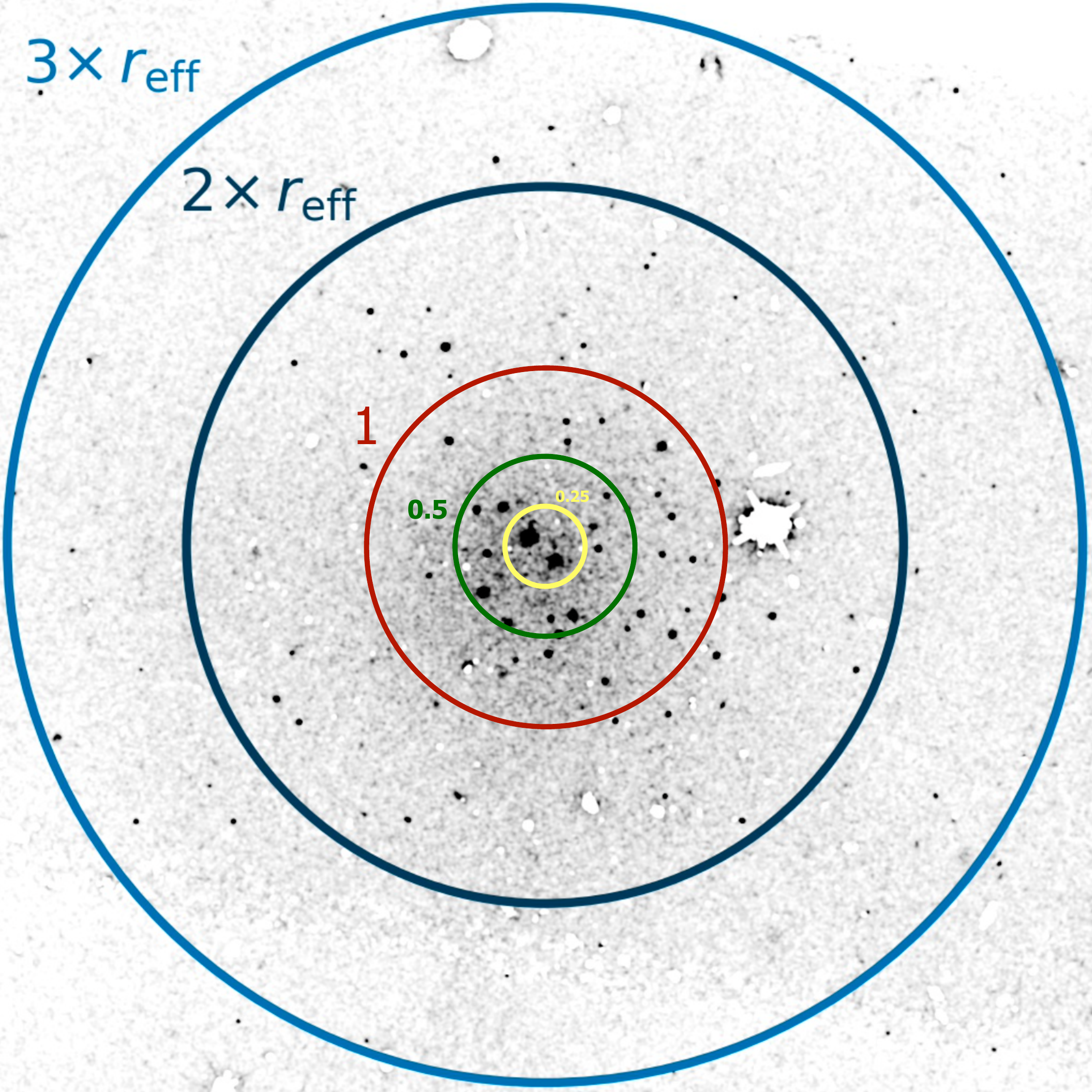}} & 
    \begin{tabular}[t]{@{}cc@{}}
        \raisebox{-\height}{\includegraphics[width=0.37\textwidth]{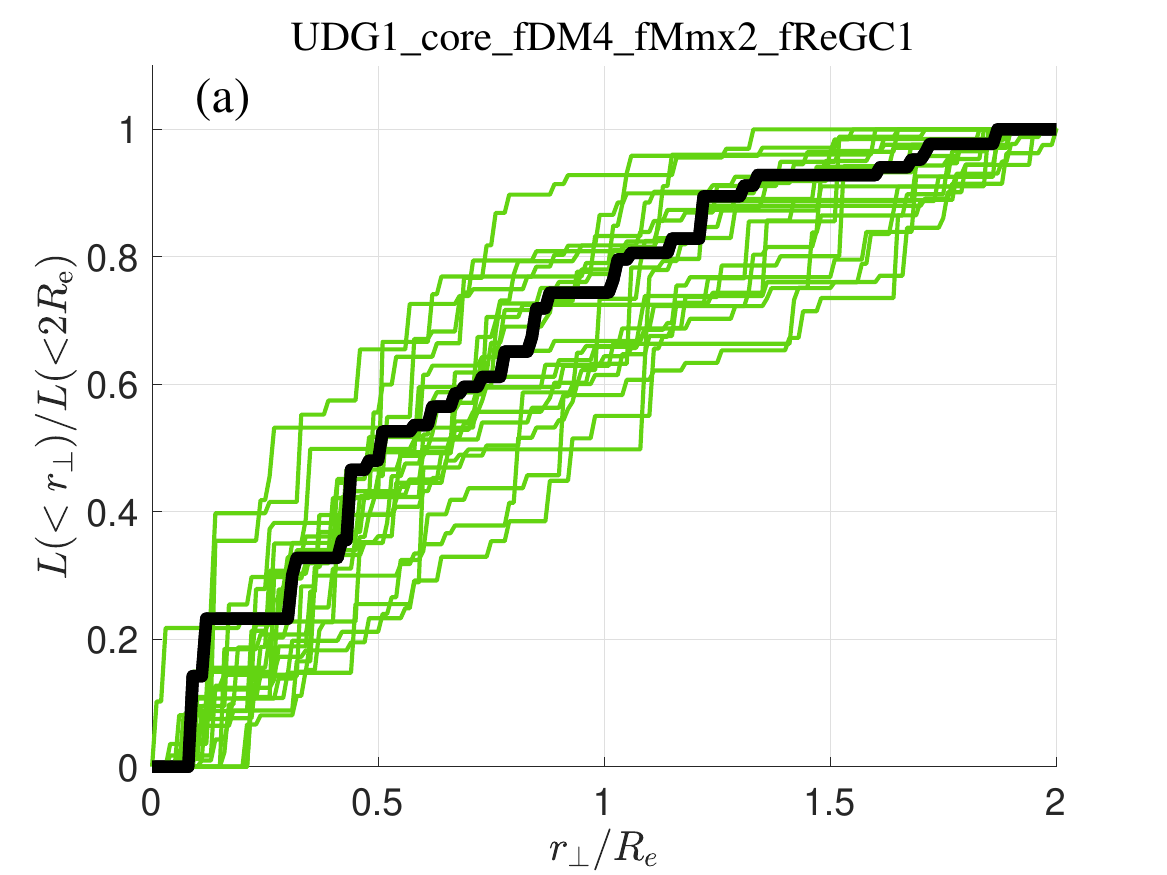}} 
        \raisebox{-\height}{\includegraphics[width=0.37\textwidth]{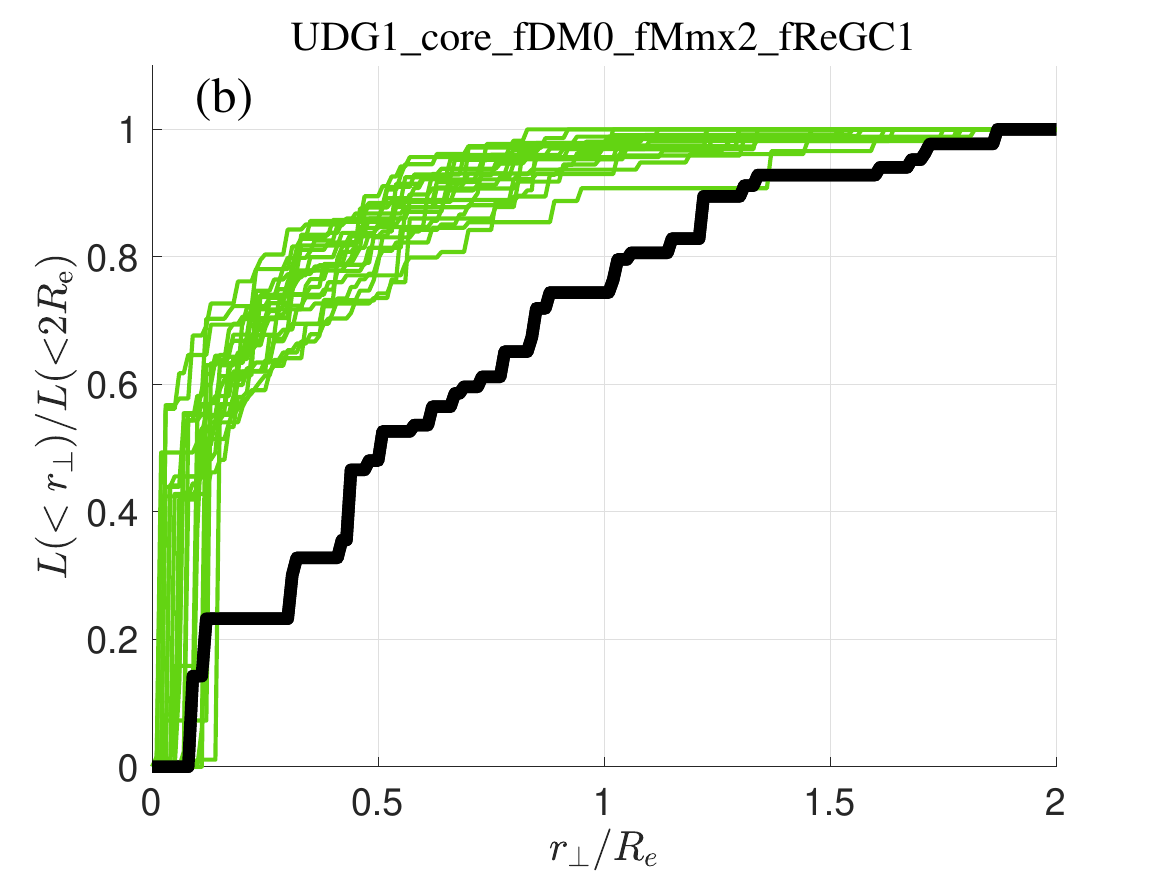}}
    \end{tabular}
\end{tabular}
\caption{Left: UDG1 with circles marking multiples of the stellar $R_{\rm e}$. Middle and right panels: GC cumulative luminosity normalized by the total GC luminosity at $r_\perp=2R_{\rm e}$, simulations (green) vs. data (black). Panel (a): a model with DM. Panel (b): a model without DM. Titles in panels (a,b) specify the halo model and GC initial conditions assumed in each case, as explained in Sec.~\ref{ss:udg1resdesc}. Here, the GC population's mass function in both panels is initialized in the same way, and the GCs have the same initial half-light radius as seen for the stellar body today.}
\label{fig:Burk_details_intro}
\end{figure*}

UDG1 is a prime example that illustrates how GC data sheds light on DM. 
Before delving into the details of the analysis, we show key results in Fig.~\ref{fig:Burk_details_intro}. The purpose of this plot is to highlight our motivation and the potential information content of the data. The left panel of Fig.~\ref{fig:Burk_details_intro} shows an Hubble Space Telescope image of UDG1\footnote{We are grateful to Shany Danieli for providing this image.}. To the right, panel (a) shows results for a DM model of the halo, and panel (b) shows a DM-free model. In each case, we run 20 simulations evolving GC orbits over 10~Gyr, and plot the GC luminosity cumulative distribution function (CDF) today, normalised to its value at projected radius $r_\perp=2R_{\rm e}$. Green lines are simulation results, thick black line is observed data. We see striking preference of the DM model over the DM-free model: a new way to infer the presence of DM, completely independent of conventional kinematics analysis. In what follows we present a systematic exploration of this data set.

\subsection{Observational data}
We start by reviewing observational results for UDG1. (A practical summary is given at the end of this section.) 
\begin{itemize}
\item {\bf \cite{muller21udg,muller2020spec}}. \cite{muller2020spec} used the IFU spectrograph MUSE to provide a spectroscopic analysis of 11 GC candidates and of the stellar body of the galaxy. 

For the stellar body, they obtain V band S\'ersic index $n=0.73\pm0.01$ and S\'ersic radius $R_{\rm e}=17.2\pm0.2''$, translating to $R_{\rm e}=2.19\pm0.02$~kpc assuming a distance $D=26.3$~Mpc. They find evidence for slight ellipticity ${\it e}=0.10\pm0.01$. The stellar body mass to light ratio was found to be $\left(M/L_V\right)_*=2^{+0.3}_{-0.1}~L_\odot/M_\odot$, with total luminosity  $L_V=0.86\times10^8~L_\odot$. 

The mass to light ratio of the stacked GC candidates was inferred to be somewhat lower than that of the stellar body, $\left(M/L_V\right)_{\rm GC}=1.6^{+0.3}_{-0.1}$ solar (but see~\cite{Forbes:2020vly} for cautionary comments regarding the uncertainty of these estimates). The velocity dispersion of the GCs was found to be $\sigma_{\rm LOS,GC}=9.4_{-5.4}^{+7.0}$~km/s, with no significant evidence for rotation.

\cite{muller21udg} used photometric HST data to identify $N_{\rm GC}=26\pm6$ within $r\leq1.75R_{\rm e}$, assuming that the spatial distribution of the GCs is well described by a Plummer profile. All 11 of the previously confirmed GCs from~\cite{muller2020spec} were detected. \cite{muller21udg} also estimated that $\sim13$ GC candidates within $1.75~R_{\rm e}$ should be background interlopers (by definition, not included in the $N_{\rm GC}=26\pm6$ that accounts for and subtracts the background estimate).\footnote{These results suggest mild tension~\citep[][]{muller21udg} between the background estimate and the galaxy's GC count, as the total number of detected GC candidates inside $r<1.75~R_{\rm e}$ was $N=49$, larger than $26+13=39$ by more than the expected Poisson noise $\sim\pm6$.}

\item {\bf \cite{Forbes2019,Forbes:2020vly}}. \cite{Forbes2019} analyzed data from the VEGAS survey and reported, for the stellar body of UDG1, $n=0.68$ and $R_{\rm e}=17.7\pm0.5''$, which they translate to $R_{\rm e}=2.14\pm0.06$~kpc assuming $D=24.89$~Mpc. They find 20 ``compact sources" associated to the galaxy, from which they infer a lower limit of 17 GCs. 

\cite{Forbes:2020vly} used IFU Keck data to provide a velocity dispersion measurement for the stellar body of UDG1, $\sigma_{\rm LOS}=17\pm2$~km/s within $r\lesssim0.5R_{\rm e}$, with no significant evidence for rotation. The total V band luminosity was found to be $L_V=0.55\times10^8~L_\odot$.

\cite{Forbes:2020vly} found a systemic recession velocity for UDG1 that is larger by 11~km/s compared to that reported by~\cite{muller2020spec}. They also measured LOS velocity to the two brightest GC candidates; their results differ from those of~\cite{muller2020spec} by 5 and 8~km/s. 

\item {\bf \cite{2022ApJ...927L..28D}} analyzed HST data (with exposure time roughly double that used in~\cite{muller21udg}) to study the GC population of UDG1. They report stellar S\'ersic index $n=0.6$ and $R_e=15.6''$, corresponding to $R_e=2$~kpc assuming $D=26.5$~Mpc. 
Within $r<2R_e$, limiting to sources with apparent magnitude $m_{F606W}<25$ (and imposing additional angular size and color cuts) they found 33 GC candidates with an estimated background of 0.7. The 20 brightest GC candidates are argued to have essentially no background contamination; this list includes the 11 spectroscopically confirmed GCs of~\cite{muller2020spec}. More faint GC candidates are also detected, but with a larger background. 

The total luminosity of the stellar body was given as $L_V=0.59\times10^8~L_\odot$. 
\end{itemize}

The projected distance of UDG1 from the brightest galaxy in the NGC5846 group~\citep[][]{Mahdavi:2005pr} was estimated at 164~kpc. At this distance, NGC5846 induces on UDG1 a tidal radius $r_t\approx6\left(\frac{10^{13}~M_\odot}{M_{\rm NGC5846}}\frac{M_{\rm UDG1}}{10^9~M_\odot}\right)^{\frac{1}{3}}~{\rm kpc}$. 
This tidal radius is roughly three times UDG1's stellar effective radius $R_e$. We do not model the tidal field of NGC5846 in our calculations, and restrict most of our attention to the system properties within $r<2R_{\rm e}$. It may be useful to revisit this approximation in subsequent work.

\cite{muller21udg} identifies the center of the smooth light profile of UDG1 at RA15:05:20, DEC1:48:45. 
\cite{Forbes:2020vly} finds RA15:05:20, DEC1:48:47. 
These estimates vary by $2''$ in declination, that is, about 10\% of $R_e$ ($\approx200$~pc). This uncertainty adds to our decision, discussed further below, to limit our quantitative analysis of the GC system to radial distance larger than $0.5R_{\rm e}\approx1$~kpc.

In summary, we use the following observational constraints for UDG1.
\begin{itemize}
\item We consider the 33 bright GC candidates from~\cite{2022ApJ...927L..28D}, neglecting background contamination. The sample was obtained with an HST F606W cut of $m_{F606W}<25$, which we translate into a V band luminosity cut,
\be L_{\rm V,GC}&>&0.5\times10^5~L_\odot.\ee
We use $D=26.5$~Mpc and follow~\cite{muller21udg} in converting HST F606W magnitudes to V band magnitudes. 

We will need to convert GC mass from the theoretical calculation into GC luminosity. Our fiducial choice is $(M/L_V)_{\rm GC}=2$ (somewhat higher than the value estimated in~\cite{muller2020spec}, but consistent with old nearby GCs~\citep[][]{2017MNRAS.464.2174B}). 

\item For the stellar body we use a S\'ersic profile with $n=0.6$, $R_e=2$~kpc, and total mass $M_*=10^8~M_\odot$.
\end{itemize}
%

\subsection{Results: GC initial radial distribution similar to stellar body}\label{ss:udg1resdesc}
We compare the predictions of DM halo models and GCIMF models to the observed distribution of GC luminosity and projected radius in UDG1. After using the GC data to reduce the viable model parameter space, we end up with a prediction for the total mass budget of UDG1, that can then be compared to kinematics data. 
In this section, the initial spatial distribution of the GCs is assumed to match the currently observed distribution of the stellar body. We revisit this assumption in subsequent sections.

\begin{figure*}
\centering
      \includegraphics[ scale= 0.245]{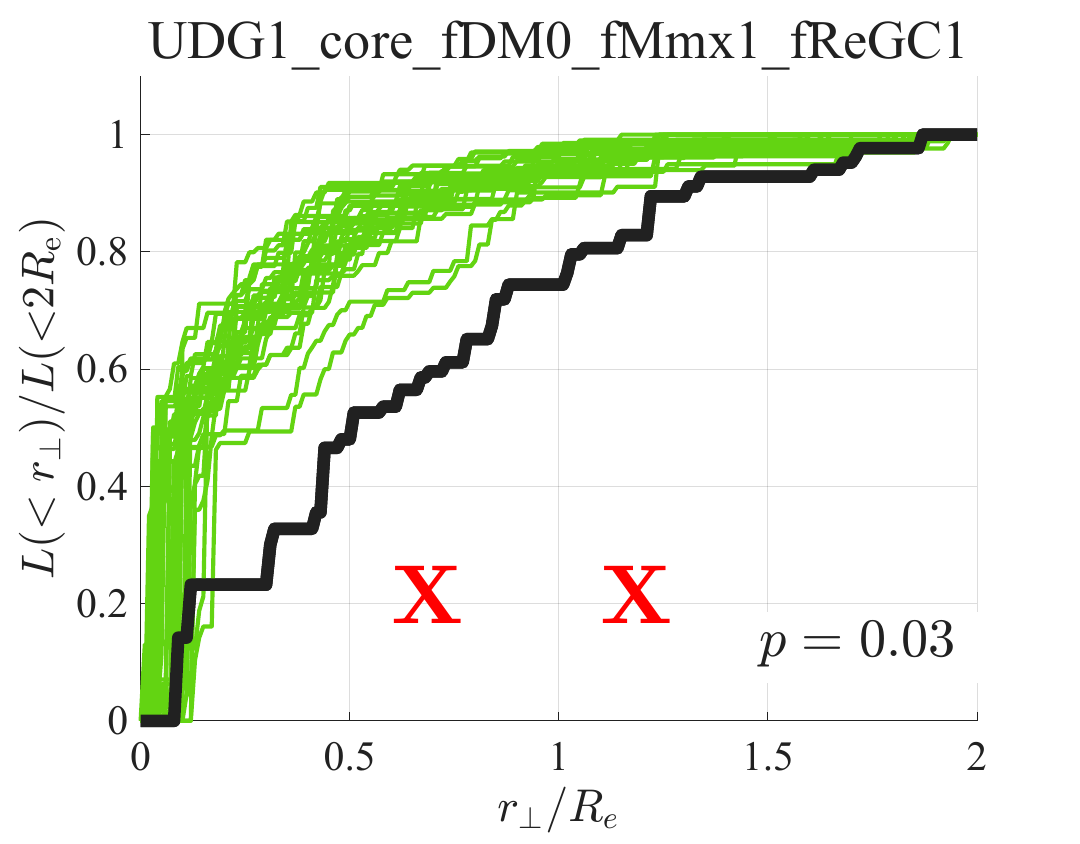} 
      \includegraphics[ scale= 0.245]{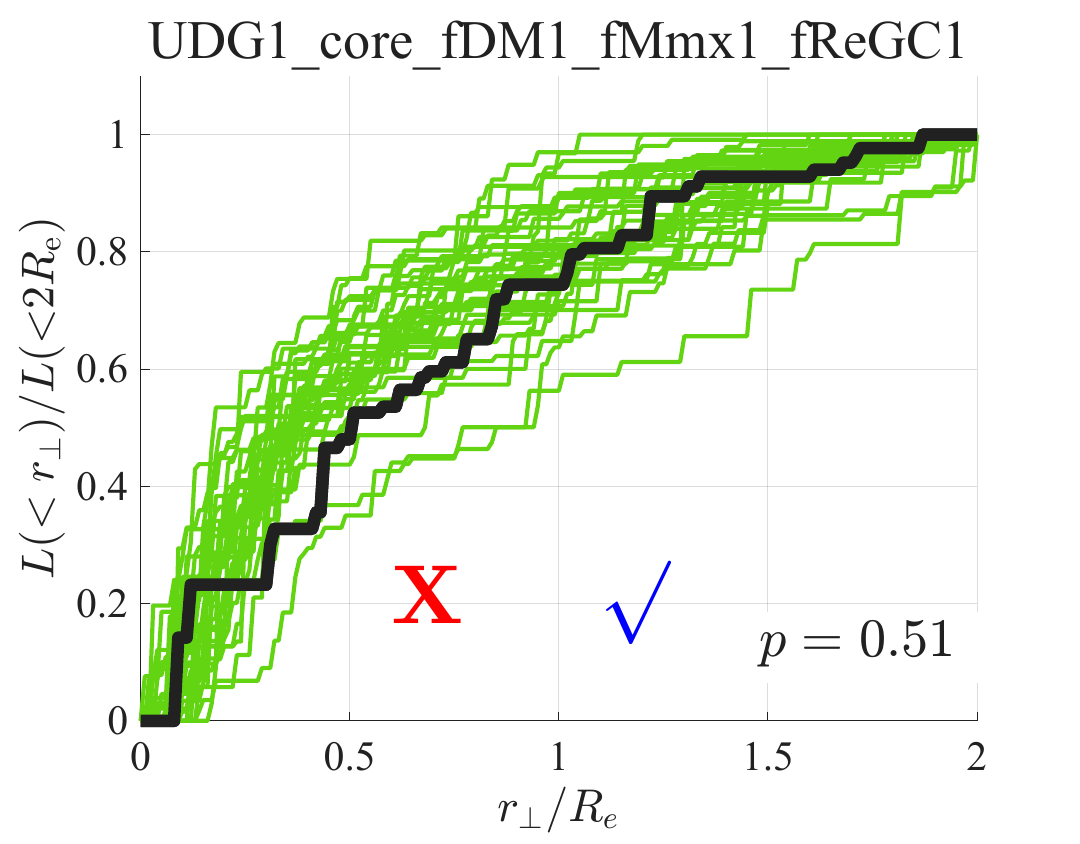} 
      \includegraphics[ scale= 0.24]{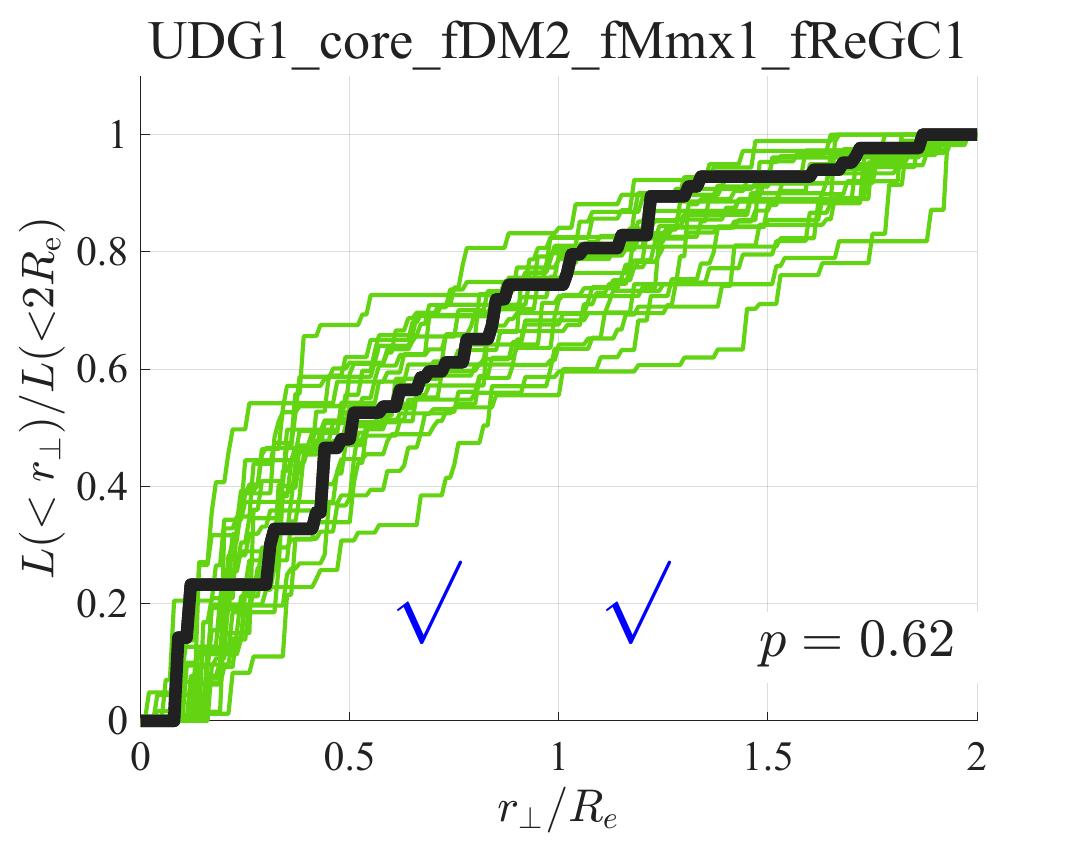}  
      \includegraphics[ scale= 0.24]{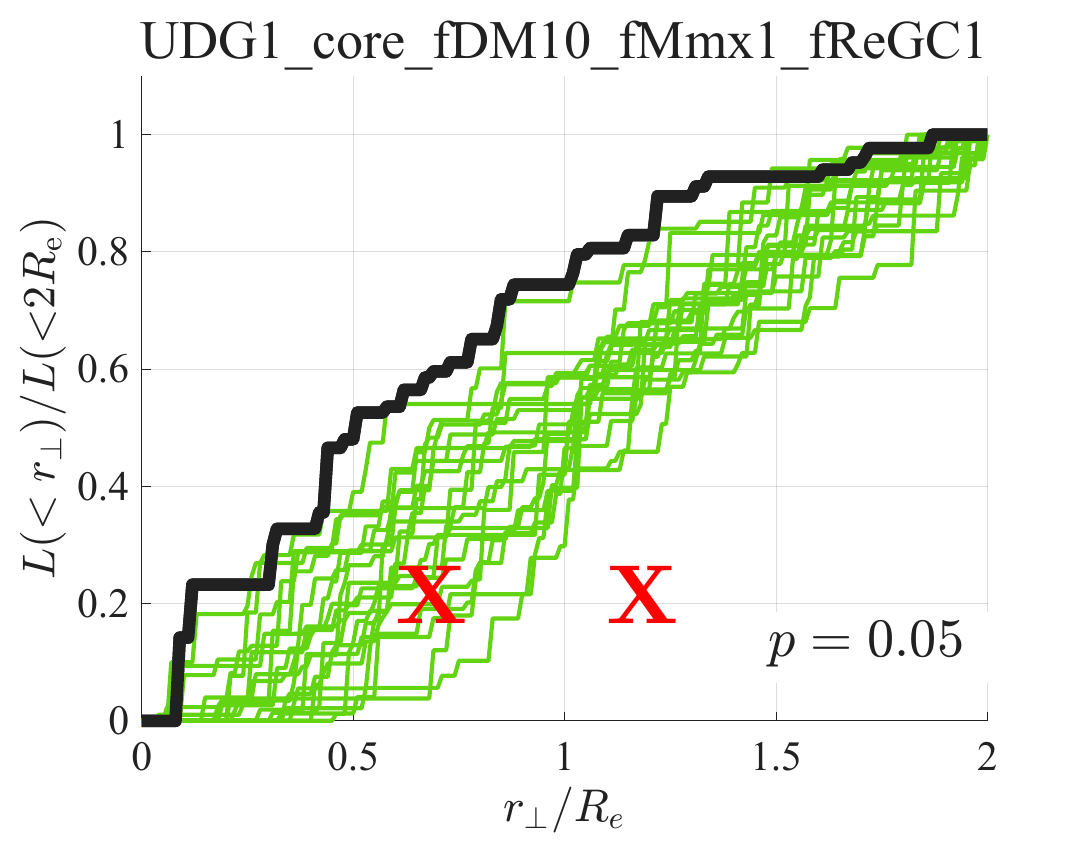} \\ 
      \includegraphics[ scale= 0.245]{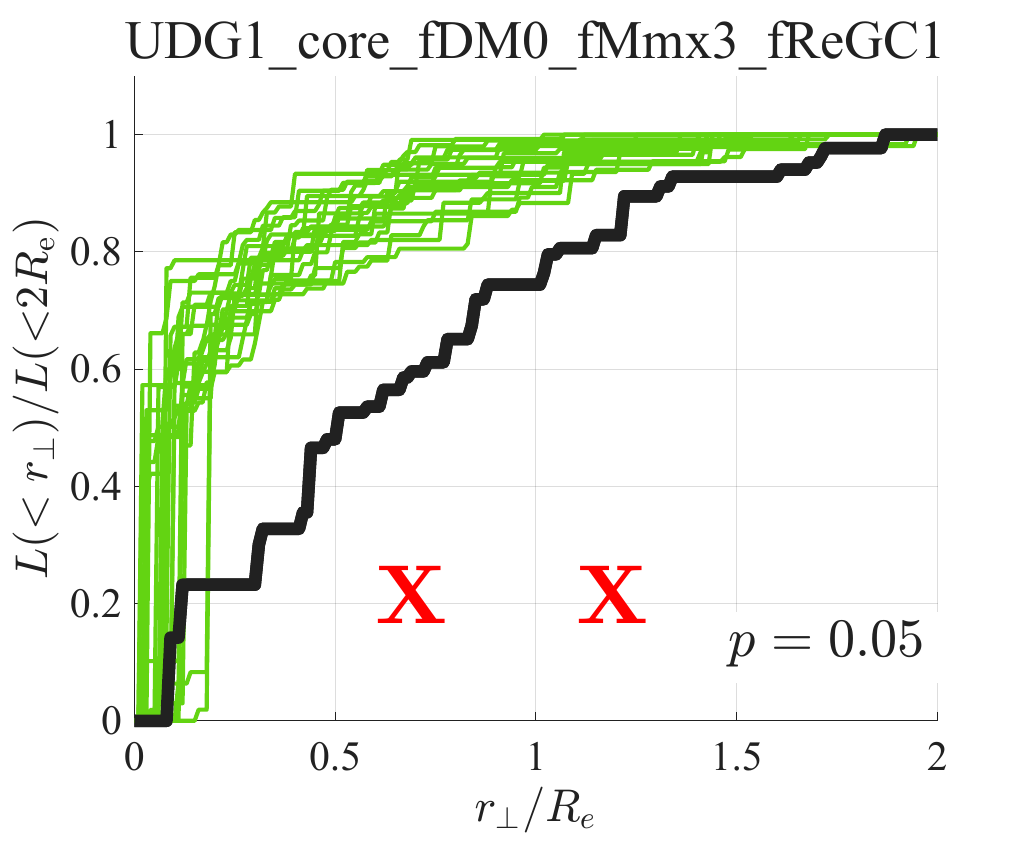} 
      \includegraphics[ scale= 0.245]{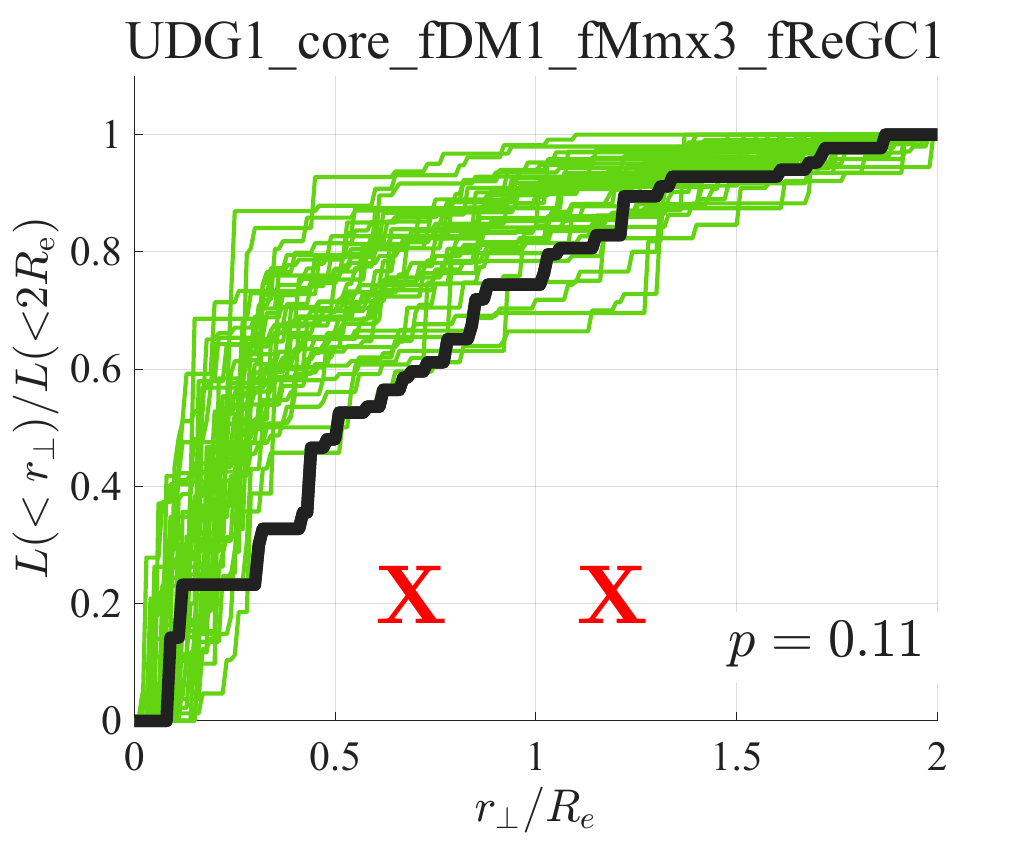}       
      \includegraphics[ scale= 0.245]{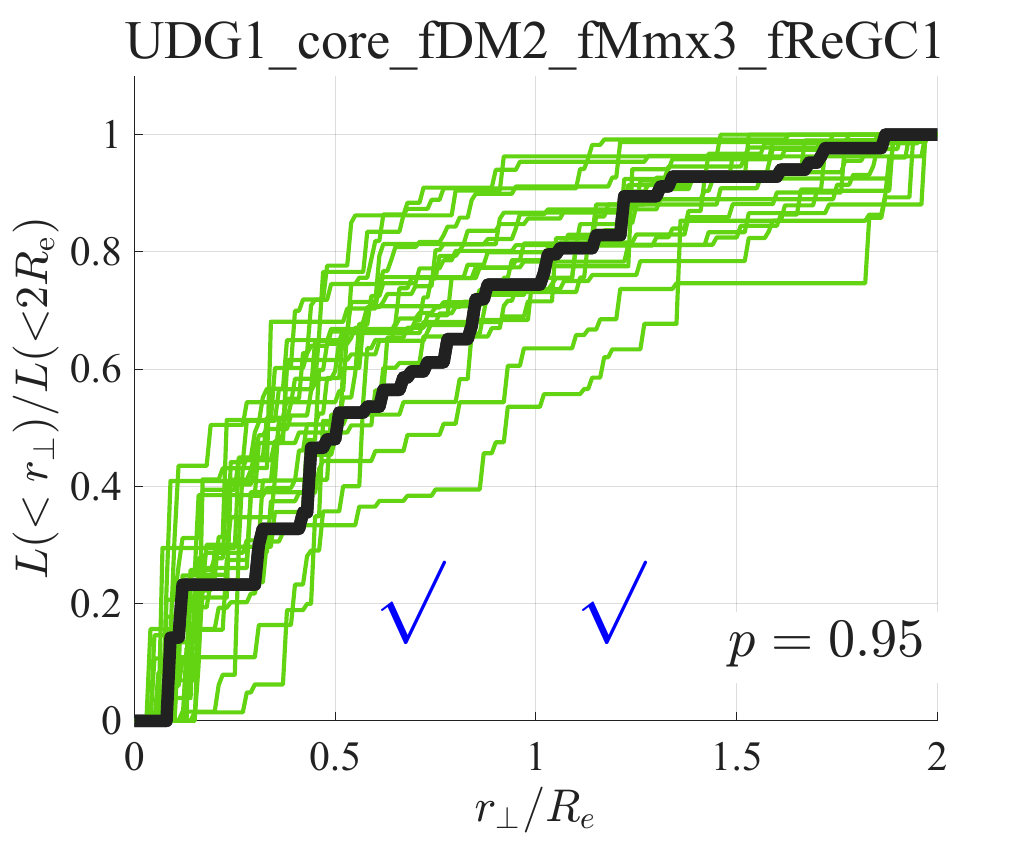}       
      \includegraphics[ scale= 0.245]{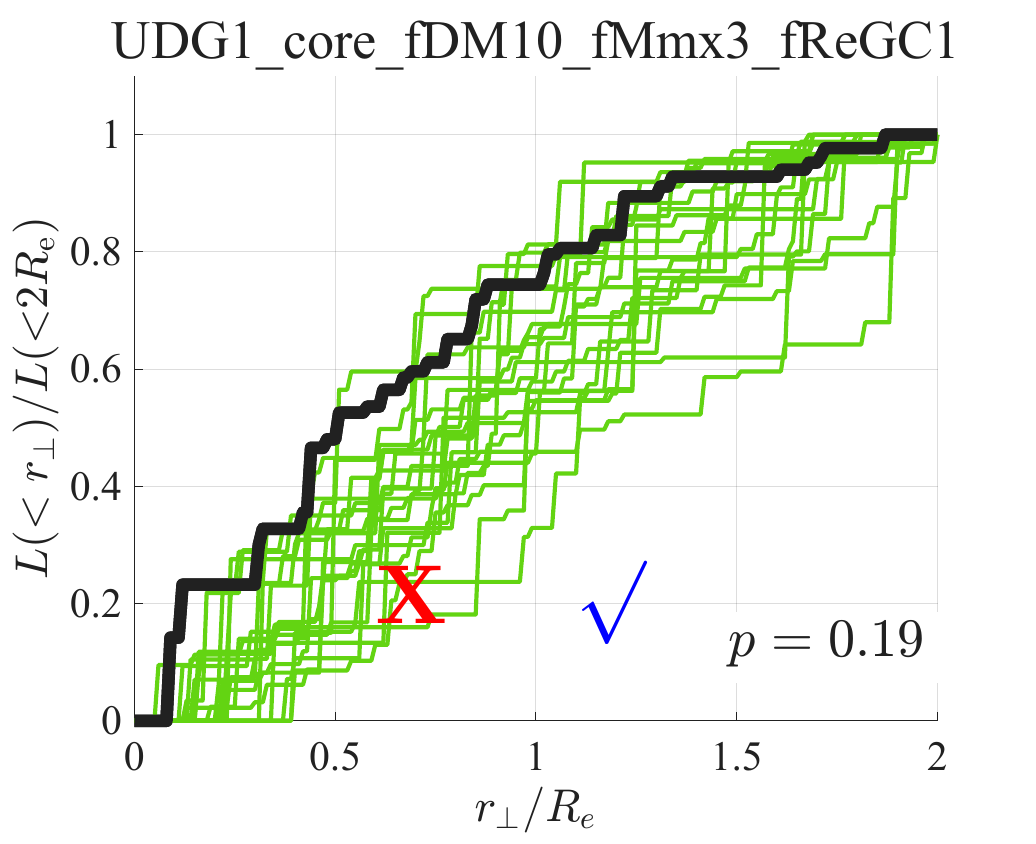}             
 \caption{Luminosity CDF: GC cumulative luminosity normalized by the total GC luminosity at $r_\perp=2R_{\rm e}$, simulations (green) vs. data (black).  Here, the DM halo follows the Burkert core profile. The DM density increases from left to right, indicated by the variable $f{\rm DM}$ in panel titles (see Sec.~\ref{ss:udg1resdesc} 
 for description). The GCIMF parameter $M_{\rm max}$ increases between the top row to the bottom row, indicated by the variable $f{\rm Mmx}$. Panels with red $\X$ (blue $\V$) are in tension (consistent) with data, where left (right) symbols refer to $r_\perp=0.5R_{\rm e}$ ($r_\perp=R_{\rm e}$). The $p$-values refer to the Anderson-Darling tail-weighted Cramér–von Mises test. 
 }
 \label{fig:Burk_LCDFs}
\end{figure*}

Fig.~\ref{fig:Burk_LCDFs} collects results for different values of the DM halo mass in the Burkert core model. We also explore different values of $M_{\rm max}$. 
The observable we use is the GC cumulative luminosity distribution (CDF) as a function of $r_\perp$, normalized to its value at $r_\perp=2R_{\rm e}$. Green lines are simulation runs, and thick black line is data. The same GC luminosity cut is applied to observed and simulated GC samples. 
We consider the normalized curve because, as explained earlier, the normalization of the GCIMF in our models is a free parameter, adjusted to reproduce the total GC luminosity. Using fractional luminosity may also ameliorate mass to light ratio uncertainties.
%

Across the different panels of Fig.~\ref{fig:Burk_LCDFs}, the DM halo density increases going from left to right, and $M_{\rm max}$ increases going from the top down, according to the name convention in the title of each panel:
\begin{itemize}
\item ``fDM0,1,2,...": the DM mass enclosed in $r<2R_{\rm e}$ is equal to (0,1,2,...) times the total stellar mass.
\item ``fMmx1,2,3": the parameter $M_{\rm max}=(\frac{1}{1.5},1,1.5)$ times the mass of the most massive GC in the sample (for UDG1 with $(M/L_V)_{\rm GC}=2$, this is ${\rm max}\,M_{\rm GC}^{\rm obs}=18.7\times10^5~M_\odot$).
\end{itemize}

We chose the points $r_\perp=0.5R_{\rm e}$ and $r_\perp=R_{\rm e}$ to define approximate consistency (or inconsistency) of the models with observational data. Models for which at least 9 out of 10 simulation runs have luminosity CDF either persistently above, or persistently below the observed value at $r_\perp=0.5R_e$ and $r_\perp=R_e$, are marked with red ${\bf \color{red}X}$. Otherwise, the model is marked with blue ${\bf \color{blue}\surd}$.

GC luminosity inside $r_\perp\ll R_e$ can also be informative, because that is where NSCs reside. However, the analysis in this case requires careful study of the uncertainty in locating the galaxy's center of light. In addition, in the inner halo GC masses are not negligible w.r.t. the enclosed halo mass, so semianalytic simulations may not capture the dynamics correctly; see Sec.~\ref{ss:calibudg}. For these reasons we mostly restrict our attention to luminosity CDF at $r_\perp\geq0.5R_{\rm e}$.

The ``${\bf\color{blue}\surd}$/${\bf\color{red}X}$" LCDF consistency criterion is, of course, a crude measure of the consistency of the models. This is appropriate because the systematic uncertainties associated with GC mass loss, mergers, beyond-Chandrasekhar DF, etc. are non-negligible. For completeness, in addition to the ${\bf\color{blue}\surd}$/${\bf\color{red}X}$ indicators we also report the p-value computed using the Anderson-Darling tail-weighted Cramér–von Mises test~\citep{AndersonDarling1952,AndersonDarling1954}, shown in the lower-right corner of each figure panel. (The formal resolution on $p$ is $\Delta p=1/(K+1)$ when the test is computed using a sample of $K$ CDFs drawn from the null model; in Fig.~\ref{fig:Burk_LCDFs}, for example, this varies between panels, in the range $\Delta p\approx0.03-0.05$.)

Our main result is that the GC luminosity CDF is only consistent with data for halo models that contain substantial DM mass. Models with too little DM (left panels in Fig.~\ref{fig:Burk_LCDFs}) predict strong DF-induced mass segregation that exceeds the gentle trend observed in the data. Models with too much DM have diminished DF and a mass segregation that is too weak to reproduce the data. Varying $M_{\rm max}$ does not change this conclusion.
\begin{figure*}
\centering
      \includegraphics[ scale= 0.23]{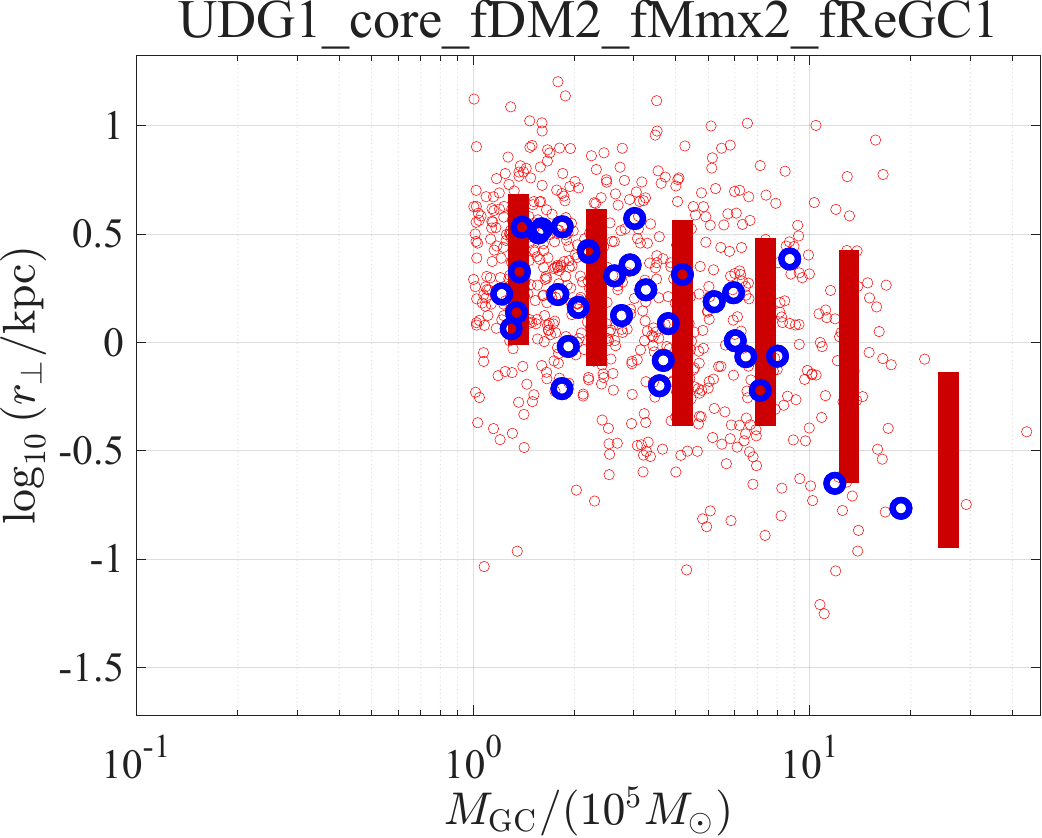}       
            \includegraphics[ scale= 0.23]{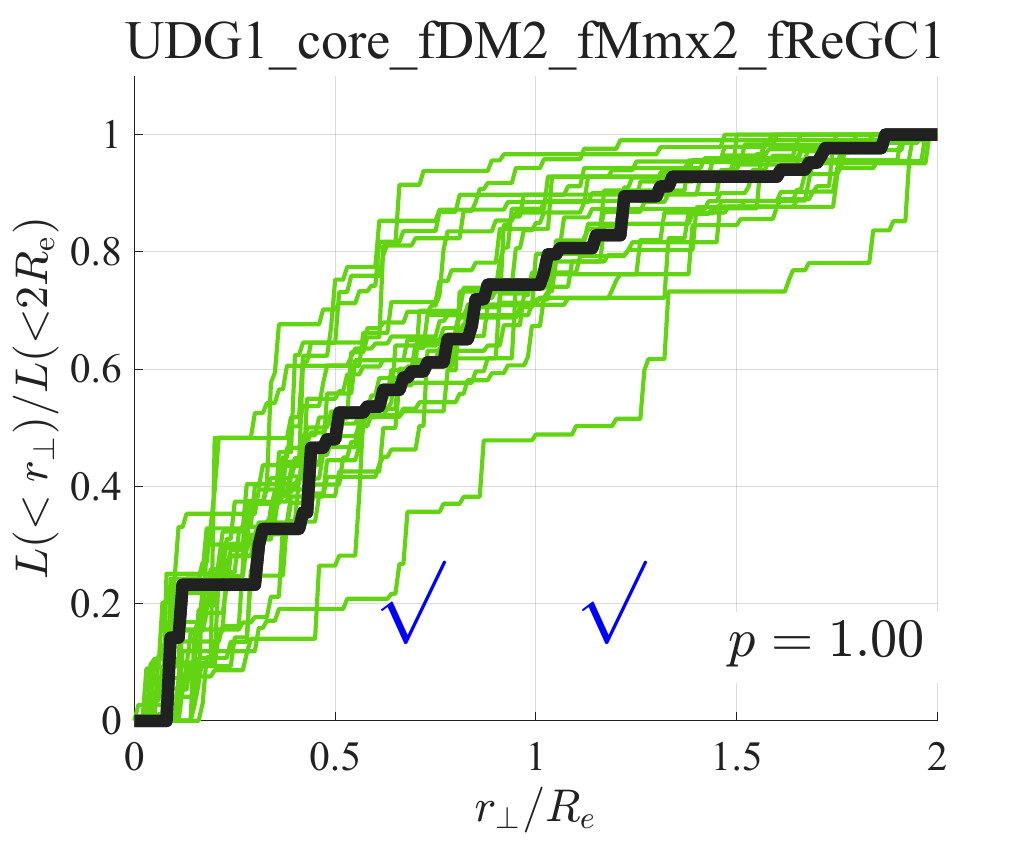}       
                  \includegraphics[ scale= 0.23]{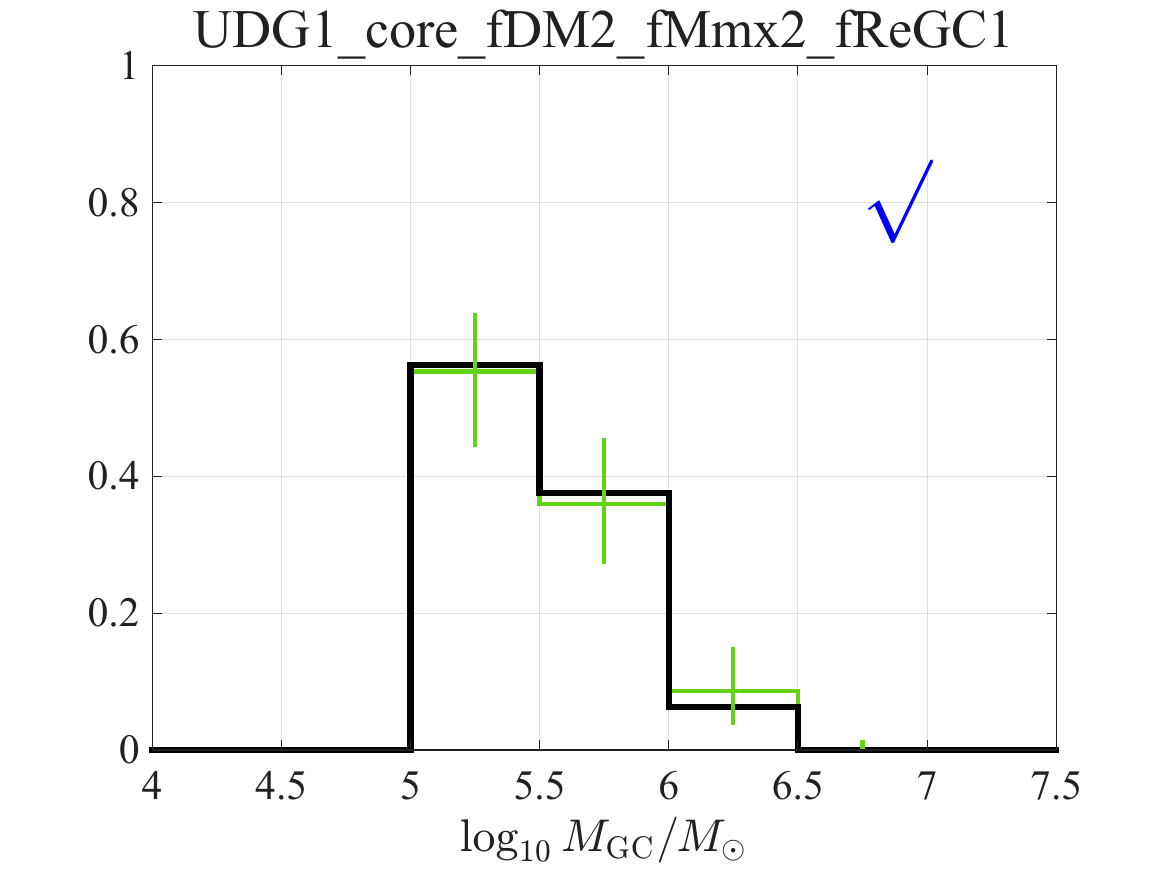}  
                  \includegraphics[ scale= 0.23]{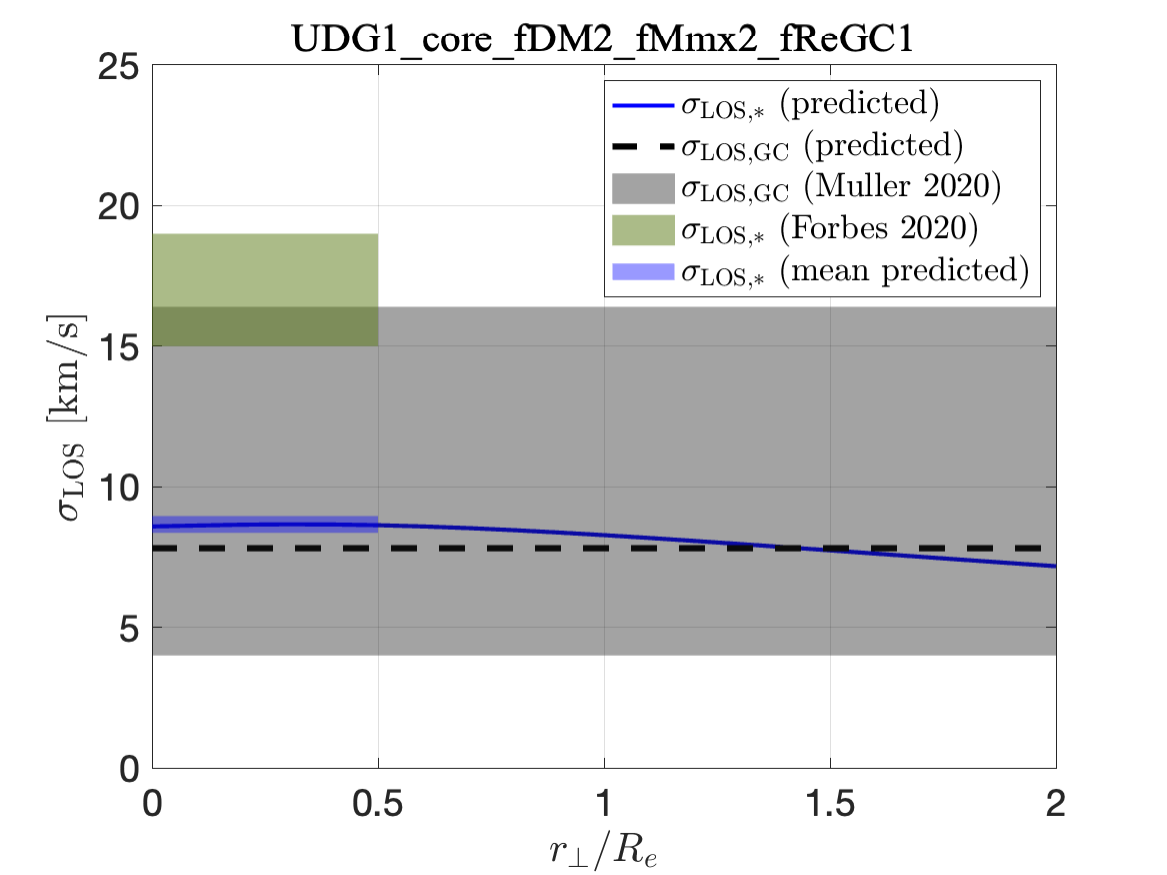} \\    
                        \includegraphics[ scale= 0.23]{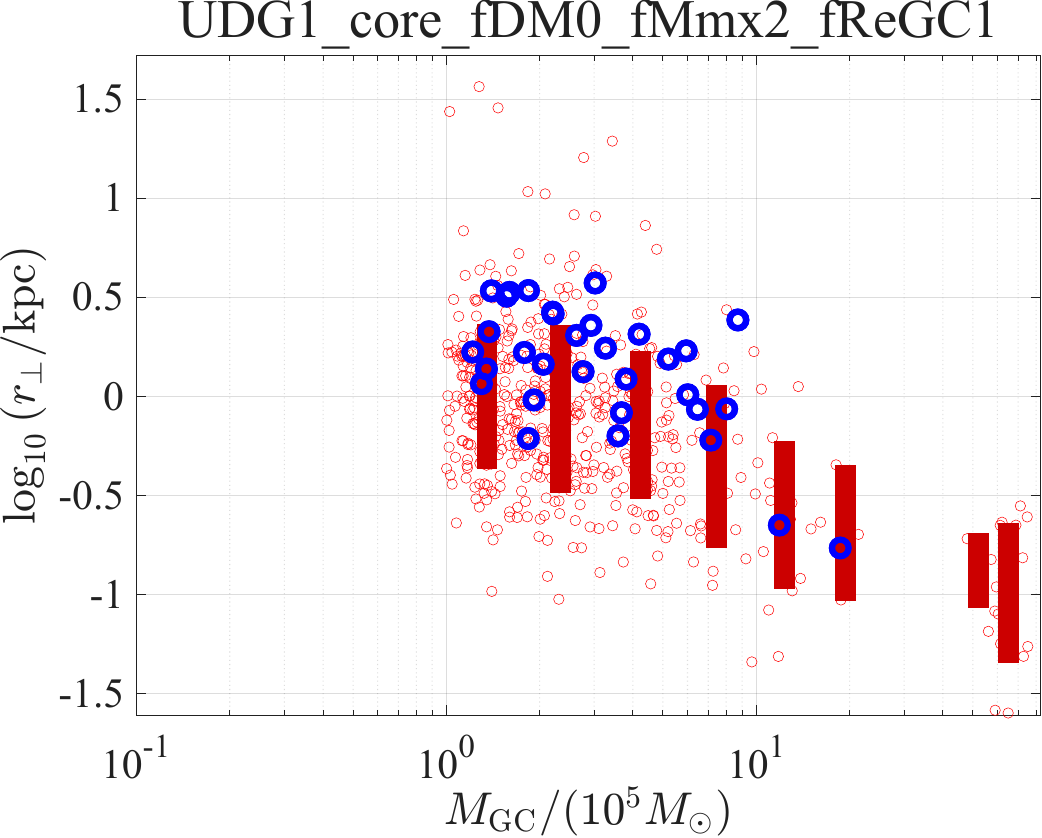}       
            \includegraphics[ scale= 0.23]{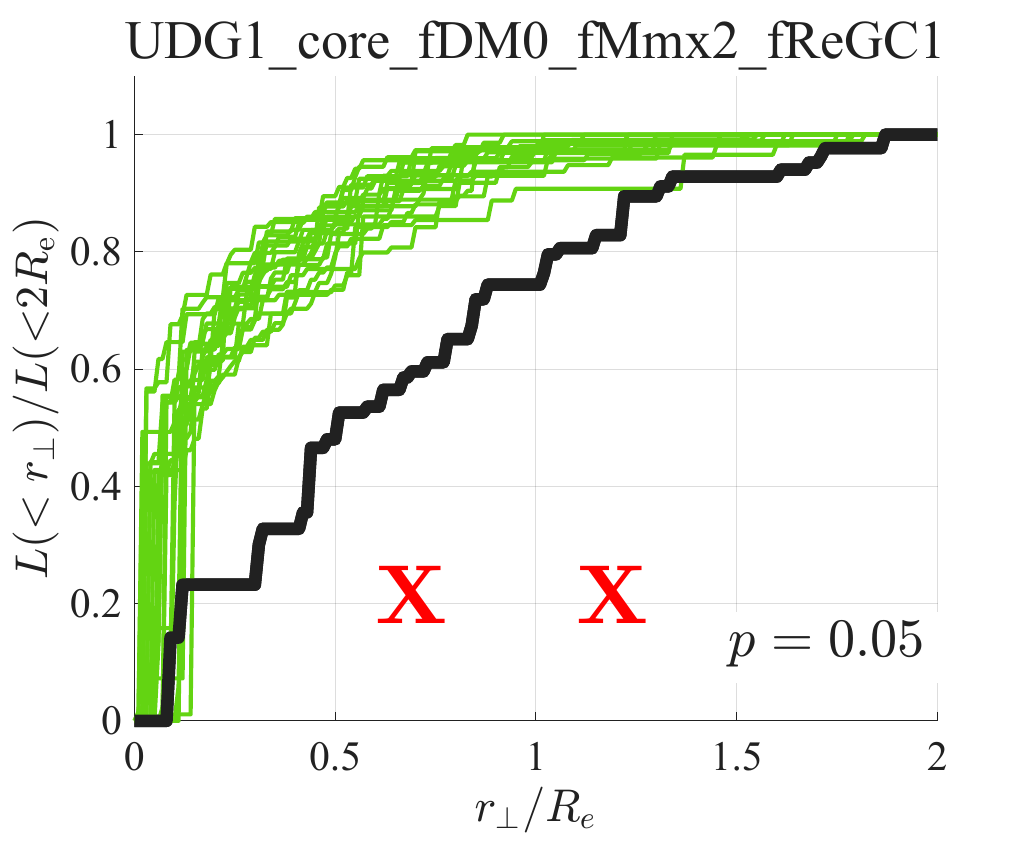}       
                  \includegraphics[ scale= 0.23]{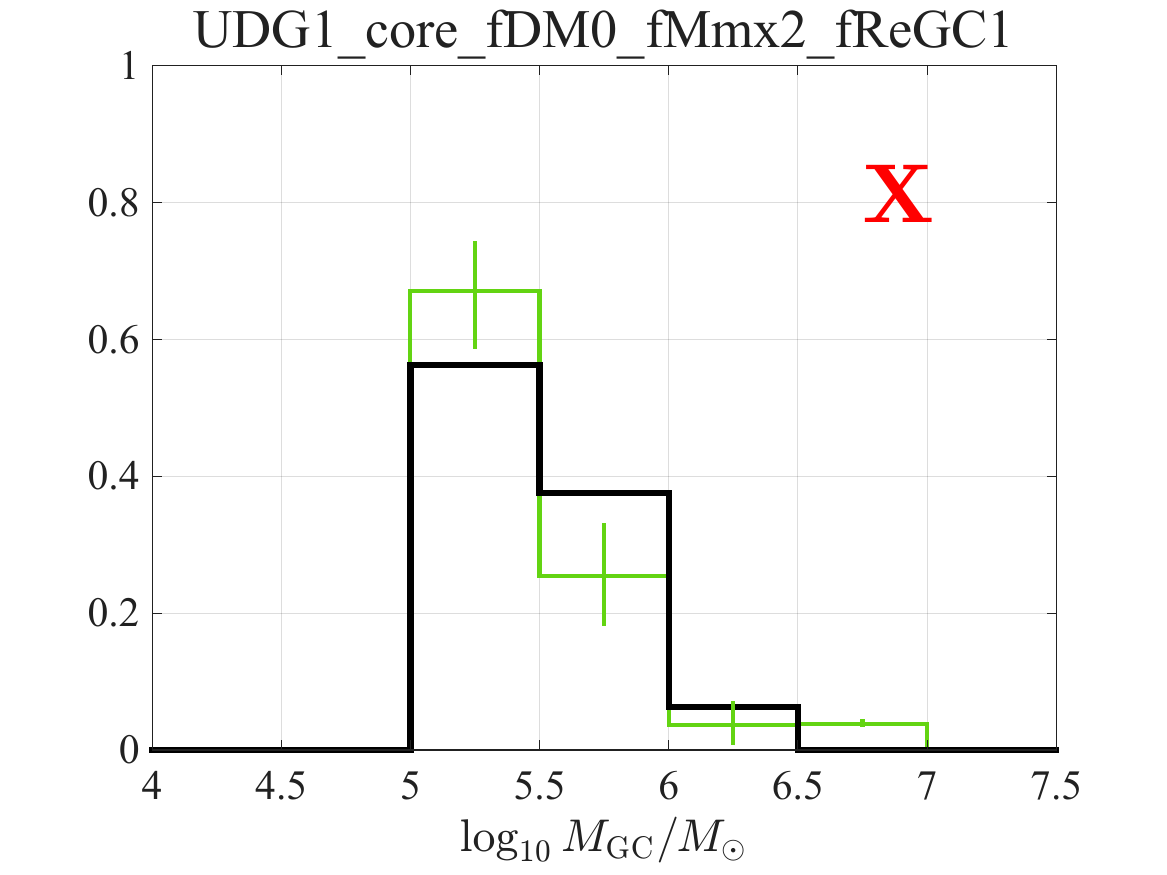}       
                  \includegraphics[ scale= 0.23]{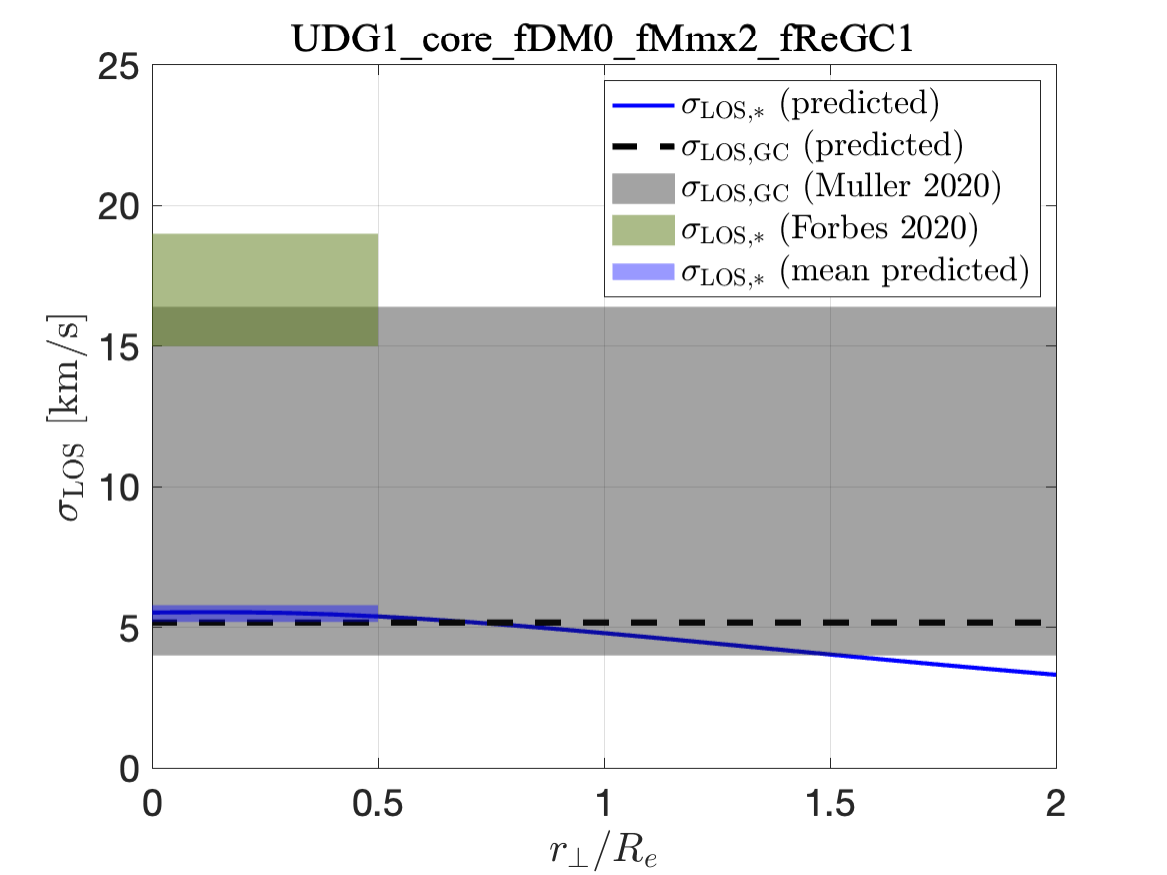}
 \caption{Detailed results for a DM core model that passes {\bf (top)} and a DM-free model that fails {\bf (bottom)} the GC luminosity test. In the left-most panels, blue circle markers show the observed projected radii and inferred mass for the 31 GCs considered in the analysis. Smaller red circles show the corresponding coordinates for GCs from the simulations (aggregate from all simulations shown together). Vertical red bars are centered at mean $\log_{10}r_\perp$ and mean $M_{\rm GC}$, bunching simulated GCs in mass bins with constant spacing of 0.1 dec. The total length of each line is twice the standard deviation of $\log_{10}r_\perp$. The second column of panels from the left shows the corresponding luminosity CDFs. The third column of panels compares the simulated GC mass function (thin green) to the observed mass function (thick black), with green error bars showing $1\sigma$ assuming Poisson statistics. The right-most panels compare simulated GC and stellar kinematics to data: blue solid line shows the $r_\perp$-dependent line of sight velocity dispersion (LOSVD) for the stellar body, predicted by the corresponding halo model; horizontal blue band shows the mean stellar LOSVD in the range $r_\perp<0.5R_{\rm e}$, to be compared with the measurement of~\cite{Forbes:2020vly} (green shade); horizontal black dashed line shows the predicted GC LOVD from the simulation, to be compared with the measurement by~\cite{muller2020spec} (grey shade).
 }
 \label{fig:Burk_details}
\end{figure*}
Fig.~\ref{fig:Burk_details} presents more detailed information for a model that passes ({top panels}), and a DM-free model that fails ({bottom panels}) the luminosity criterion. 
The left panels show a scatter plot of GC projected radius vs. mass, as suggested by \cite{Bar:2022liw} to highlight DF-induced mass segregation. Red circles show stacked GCs from all simulations. Vertical lines are centered at mean $\log_{10}r_\perp$ and mean $M_{\rm GC}$, bunching GCs in mass bins with constant logarithmic  spacing of 0.1 dec. The total length of each line is twice the standard deviation of $\log_{10}r_\perp$. Bins with less than 2 GCs (from all simulations combined) do not have a vertical line. Blue circles show unbinned observed data. 
Second-column panels repeat the luminosity CDF from Fig.~\ref{fig:Burk_LCDFs}. 
Third-column panels show the normalized GCMF. Here we add another consistency criterion: calculating $\chi^2$ from the data, prediction, and error bars in the plot (obtained from run-to-run scatter), models with $\chi^2/{\rm dof}>9$, where dof=number of mass bins in which the simulation variance is non-vanishing (e.g., dof=3 in the top row, and dof=4 in the bottom), are marked by red ${\bf \color{red}X}$. Otherwise, the model is marked with blue ${\bf \color{blue}\surd}$. 
Finally, the right-most panels show LOSVD data and predictions, discussed further below.

In the bottom row of Fig.~\ref{fig:Burk_details} the model predicts NSC formation, seen as the bunch of red circles at large $M_{\rm GC}$ and small $r_\perp$ in the left panel. If the GC initial radial distribution follows that of the stars, then a DM-free model of UDG1 always contains an NSC that is too massive to match the data. 

\begin{figure*}
\centering
      \includegraphics[ scale= 0.24]{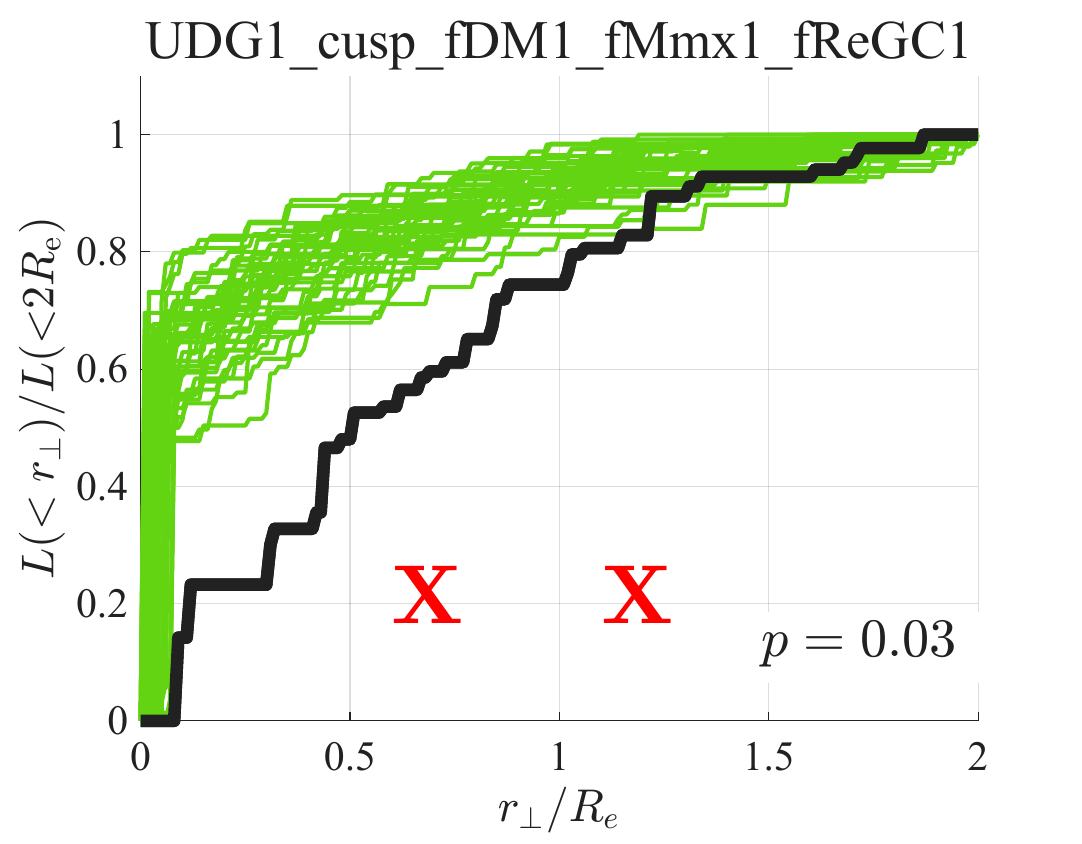}       
     \includegraphics[ scale= 0.24]{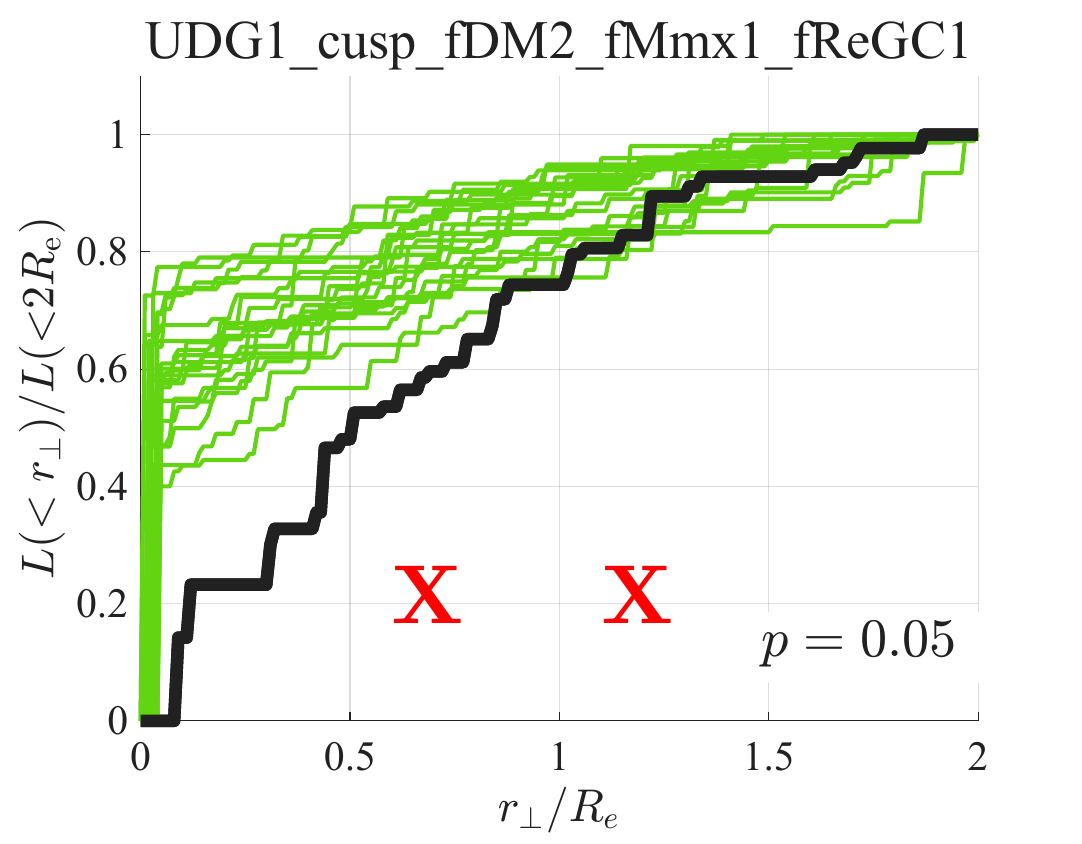}             
      \includegraphics[ scale= 0.24]{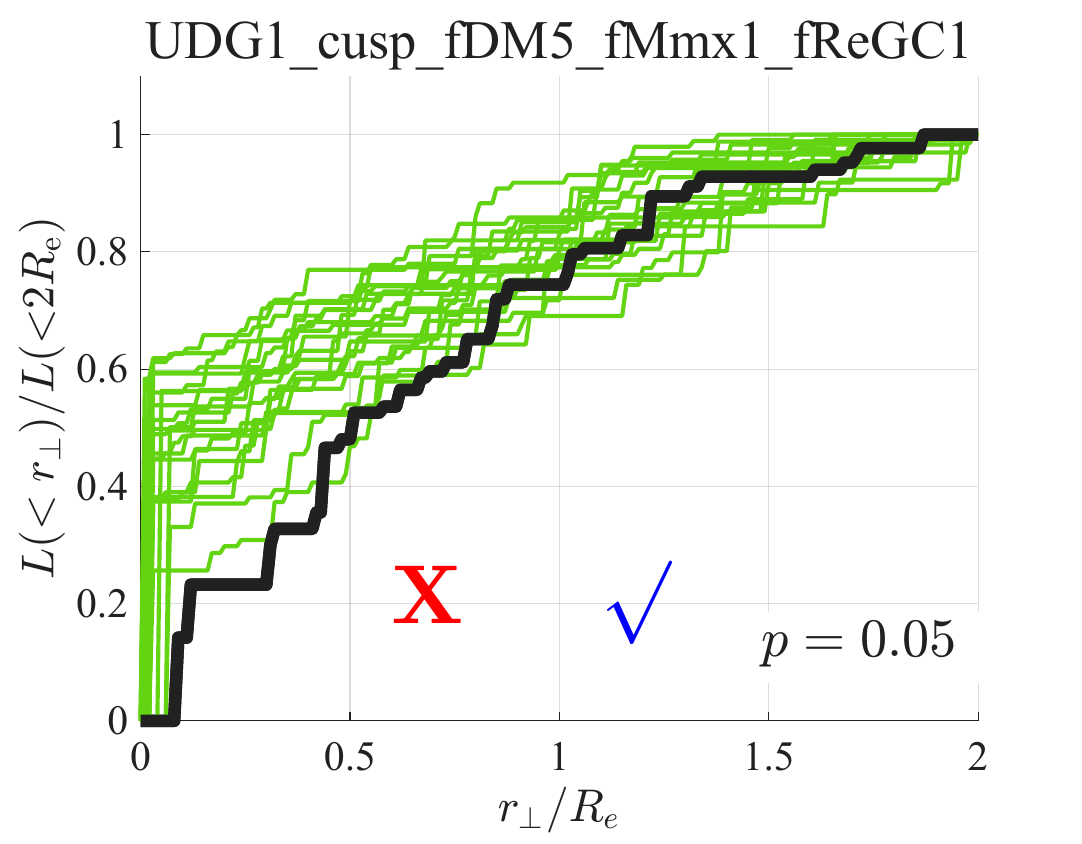} 
      \includegraphics[ scale= 0.24]{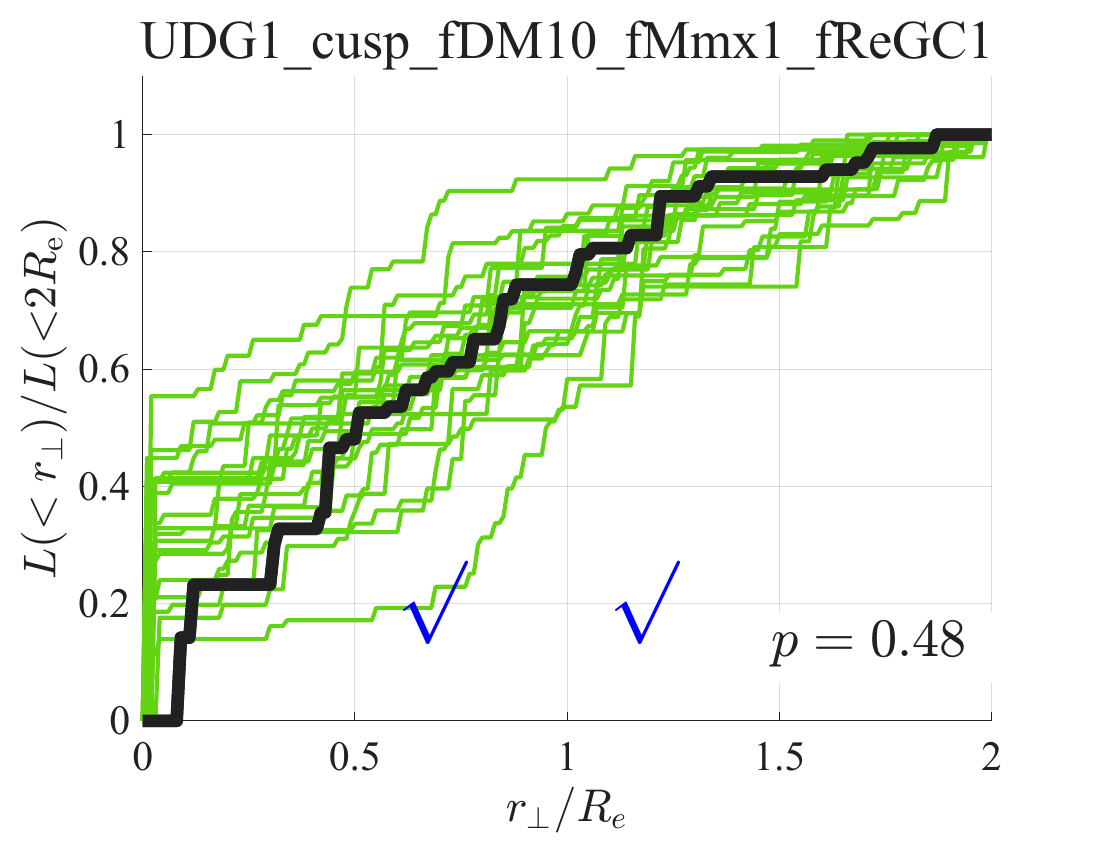} \\      
      \includegraphics[ scale= 0.245]{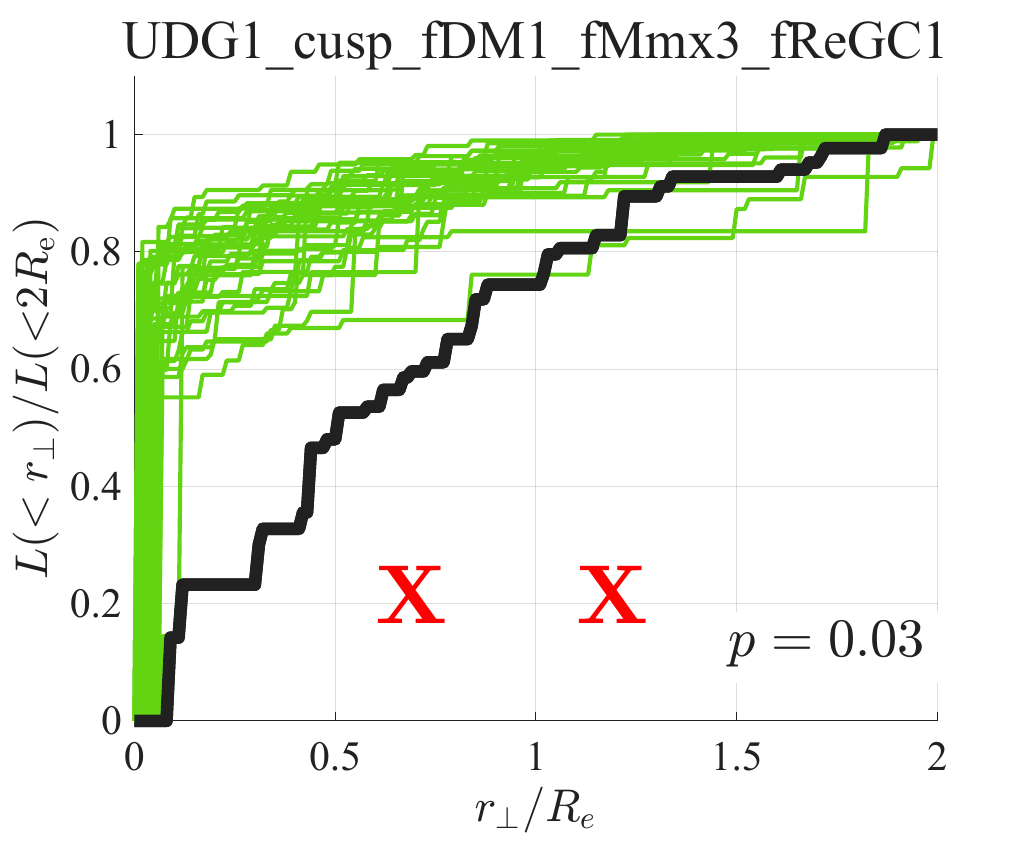}       
      \includegraphics[ scale= 0.245]{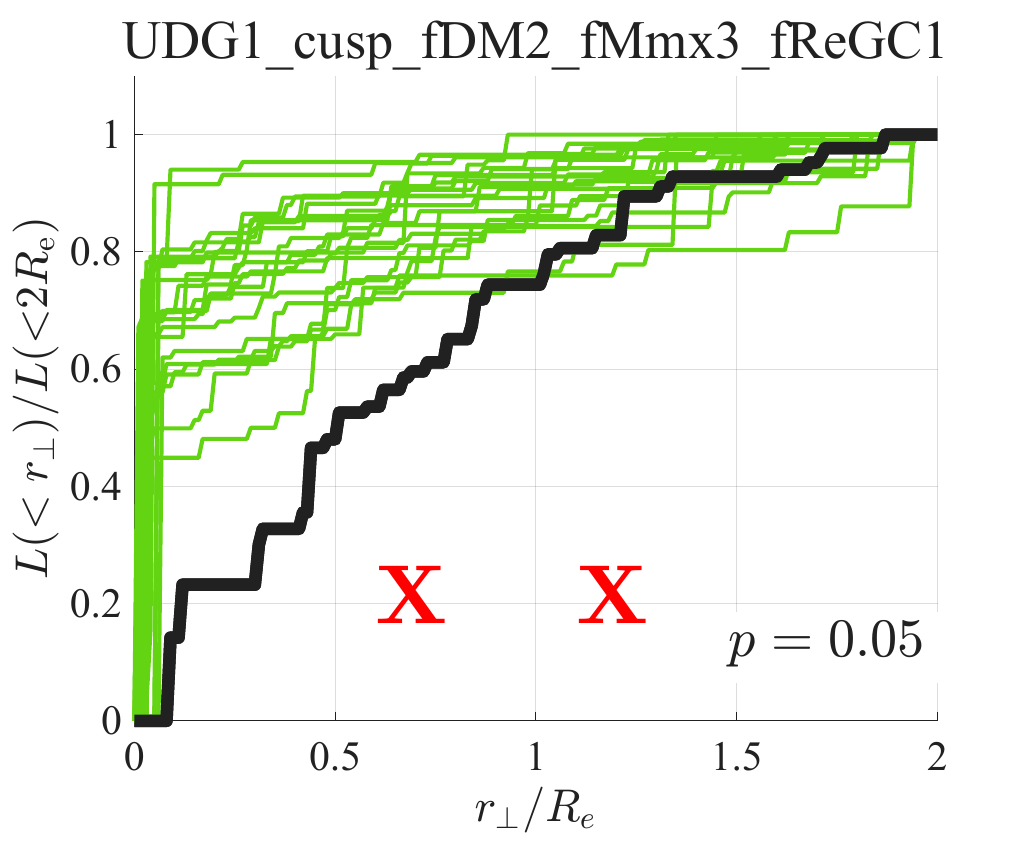}             
      \includegraphics[ scale= 0.245]{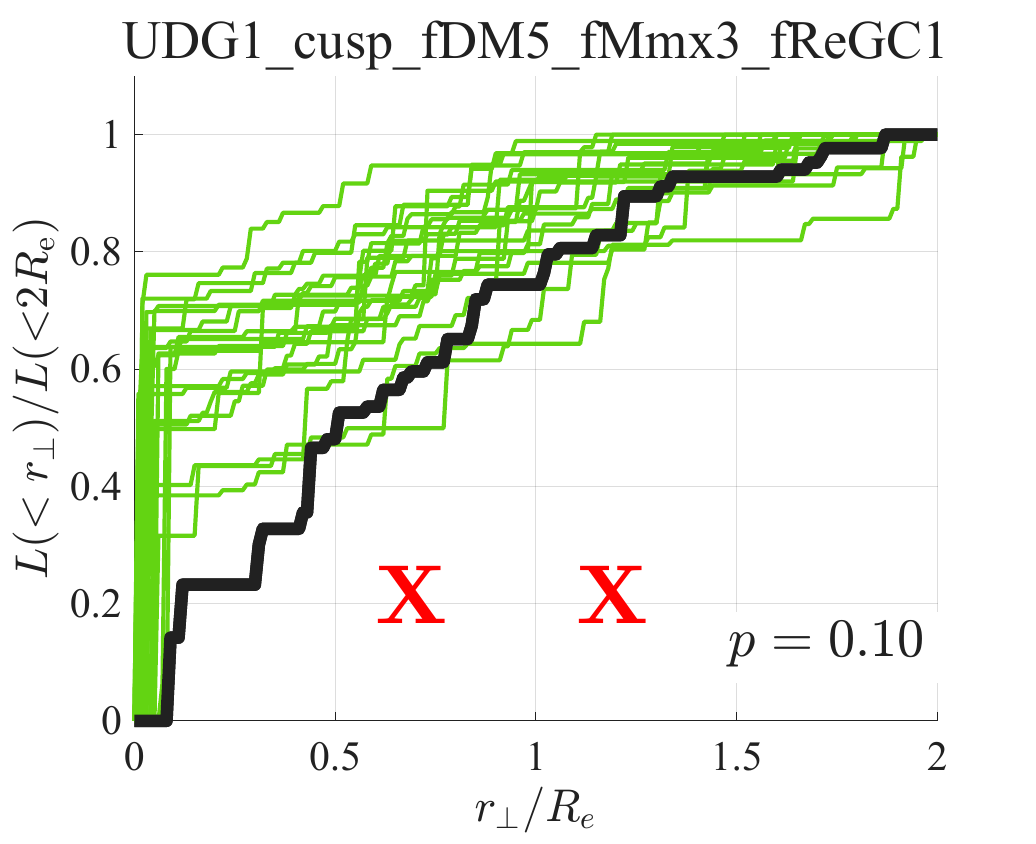} 
       \includegraphics[ scale= 0.245]{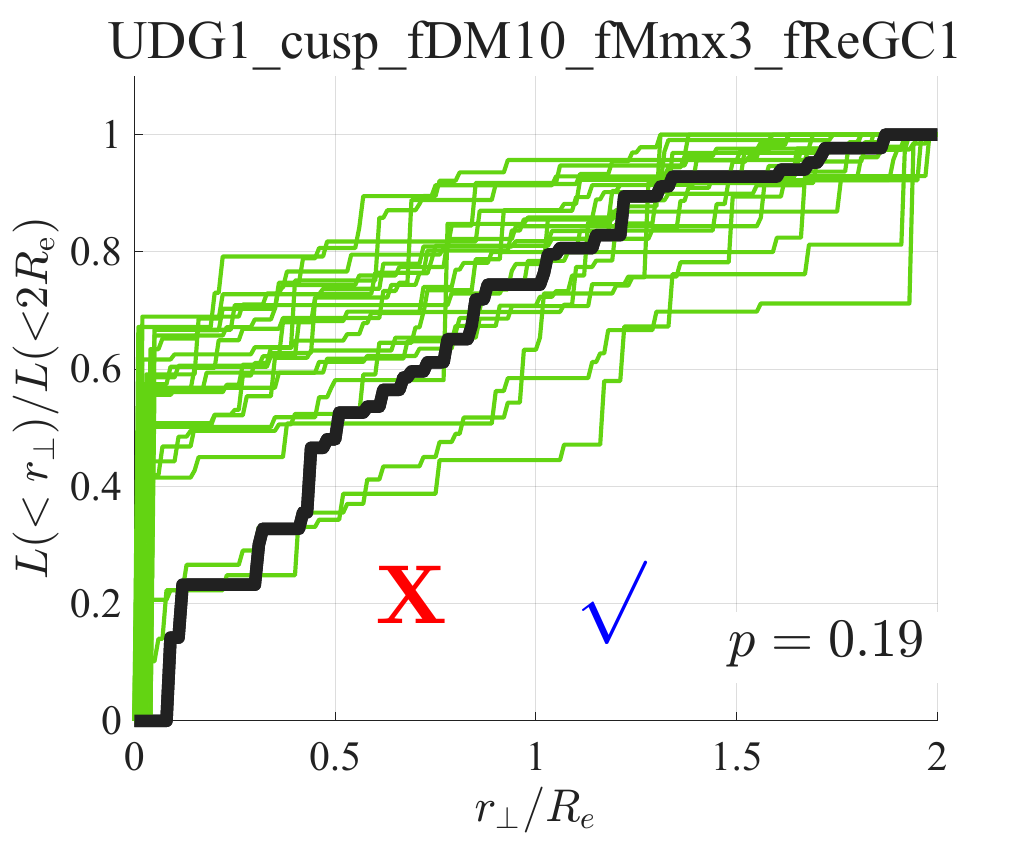}     
      \caption{Similar to Fig.~\ref{fig:Burk_LCDFs}, but using the NFW cusp DM halo model.}
 \label{fig:NFW_LCDFs}
\end{figure*}

\begin{figure*}
\centering
      \includegraphics[ scale= 0.23]{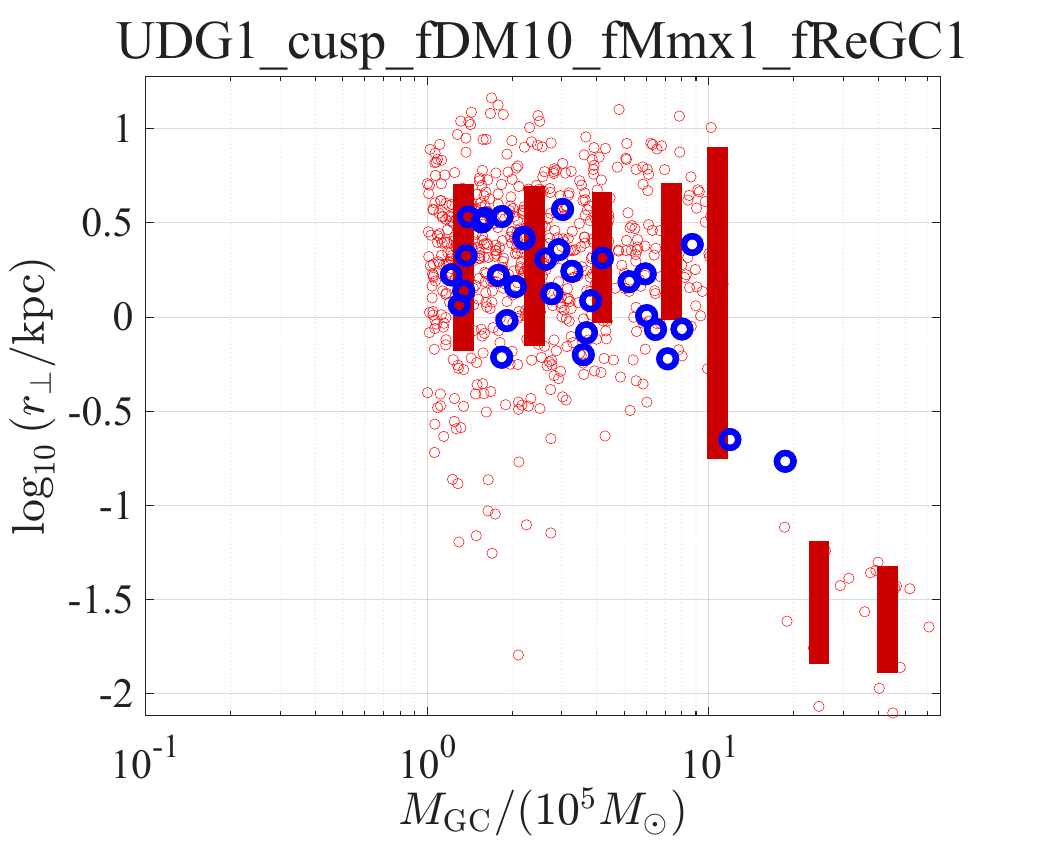}       
            \includegraphics[ scale= 0.23]{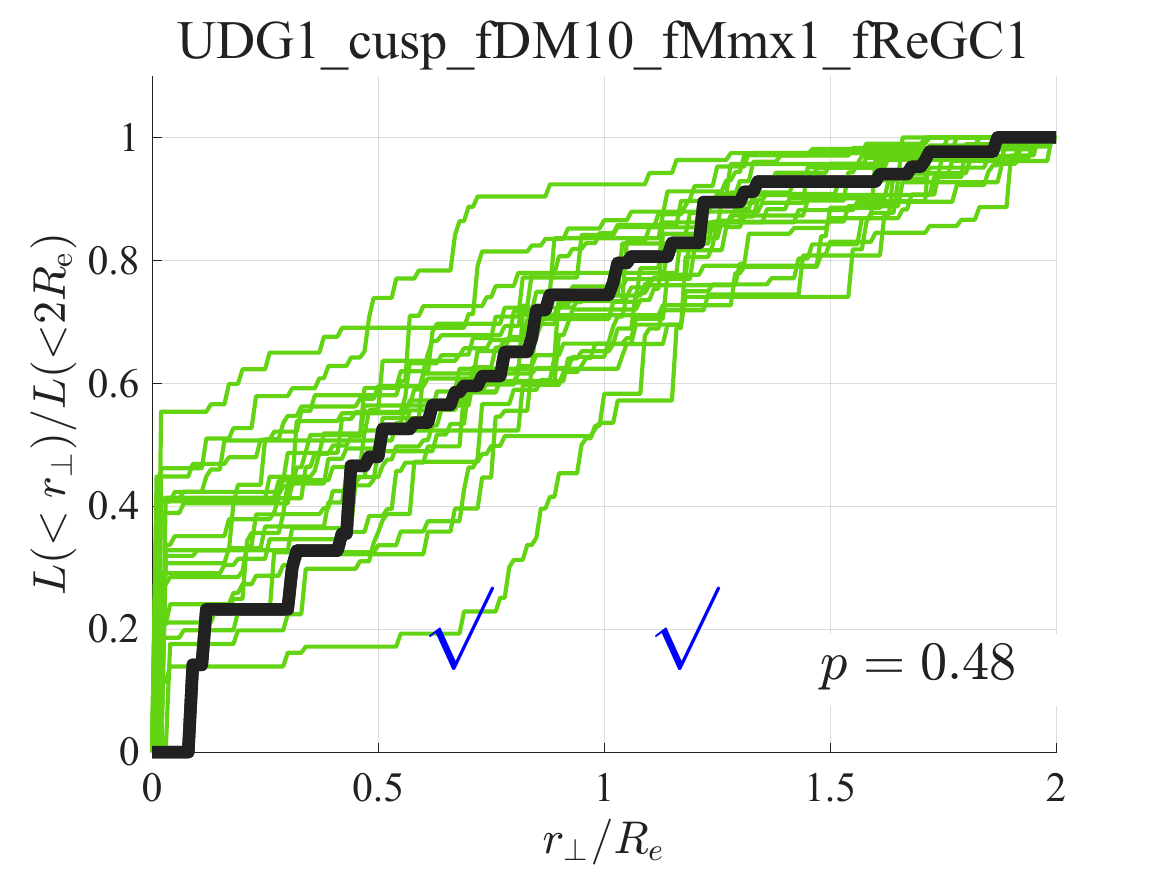}       
                  \includegraphics[ scale= 0.23]{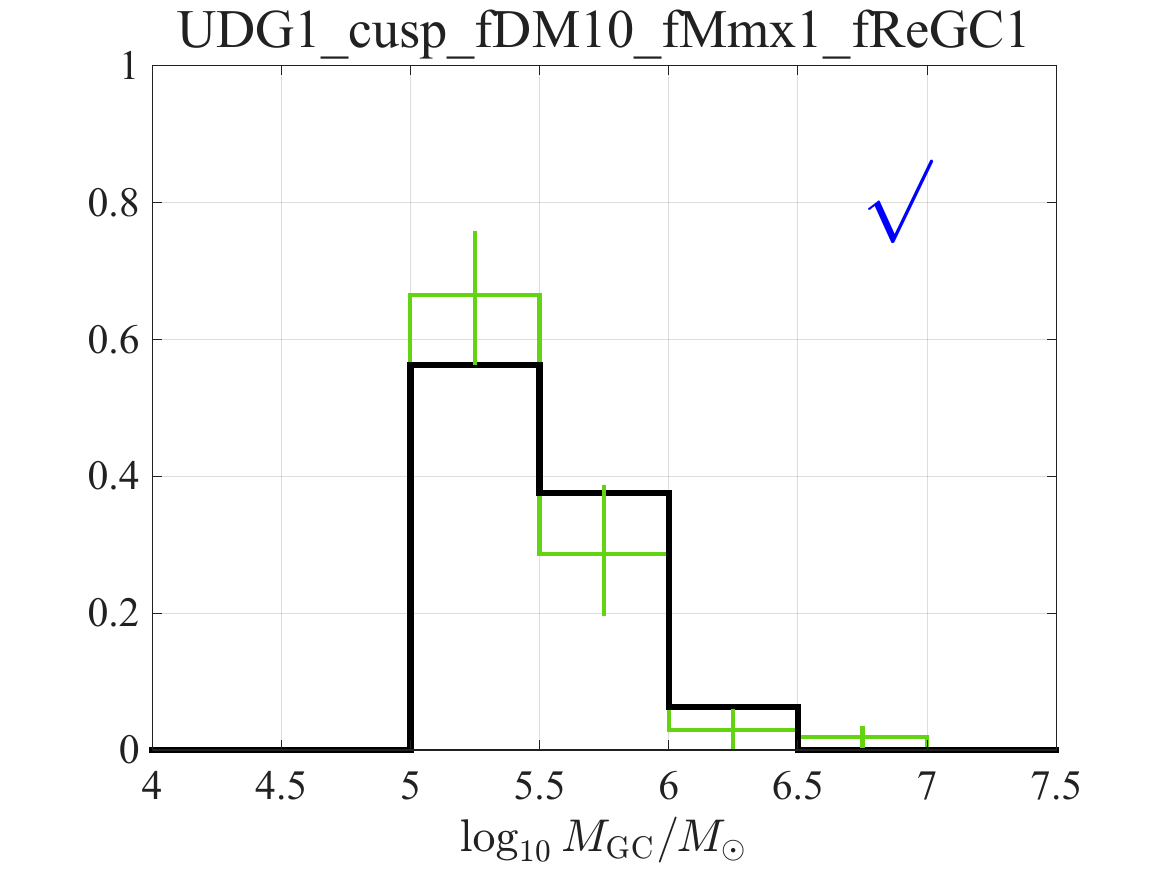}
                  \includegraphics[ scale= 0.23]{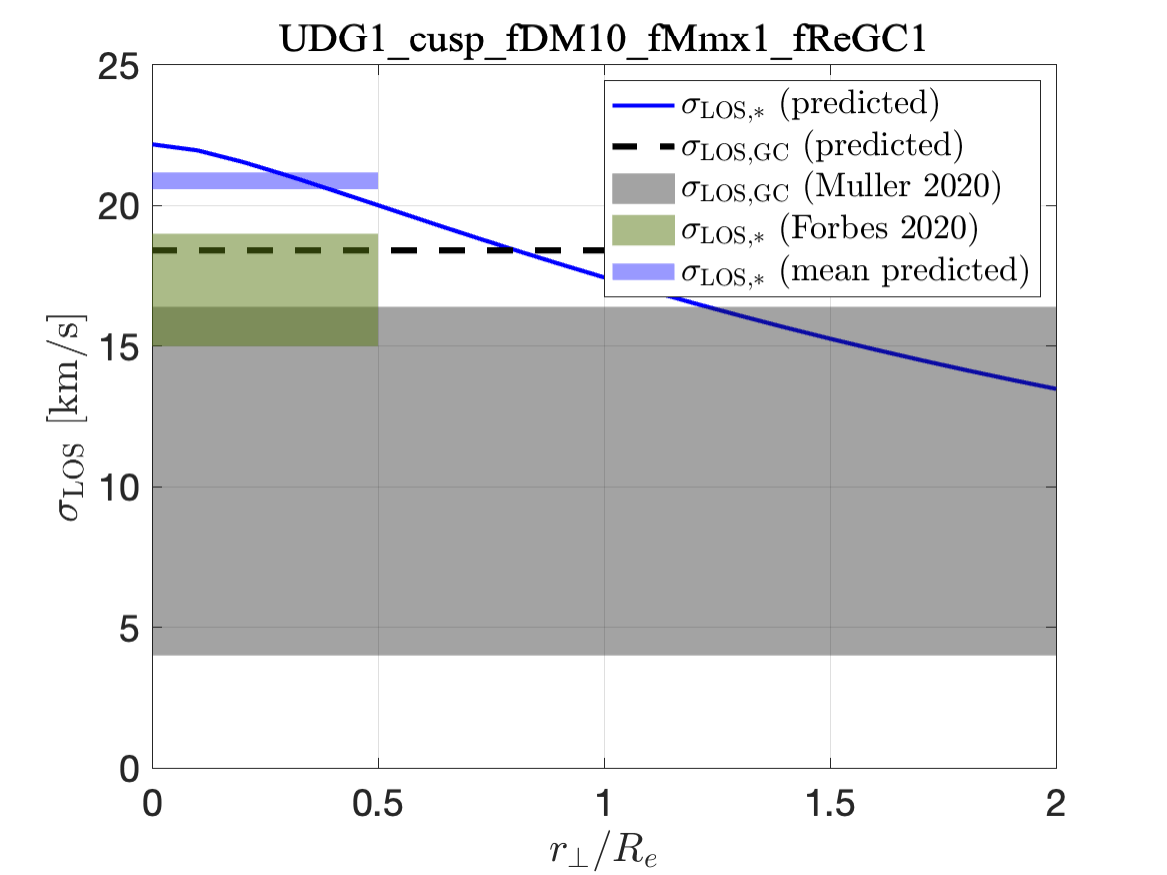}   \\    
                        \includegraphics[ scale= 0.23]{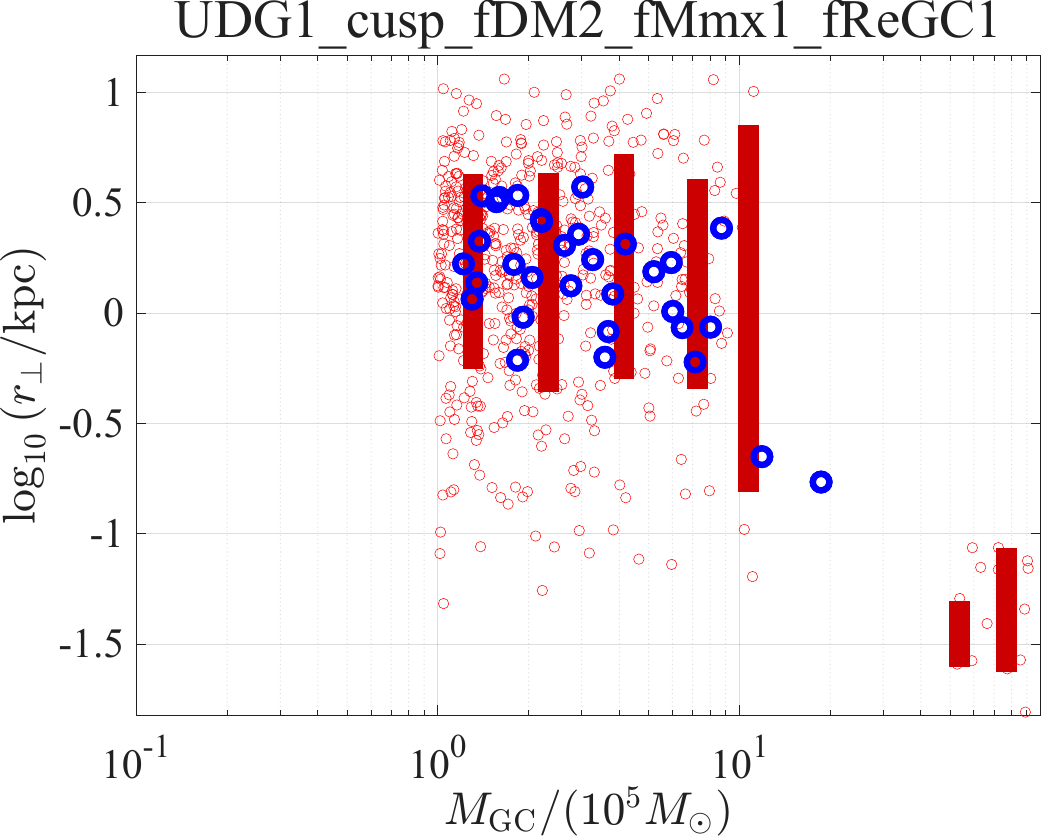}       
            \includegraphics[ scale= 0.23]{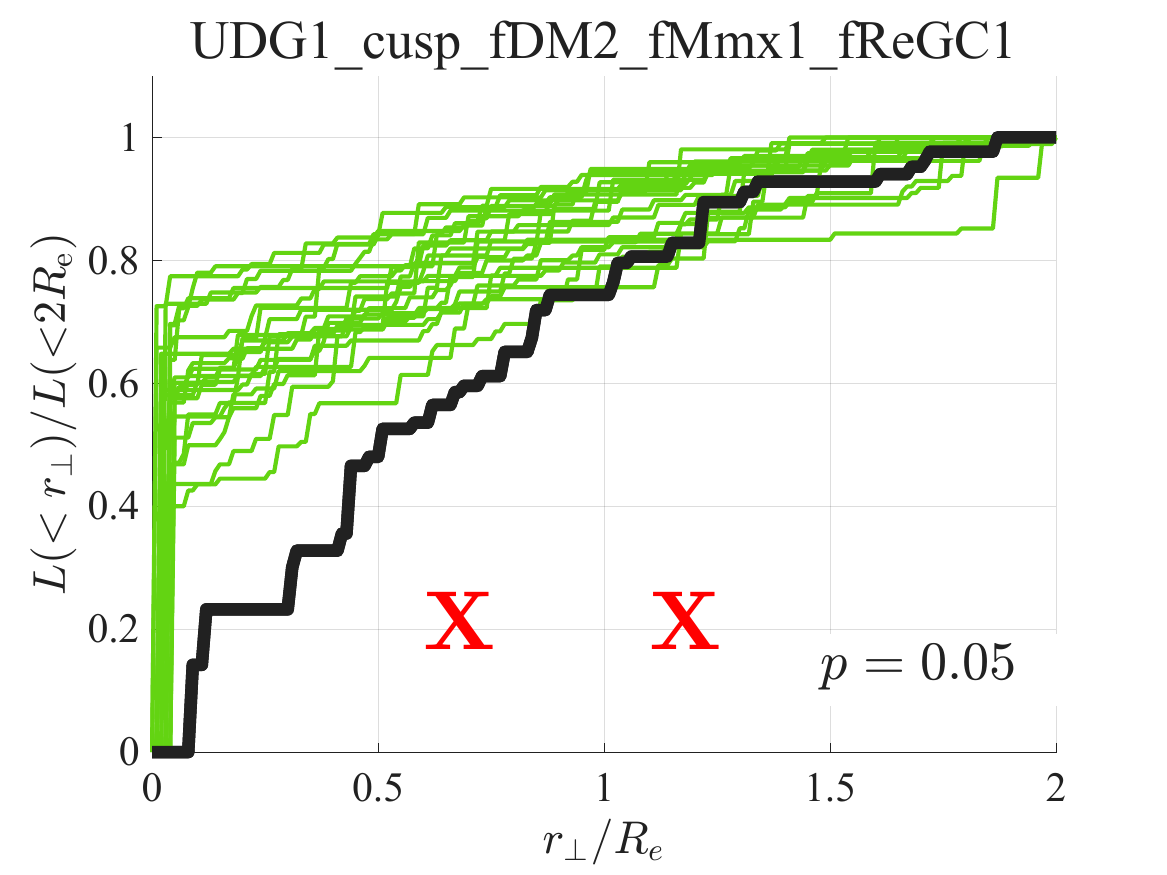}       
                  \includegraphics[ scale= 0.23]{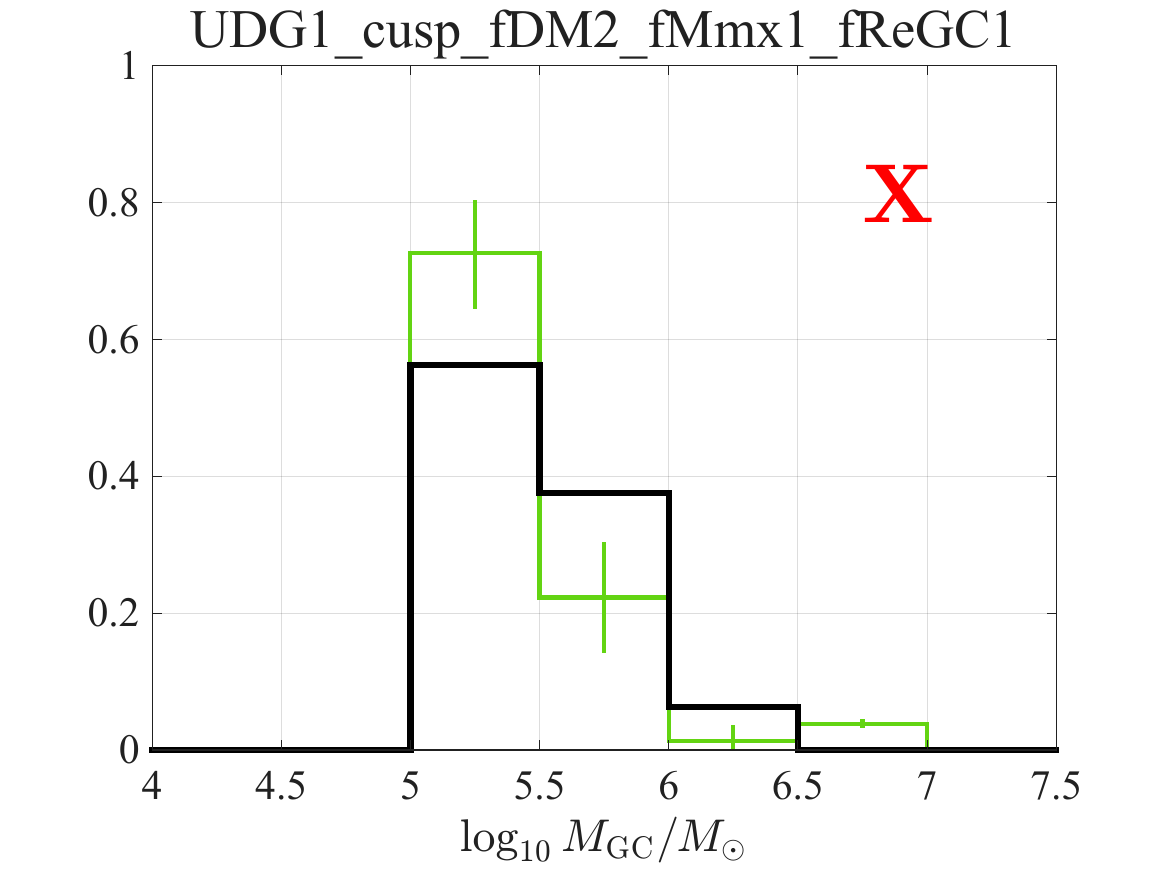}       
                  \includegraphics[ scale= 0.23]{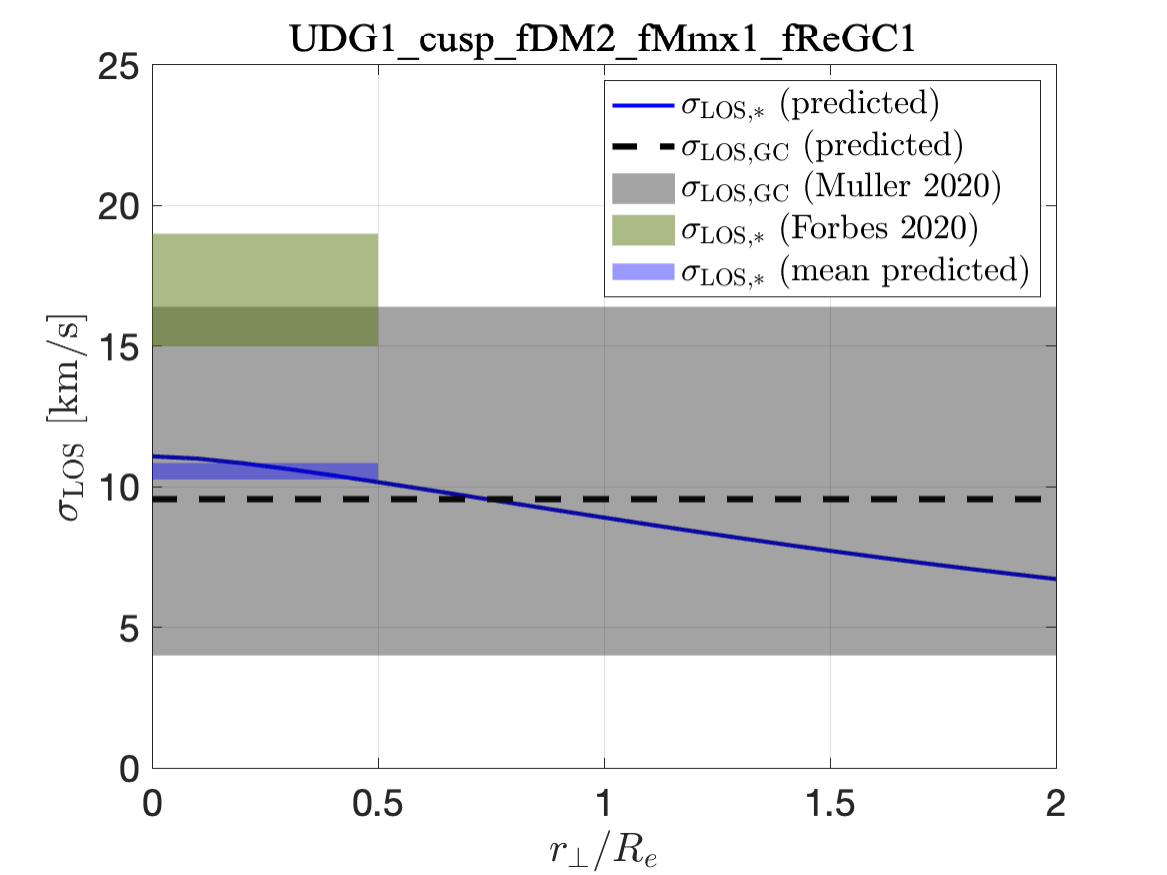}
 \caption{Similar to Fig.~\ref{fig:Burk_details}, but using the NFW cusp DM halo model.}
 \label{fig:NFW_details}
\end{figure*}

%
Figs.~\ref{fig:NFW_LCDFs} and~\ref{fig:NFW_details} repeat the analysis for the NFW cusp model. (We do not repeat the DM-free scenario.) The model in the top row of Fig.~\ref{fig:NFW_details} passes the luminosity CDF criterion, and the model at the bottom fails it. The left panels of Fig.~\ref{fig:NFW_details} show that both models predict NSCs (often for the top, and always for the bottom). In these models the maximal mass in the GCIMF is $0.66$ that of the most massive GC observed (indicated by ``fMmx1" in panel titles), so NSCs come from mergers. 

Because of the tendency of the NFW model to produce NSCs, it is important to verify that our results are not overly sensitive to our crude treatment of mergers. In App.~\ref{a:nomerge} we repeat the calculation of key results of the NFW model, with GC mergers turned off altogether in the simulations. We find that the luminosity CDF at $r_\perp\geq0.5R_{\rm e}$ remains sufficiently robust to this change.
%

\subsection{Results: GC initial radial distribution stretched compared to stellar body}
\begin{figure*}
\centering
      \includegraphics[ scale= 0.26]{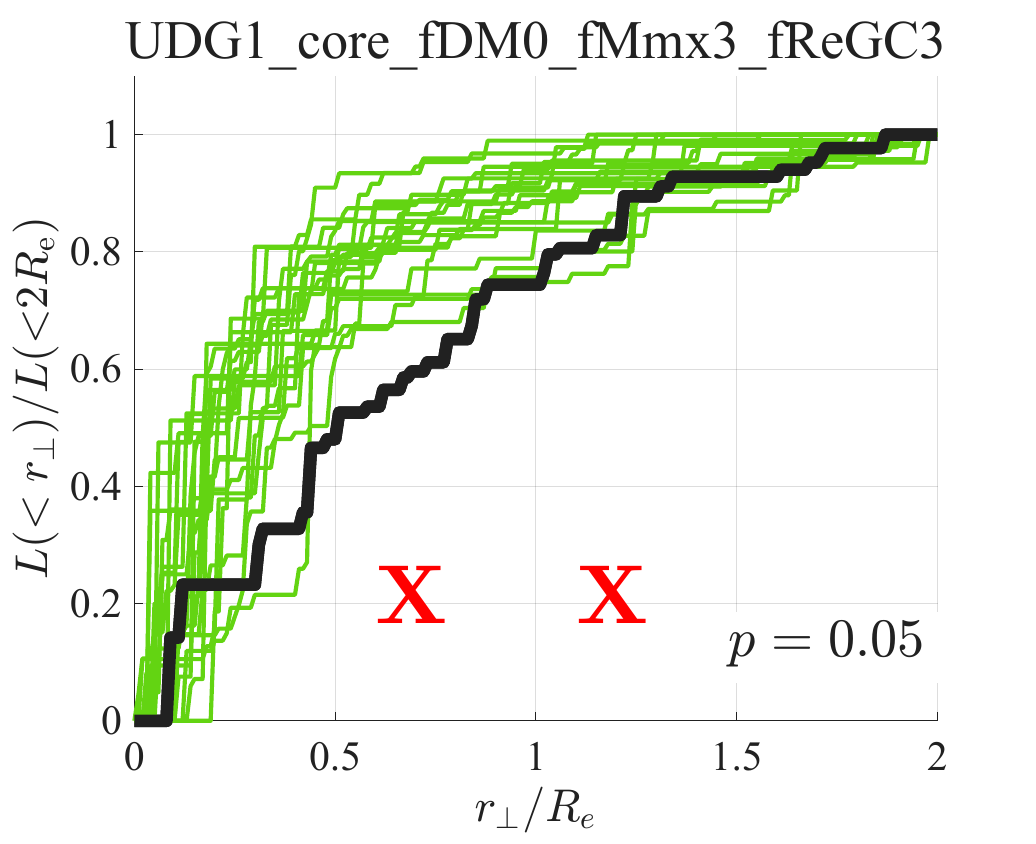} 
            \includegraphics[ scale= 0.26]{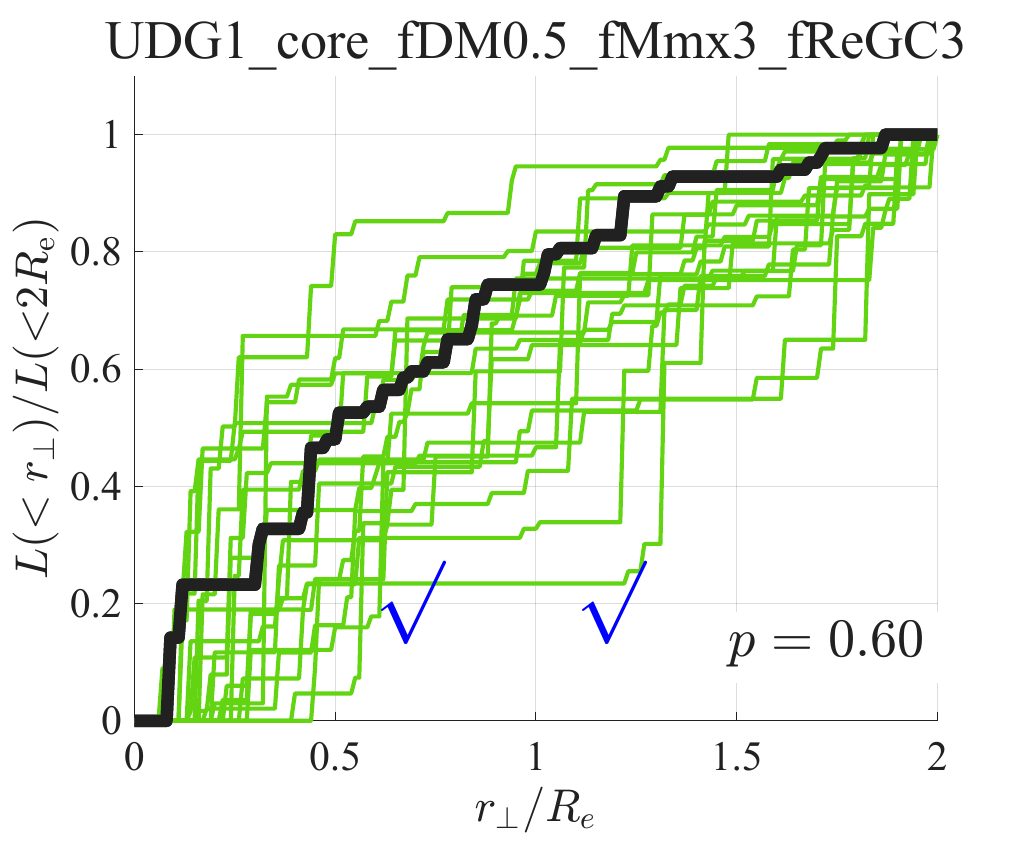} 
      \includegraphics[ scale= 0.26]{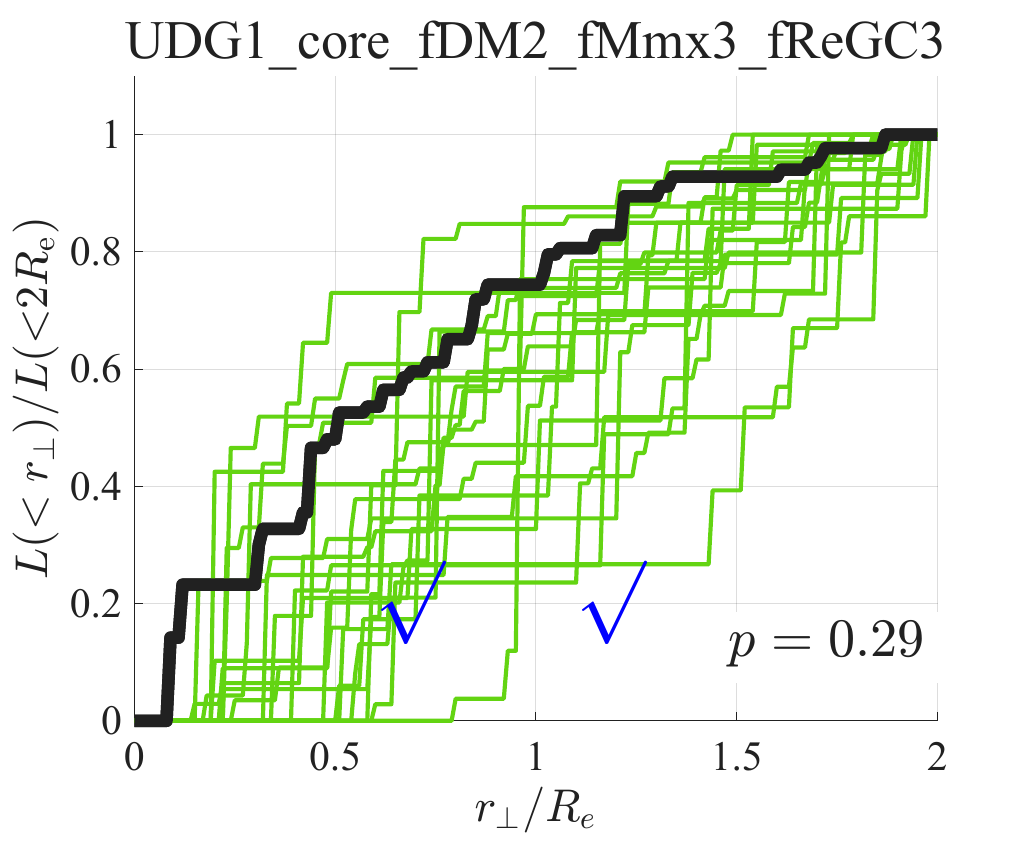}       \\
                  \includegraphics[ scale= 0.26]{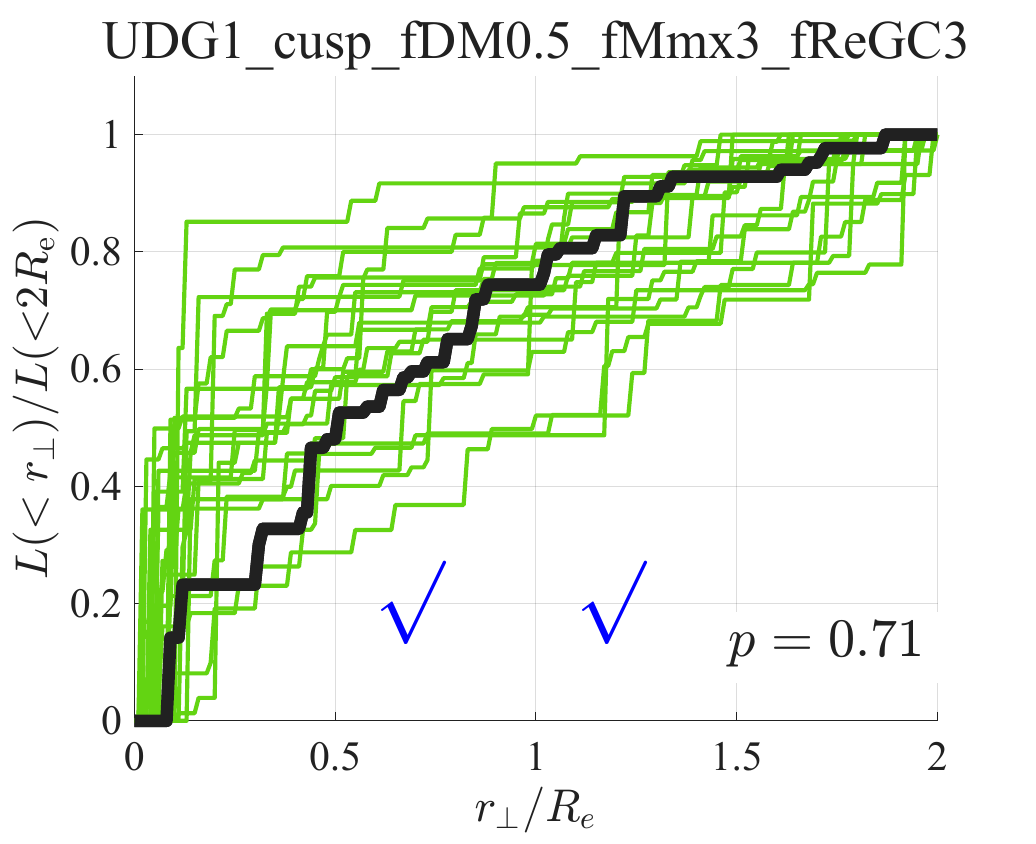}       
      \includegraphics[ scale= 0.26]{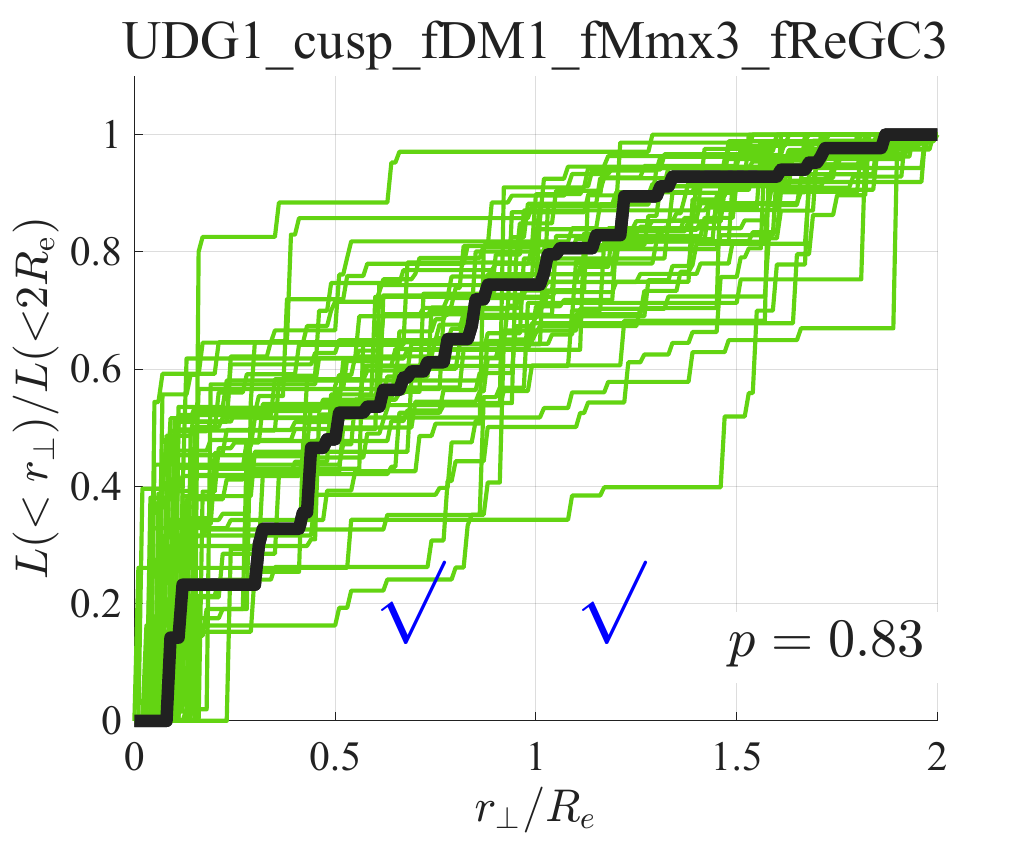}                 
      \includegraphics[ scale= 0.25]{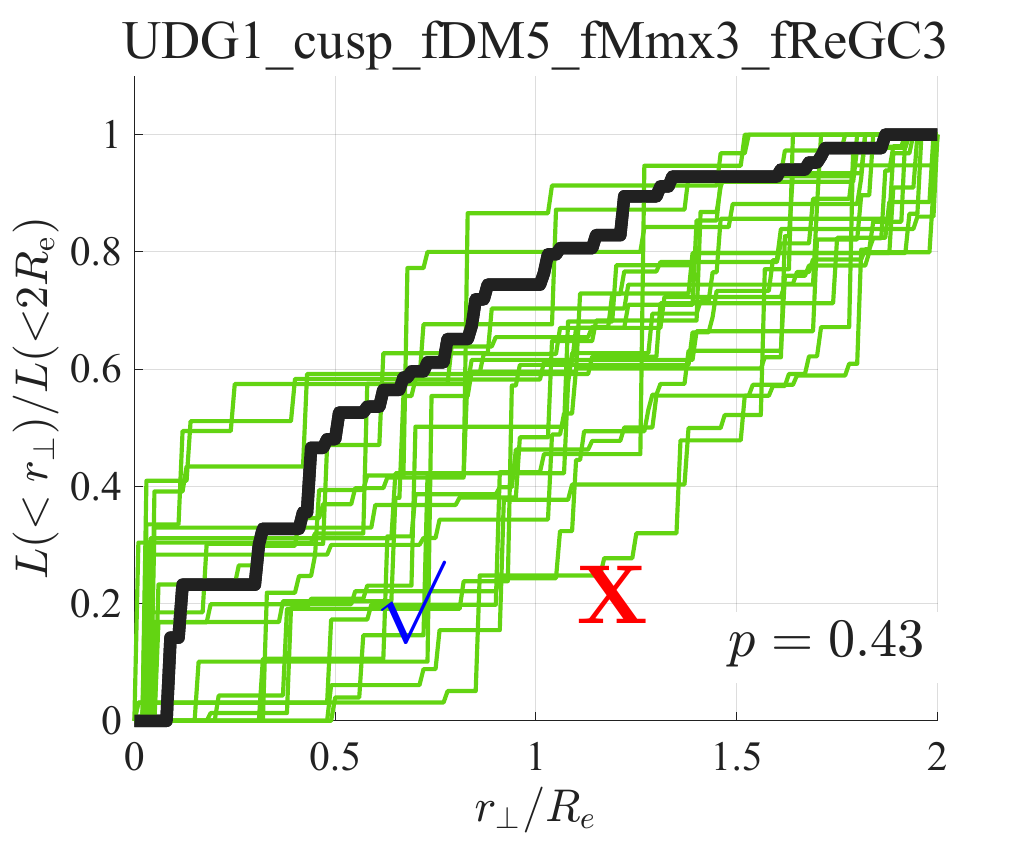} 
 \caption{Similar to Figs.~\ref{fig:Burk_LCDFs} and~\ref{fig:NFW_LCDFs}, but with GC initial radial distribution stretched by a factor of 3 w.r.t. the stellar body. 
 }
 \label{fig:Burk_LCDFs_stretch3}
\end{figure*}

We now consider the scenario in which the initial radial distribution of GCs is extended (``stretched") w.r.t. the currently observed distribution of the stellar body. Fig.~\ref{fig:Burk_LCDFs_stretch3} 
repeats a similar analysis as in Figs.~\ref{fig:Burk_LCDFs} and~\ref{fig:NFW_LCDFs}, but with initial GC radial distribution stretched by a factor of 3 w.r.t. the stellar body (indicated in panel titles by ``fReGC3"). Namely, the GCs start their evolution with a S\'ersic distribution with $R_{\rm e}=6$~kpc. We keep the S\'ersic index at $n=0.6$. Note that a factor of 3 stretch brings the initial GC effective radius close to the tidal radius of UDG1~\citep[][]{Bar:2022liw}, a consideration that deserves further study, but is outside of our current scope. 
%

We find that stretched initial conditions lead to large realization-to-realization spread in the GC distribution today. For example, the cusp model with fDM5 and Mmx3 predicts GC luminosity CDF in the range $L(0.5R_{\rm e})/L(2R_{\rm e})\approx(0.5-0.8)$ if GCs initially follow field stars (fReGC1), and $L(0.5R_{\rm e})/L(2R_{\rm e})\approx(0-0.6)$ if the GCs are initially stretched (fReGC3). 

Stretched initial conditions for GCs counteract the DF-induced decay of the orbits caused by a low-dispersion halo, allowing  lower halo mass. Nevertheless, even with a factor of three stretch (``fReGC3"), a DM-free model is still excluded in our fiducial simulations. The dynamical evidence for DM in UDG1 therefore seems quite convincing, although not bullet-proof: we show in Sec.~\ref{ss:udg1ressum} (see also App.~\ref{a:massloss}) that the combination of a factor of 3 initial radial stretch and a low GC mass loss rate does allow a DM-free model to fit the data.

\subsection{Comparison with stellar kinematics}\label{ss:kin}
\begin{figure*}
\centering
      \includegraphics[ scale= 0.24]{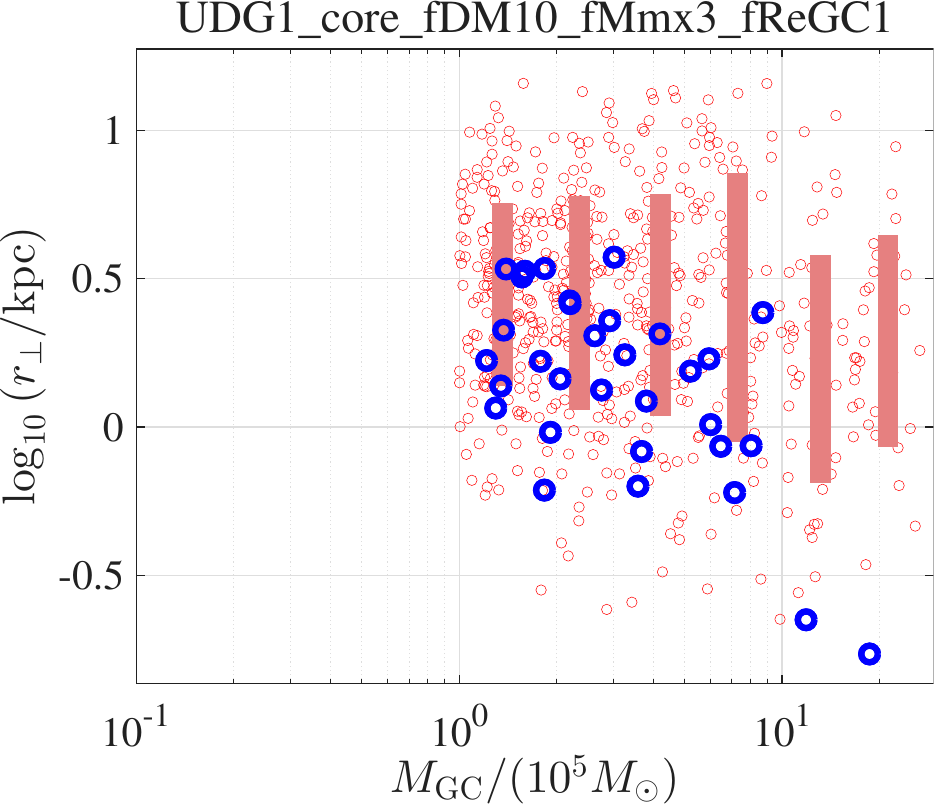}       
            \includegraphics[ scale= 0.23]{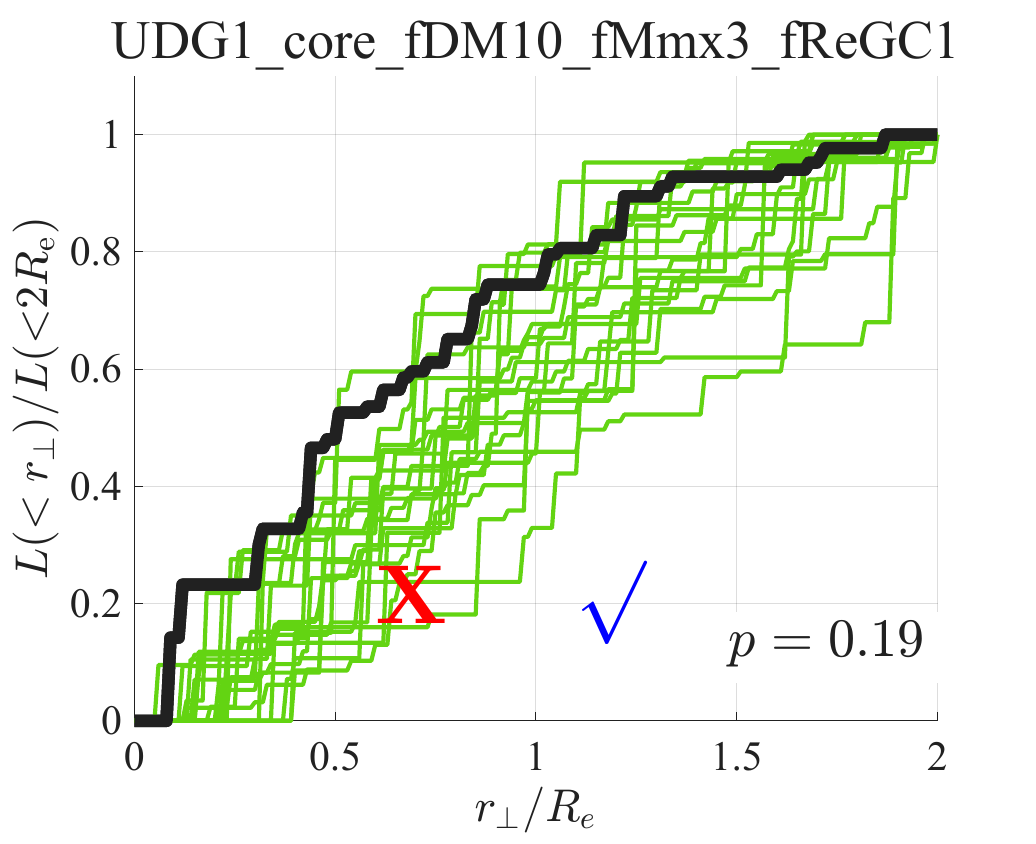}       
                        \includegraphics[ scale= 0.23]{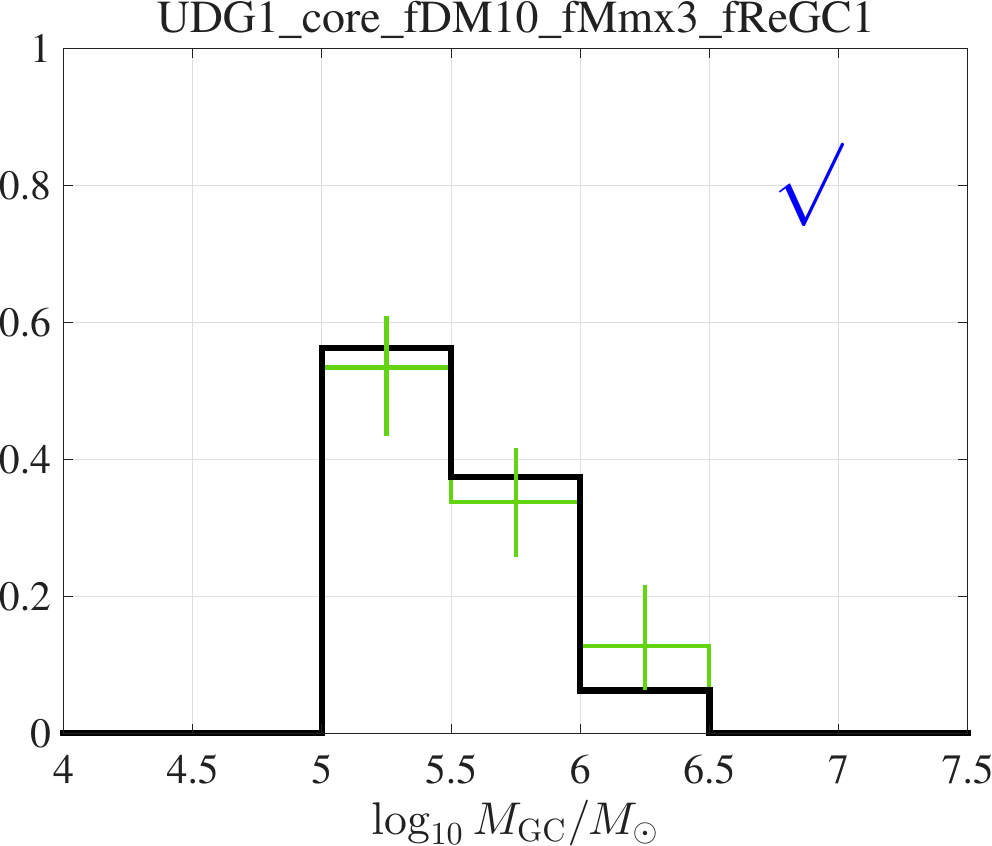}       
                  \includegraphics[ scale= 0.23]{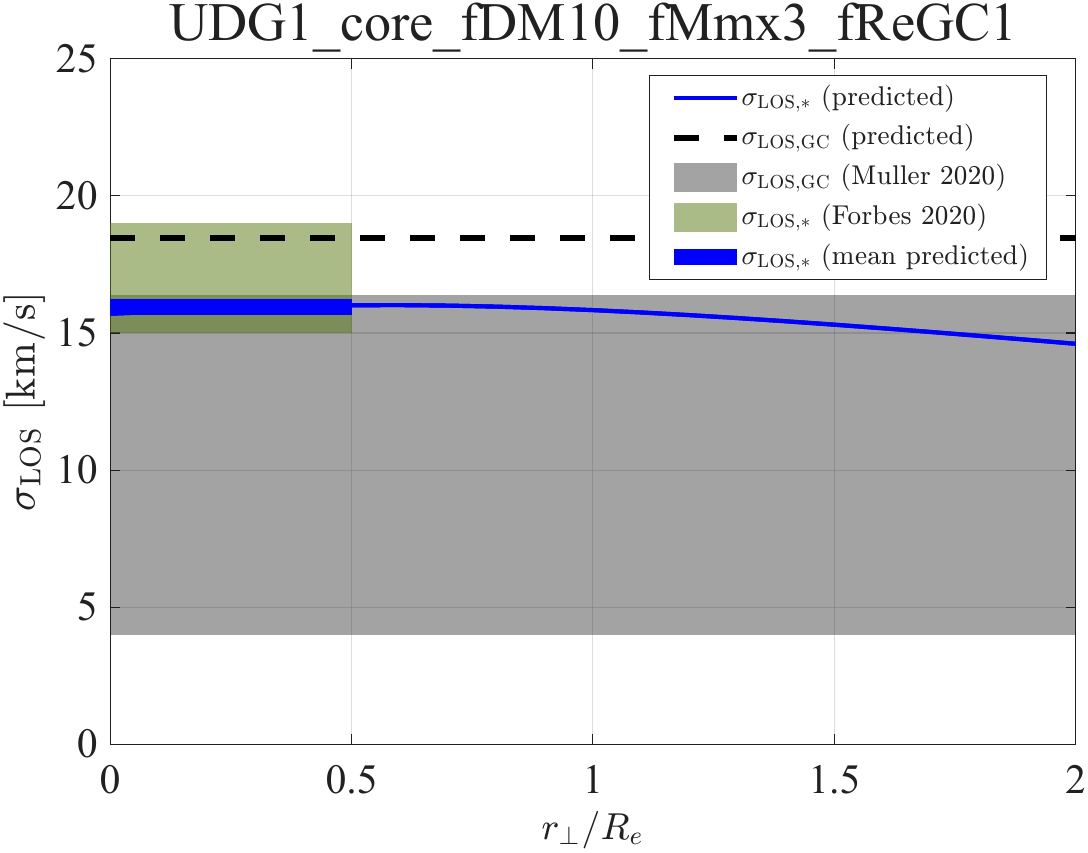}        \\
                        \includegraphics[ scale= 0.24]{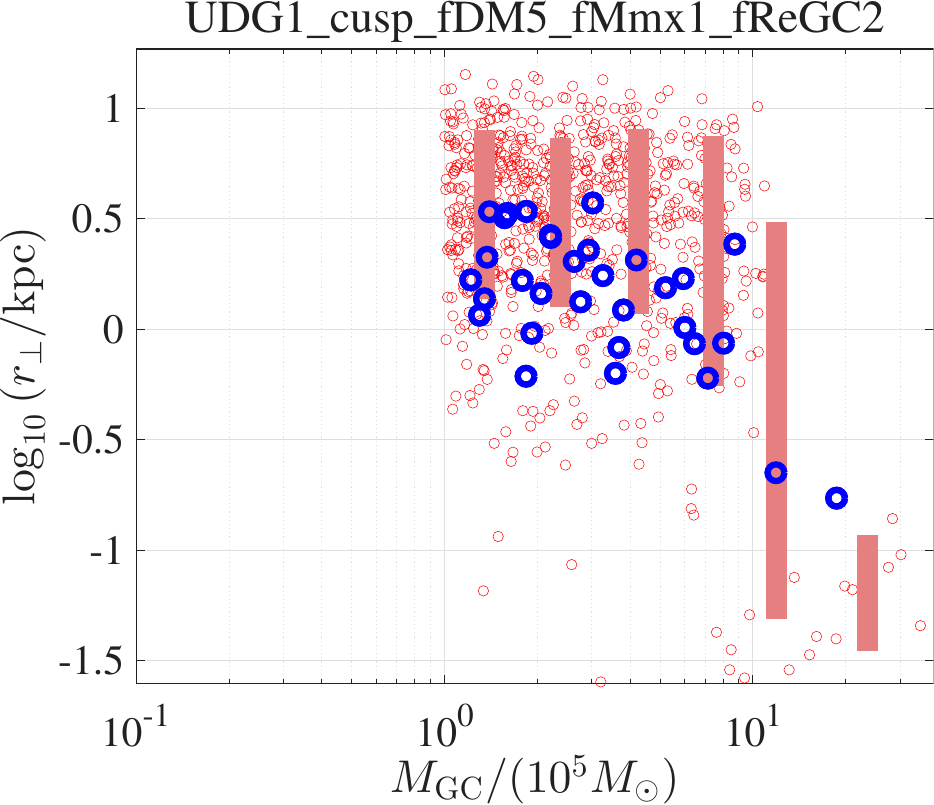}       
            \includegraphics[ scale= 0.23]{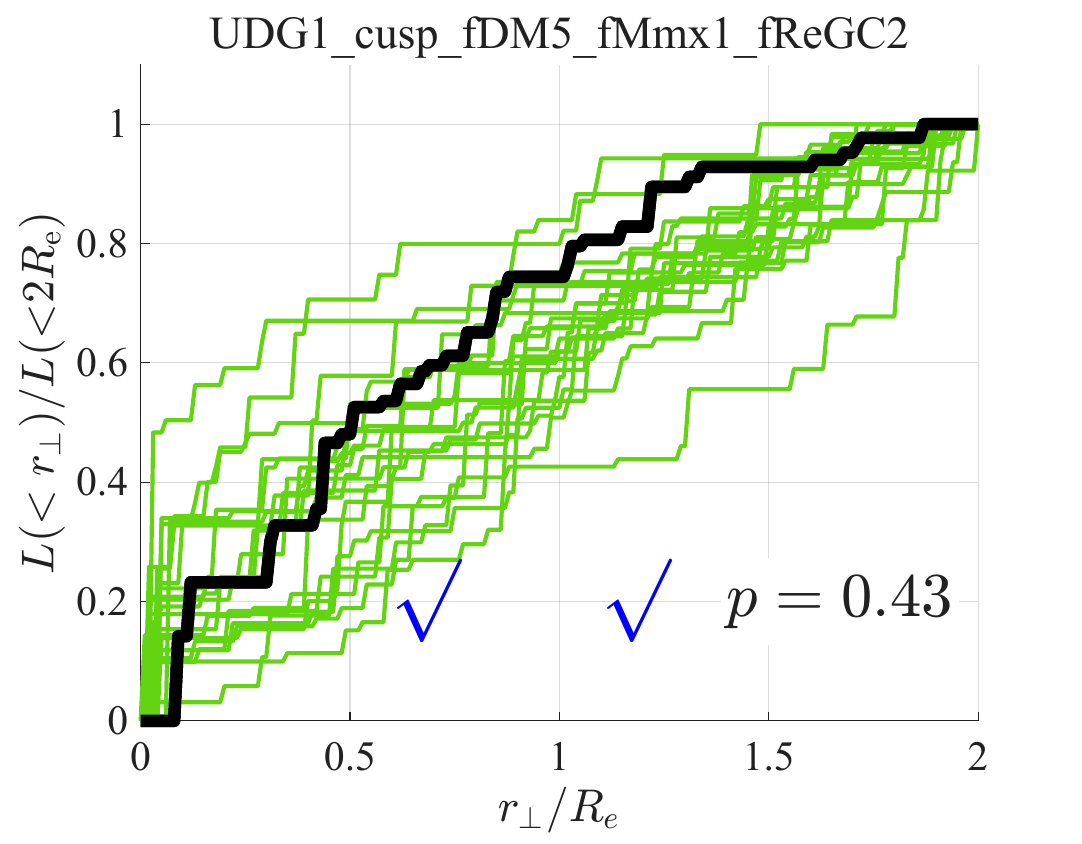}       
                        \includegraphics[ scale= 0.23]{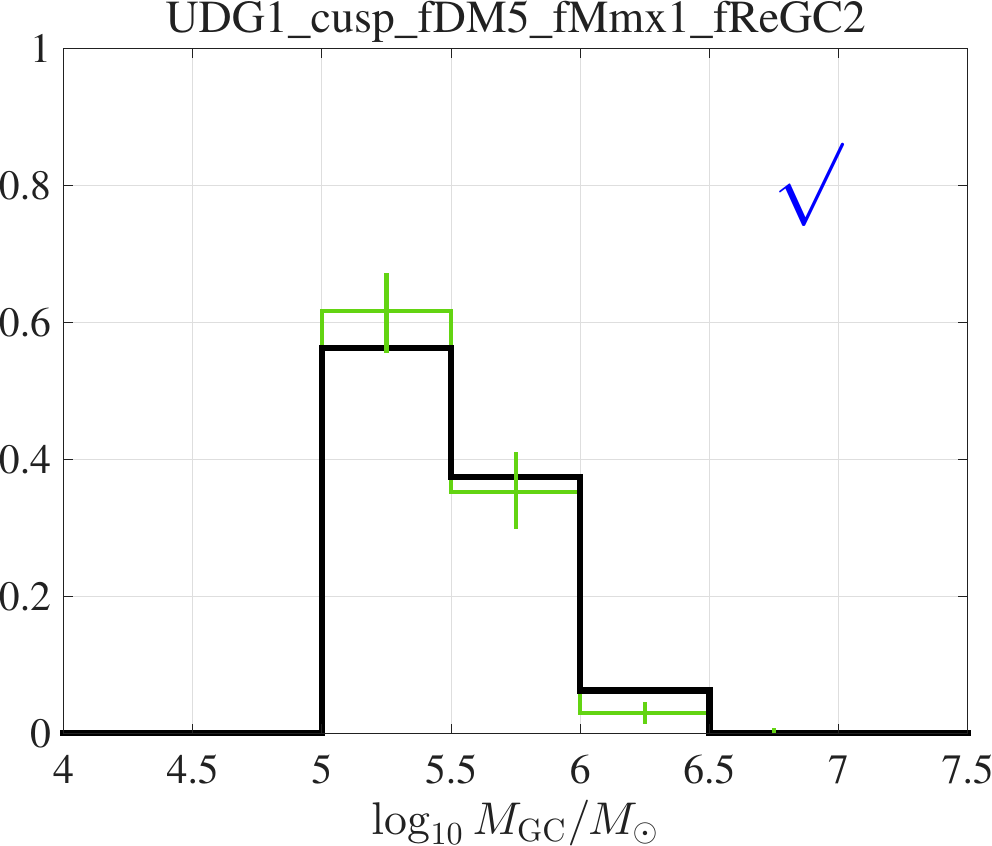}       
                  \includegraphics[ scale= 0.23]{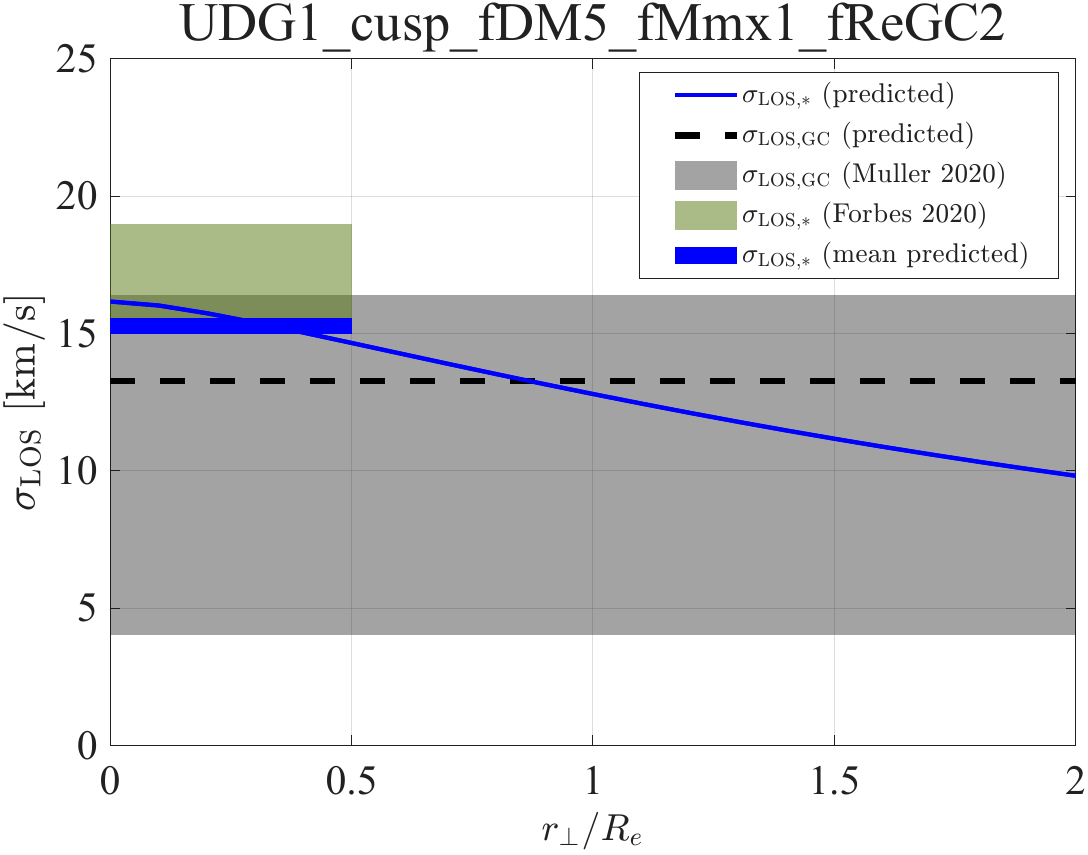}         
 \caption{GC observables predictions for core (top) and cusp (bottom) halo models consistent with stellar kinematics data (\cite{Forbes:2020vly}).}
 \label{fig:details_kin2}
\end{figure*}

The right-most panels of Figs.~\ref{fig:Burk_details} and~\ref{fig:NFW_details} compare data and predictions for stellar and GC LOSVD. 
The bottom row of panels in Fig.~\ref{fig:Burk_details} is a DM-free model: it strongly fails the GC data comparison (notably the LCDF plot in the second panel), and also undershoots the LOSVD data (right-most panel). The top row in Fig.~\ref{fig:Burk_details} roughly represents the minimal DM content compatible with GC data, assuming that GC initial conditions follow the stars. The model's stellar LOSVD prediction (blue shaded patch) is still lower than the measurement of \cite{Forbes:2020vly} (green shade): this mismatch in $\sigma_{\rm LOS,*}$ means that the GC-derived lower bound on the DM halo is weaker than (but consistent with) that derived from stellar kinematics. The core model can become consistent with kinematics data with a more massive DM halo, as we illustrate using an fDM10 core model in Fig.~\ref{fig:details_kin2} (top row, right panel). However, the larger DM halo mass, motivated by the LOSVD data, causes reduced DF and tends to undershoot the LCDF (second panel). The mismatch is not extreme, with $p=0.19$, but it  indicates some tension for the core model when GC and LOSVD data are considered in tandem.

Fig.~\ref{fig:NFW_details} shows GC and LOSVD comparisons for the cusp halo model. Here, again, the bottom row with a low DM content (fDM2) fails both GC and LOSVD comparisons. The top row with a substential DM halo (fDM10) captures GC data well but slightly overshoots the stellar and GC LOSVD data of \cite{Forbes:2020vly} and \cite{muller2020spec}. This mild tension  can be remedied with moderate variation of the initial conditions, as we demonstrate in the bottom row of Fig.~\ref{fig:details_kin2} with a lower DM halo mass and moderate GC radial stretch. 
%

\begin{figure*}
\centering
                       \includegraphics[ scale= 0.3]{UDG1_cusp_fDM5_fMmx1_fReGC1_LCDFnorm_cx}      
            \includegraphics[ scale= 0.3]{UDG1_cusp_fDM5_fMmx1_fReGC2_LCDFnorm}       
             \includegraphics[ scale= 0.3]{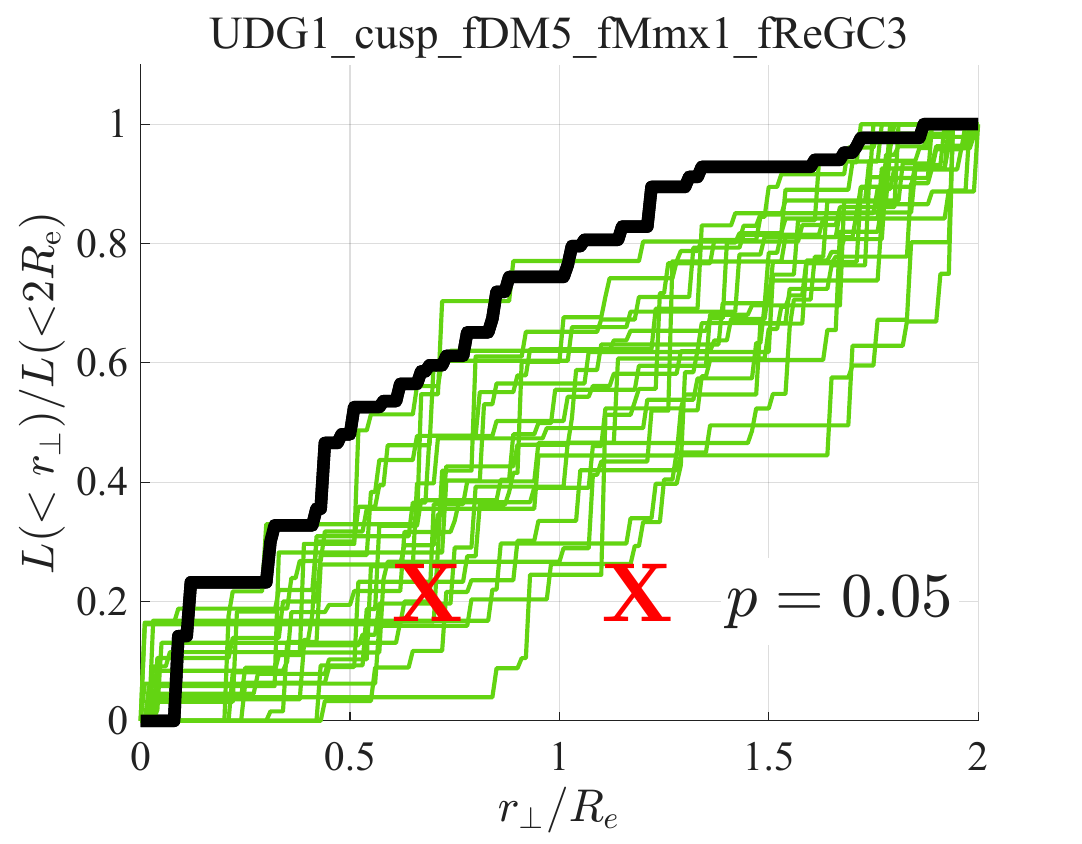} \\
 \caption{Complementarity between GC morphology and stellar kinematics (NFW cusp model): effect of initial GC distribution stretch.
 }
 \label{fig:details_kin3}
\end{figure*}

Fig.~\ref{fig:details_kin3} demonstrates how varying the GC radial stretch for the cusp fDM5 model brings the GC LCDF out of, or into, compatibility with the data. All three panels have the same underlying DM halo and thus the same stellar LOSVD. The GC LOSVD prediction varies slightly across the panels but stays compatible with the data. 
%

\subsection{UDG1: discussion and summary tables}\label{ss:udg1ressum}
\begin{figure*}
\centering
      \includegraphics[scale=0.45]{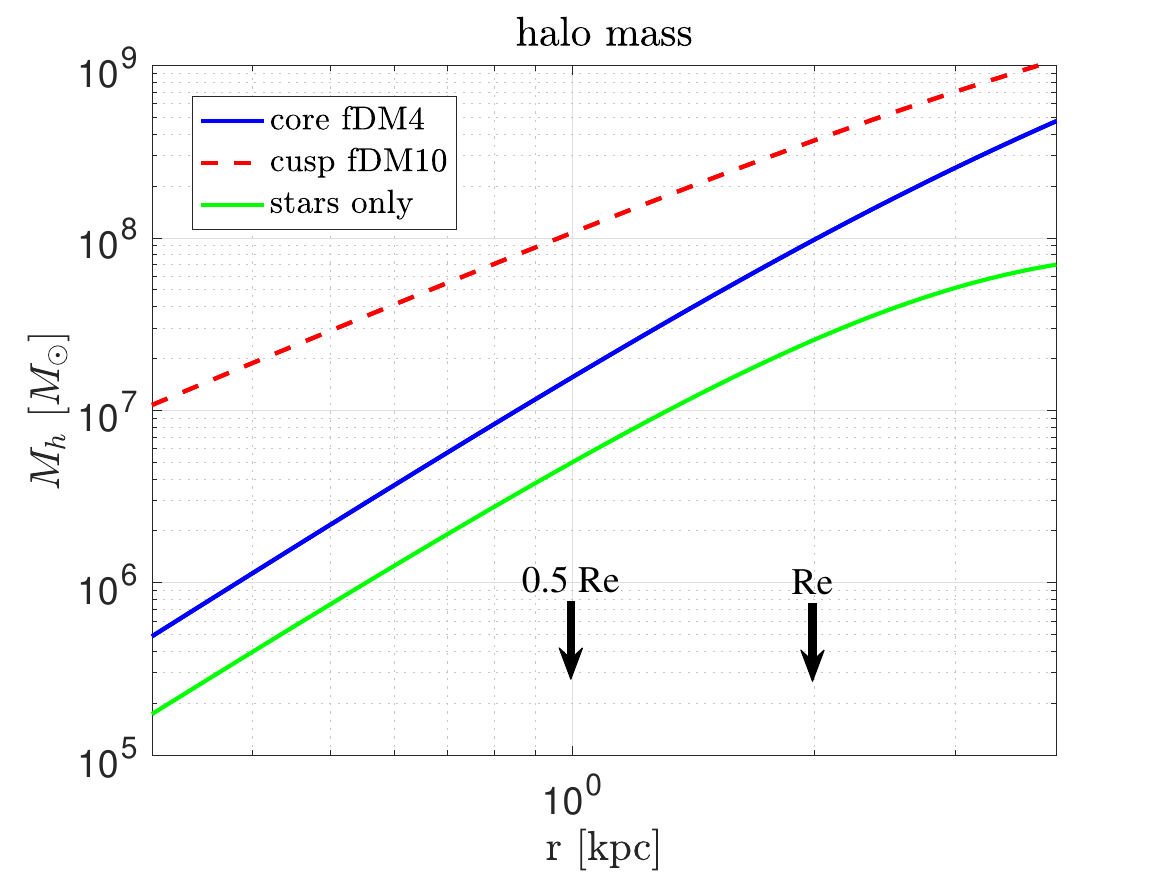}
            \includegraphics[scale=0.45]{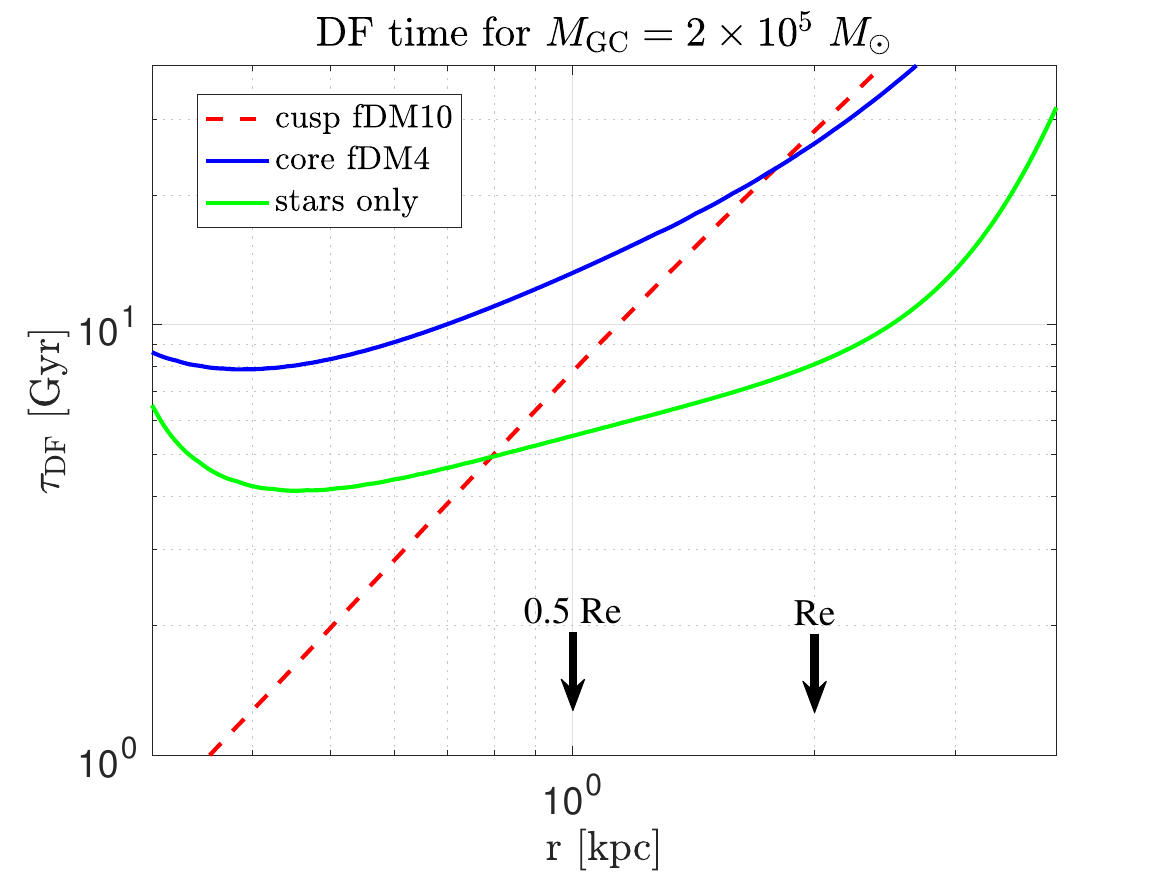}
 \caption{Enclosed mass profile (left) and DF time (right) for halo models discussed in the text. Arrows highlight 0.5 and 1 times the stellar effective radius, which roughly delimits the most important region of the halo for the purpose of the GC analysis.
 }
 \label{fig:tDF}
\end{figure*}

The left panel of Fig.~\ref{fig:tDF} shows mass models of UDG1. The DM density in the core (cusp) model is 4 (10) times the stellar density at $r=2R_{\rm e}$. Both models fit well the GC distribution data; as discussed in Sec.~\ref{ss:kin}, the core model undershoots the stellar LOSVD data, while the cusp model slightly overshoots it. 

The right panel of Fig.~\ref{fig:tDF} shows the local DF time for a $M_{\rm GC}=2\times10^5~M_\odot$ GC on a circular orbit. In the NFW profile $\tau_{\rm DF}\propto r^\beta$ with $1.85\lesssim\beta\lesssim2$~\citep[][]{Bar:2021jff}. In the Burkert profile $\tau_{\rm DF}$ saturates at small $r$. The saturation occurs also in the star-only profile, but at shorter $\tau_{\rm DF}$. 
Using the circular velocity for $V$, the parametric scaling of $\tau_{\rm DF}$ with halo mass is roughly $\tau_{\rm DF}\propto M_h^{0.5}$. The precise scaling depends on the details of the velocity distribution of the halo particles, but the point remains that a more massive halo exhibits weaker DF. 

The fact that DF becomes inefficient for a massive halo means that using GC mass segregation to identify the imprint of DM is a delicate game, because the signal disappears if the galaxy contains too much DM. The case of UDG1 may be  fortunate in this respect~\citep{Liang:2023ryi}: as the right panel of Fig.~\ref{fig:tDF} shows, the stars-only model yields $\tau_{\rm DF}$ that is about a factor of two shorter than the typical ages of the GCs inside the stellar half-light radius, predicting significant mass segregation for GCs with orbits inwards of $r\approx R_{\rm e}$. 
At the same time, DM models that are consistent with stellar kinematics yield $\tau_{\rm DF}$ at $r\approx R_{\rm e}$ that is slightly larger than the probable age of the system, producing mild mass segregation consistent with the observations.

A summary of the luminosity CDF criterion at $r_\perp=0.5R_{\rm e}$ and $r_\perp=R_{\rm e}$ for most of our semianalytic simulations is given in Tab.~\ref{tab:UDG1summarycore}. In each entry of the table, the first and second marker ($\X$ or $\V$) denotes success or failure in the luminosity CDF criterion at $r_\perp=0.5R_{\rm e}$ and $r_\perp=R_{\rm e}$, respectively.

Tab.~\ref{tab:UDG1summarycoreMdot01} repeats the analysis using GC mass loss rate $(dM_{\rm GC}/dt)_{\rm iso}$ that is a factor of ten below that in Eq.~(\ref{eq:Mdotiso}). The results are mostly similar to the fiducial implementation, as long as GC initial stretch is not large; in particular, for stretch of 1 or 2, the lowest allowed DM content in the low mass loss runs is the same as in the fiducial runs. In the case of stretch 3, however, the DM-free model becomes acceptable at low GC mass loss. App.~\ref{a:massloss} shows more comparisons between the fiducial and low mass loss rate results, demonstrating that the main effect of the low mass loss rate is to increase the statistical spread of the evolved GC distributions.
%
%
\begin{table*}[htb!]
\centering
\caption{Results summary: UDG1, fiducial models (see Sec.~\ref{ss:udg1resdesc}). ``Stretch" specifies initial GC radial distribution (1,2,3 times the current stellar body distribution scale radius). $fDM$ specifies the ratio between the DM halo mass at $r=3R_{\rm e}$ to the total stellar mass in UDG1. $fMmx$ specifies the maximal GC mass in the GCIMF, with the flag $fMmx=1,2,3$ corresponding to $\frac{1}{1.5},1,1.5$ times the largest GC mass observed today. In each entry of the table, the first and second marker ($\X$ or $\V$) denotes success or failure in the luminosity CDF criterion at $r_\perp=0.5R_{\rm e}$ and $r_\perp=R_{\rm e}$, respectively.}\label{tab:UDG1summarycore}
\begin{tabular}{|*{19}{c|}}
\hline
\multicolumn{19}{|c|}{Core}\\ \hline
&\multicolumn{6}{|c|}{ stretch: 1}&\multicolumn{6}{|c|}{ stretch: 2}&\multicolumn{6}{|c|}{ stretch: 3}\\ \hline
fDM & 0 & 0.5 & 1 & 2  & 5 & 10 & 0 & 0.5 & 1 & 2  & 5 & 10 & 0 & 0.5 & 1 & 2  & 5 & 10\\ \hline
fMmx1 & $\X\X$ & $\X\X$ & $\X\V$ & $\V\V$ &  $\X\V$ & $\X\X$ & $\X\X$ & $\V\V$ & $\V\V$ & $\X\V$ &  $\X\X$ & $\X\X$ & $\X\X$ & $\V\V$ & $\X\X$ & $\X\X$ &  $\X\X$ & $\X\X$\\
fMmx2 & $\X\X$ & $\X\X$ & $\X\V$ & $\V\V$  & $\V\X$ & $\X\X$ & $\X\X$ & $\V\V$ & $\V\V$ & $\X\V$ &  $\X\X$ & $\X\X$ & $\X\X$ & $\V\V$ & $\X\V$ & $\X\X$ &  $\V\X$ & $\X\X$\\
fMmx3 & $\X\X$ & $\X\X$ & $\X\X$ & $\V\V$ & $\V\V$ & $\X\V$ & $\X\X$ & $\V\V$ & $\V\V$ & $\V\V$ &  $\X\X$ & $\X\X$ & $\X\X$ & $\V\V$ & $\V\V$ & $\V\V$ &  $\X\X$ & $\X\X$\\
\hline
\multicolumn{19}{|c|}{Cusp}\\ \hline
&\multicolumn{6}{|c|}{ stretch: 1}&\multicolumn{6}{|c|}{ stretch: 2}&\multicolumn{6}{|c|}{ stretch: 3}\\ \hline
fDM & 0 & 0.5 & 1 & 2 & 5 & 10 & 0 & 0.5 & 1 & 2 & 5 & 10 & 0 & 0.5 & 1 & 2 & 5 & 10\\ \hline
fMmx1 & $\X\X$ & $\X\X$ & $\X\X$ & $\X\X$ & $\X\V$ & $\V\V$ & $\X\X$ & $\X\V$ & $\V\V$ & $\V\V$ & $\V\V$ & $\X\X$ & $\X\X$ & $\V\V$ & $\X\X$ & $\X\X$ & $\X\X$ & $\X\X$\\
fMmx2 & $\X\X$ & $\X\X$ & $\X\X$ & $\X\X$ & $\X\V$ & $\X\V$ & $\X\X$ & $\X\V$ & $\X\V$ & $\V\V$ & $\V\V$ & $\X\X$ & $\X\X$ & $\V\V$ & $\V\V$ & $\V\V$ & $\V\V$ & $\V\X$\\
fMmx3 & $\X\X$ & $\X\X$ & $\X\X$ & $\X\X$ & $\X\X$ & $\X\V$ & $\X\X$ & $\V\V$ & $\V\V$ & $\X\V$ & $\V\V$ & $\V\V$ & $\X\X$ & $\V\V$ & $\V\V$ & $\V\V$ & $\V\X$ & $\X\X$\\
\hline
\end{tabular}
%
%
\caption{Results summary: UDG1, reduced mass loss rate ($dM_{\rm GC}/dt$ 0.1 of fiducial)}\label{tab:UDG1summarycoreMdot01}
\begin{tabular}{|*{19}{c|}}
\hline
\multicolumn{19}{|c|}{Core}\\ \hline
&\multicolumn{6}{|c|}{ stretch: 1}&\multicolumn{6}{|c|}{ stretch: 2}&\multicolumn{6}{|c|}{ stretch: 3}\\ \hline
fDM & 0 & 0.5 & 1 & 2 & 5 & 10 & 0 & 0.5 & 1 & 2 & 5 & 10 & 0 & 0.5 & 1 & 2 & 5 & 10\\ \hline
fMmx1 & $\X\X$ & $\X\X$ & $\X\V$ & $\V\V$ & $\X\V$ & $\X\X$ & $\X\X$ & $\V\V$ & $\V\V$ & $\V\V$ & $\X\X$ & $\X\X$ & $\V\V$ & $\V\V$ & $\V\V$ & $\X\X$ & $\X\X$ & $\X\X$\\
fMmx2 & $\X\X$ & $\X\X$ & $\X\V$ & $\V\V$ & $\X\V$ & $\V\X$ & $\X\X$ & $\V\V$ & $\V\V$ & $\X\X$ & $\V\V$ & $\X\X$ & $\X\X$ & $\V\V$ & $\V\V$ & $\X\V$ & $\X\X$ & $\X\X$\\
fMmx3 & $\X\X$ & $\X\X$ & $\X\V$ & $\V\V$ & $\V\V$ & $\X\V$ & $\X\X$ & $\V\V$ & $\V\V$ & $\V\V$ & $\X\X$ & $\X\X$ & $\X\X$ & $\V\V$ & $\V\V$ & $\V\V$ & $\X\V$ & $\X\X$\\
\hline
\multicolumn{19}{|c|}{Cusp}\\ \hline
&\multicolumn{6}{|c|}{ stretch: 1}&\multicolumn{6}{|c|}{ stretch: 2}&\multicolumn{6}{|c|}{ stretch: 3}\\ \hline
fDM & 0 & 0.5 & 1 & 2 & 5 & 10 & 0 & 0.5 & 1 & 2 & 5 & 10 & 0 & 0.5 & 1 & 2 & 5 & 10\\ \hline
fMmx1 & $\X\X$ & $\X\X$ & $\X\X$ & $\X\X$ & $\V\V$ & $\V\V$ & $\X\X$ & $\V\V$ & $\V\V$ & $\V\V$ & $\V\V$ & $\X\X$ & $\V\V$ & $\V\V$ & $\V\V$ & $\V\X$ & $\X\X$ & $\X\X$\\
fMmx2 & $\X\X$ & $\X\X$ & $\X\X$ & $\X\X$ & $\X\X$ & $\V\V$ & $\X\X$ & $\X\V$ & $\V\V$ & $\V\V$ & $\V\V$ & $\V\V$ & $\X\X$ & $\V\V$ & $\V\V$ & $\V\V$ & $\V\V$ & $\X\X$\\
fMmx3 & $\X\X$ & $\X\X$ & $\X\X$ & $\X\X$ & $\X\X$ & $\V\V$ & $\X\X$ & $\X\V$ & $\V\V$ & $\V\V$ & $\V\V$ & $\V\V$ & $\X\X$ & $\V\V$ & $\V\V$ & $\V\V$ & $\V\V$ & $\V\X$\\
\hline
\end{tabular}
\end{table*}

\FloatBarrier
\section{Results: Fornax dSph}\label{s:fornax}
The GC distribution of the Fornax dSph attracted many studies~\citep[][]{Tremaine1976a,Hernandez:1998hf,ohlinricher2000,Lotz:2001gz,Goerdt2006,Cowsik:2009uk,angus2009resolving,Cowsik:2009uk,Cole2012,Kaur2018,Hui:2016ltb,Leung_2019,boldrini2020embedding,Berezhiani2019,Hartman:2020fbg,Bar-Or:2018pxz,Lancaster2020,Meadows20}. With only six GCs, statistics is too low for detailed analysis, but there are still interesting constraints. 
As we show below, it is a striking fact that if Fornax did not have a massive DM halo, its stellar distribution should have been so cold that its GC population would plunge into the center in a very short time. This comment could seem out of place given that Fornax's stellar LOSVD was measured long ago (see, e.g.~\cite{Read:2018fxs}) and it is not cold, that is, we know from kinematics that it does contain DM; however, the point is that the stellar surface brightness profile and the GC morphology {\it by themselves} give a constraint {\it independent of kinematics}. 

%

\subsection{Observational data}
Observational information for Fornax's GCs was summarised in \cite{Bar:2021jff}, reproduced in Tab.~\ref{tab:Fornaxinfo}. The last column quotes an estimate of $\tau_{\rm DF}$ for each GC at its observed position. For the stellar body of Fornax we assume $M_*=4.3\times10^7~M_\odot$~\citep[][]{deBoer:2012py}, and use S\'ersic index $n=0.8$ and $R_{\rm e}=668$~pc~\citep[][]{DES:2018jtu}.
\begin{table*}
	\centering
	\caption{Some details of Fornax's six GCs. Reproduced from \cite{Bar:2021jff}. The last column quotes an estimate of $\tau_{\rm DF}$ for each GC. Reference key: {\it a}: \cite{de2016four}; {\it b}: \cite{Lauberts1982}; {\it c}: \cite{Letarte2006}; {\it d}: \cite{Mackey2003a}; {\it e}: \cite{Hendricks_2014}; {\it f}: \cite{Morrison2001}; {\it g}: \cite{Skrutskie2006}; {\it h}: \cite{Larsen_2012}; {\it i}: \cite{wang2019rediscovery}; {\it j}: \cite{Shao:2020tsl}.}\label{tab:Fornaxinfo}
	\begin{tabular}{ c| c c c| c}
	\hline
		& $ M_{\rm GC}\;[10^5M_{\odot}]$ & $ r_\perp [{\rm kpc}]$ & Refs. & {\color{blue}$ \tau_{\rm DF}~[{\rm Gyr}] $}\\ \hline
		GC1 & $ 0.42\pm 0.10 $ & $ 1.73\pm 0.05 $  & 
		{\it a,b,c,d,e}& {\color{blue}$ 119 $}\\  
		GC2 & $ 1.54\pm 0.28 $ & $ 0.98\pm 0.03 $ & 
		{\it a,c,d,f}& {\color{blue}$ 14.7 $}\\
		GC3 & $ 4.98\pm 0.84 $ & $ 0.64\pm 0.02$  & 
		{\it a,d,g,h} & {\color{blue}$ 2.63 $}\\
		GC4 & $ 0.76 \pm 0.15 $ & $ 0.154\pm 0.014 $ & 
		{\it a,d,g,h,} & {\color{blue}$ 0.91 $}\\
		GC5 & $ 1.86\pm 0.24 $ & $ 1.68\pm 0.05 $  & 
		{\it a,d,e,g,h} & {\color{blue}$ 32.2 $} \\ 
		GC6 & $ \sim 0.29 $ & $ 0.254\pm 0.015 $ & 
		{\it i,j} & {\color{blue}$ 5.45 $}\\
		\hline
	\end{tabular}	
\end{table*}
%

\subsection{Results: GC initial radial distribution similar to stellar body}
\begin{figure*}
\centering
      \includegraphics[ scale= 0.25]{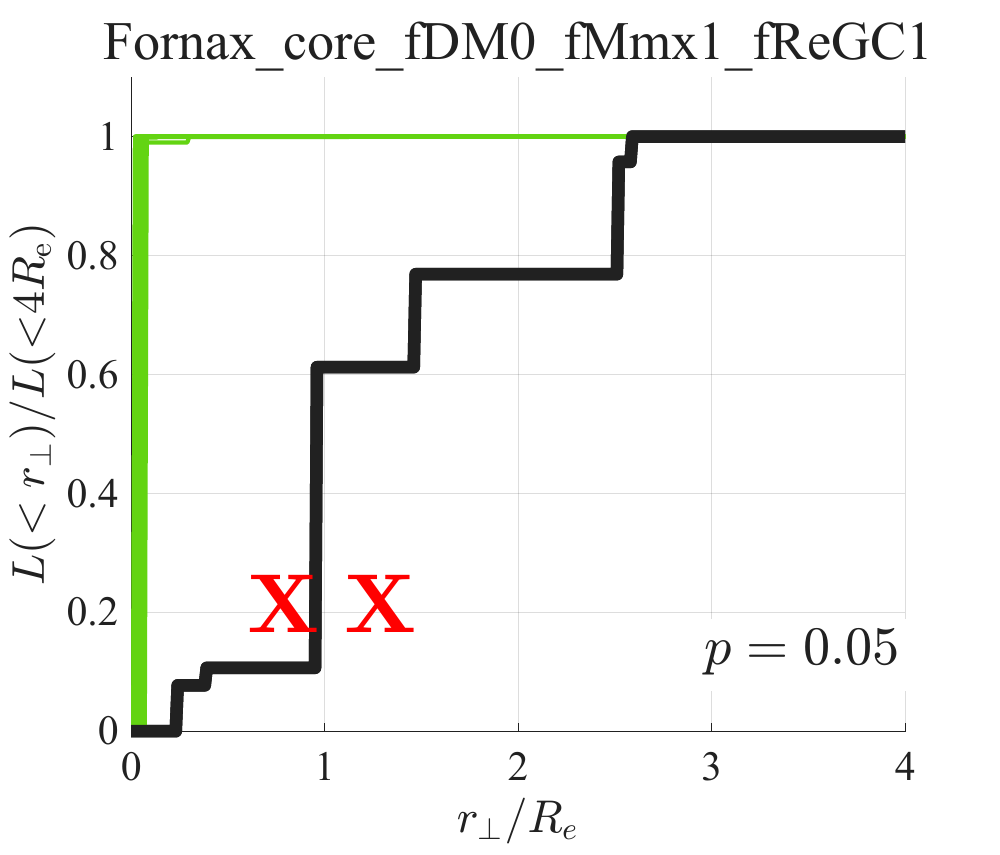} 
     \includegraphics[ scale= 0.25]{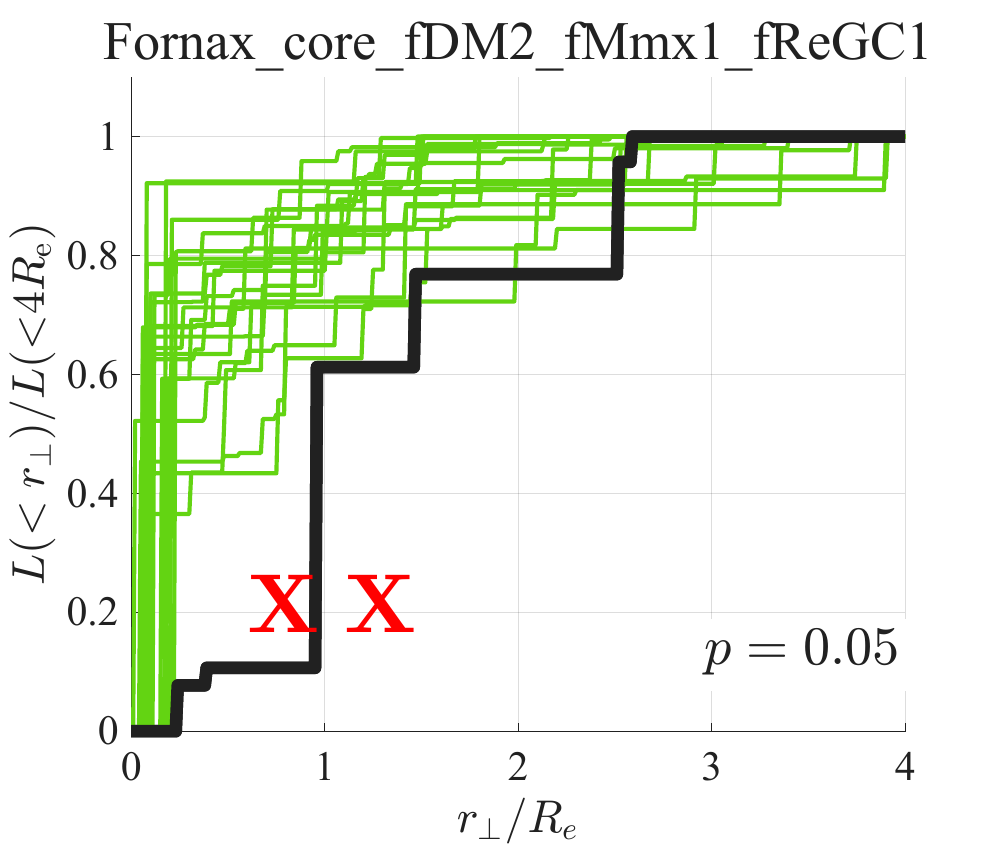} 
          \includegraphics[ scale= 0.25]{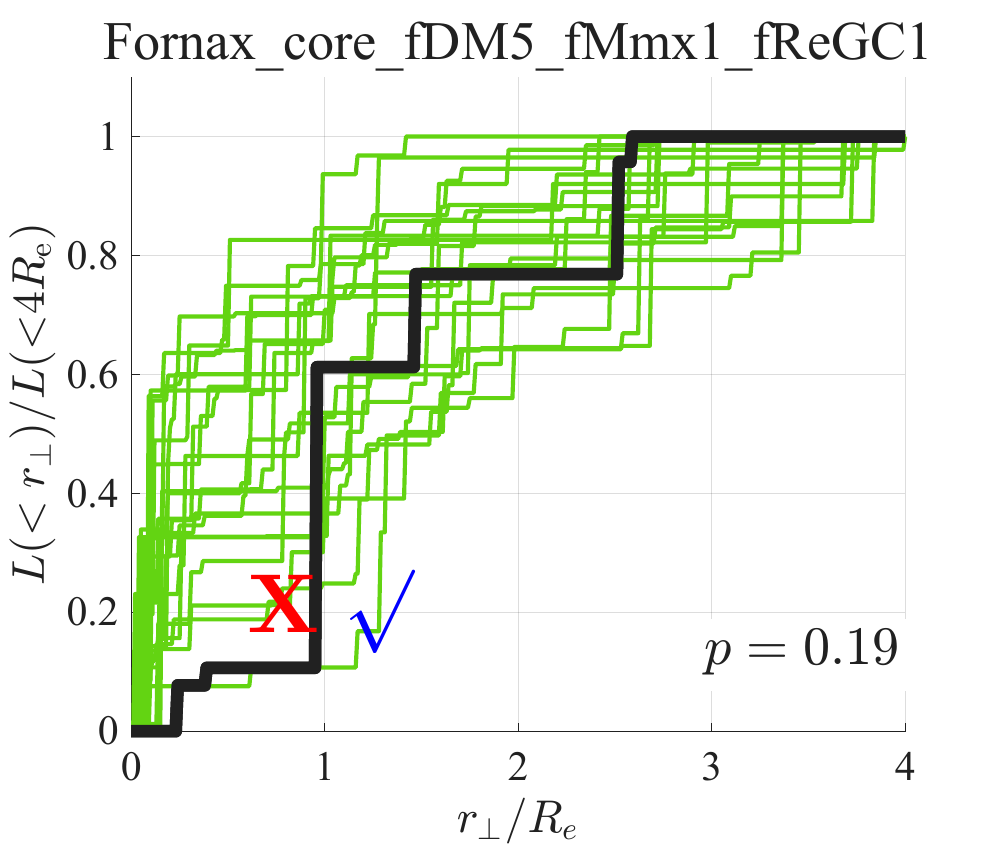} 
      \includegraphics[ scale= 0.25]{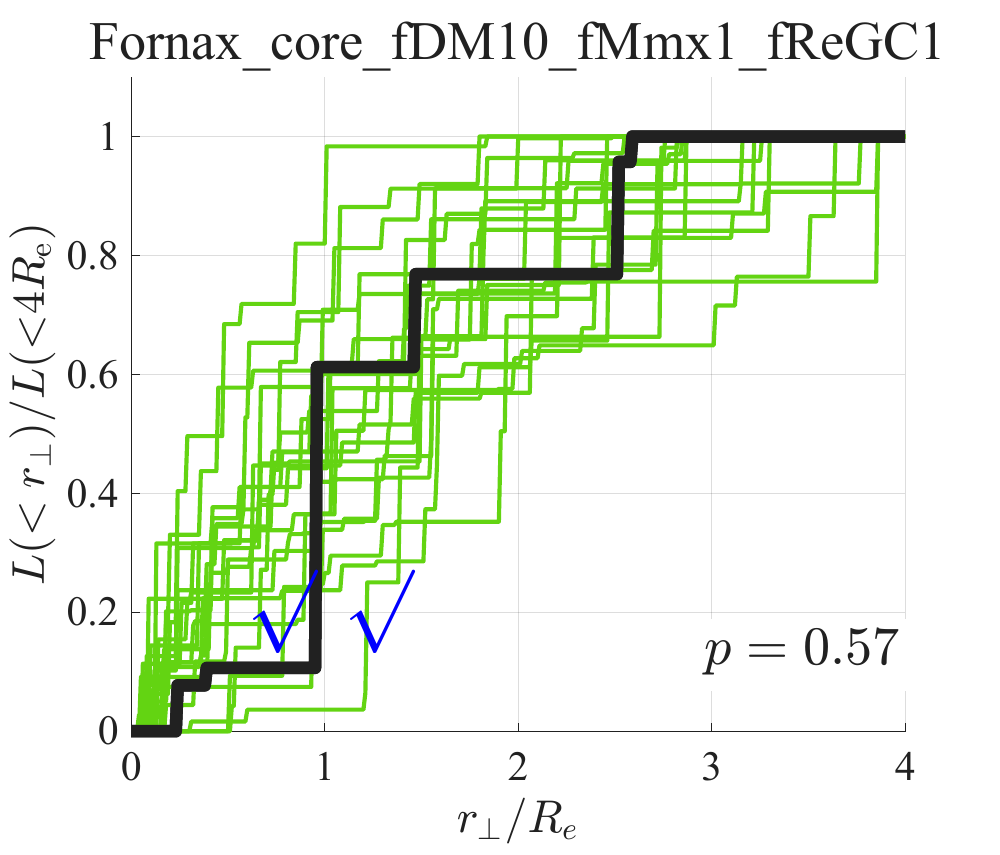} \\
            \includegraphics[ scale= 0.25]{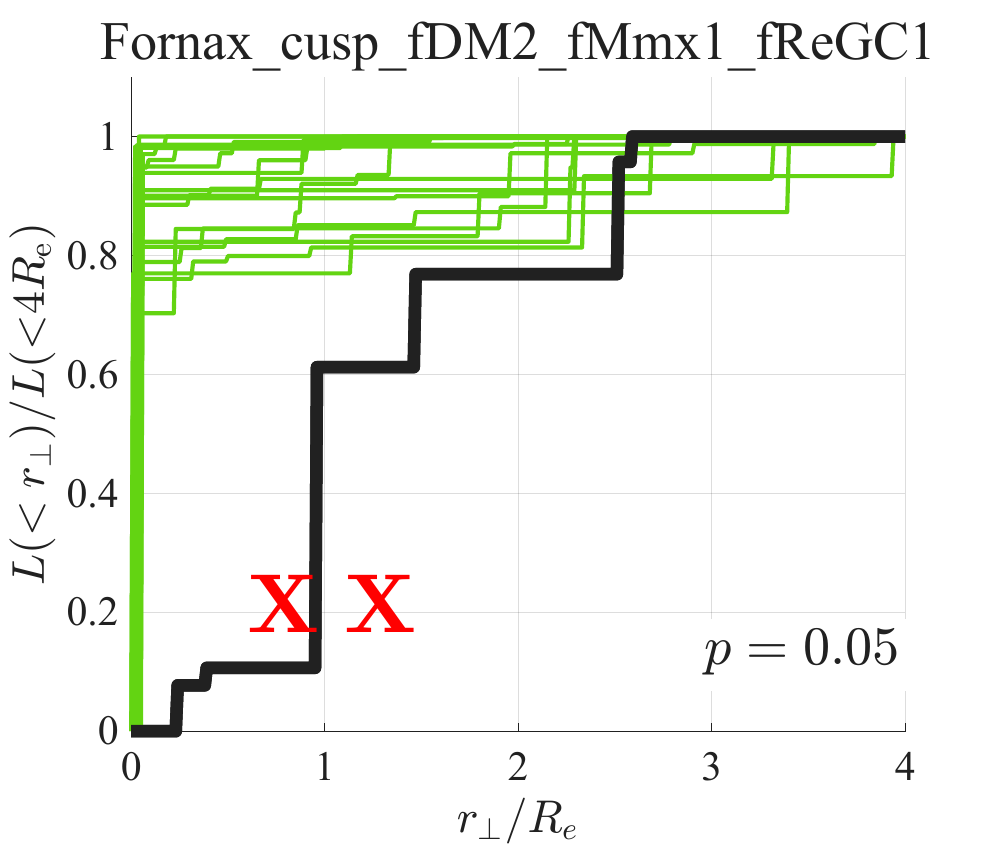}       
     \includegraphics[ scale= 0.25]{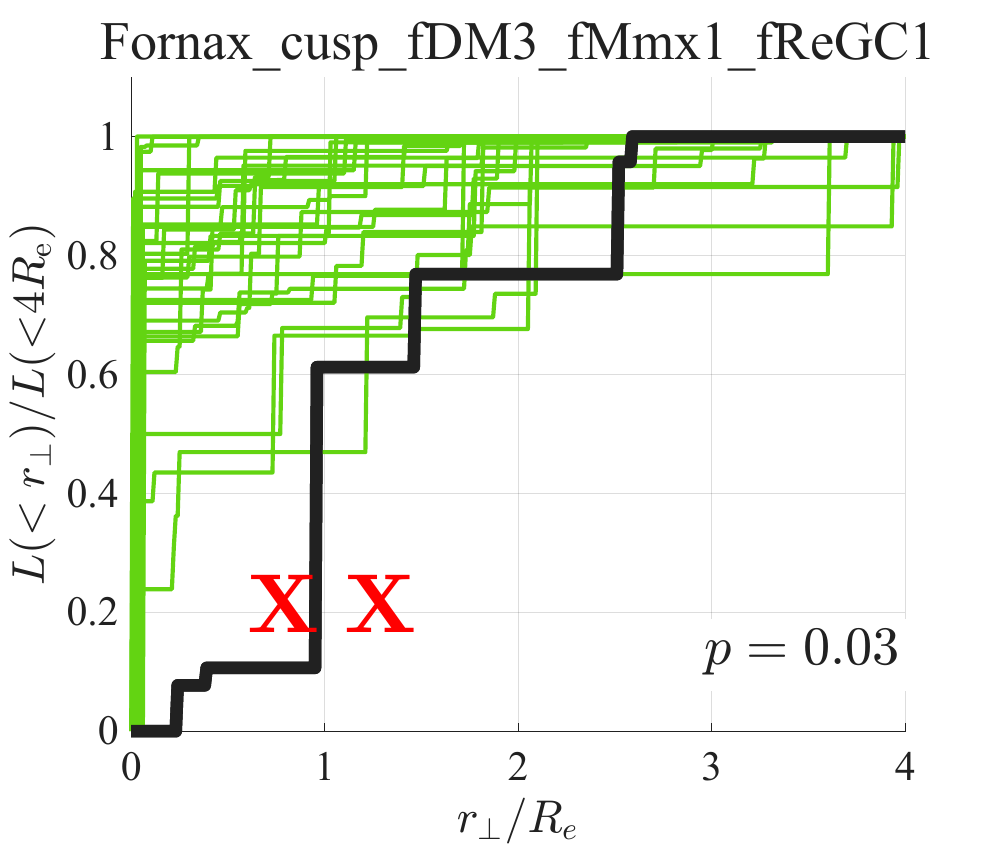} 
          \includegraphics[ scale= 0.25]{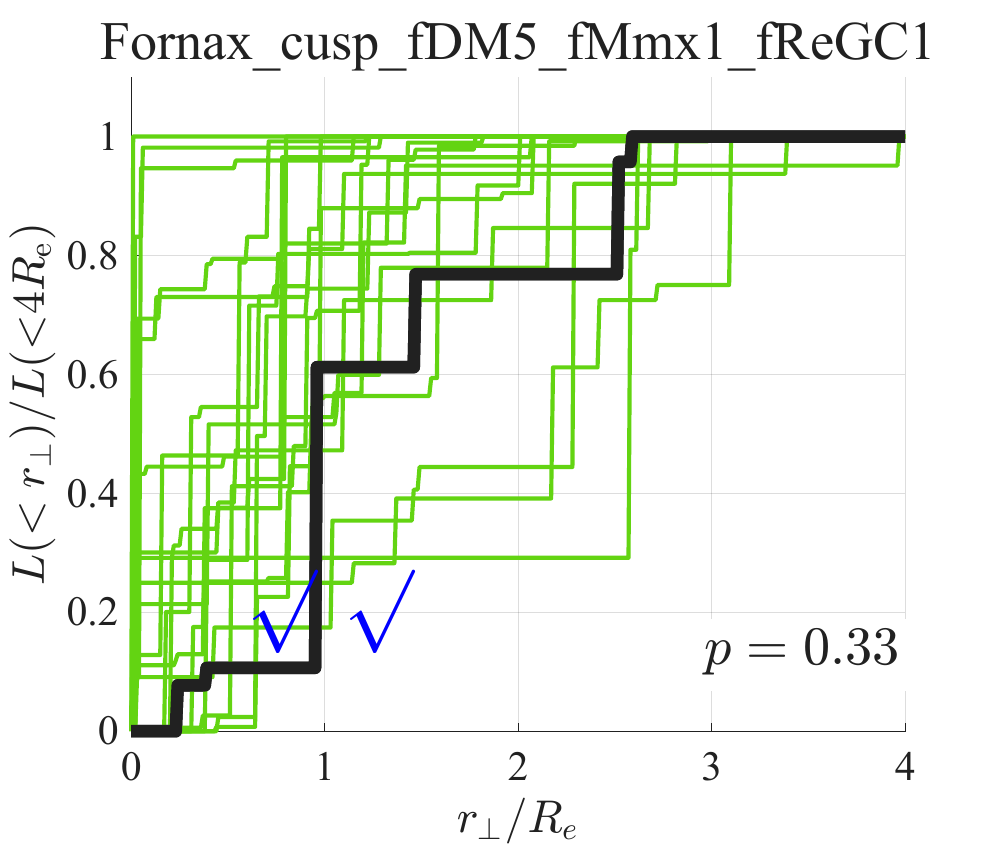} 
      \includegraphics[ scale= 0.25]{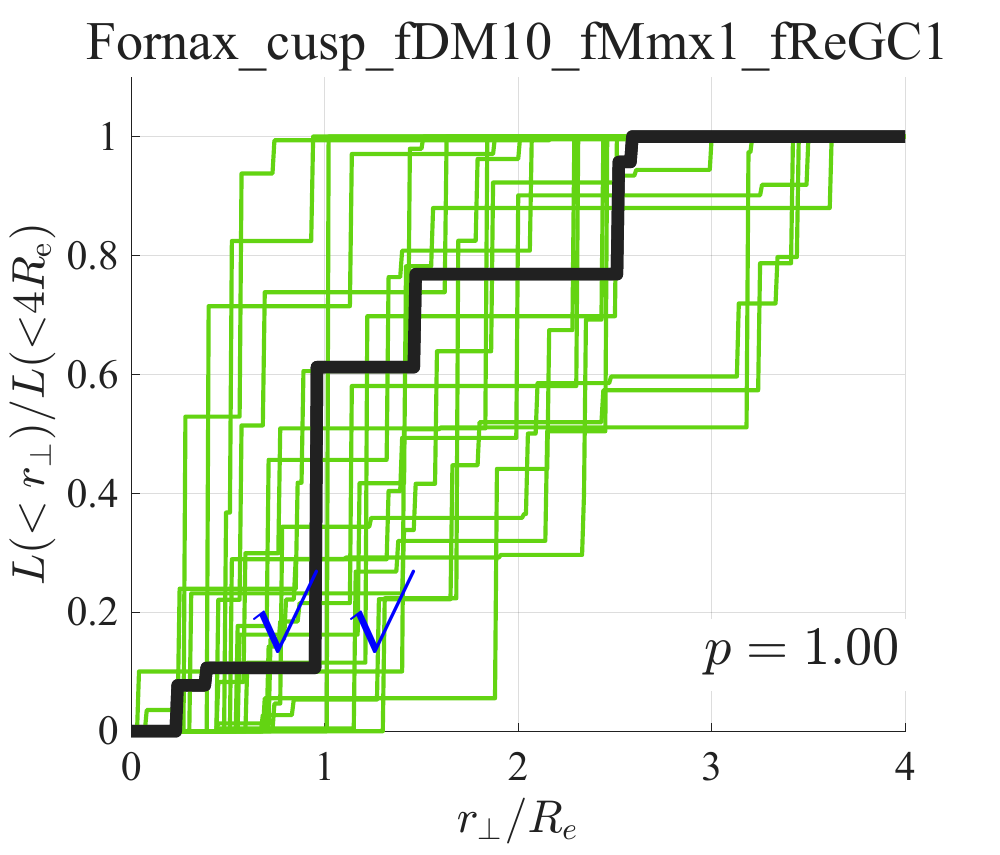} 
 \caption{Luminosity CDF for Burkert ({top}) and NFW ({bottom}) models of the Fornax dSph.  
 }
 \label{fig:LCDFs_Fornax}
\end{figure*}

We start assuming that the initial distribution of Fornax's GCs followed the currently distribution of the stellar body. Fig.~\ref{fig:LCDFs_Fornax} shows luminosity CDF for core ({top panels}) and cusp ({bottom panels}) models, normalized to the total GC luminosity at $r_\perp=4R_{\rm e}$. 
The GC morphology requires Fornax to host a massive DM halo.
%

\subsection{Results: GC initial radial distribution stretched compared to stellar body}
\begin{figure*}
\centering
      \includegraphics[ scale= 0.25]{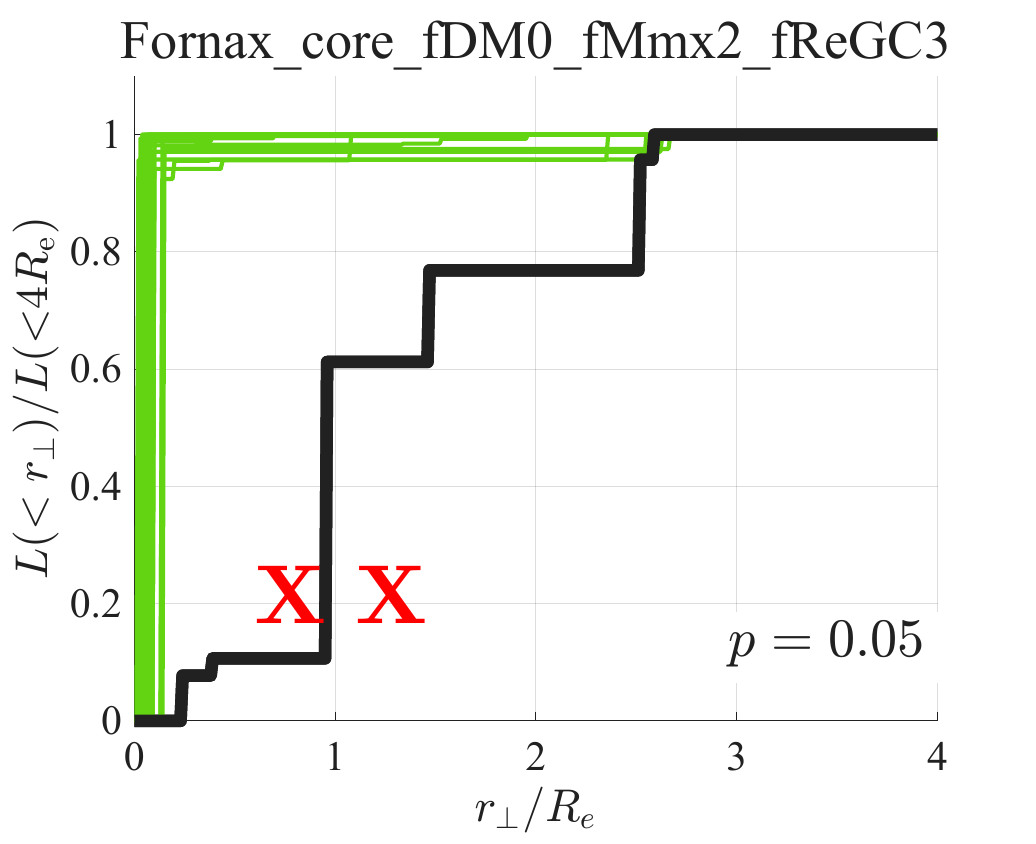} 
     \includegraphics[ scale= 0.25]{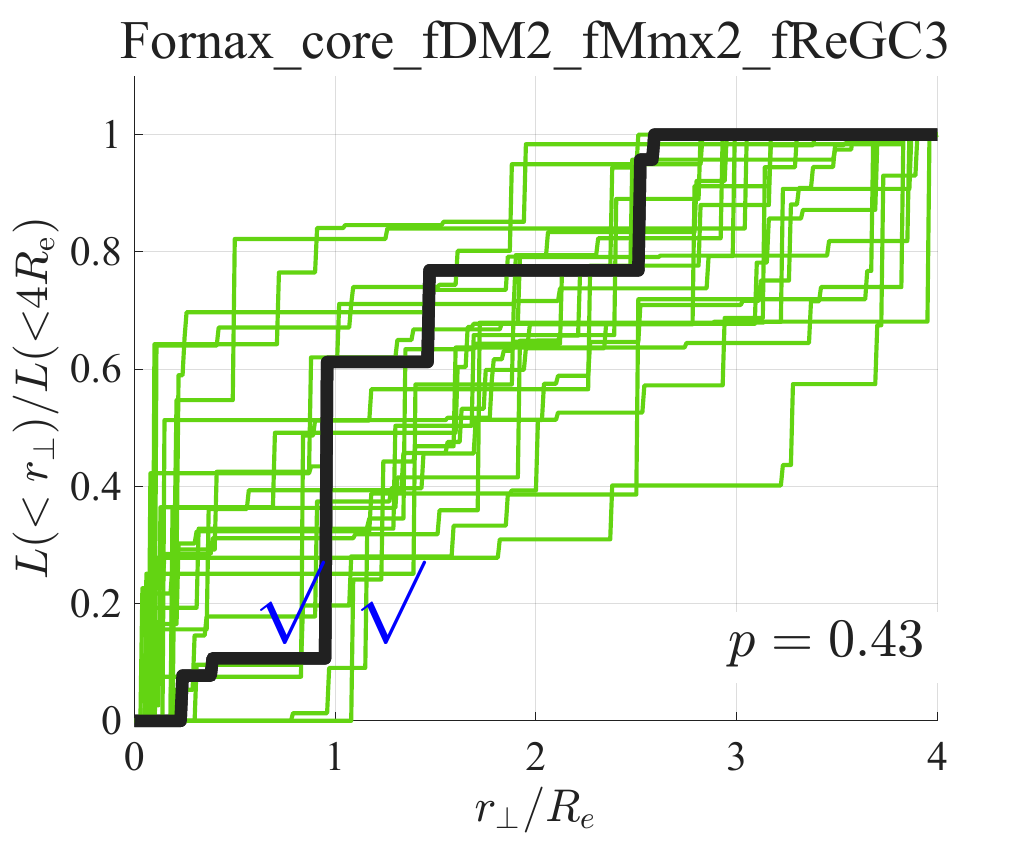} 
          \includegraphics[ scale= 0.25]{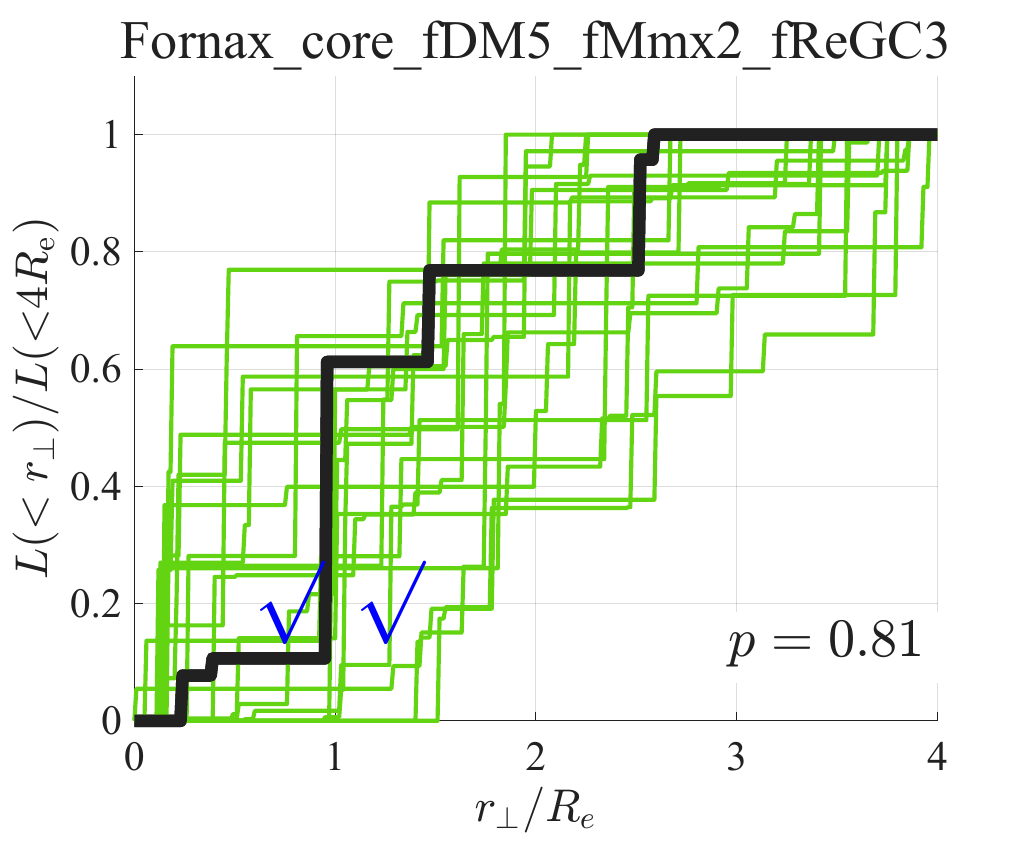} 
      \includegraphics[ scale= 0.25]{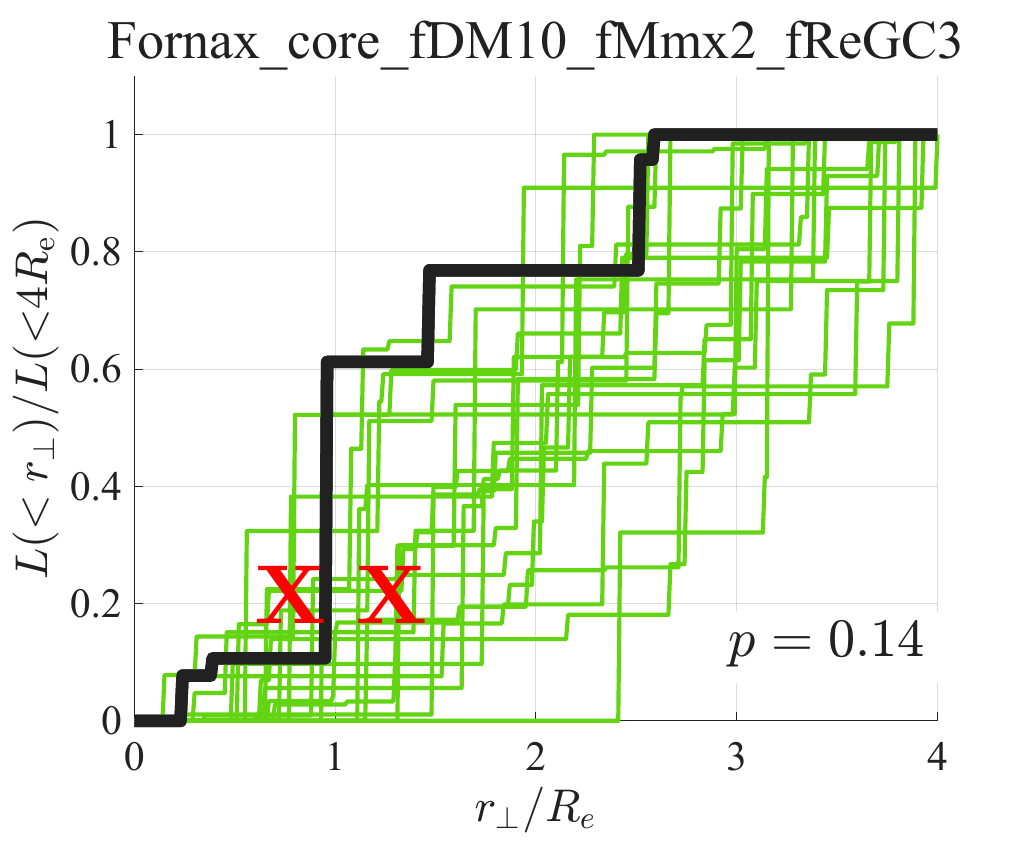} \\      
      \includegraphics[ scale= 0.25]{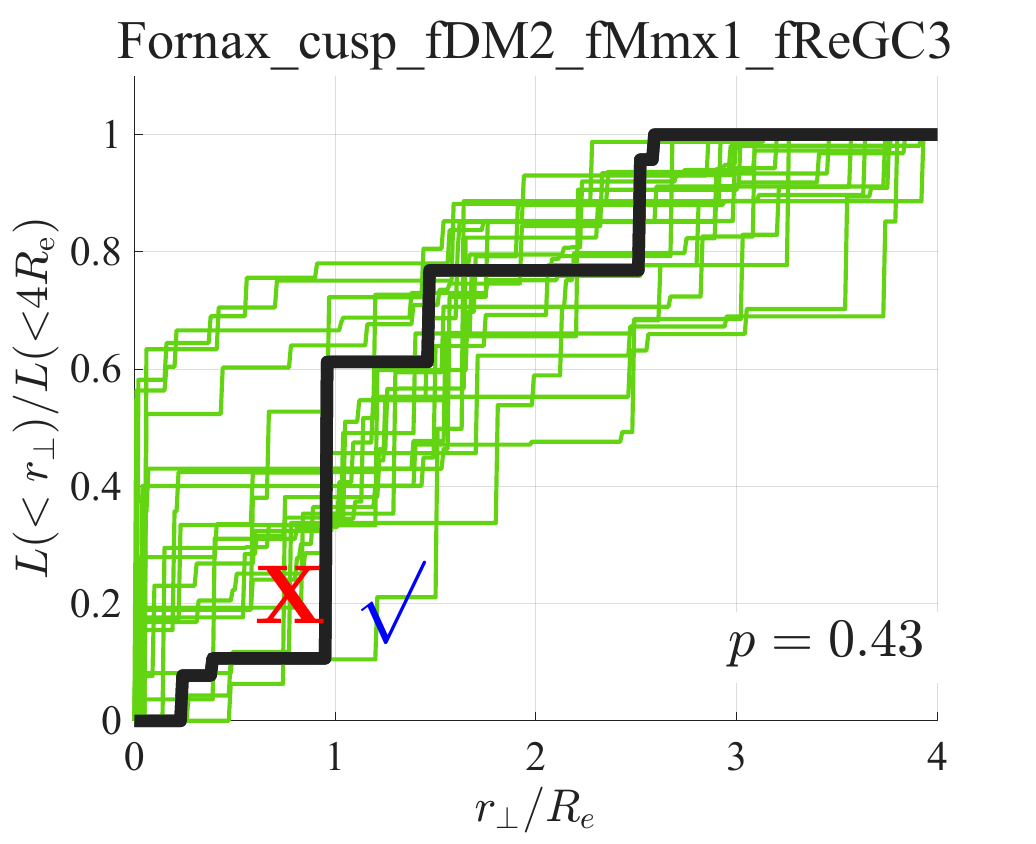}       
      \includegraphics[ scale= 0.25]{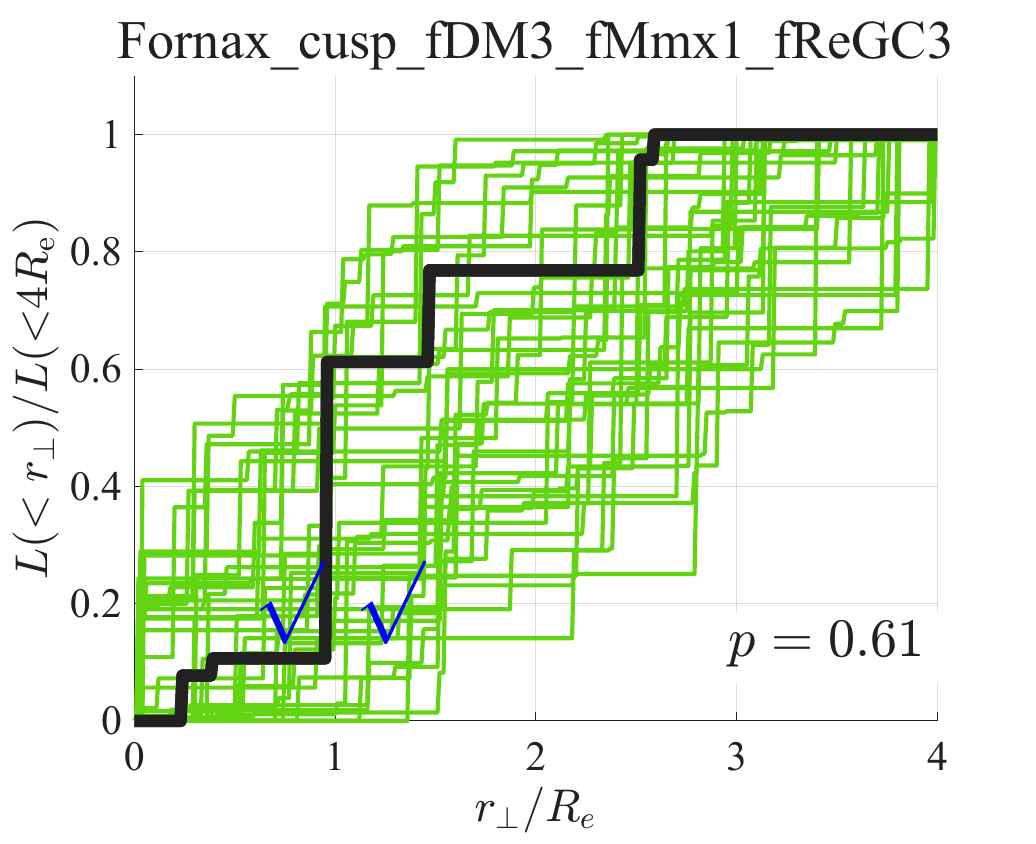} 
      \includegraphics[ scale= 0.25]{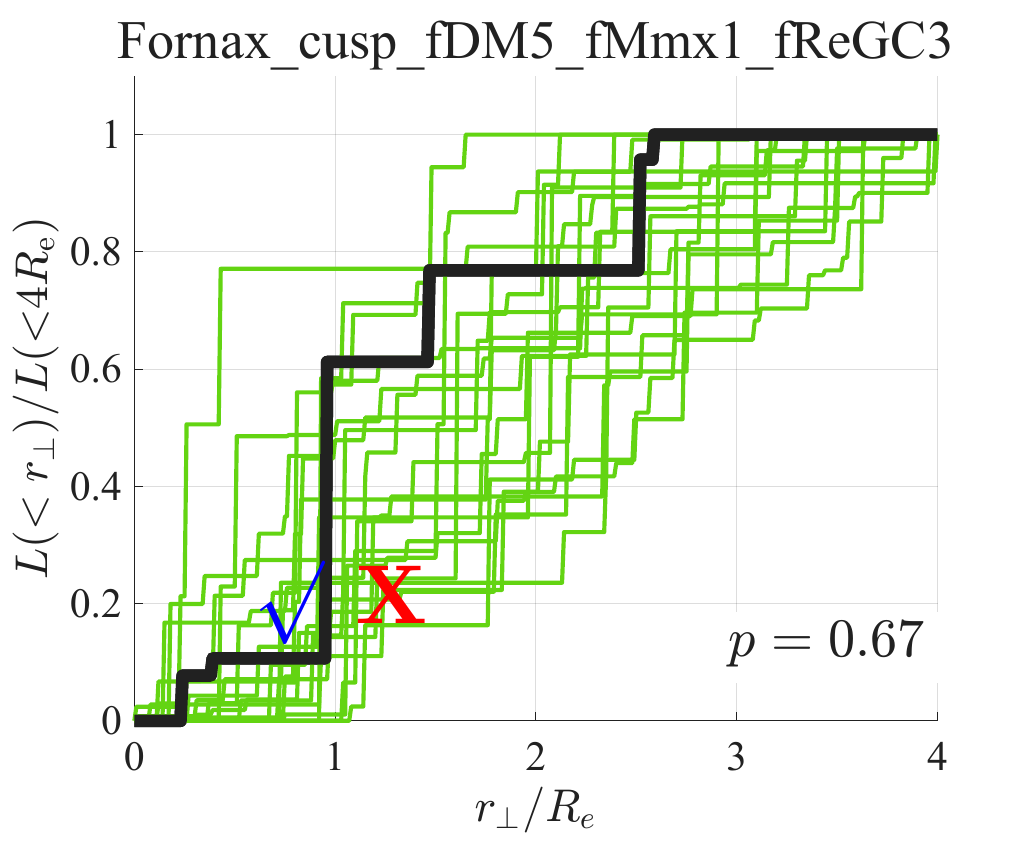} 
      \includegraphics[ scale= 0.25]{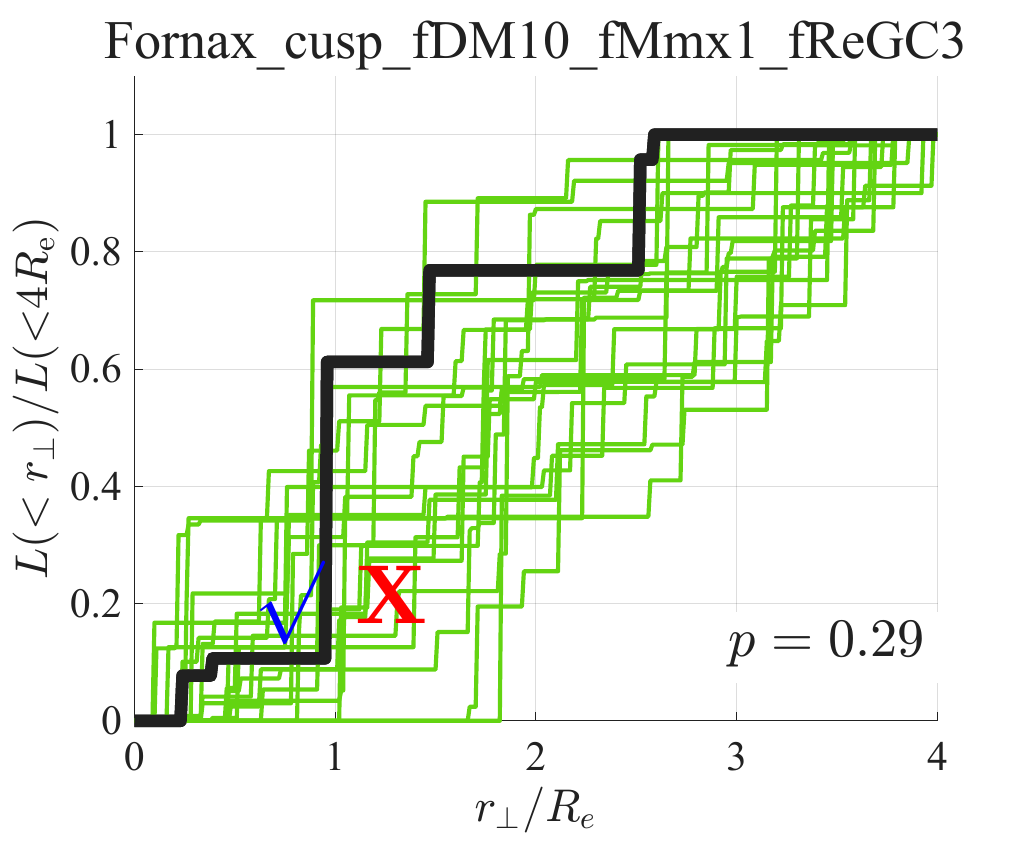}             
 \caption{Luminosity CDF for Burkert ({top}) and NFW ({bottom}) models of the Fornax dSph. Here, the initial GC radial distribution is stretched by a factor of 3 compared to the stellar body.
 }
 \label{fig:LCDFs_Fornax_stretch}
\end{figure*}

Fig.~\ref{fig:LCDFs_Fornax_stretch} shows the results obtained assuming that Fornax GCs start their lives with radial distribution stretched by a factor of 3 w.r.t. the stellar body. Initial stretch shifts the allowed DM model to lower halo mass. We will see that consistency with both GC morphology and stellar kinematics data requires some initial GC stretch.
%

\subsection{Comparison with stellar kinematics}
\begin{figure*}
\centering
      \includegraphics[ scale= 0.3]{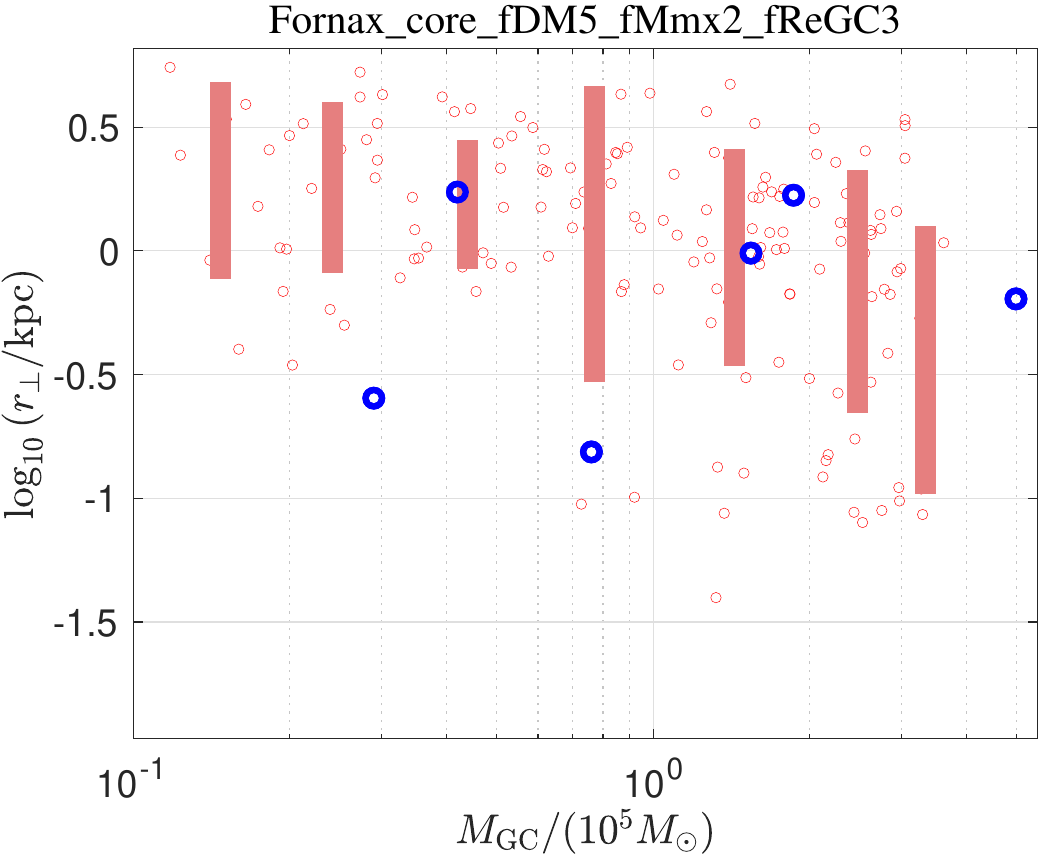}       
            \includegraphics[ scale= 0.3]{Fornax_core_fDM5_fMmx2_fReGC3_LCDFnorm_cx}       
                  \includegraphics[ scale= 0.3]{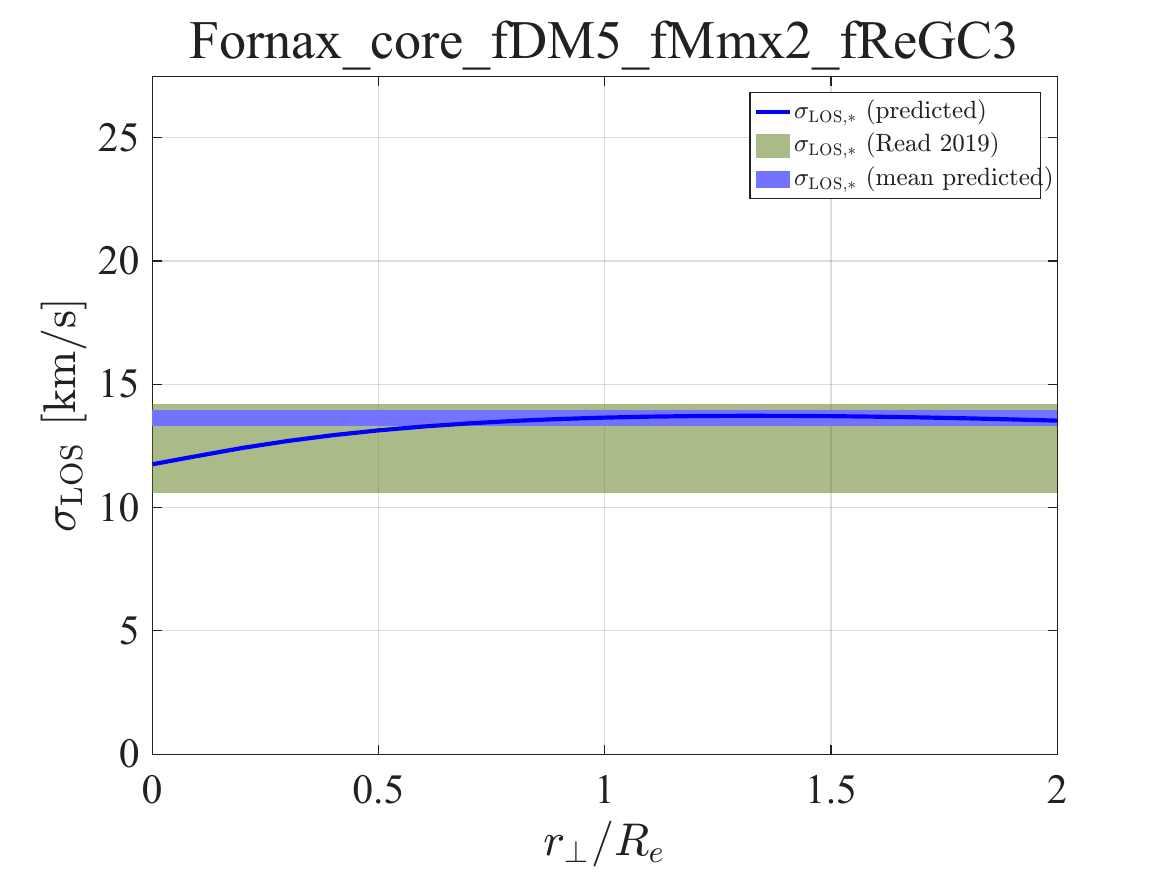}   \\    
      \includegraphics[ scale= 0.3]{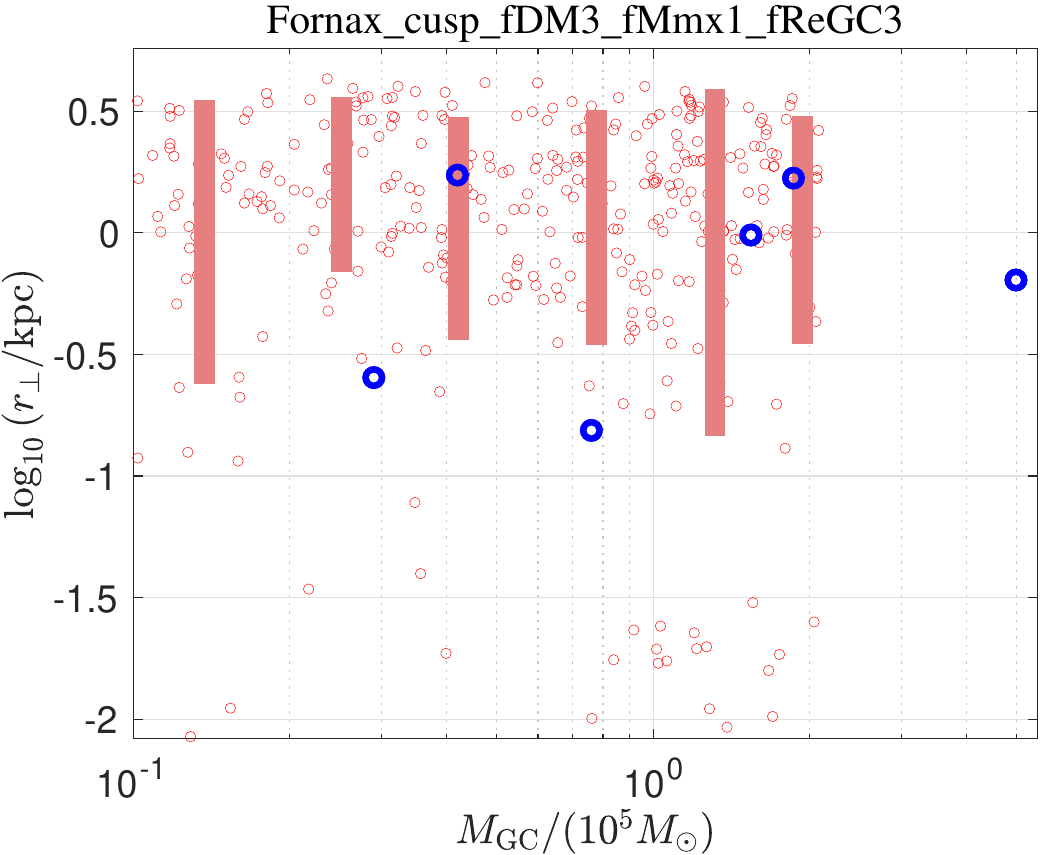}       
            \includegraphics[ scale= 0.3]{Fornax_cusp_fDM3_fMmx1_fReGC3_LCDFnorm_cx}       
                  \includegraphics[ scale= 0.3]{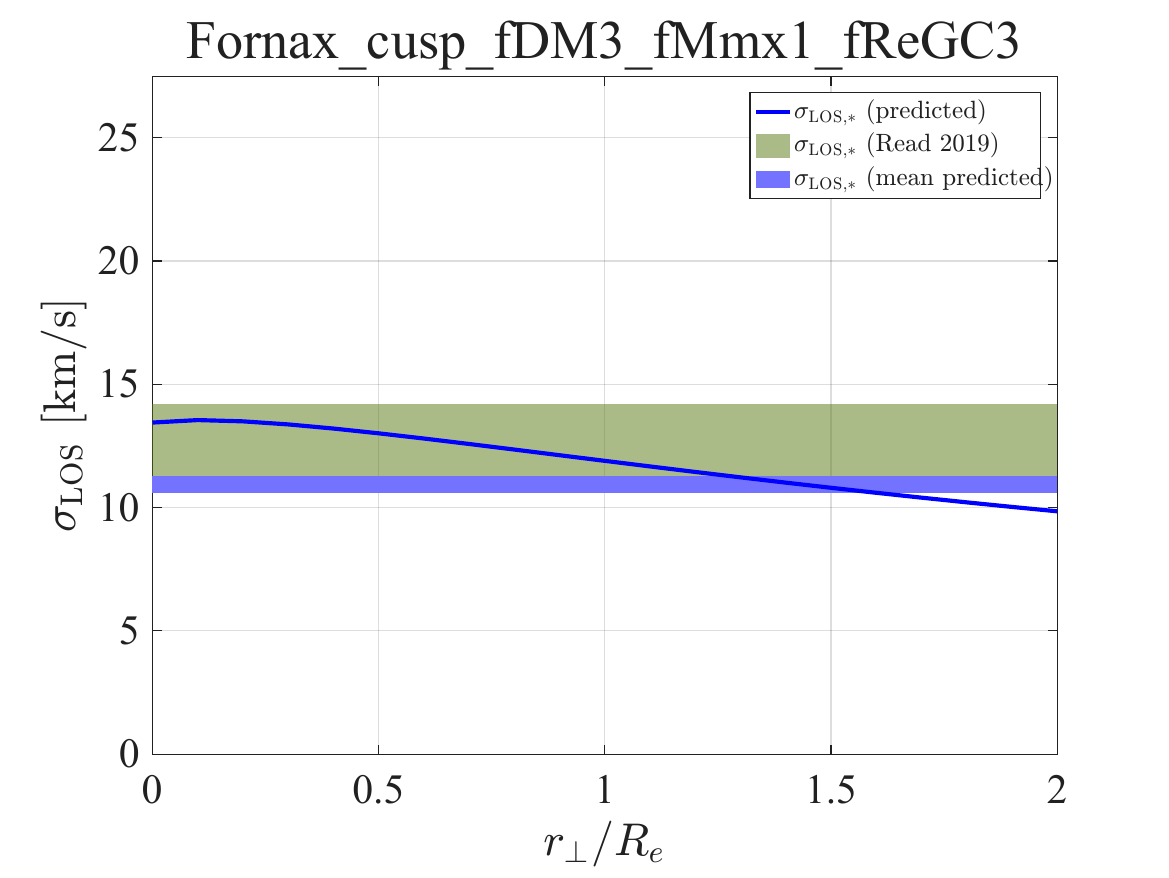}   
 \caption{Comparison of GC data analysis with stellar kinematics data, for Burkert ({top}) and NFW ({bottom}) models of the Fornax dSph.
 }
 \label{fig:details_kin_Fornax}
\end{figure*}

In Fig.~\ref{fig:details_kin_Fornax} we demonstrate consistency between models addressing Fornax's GC morphology and stellar kinematics data. The top (bottom) panels show a core (cusp) model. 
The green shaded region shows kinematics data from \cite{Read:2018fxs}.

\begin{figure*}
\centering
                         \includegraphics[ scale= 0.3]{Fornax_cusp_fDM3_fMmx1_fReGC1_LCDFnorm_cx}
             \includegraphics[ scale= 0.3]{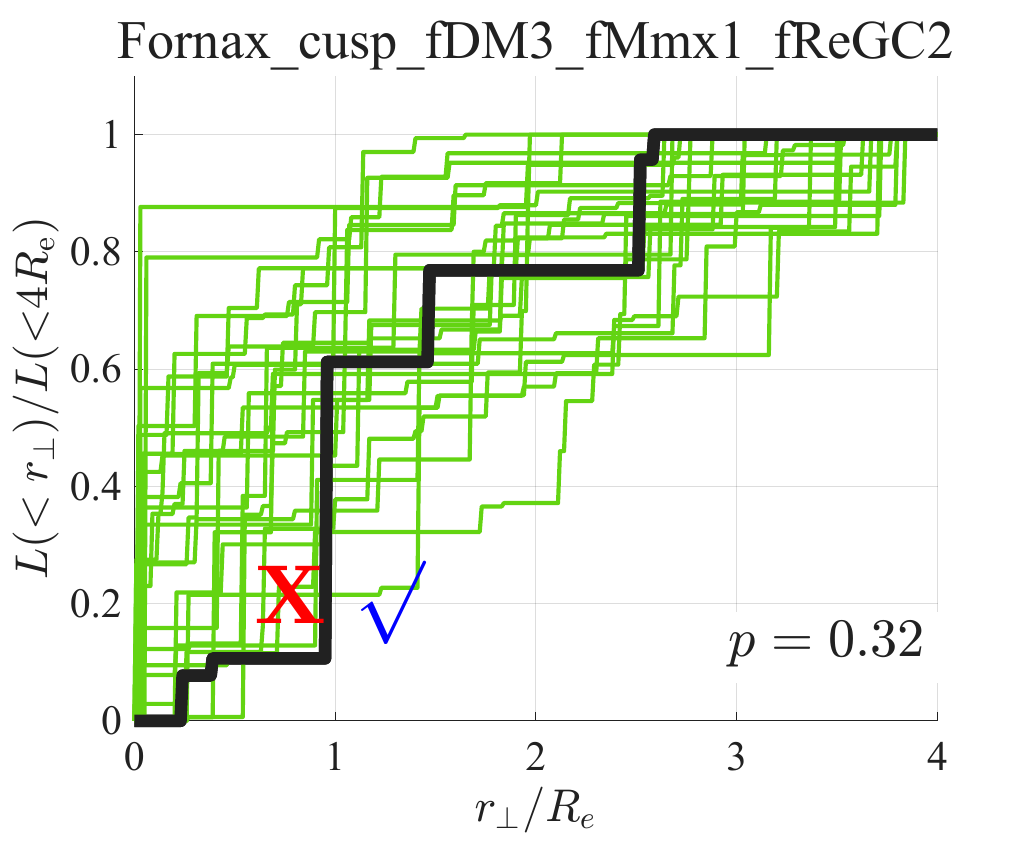}
              \includegraphics[ scale= 0.3]{Fornax_cusp_fDM3_fMmx1_fReGC3_LCDFnorm_cx} 
 \caption{Complementarity between stellar kinematics and GC morphology data, for the NFW model of the Fornax dSph.}
 \label{fig:details_kin_Fornax2}
\end{figure*}

In both rows of Fig.~\ref{fig:details_kin_Fornax}, simultaneous consistency with stellar kinematics and GC morphology data required us to assume initial GC stretch: fReGC3 in this case. We explore this point further in Fig.~\ref{fig:details_kin_Fornax2}, comparing results for different stretch for a cusp DM model. The scenario with initial GC distribution aligned with the current stellar body is disfavored by the combination of GC morphology and stellar LOSVD. This observation may be related to Fornax's tidal history as a dwarf satellite of the Milky Way~\citep[][]{Shao:2020tsl}.
%

\subsection{Fornax: summary tables}\label{ss:fornaxressum}
Tab.~\ref{tab:Fornaxsummarycore} summarizes the luminosity CDF criteria for core and cusp halo models.
\begin{table*}[htb!]
\centering
\caption{Results summary: Fornax}\label{tab:Fornaxsummarycore}
\begin{tabular}{|*{13}{c|}}
\hline
\multicolumn{13}{|c|}{ Core}\\ \hline
&\multicolumn{4}{|c|}{ stretch: 1}&\multicolumn{4}{|c|}{ stretch: 2}&\multicolumn{4}{|c|}{ stretch: 3}\\ \hline
fDM & 0 & 2 & 5 & 10 & 0 & 2 & 5 & 10 & 0 & 2 & 5 & 10\\ \hline
fMmx1 & $\X\X$ & $\X\X$ & $\X\V$ & $\V\V$ & $\X\X$ & $\X\V$ & $\V\X$ & $\V\X$ & $\X\X$ & $\X\V$ & $\V\X$ & $\V\X$\\
fMmx3 & $\X\X$ & $\X\X$ & $\X\X$ & $\V\V$ & $\X\X$ & $\X\V$ & $\V\V$ & $\V\X$ & $\X\X$ & $\V\V$ & $\V\V$ & $\V\V$\\
\hline
\multicolumn{13}{|c|}{Cusp}\\ \hline
&\multicolumn{4}{|c|}{ stretch: 1}&\multicolumn{4}{|c|}{ stretch: 2}&\multicolumn{4}{|c|}{ stretch: 3}\\ \hline
fDM & 2 & 3 & 5 & 10 & 2 & 3 & 5 & 10 & 2 & 3 & 5 & 10\\ \hline
fMmx1 & $\X\X$ & $\X\X$ & $\V\V$ & $\V\V$ & $\X\V$ & $\X\V$ & $\V\V$ & $\V\X$ & $\X\V$ & $\V\V$ & $\V\X$ & $\V\X$\\
fMmx3 & $\X\X$ & $\X\V$ & $\V\V$ & $\V\V$ & $\X\X$ & $\X\V$ & $\V\V$ & $\V\V$ & $\X\V$ & $\V\V$ & $\V\V$ & $\V\V$\\
\hline
\end{tabular}
\end{table*}
%

\FloatBarrier
\section{Results: UDG-DF44}\label{s:df44}
\cite{Saifollahi:2022yyb} presented a detailed analysis of the GC distribution in six UDGs in the Coma cluster. 
Of these, we select UDG-DF44, for which stellar kinematics measurements also exist~\citep[][]{vanDokkum:2019fdc}, as a case study.

\subsection{Observational data}
Adopting the central values reported in~\cite{Saifollahi:2022yyb}, we take the total stellar mass of DF44 as $M_*=2.1\times10^8~M_\odot$, with Se\'rsic index $n=0.77$ and radius $R_{\rm e}=3.9$~kpc. From the GC catalog presented in~\cite{Saifollahi:2022yyb}, we include GCs with $m_{814}<28$. For this sample, the background estimate inside $r_\perp<1.5R_{\rm e}$ is about one. We model the background by drawing fake GCs from the reported background histogram, and adding them to the simulation final state.

The GC catalog of \cite{Saifollahi:2022yyb} contains two very bright GC candidates, that pass the selection criteria but are clear outliers in the GCMF\footnote{In fact, they fall outside of the axes of the GCLF in Fig.~11 of~\cite{Saifollahi:2022yyb}.}. Neither of these GCs is located near the galaxy's center of light. We include these GCs in our analysis. 

\subsection{Results}
\begin{figure*}
\centering
      \includegraphics[ scale= 0.24]{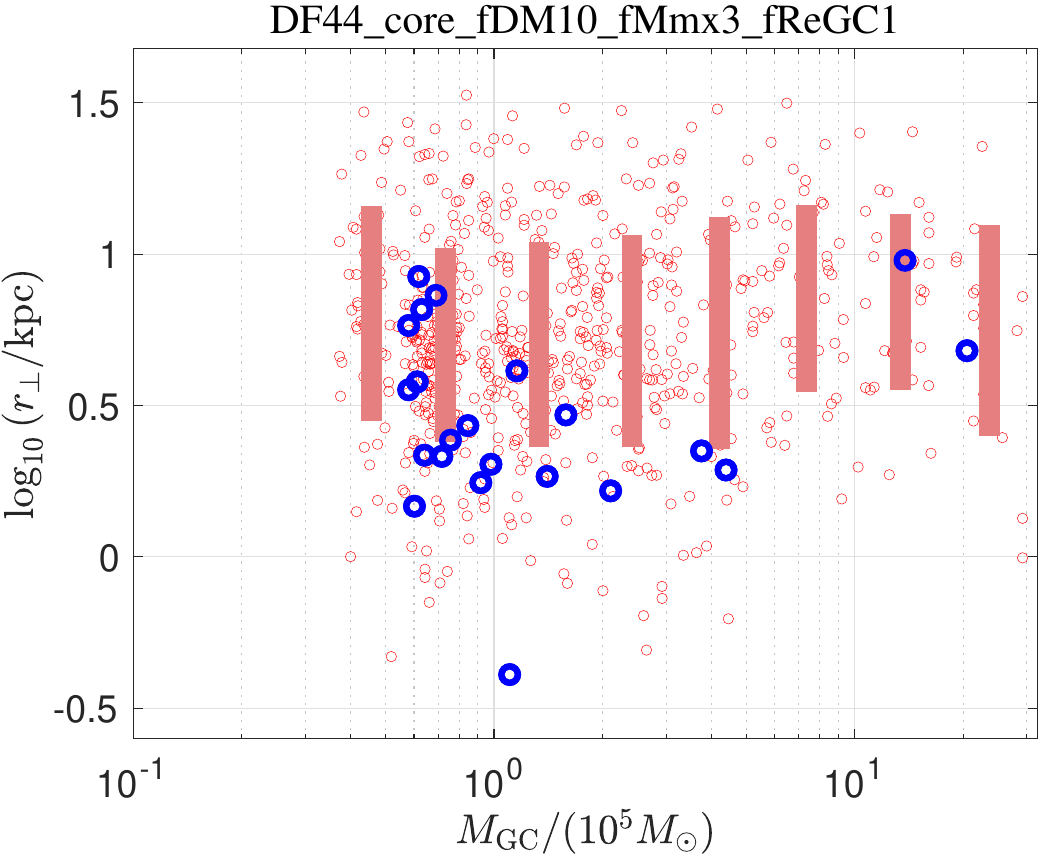}       
            \includegraphics[ scale= 0.24]{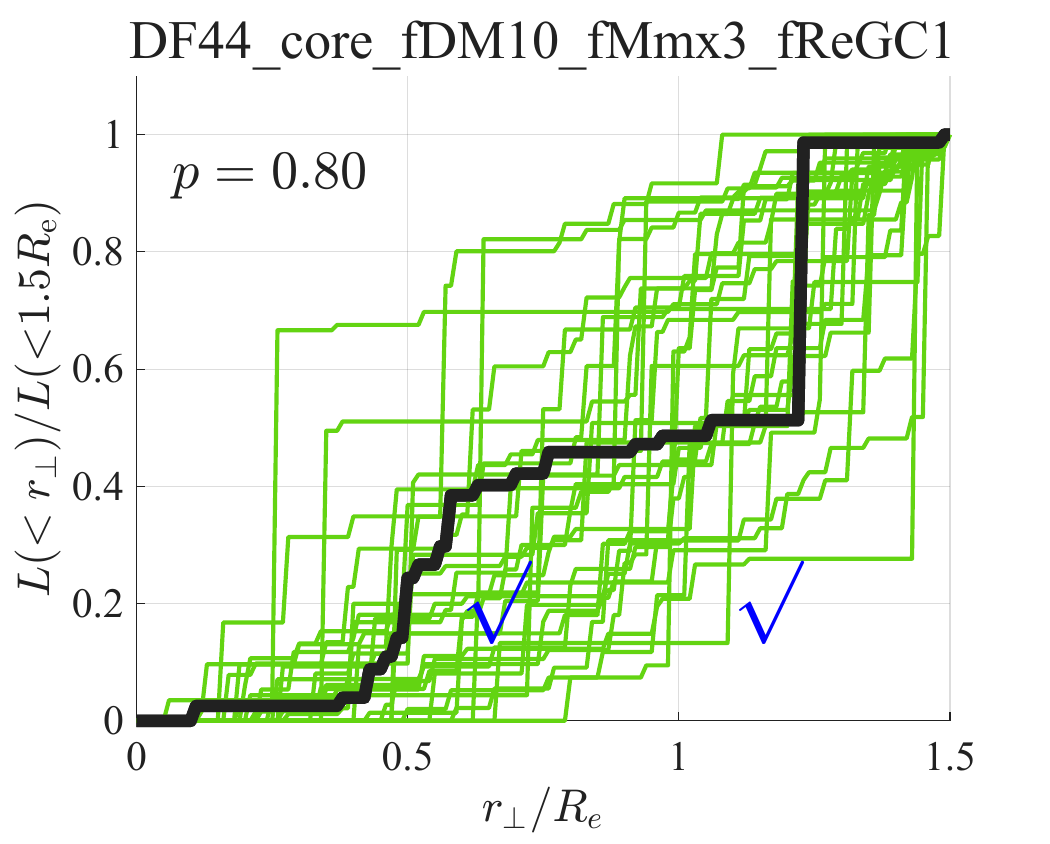}       
                  \includegraphics[ scale= 0.24]{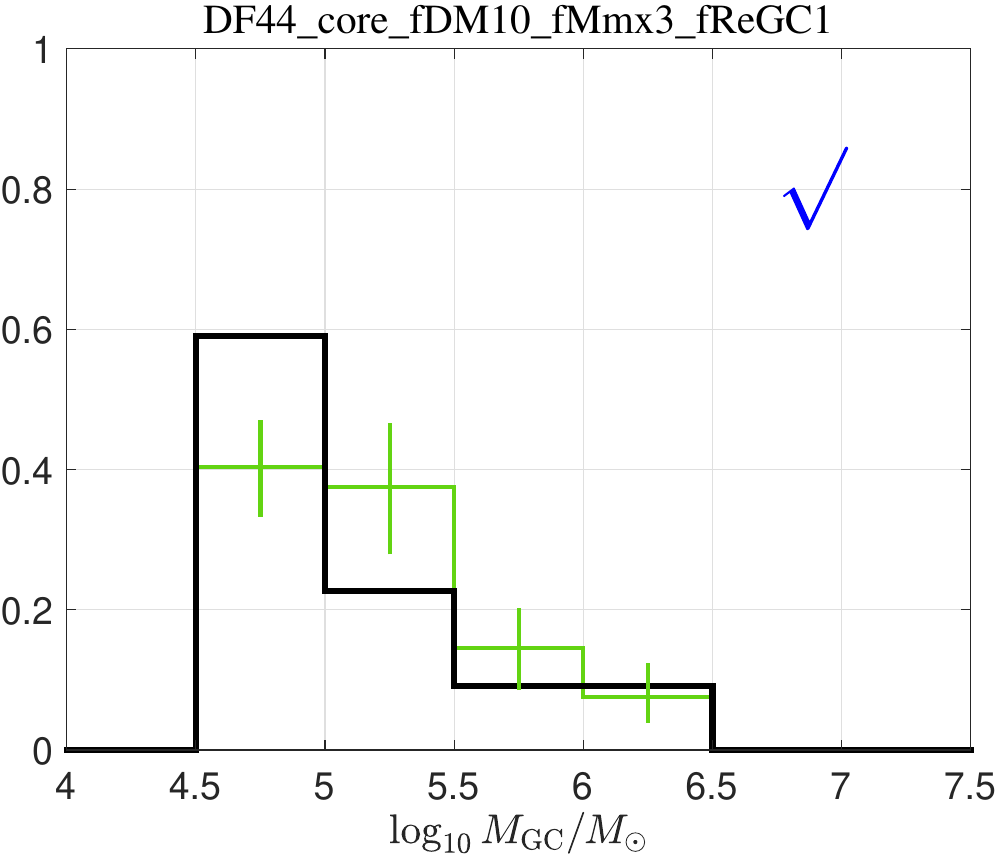}   
                              \includegraphics[ scale= 0.275]{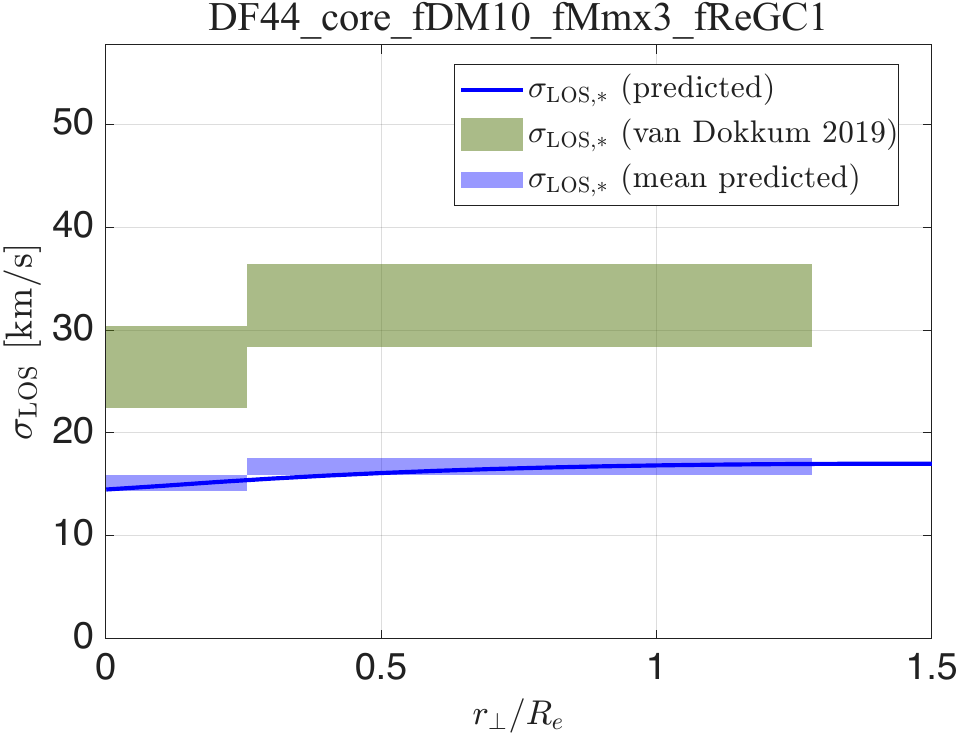}       \\    
      \includegraphics[ scale= 0.24]{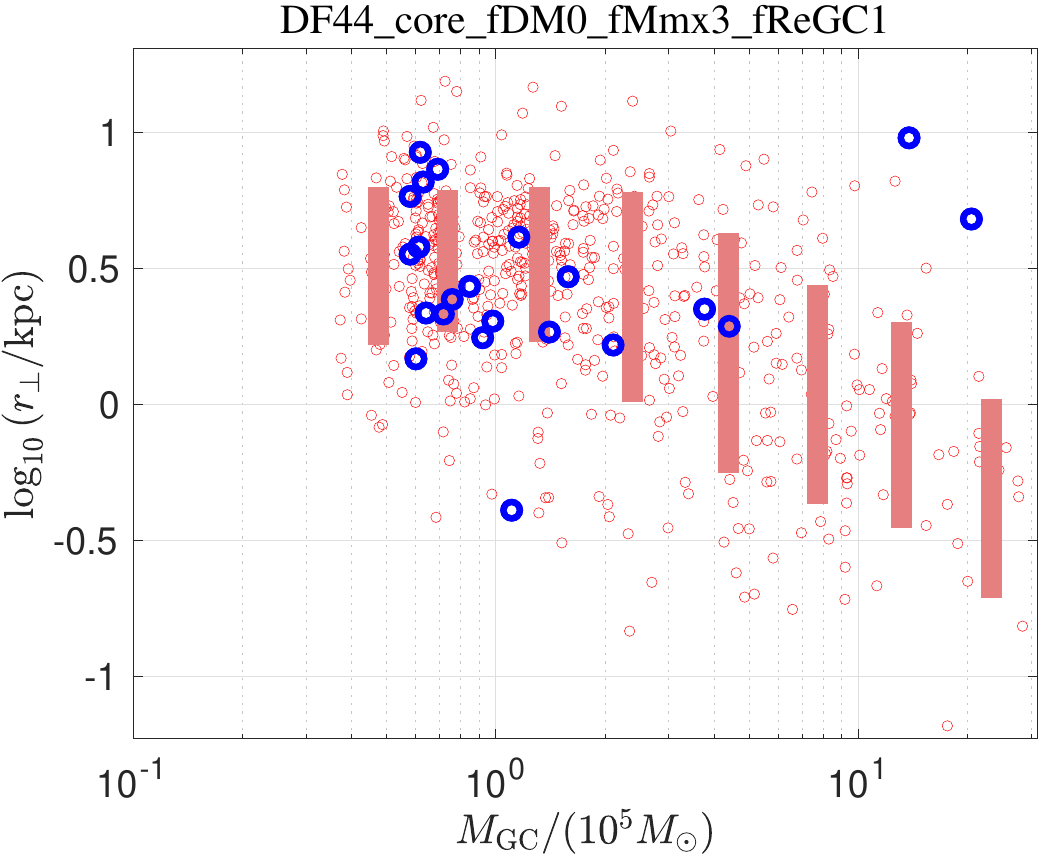}       
            \includegraphics[ scale= 0.24]{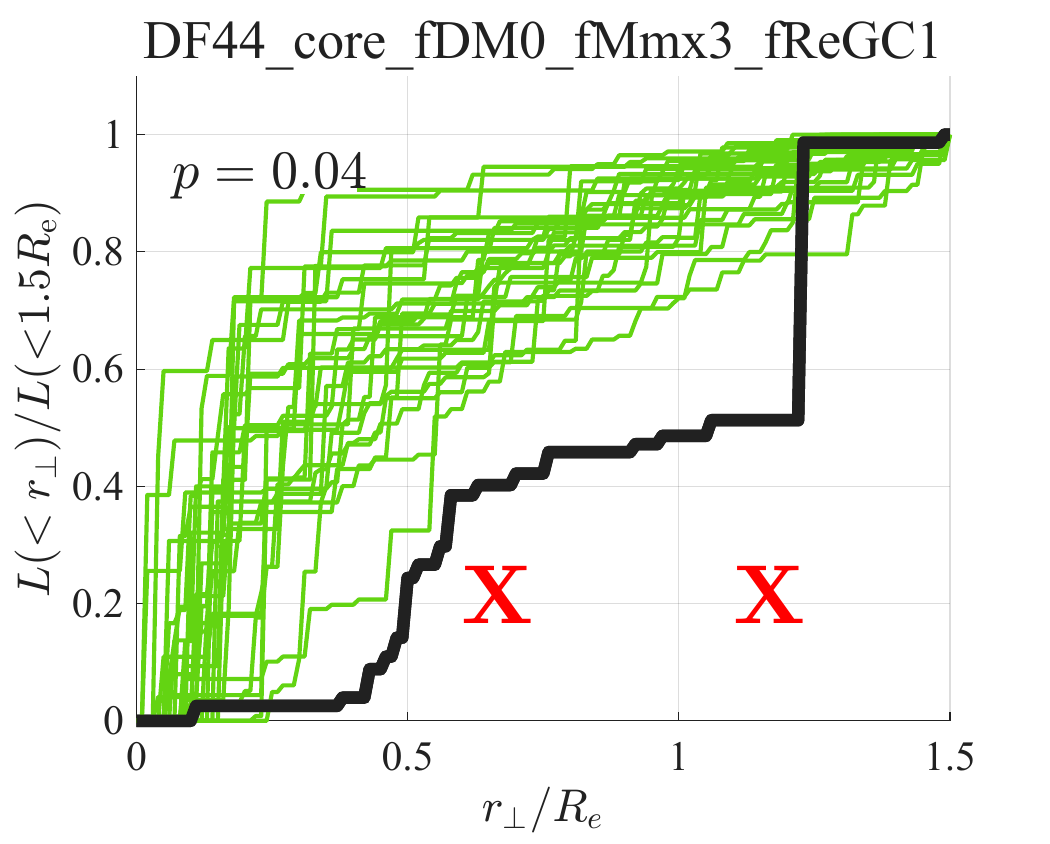}       
                  \includegraphics[ scale= 0.24]{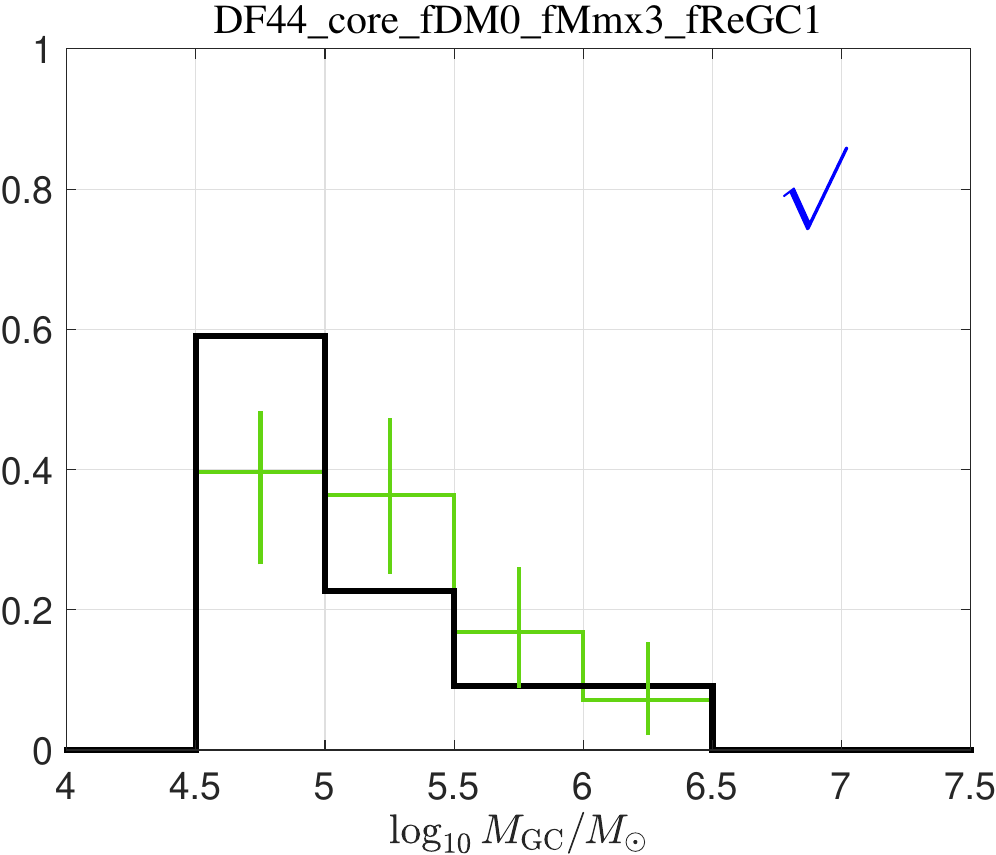}
                                    \includegraphics[ scale= 0.275]{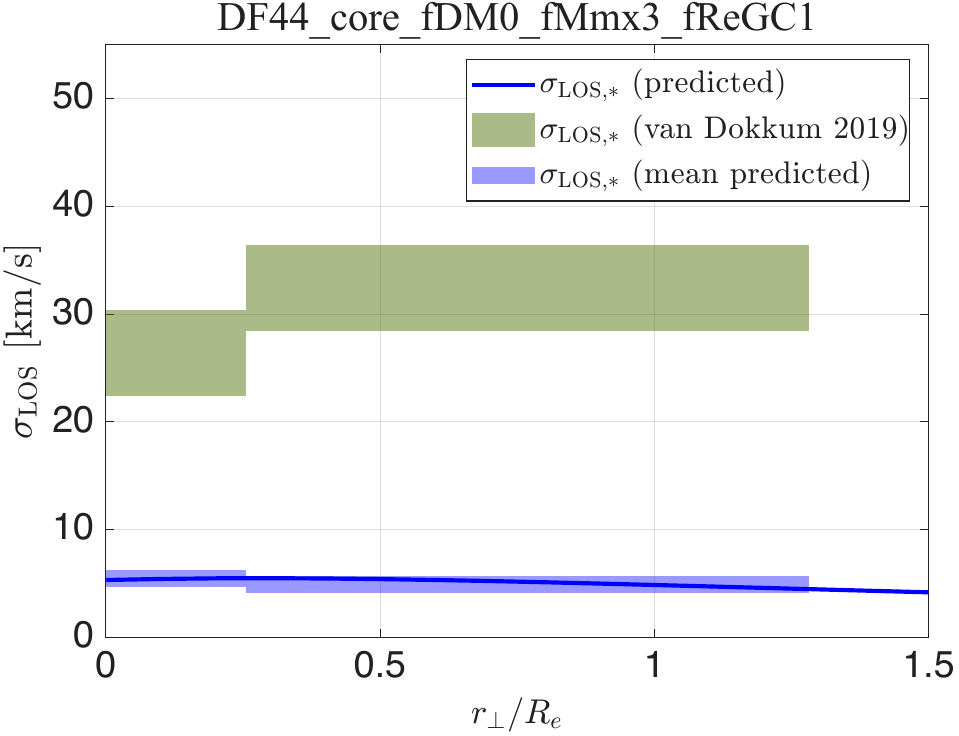}
 \caption{Detailed results for a core model that pass {\bf (top)} and a model that fails {\bf (bottom)} the GC cumulative luminosity test. {\bf DF44 galaxy.}
 }
 \label{fig:DF44_Burk_details}
\end{figure*}

\begin{figure*}
\centering
      \includegraphics[ scale= 0.24]{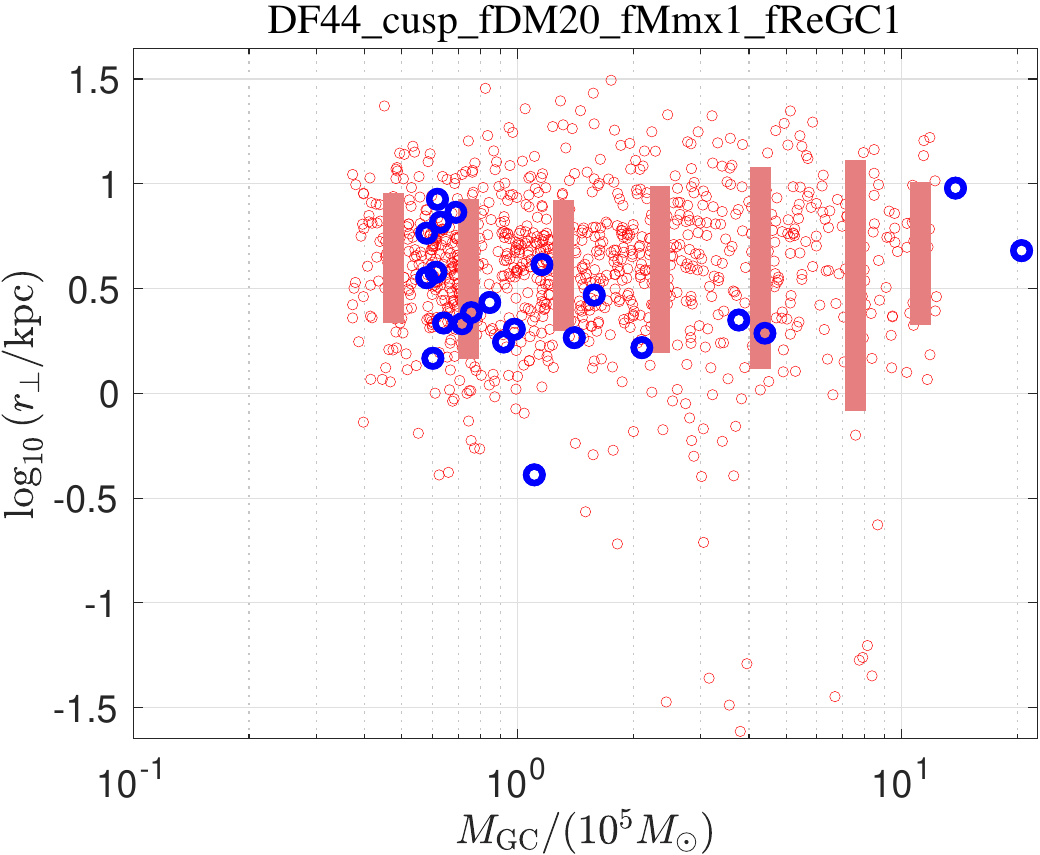}       
            \includegraphics[ scale= 0.24]{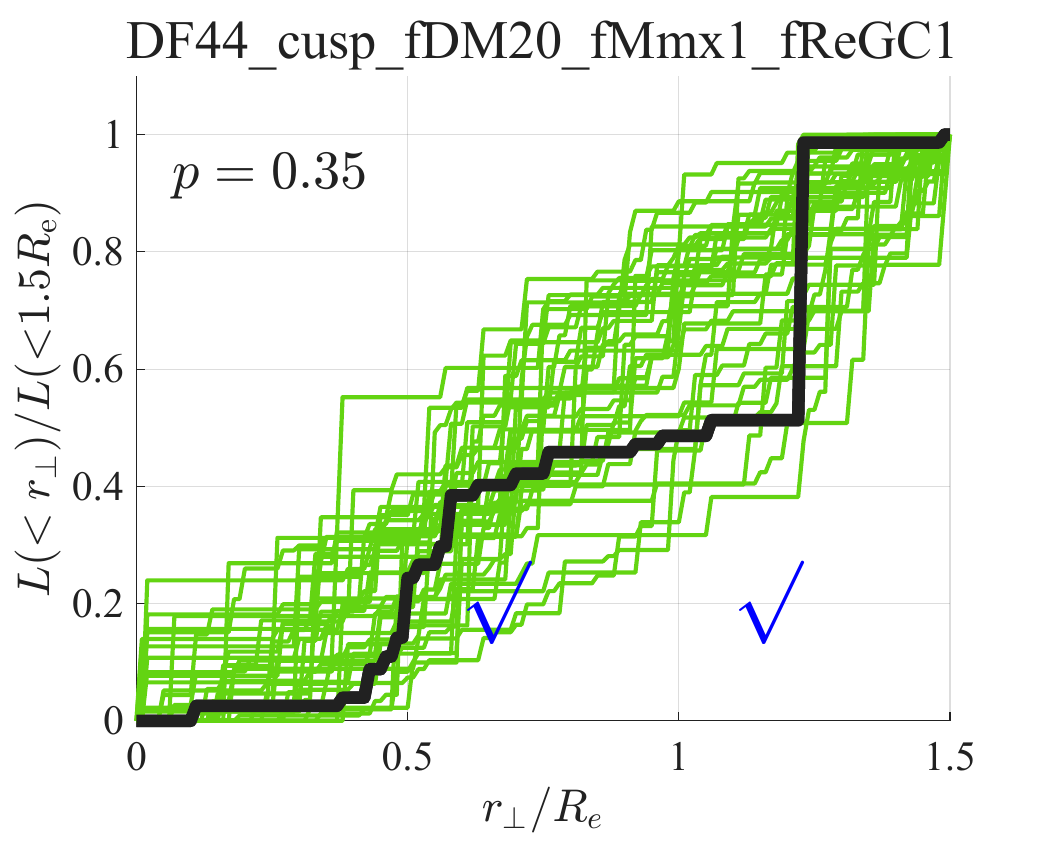}       
                  \includegraphics[ scale= 0.24]{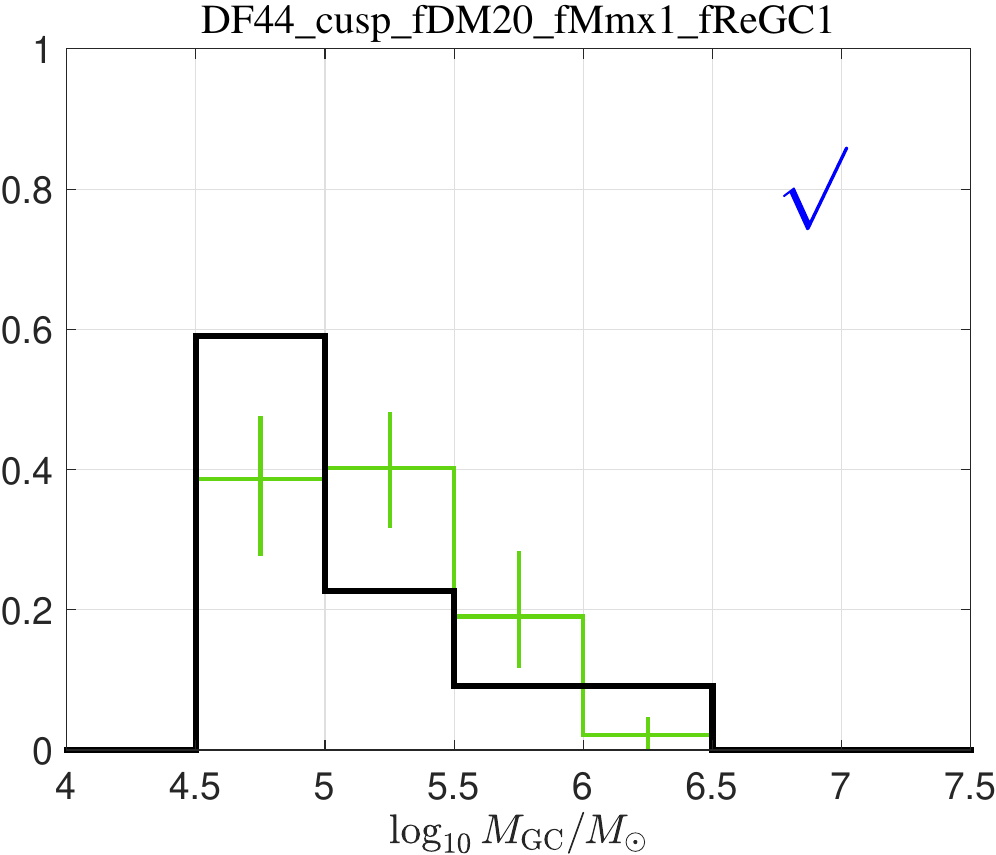}   
                              \includegraphics[ scale= 0.275]{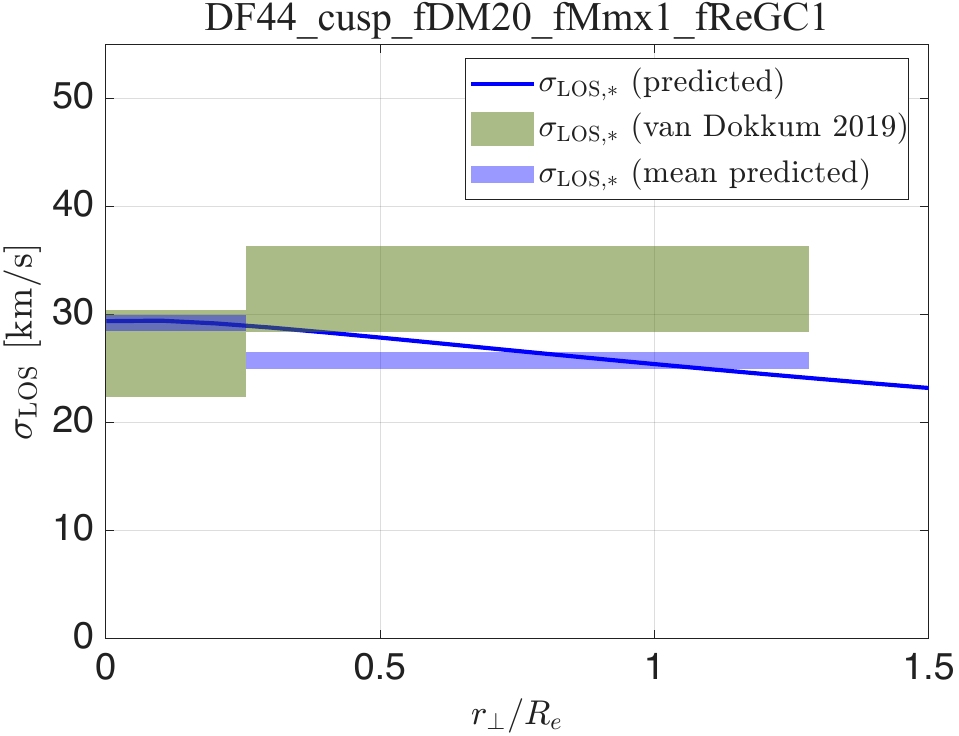}       \\    
      \includegraphics[ scale= 0.24]{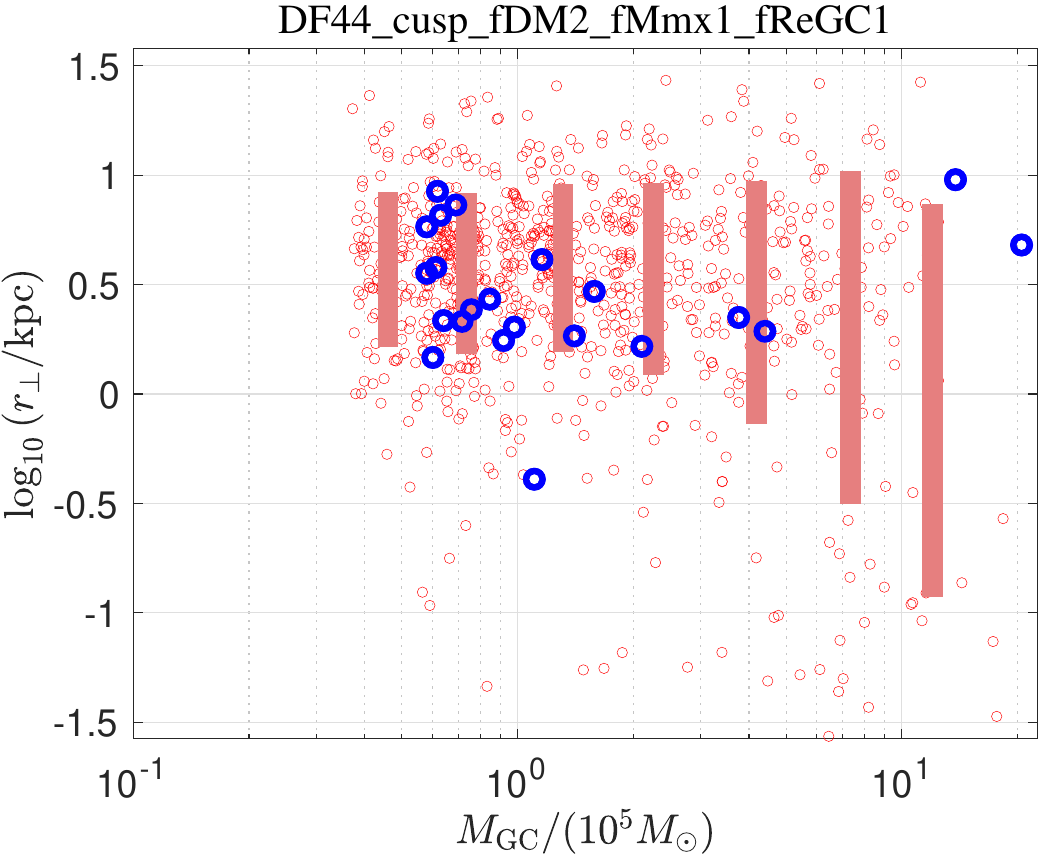}       
            \includegraphics[ scale= 0.24]{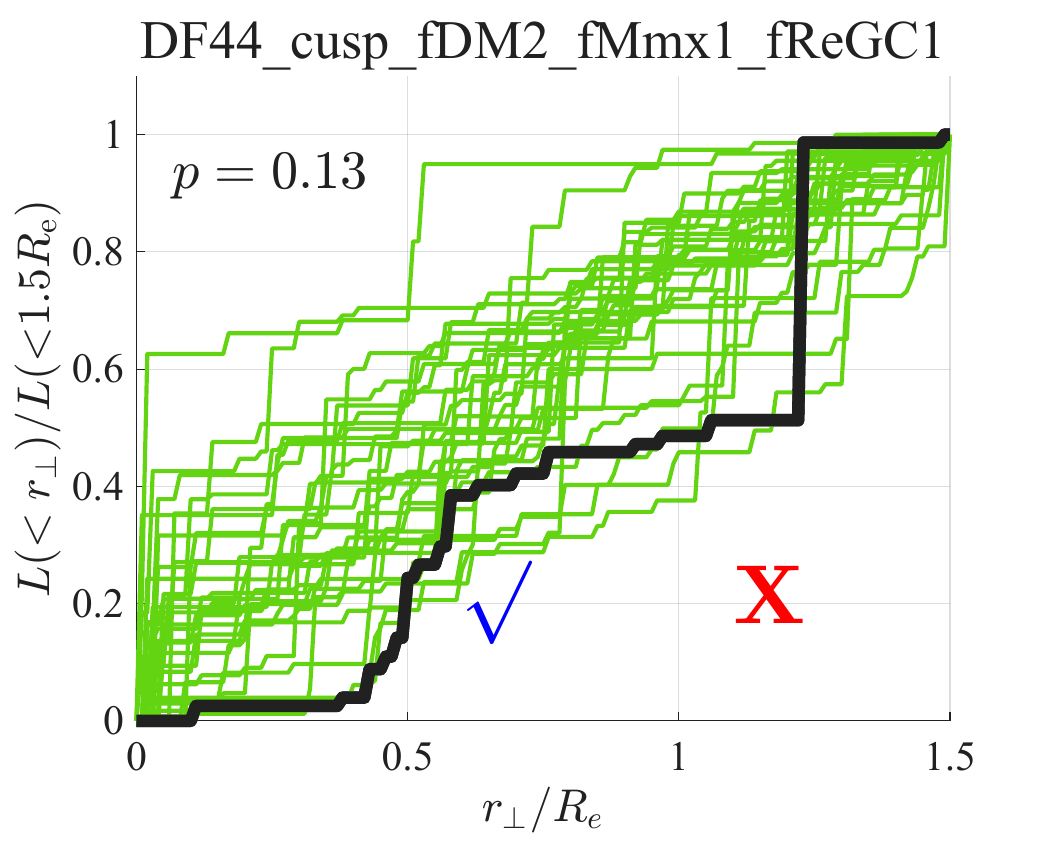}       
                  \includegraphics[ scale= 0.24]{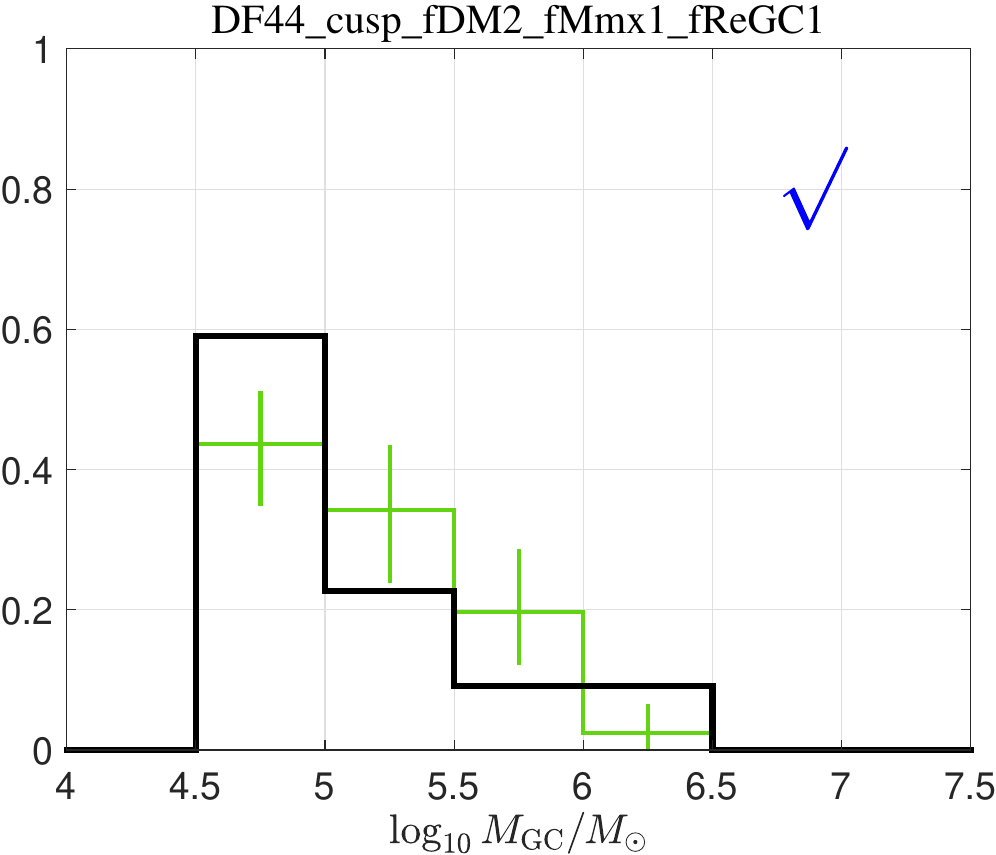}
                                    \includegraphics[ scale= 0.275]{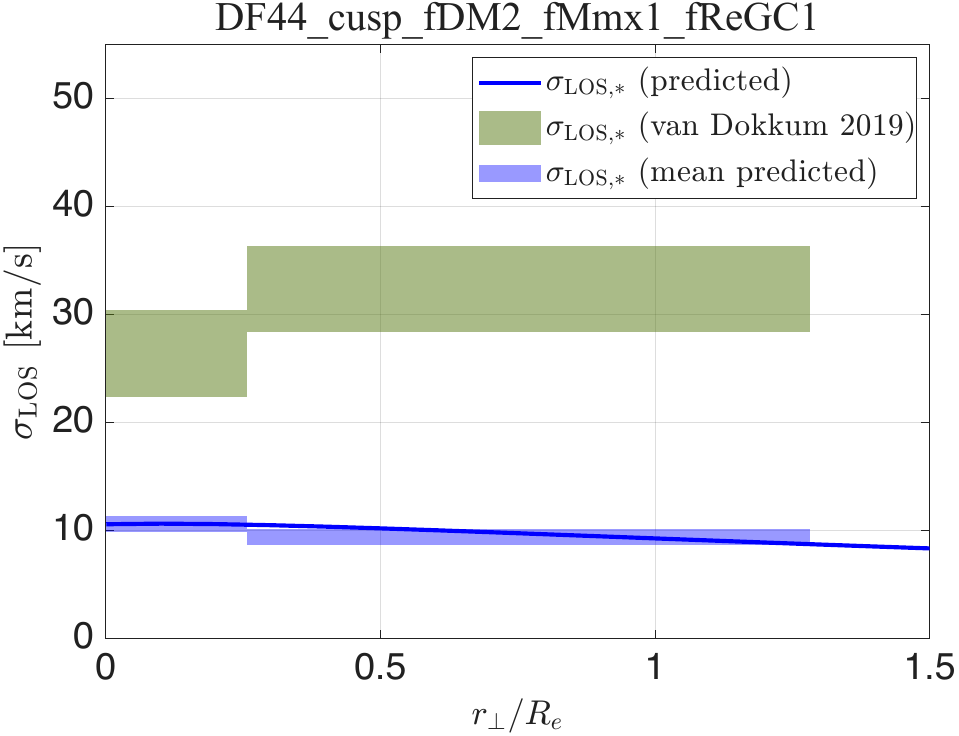}
 \caption{Detailed results for a cusp model that pass {\bf (top)} and a model that fails {\bf (bottom)} the GC cumulative luminosity test.  {\bf DF44 galaxy.}
 }
 \label{fig:DF44_NFW_details}
\end{figure*}

Figs.~\ref{fig:DF44_Burk_details} and~\ref{fig:DF44_NFW_details} show detailed results for core and cusp models, comparing scenarios that pass and scenarios that fail the GC luminosity CDF test. 

\begin{figure*}
\centering
      \includegraphics[ scale= 0.24]{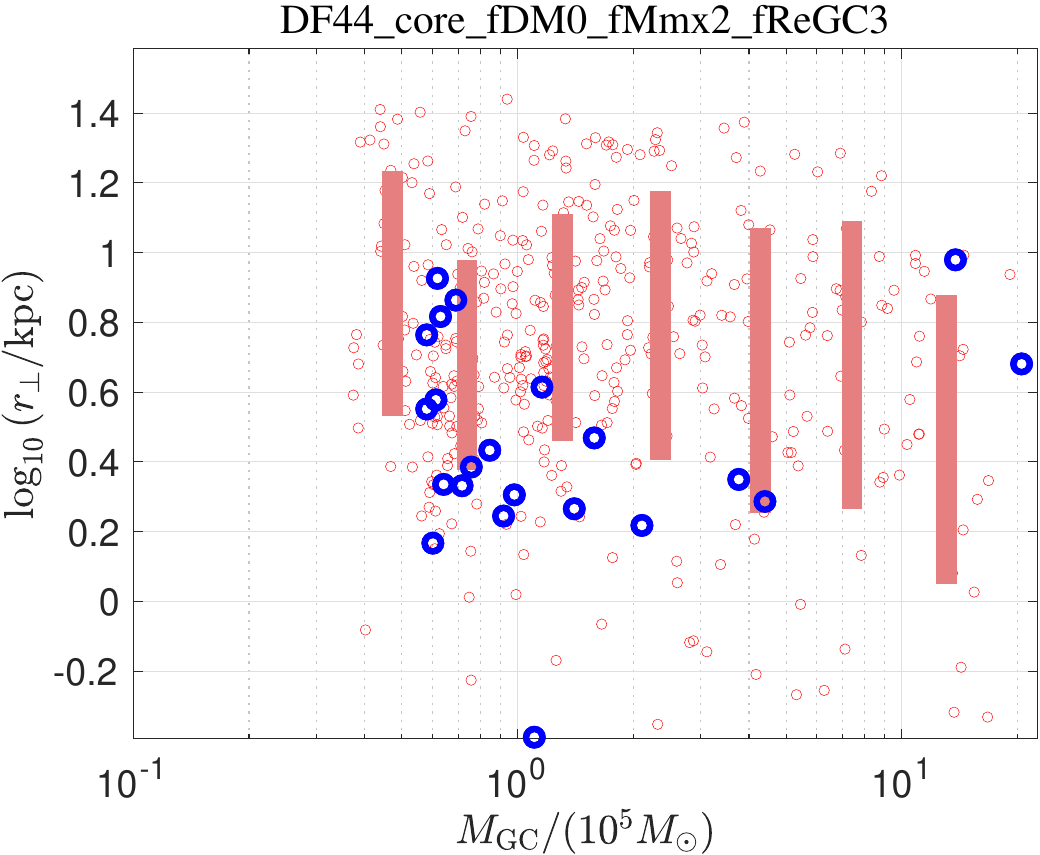}       
            \includegraphics[ scale= 0.24]{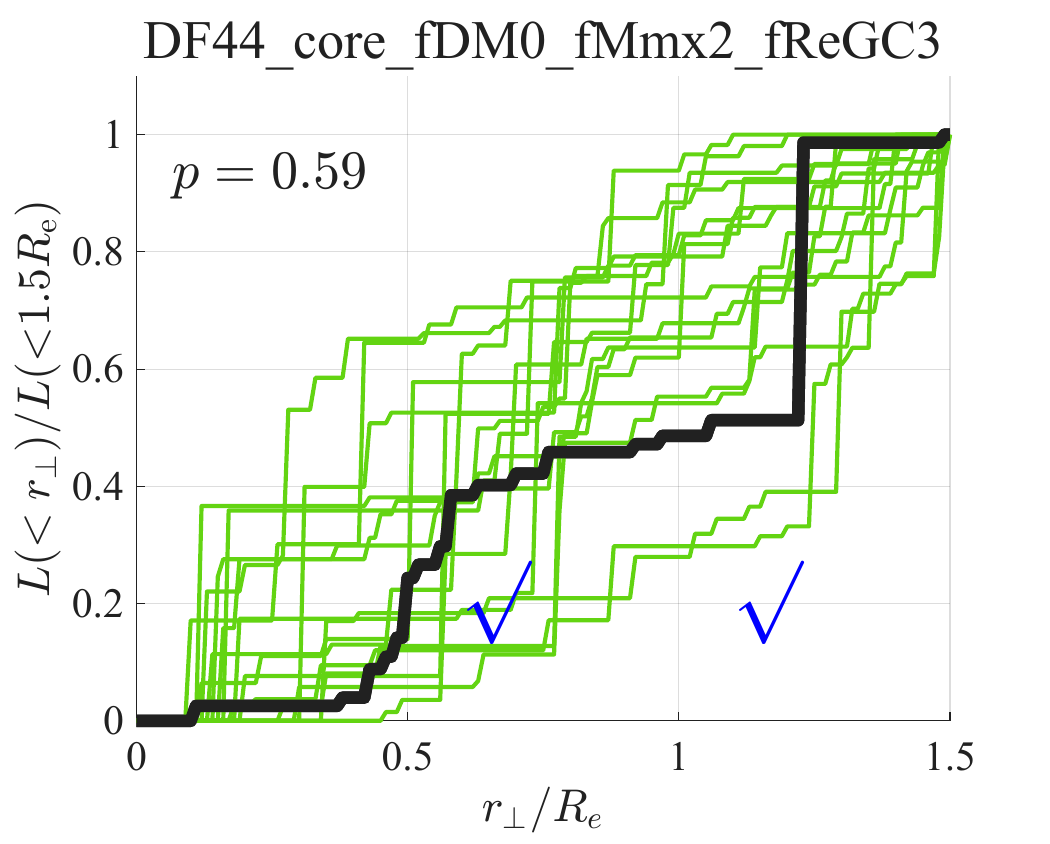}       
                  \includegraphics[ scale= 0.24]{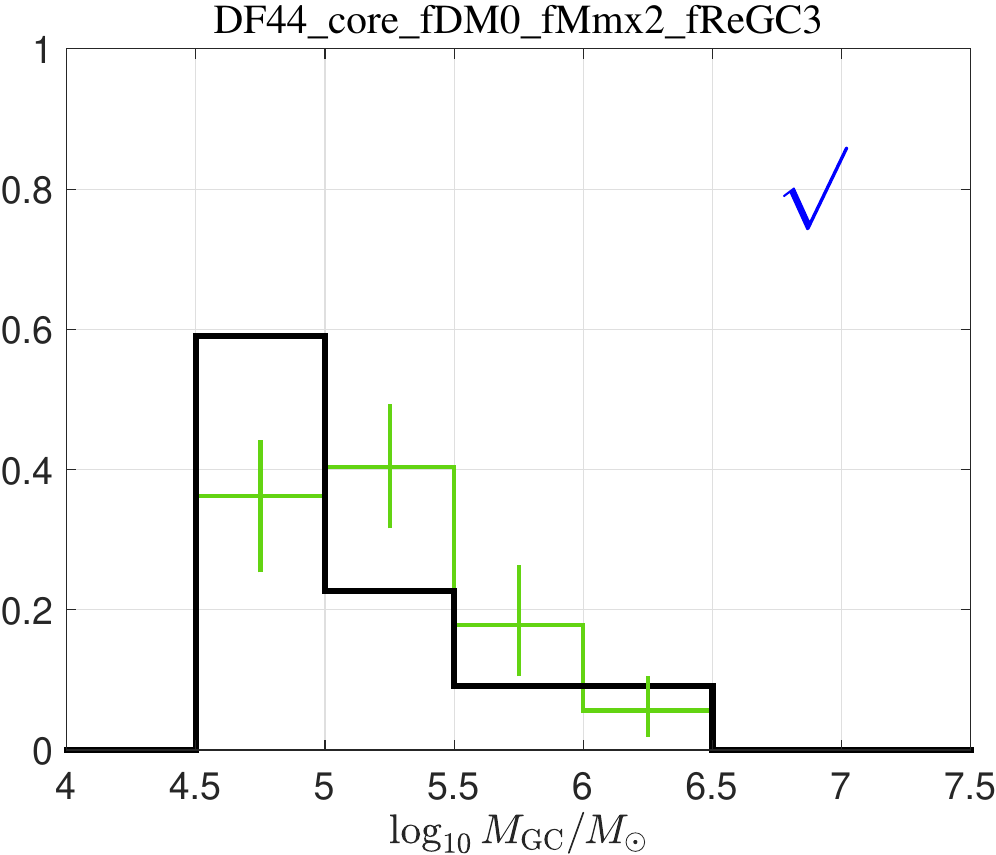}   
                              \includegraphics[ scale= 0.285]{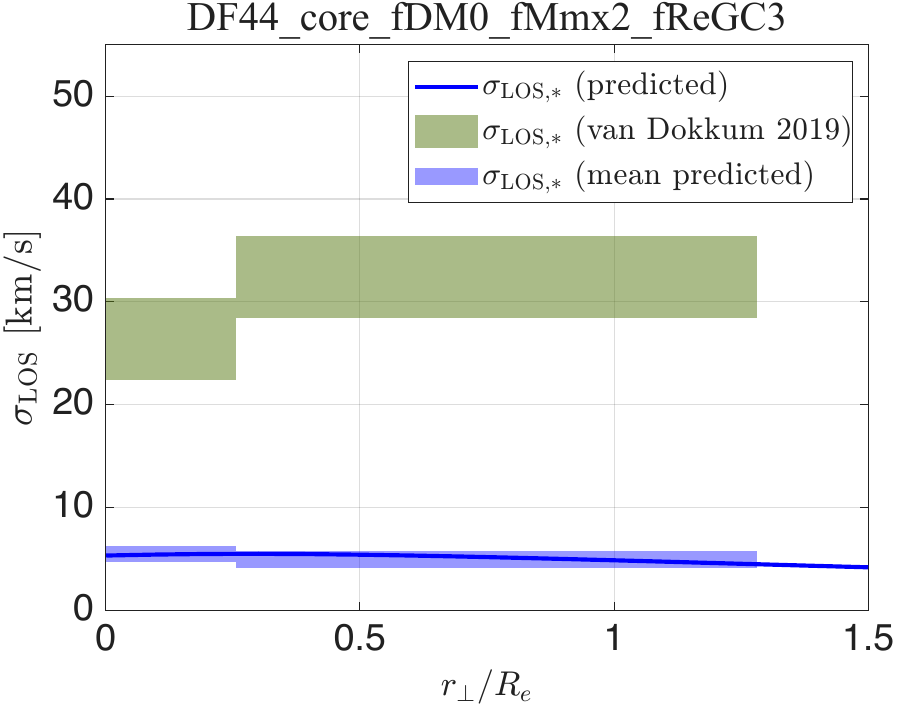}
 \caption{Detailed results for a DM-free model of DF44, that passes the GC cumulative luminosity test.
 }
 \label{fig:DF44_NoDM}
\end{figure*}

Notably, the GC data of DF44 becomes roughly consistent with DM-free models if we allow significant initial stretch. We show an example in Fig.~\ref{fig:DF44_NoDM} (see also Tab.~\ref{tab:DF44summarycore}). Although the model in Fig.~\ref{fig:DF44_NoDM} survives our luminosity criterion, it shows tension in the mass-radius scatter plot (left panel) and could perhaps be disfavored in a more detailed analysis.
%

%

\subsection{DF44: summary tables}\label{ss:df44ressum}
A summary of the luminosity CDF criterion at $r_\perp=0.5R_{\rm e}$ and $r_\perp=R_{\rm e}$ for most of our semianalytic simulations is given in Tab.~\ref{tab:DF44summarycore}.
\begin{table*}[htb!]
\centering
\caption{Results summary: DF44 core}\label{tab:DF44summarycore}
\begin{tabular}{|*{16}{c|}}
\hline
\multicolumn{16}{|c|}{Core}\\ \hline
&\multicolumn{5}{|c|}{ stretch: 1}&\multicolumn{5}{|c|}{ stretch: 2}&\multicolumn{5}{|c|}{ stretch: 3}\\ \hline
fDM & 0 & 1 & 2 & 5 & 10 & 0 & 1 & 2 & 5 & 10 & 0 & 1 & 2 & 5 & 10\\ \hline
fMmx1 & $\X\X$ & $\V\X$ & $\V\V$ & $\V\V$ & $\V\V$ & $\X\X$ & $\V\V$ & $\X\V$ & $\V\V$ & $\V\V$ & $\X\X$ & $\V\V$ & $\V\V$ & $\V\V$ & $\V\V$\\
fMmx3 & $\X\X$ & $\V\X$ & $\V\V$ & $\V\V$ & $\V\V$ & $\X\X$ & $\V\V$ & $\V\V$ & $\V\V$ & $\V\V$ & $\V\V$ & $\V\V$ & $\V\V$ & $\X\V$ & $\V\V$\\
\hline
\multicolumn{16}{|c|}{Cusp}\\ \hline
&\multicolumn{5}{|c|}{ stretch: 1}&\multicolumn{5}{|c|}{ stretch: 2}&\multicolumn{5}{|c|}{ stretch: 3}\\ \hline
fDM & 0 & 1 & 2 & 5 & 10 & 0 & 1 & 2 & 5 & 10 & 0 & 1 & 2 & 5 & 10 \\ \hline
fMmx1 & $\X\X$ & $\X\V$ & $\V\X$ & $\V\X$ & $\V\V$ & $\X\X$ & $\V\V$ & $\V\V$ & $\V\V$ & $\V\V$ & $\X\X$ & $\X\V$ & $\V\V$ & $\V\V$ & $\V\V$ \\
fMmx3 & $\X\X$ & $\X\X$ & $\X\X$ & $\V\X$ & $\X\V$ & $\X\X$ & $\V\V$ & $\V\V$ & $\V\V$ & $\V\X$ & $\V\V$ & $\V\V$ & $\V\V$ & $\V\V$ & $\V\V$ \\
\hline
\end{tabular}
\end{table*}
%

\FloatBarrier
\section{Discussion}\label{s:disc}
{
%
%
%

For UDG1, our results in Sec.~\ref{s:udg1} suggest that a DM halo is needed for a reasonable description of the GC system. This conclusion remains robust under varying GCIMF and initial spatial stretch, within what we think are reasonable limits. 
We also compared the GC analysis to standard stellar kinematics~\citep{muller2020spec,muller21udg,Forbes:2020vly,2022ApJ...927L..28D}, finding models that reproduce the GC system while being consistent with kinematics (we briefly comment below on the recent measurements of \cite{2025MNRAS.539..674H}).

For Fornax, Sec.~\ref{s:fornax} showed that the GC system -- despite its meager statistics -- provides strong indication of a DM halo. This point may seem moot given that stellar kinematics data, available now for many years, already famously implies that Fornax is DM-dominated. However, again, DF is a qualitatively different test of the DM paradigm. In past and recent years, the GC system of Fornax attracted attention for this same reason. Naive estimates of the DF time for some of Fornax's GCs are much shorter than their age~\citep[][]{Tremaine1976a,Hernandez:1998hf,ohlinricher2000,Lotz:2001gz,Goerdt2006,Cowsik:2009uk,angus2009resolving,Cole2012,Kaur2018,Hui:2016ltb,Leung_2019,boldrini2020embedding,Berezhiani2019,Hartman:2020fbg,Bar-Or:2018pxz,Lancaster2020,Meadows20}, leading to suggestions of exotic properties of DM (see \cite{Bar:2021jff} for review of ideas). Our results show that exotic models of DM are not needed and that the minimal theory of DM-induced DF can explain the GC data, consistent with stellar kinematics, provided that the initial conditions of the Fornax GC system are somewhat stretched w.r.t. its current stellar body. This conclusion is consistent with that of \cite{Shao:2020tsl}. We also found that both cusp and core DM halo models are consistent with the GC data, as long as an initial radial stretch is permitted.

In Sec.~\ref{s:df44} we considered UDG-DF44, which hosts massive GC candidates far from its center of light, naively defying the premise of mass segregation. Nevertheless, we found that reasonable DM halo models and GC initial conditions can address the data. The feature that lets DF44 ``walk away" with massive off-center GCs, is its excessive diffusivity: its stellar radius is twice larger than that of UDG1, at similar total luminosity. Thus DF44 does not look like a breakdown of our framework, although our framework is not particularly predictive for such a diffuse system and the GC-based constraints we derive are much weaker than those obtained from kinematics.


A few more remarks are in order. 
An interesting target for our analysis would be UDG-DF2, that was argued in~\cite{Danieli_2019,Shen_2021} to lack DM altogether. \cite{Dutta_Chowdhury_2019,Dutta_Chowdhury_2020} studied the spatial distribution of GCs in UDG-DF2, and concluded that it could be consistent with a DM-free halo, if the initial distribution of GCs was extended w.r.t. the stellar body. The tools we developed are applicable to the case of DF2, but we defer this study to a separate work.

Our analysis neglected perturbations like external tides or baryonic feedback (for an attempt to include such effects, see~\cite{2024A&A...690A.119B}). If the galaxies we considered go through complex dynamical evolution, then our analysis may need to be revisited. Feedback, mergers, and tidal perturbations were indeed suggested for at least some UDGs~\citep[][]{DiCintio:2016ehs,vanDokkum:2022zdd,2023ApJ...954L..39F,2024arXiv240101931F}. 

We ignored halo rotation. The impact of rotation on DF may be important for GC dynamics in disc galaxies, which is an interesting problem, among other reasons, because such systems may represent pristine environments where galaxy  mergers could be less important than for spheroidal galaxies. At this point, however, we restricted our attention to galaxies that appear mostly pressure supported. 

Another systematic uncertainty, that we did not cover, is in the assessment of GC mass-to-light ratio. We left this aside for now because we find the problem of the information content of the GC system, coupled to the uncertain initial conditions, rather overwhelming by itself, and we decided to explore these aspects of the problem first. We will investigate mass-to-light uncertainties in subsequent work.

It is worth noting that our analysis of the observational data did not make use of the full information content of the GC system. We considered integrated quantities: the luminosity CDF and the GCMF. The first integrates the GCMF in some spatial region, and the second integrates over spatial positions in some mass range. More information is stored in the statistics of the two-dimensional distribution in GC mass and position; this is the point of mass segregation~\citep{Bar:2022liw}. Even more information is available when coupling the above with LOS velocity measurements. While we tried to exhibit some of this information in our results, we did not fully use it to differentiate between models.

Finally, a different line of work aims to infer dark matter halo properties from GC mass and number counts in galaxies and clusters of galaxies, based on empirical statistical correlations, or comparison with cosmological simulations (see, e.g.~\cite{Saifollahi:2022yyb,Diego:2023jed,2023ApJ...950..179D,2009hst..prop11710B,2021MNRAS.505.5815V,2023MmSAI..94c..47V}). Our work, in comparison, was a case-by-case study, attempting to remain agnostic w.r.t. initial conditions. Ultimately, the information from both lines of study should be combined. 

\subsection{Comparison to previous work}\label{ss:comp}
Our results can be compared to previous work that highlighted DM-dominated DF. We now discuss the most relevant examples from the literature.

\begin{itemize}
\item
\cite{Bar:2022liw} used semianalytic simulations and found evidence for DM-induced mass segregation in UDG1, previously speculated by~\cite{muller21udg}. Our work here used a similar semianalytic code, but extended the analysis considerably. First, N-body simulations allowed us to test the semianalytic code. Second, we systematically explored the effect of the initial conditions of GC radial and mass distribution, and surveyed a large set of DM halo models, while \cite{Bar:2022liw} considered one DM halo model representative for each choice of core or cusp. Third, we presented new ways to compare the models to data, notably via the LCDF that captures realization scatter while maintaining inter-realization correlations. \cite{Bar:2022liw} found a lower GC merger probability than we tend to: for example, their mean merger probability for the stars-only model was $\sim0.8$, while our results for a similar model  predict one or more. We think that the difference is mainly due to our simulations evolving a larger initial number of GCs, drawn from an GCIMF and evolved into GCMF accounting for observational luminosity cut, facilitating mergers among low-mass GCs. These would not occur in \cite{Bar:2022liw} that mostly evolved only the currently observed GCs. Despite the different merger efficiency, the results of \cite{Bar:2022liw} were roughly consistent with ours: their core model is comparable to an $fDM5$ or $fDM6$ model in our definitions, and their cusp model to our $fDM4$ or $fDM5$, compatible with the allowed or marginally allowed entries in Tab.~\ref{tab:UDG1summarycore}. 
%
\item
\cite{Liang:2023ryi} also analyzed DM-induced DF in UDG1 based on semianalytic calculations. The details of their modeling, however, were rather different to ours. \cite{Liang:2023ryi} did not model GC mergers, instead simply assuming that all GCs that arrive in $r<100$~pc join an NSC. Their simulations were described as starting with an ``arbitrarily large initial number of GCs". Another difference to our work was their model of GC evolution and mass loss. 
As noted earlier, their mass loss rate for an isolated GC is smaller by a factor of $\sim$20 than the corresponding rate in our work, Eq.~(\ref{eq:Mdotiso}) (adopted from \cite{Gnedin:2013cda}).\footnote{We thank the authors of \cite{Liang:2023ryi} for a discussion on this point.} The main difference can be traced to two points: (i) a factor of 2.5 increase in loss rate, attributed in \cite{Gnedin:2013cda} to multi-mass models of GC stellar populations~\citep[]{2011MNRAS.413.2509G}, is not used in \cite{Liang:2023ryi}; and (ii) different estimate of the half-mass relaxation time at a given GC mass, which, in turn, stems from different estimates of GC central densities. 
Indeed, \cite{Gnedin:2013cda} assumed GC half-mass mean density of $\rho_h=10^3M_\odot/{\rm pc}^3$ at $M_{\rm GC}\leq10^5~M_\odot$, to be compared with $\rho_h\approx(35-185)~M_\odot/{\rm pc}^3$ at $M_{\rm GC}=10^5~M_\odot$ that can be gleaned from \cite{Liang:2023ryi}'s Figs.3 and 5 (with Fig.5, representing UDG1 GCs, tending towards lower $\rho_h$). The difference in $\rho_h$ translates to a factor of $\sim2.3-5.3$ in mass loss rate at $M_{\rm GC}=10^5M_\odot$. 

Low mass loss rate requires stronger DF at fixed GC radial distribution today~\citep[][]{Bar:2022liw,Liang:2023ryi}, and numerical exploration in App.~\ref{a:massloss} shows that it also increases the statistical spread of the evolved GC radial distribution compared to its spread at larger mass loss rate. 

Note that the mass loss rate adopted in~\cite{Liang:2023ryi}, coupled to a power-law GCIMF \citep{Gnedin:2013cda}), would lead to $M_{\rm GC,peak}\approx10^4~M_\odot$, an order of magnitude lower than the peak inferred from observations. Thus, if this mass loss rate estimate is indeed correct (see also \cite{Gieles:2008ew}), it would imply that the power-law GCIMF model, and perhaps the basic association of GCs with evolved YMCs~\citep[][]{Kravtsov:2003sm,Gnedin:2013cda,Krumholz2019,Hughes:2021got} may need to be revised. In \cite{Liang:2023ryi} the power-law GCIMF was modified via exponential truncation at an upper and lower scale mass $M_{\rm max}$ and $M_{\rm min}$~\citep{2019MNRAS.488.3972T}, with $M_{\rm min}=10^{5.5}~M_\odot$ effectively implementing the GCMF break by hand already from the GCIMF. This procedure may leave open the question of the similarity of the GCMF among galaxies in different environments. 

Moving to galaxy halo models, \cite{Liang:2023ryi}'s UDG1 stellar mass was taken to be $\approx2\times10^{8}~M_\odot$, twice larger than in our analysis. In addition, while in this paper we set the DM halo radius parameter to a fixed value of $3R_{\rm e}$ (namely, e.g., 6~kpc for UDG1; see Sec.~\ref{s:model}), \cite{Liang:2023ryi} varied the halo  radius as a fit parameter. This approach would be important for drawing detailed conclusions about the halo profile, but varying the halo scale radius likely requires N-body simulations to test deviations from semianalytic approximations on a case by case basis. Without such tests, semianalytic simulations with varying scale radius could introduce a bias that is difficult to model. We also note that the statistical inference in \cite{Liang:2023ryi} does not show convergence: both for the NFW and Burkert DM halo models, their posterior distributions for the halo scale radius are bounded from above only by the prior (Figs.~4 and ~6 {\it there}). 

Finally, \cite{Liang:2023ryi} considered GC half-light radii as an additional observational input. However, GCs in UDG1 are resolved by a few HST pixels at best \citep{2022ApJ...927L..28D}. 
We therefore opted not to include this information without careful study of systematics.

Despite the differences in modeling, \cite{Liang:2023ryi}'s results support our conclusion that UDG1 is a prime example of DM-induced DF. Their fit results are compatible with ours; for example, their NFW halo mass at $r= 3R_{\rm e}$ is roughly in the range $M_h(6~{\rm kpc})\approx10^8-10^9~M_\odot$, which covers our models $fDM1-fDM10$, compatible with data with some GC stretch, and even more compatible if we adjust the GC mass loss closer to that in \cite{Liang:2023ryi} (see Tabs.~\ref{tab:UDG1summarycore} and~\ref{tab:UDG1summarycoreMdot01}).

\item
\cite{Modak:2022kxv} studied the galaxy UGC7369, using semianalytical simulations similar to these here and in \cite{Bar:2022liw}. The main discussion focused on the brightest GC in UGC7369 that may be understood as an NSC product of DF-induced GC mergers. With this understanding, the GC system in that galaxy also presents evidence of DM, and \cite{Modak:2022kxv} argued further that the DM halo is likely to be a cusp. Our numerical experiments in the current work suggest that uncertainties related to GC mass loss, merger dynamics, and initial conditions deserve careful exploration if one aims to constrain the halo profile. Given our assessment of the uncertainties, we did not claim convincing discrimination of core vs. cusp for any of the three galaxies we studied. Once we investigate GC merger dynamics more thoroughly via N-body simulations, it would be interesting to revisit and extend the analysis of \cite{Modak:2022kxv}, and consider the general problem of NSC statistics in dwarf galaxies. 

\item 
Finally, a recent re-analysis of UDG1's GC velocity dispersion was reported in \cite{2025MNRAS.539..674H}. The velocity dispersion of 20 (nominally high-confidence) GC candidates was measured using the Keck Cosmic Web Imager (KCWI)~\citep{2018ApJ...864...93M}, finding a result of $\sigma_{\rm GC}=29.8^{+6.4}_{-4.9}$~km/s, in strong disagreement with the previous result of \cite{muller2020spec}, $\sigma_{\rm GC}=9.4_{-5.4}^{+7.0}$~km/s. Interestingly, however, the large $\sigma_{\rm GC}$ result of \cite{2025MNRAS.539..674H} is driven by the measurements from the faintest $\sim6$ GCs in the sample (see Fig.7 {\it there}). Restricting to the 10-14 brightest GCs (including the sample of 11 GCs used in the MUSE data of \cite{muller2020spec}) the authors find lower $\sigma_{\rm GC}$, consistent with that of \cite{muller2020spec}. 

High-mass DM halo models that we considered in this work can, in principle, be consistent with a high GC velocity dispersion alongside the observed radial distribution; see, e.g., the $fDM10$ NFW model in Fig.~\ref{fig:NFW_details}. This said, given the observational discrepancies noted in \cite{2025MNRAS.539..674H}, and the puzzle presented there w.r.t. the dihcotomic high- vs. low-mass GC velocities, we believe that an attempt to fit the kinematics results with GC dynamical models may be premature. 
\end{itemize}
}

\section{Summary}\label{s:sum}
We considered the information content of the projected radial distribution and mass function of globular clusters (GCs) in dwarf and ultradiffuse galaxies. Using a semianalytic implementation of dynamical friction (DF), tested against live-halo (but point-mass GC) N-body simulations, we conducted thousands of simulations of GC systems, exploring a range of initial conditions and models of the dark matter (DM) halo. 

We focused on three galaxies: UDG1 (Sec.~\ref{s:udg1}), which shows convincing positive evidence of mass segregation; the Fornax dwarf spheroidal (Sec.~\ref{s:fornax}), a Milky Way satellite with a GC system that is old in comparison with naive estimates of its DF time scale; and the Coma cluster UDG-DF44 (Sec.~\ref{s:df44}).

GC distributions provide a dynamical test of the DM paradigm, observationally independent and theoretically distinct from the more familiar gas and stellar kinematics analyses. The study involves systematic uncertainties, notably due to the initial conditions around the time of galaxy and GC formation. Other theoretical uncertainties, like the rate of GC mass loss and the detailed dynamics of GC mergers, are less important for the main results (examined in App.~\ref{a:checks}), and moreover can be tackled with additional dedicated simulations. We plan to perform such simulations using live GCs; this will be important for pinning down the formation of nuclear star clusters, a promising diagnostic of DM models. Given the uncertainties we do not expect that GC morphology could replace stellar kinematics as a detailed quantitative tracer of DM. However, we do find that GC-rich dwarf galaxies like UDG1, and even systems with only a handful of GCs like the Fornax dSph, provide  compelling evidence for massive DM halos.  

Considering the DF or mass segregation phenomenon itself, the main effect of a massive DM halo is to produce high velocity dispersion for halo particles, rendering DF {\it less efficient} than it would be if the halo contained only stellar mass and thus lower dispersion. Indeed, the basic observation is that DM-free models of UDG1 and Fornax predict rapid contraction of the GC system, leading to over-pronounced mass segregation. In this sense the DM signal we are after is a null signal: the absence of strong segregation. 
That said, the subtle positive hint for mass segregation in UDG1 seems like a rare and exciting evidence of beyond-mean field gravitational dynamics of DM, because both our GC analysis, and the kinematics data, suggest the galaxy is DM-dominated, thus the DF we see there is mostly DM-induced. It is also noteworthy, and nontrivial, that the velocity dispersion predicted by our GC analysis is broadly consistent with the kinematics data.

\acknowledgments
We thank Nitsan Bar, Shany Danieli, Fangzhou Jiang, Jinning Liang, Ignacio Trujillo, Teymoor Saifollahi, Mireia Montes Quiles, and Yossi Nir for useful discussions and comments. This research was supported by    Israel Science Foundation grant 1784/20, by MINERVA grant 714123, and by the EU UNDARK project 101159929. 

\bibliography{ref}
\bibliographystyle{aasjournal}

\begin{appendix}

\section{Testing the semianalytic dynamical friction implimentation with N-body simulations}\label{a:calib}
\subsection{Single GC tests, and implementation of core-stalling: comparison to the simulations of Inoue 2009}\label{ss:Inoue}
We start with a setup similar to that studied by \cite{Inoue:2009wd} to investigate core stalling.  
We calculate the orbit of a GC starting on a circular orbit at $r=0.75$~kpc in a halo with Burkert density profile (Eq.~(\ref{eq:burk})).
The halo parameters are $R_0=1$~kpc, $\rho_0=0.1~M_\odot/{\rm pc}^3$, as chosen in~\cite{Inoue:2009wd}. 

\cite{Inoue:2009wd} used $N=10$~M with a Plummer softening length of 3~pc and tree code opening angle parameter $\theta=0.5^o$. 
Our fiducial GONBY simulations use the same opening angle, and employ a reflective boundary $R_{\rm sim}=1$~kpc. We run simulations with different values of $N=100$K, $N=1$M, and different softening lengths ranging from 3~pc to 15~pc. We also test a different $R_{\rm sim}=10$~kpc, to check convergence. 
\cite{Inoue:2009wd} did not use a reflective boundary, but cut-off the halo profile at $R_{\rm max}\sim10$~kpc. To obtain a fair comparison of the effective number of particles in the GONBY simulations, using $R_{\rm sim}=1$~kpc, we should scale the GONBY $N$ as $N_{\rm eff}\sim\frac{M(10~{\rm kpc})}{M(1~{\rm kpc})}N\approx12.7N$, where $M(r)$ is the halo mass contained in radius $r$. The $N=1$M GONBY simulation has a particle mass of $1.6\times10^2~M_\odot$, to be compared with the particle mass $2.44\times10^2~M_\odot$ used by the $N=10$M simulation of~\cite{Inoue:2009wd}, consistent with the halo density radial scaling.

\begin{figure}[hbp!]
\centering
      \includegraphics[scale=0.3]{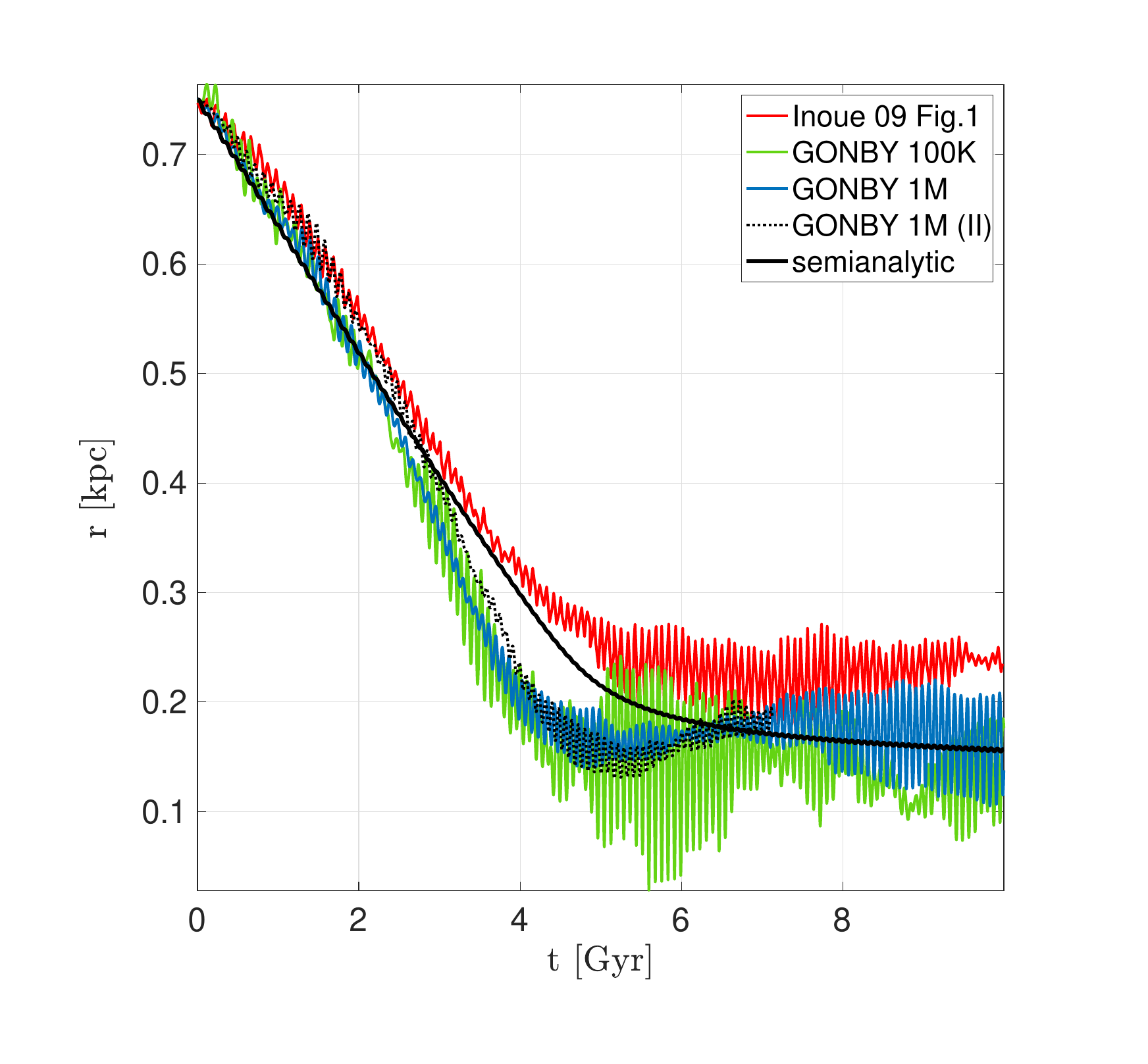}
            \includegraphics[scale=0.3]{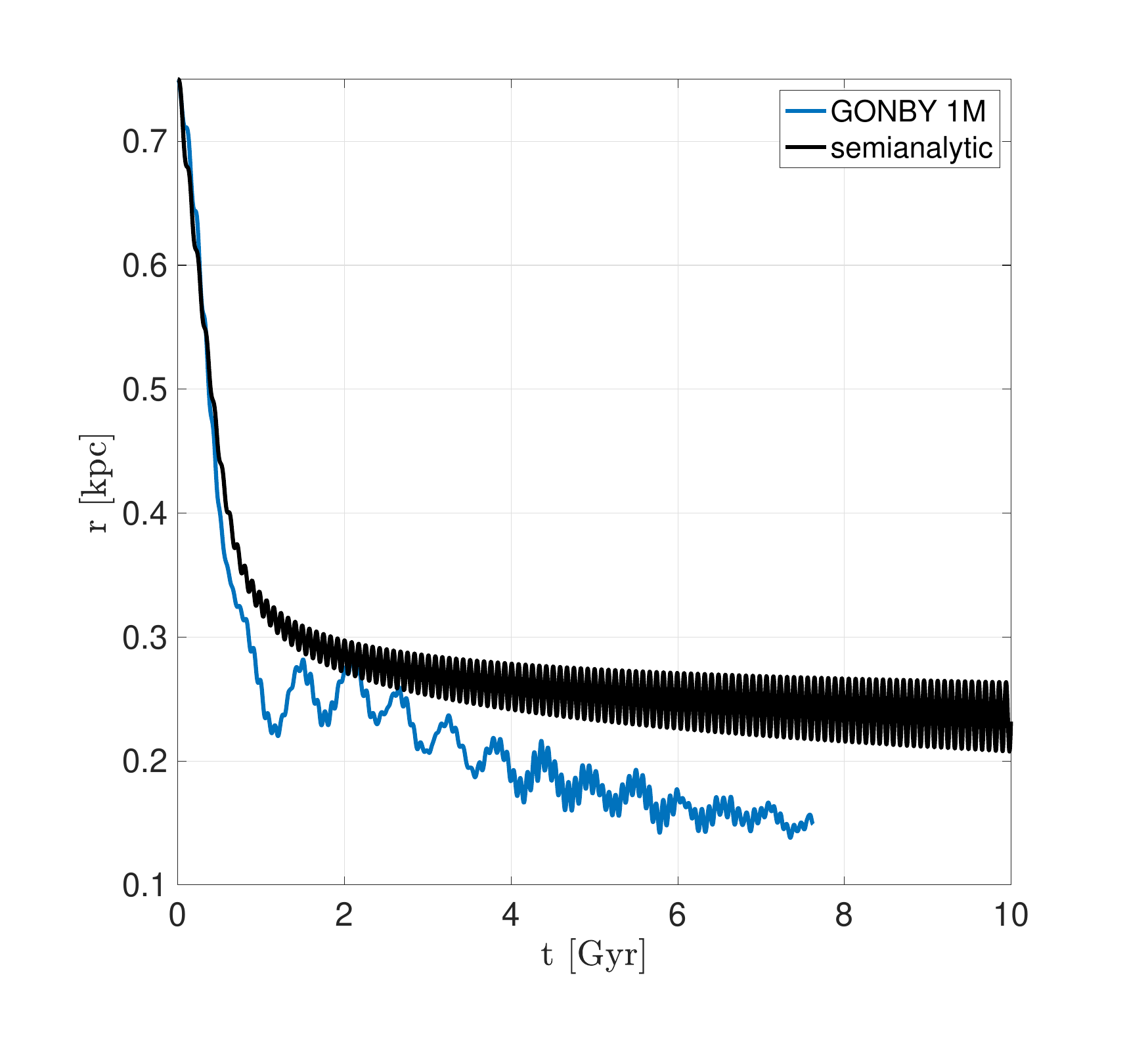}
 \caption{N-body simulations (solid colored and dotted black lines) compared with semianalytic integration (solid black line), showing the orbital decay of a GC with $M_{\rm GC}=2\times 10^5~M_\odot$ (left) and $M_{\rm GC}=1.6\times 10^6~M_\odot$ (right). The GC is initialized on a circular orbit at $R_{\rm GC}(t=0)=0.75$~kpc. The halo parameters are similar to those chosen in \cite{Inoue:2009wd}. On the left we show GONBY runs with 100K (green) and 1M (blue and dotted black) halo particles; two realizations are shown for the case of 1M particles, to illustrate realization variance. The red line shows the result obtained by \cite{Inoue:2009wd}. 
 }
 \label{fig:Inoue}
\end{figure}

The {left panel} of Fig.~\ref{fig:Inoue} shows results for GC mass $M=2\times10^5~M_\odot$. Red line shows the result of \cite{Inoue:2009wd}. Green and blue show GONBY results with $N=100$K and $N=1$M particles, respectively, employing $R_{\rm sim}=1$~kpc and a softening length of 7~pc. Dotted black shows an $N=1$M GONBY run with a softening length of 3~pc (``GONBY 1M (II)" in the legend).

Our results are close but not identical to those of \cite{Inoue:2009wd}, with core stalling occurring at a slightly smaller radius in our simulations. Our results are not sensitive to changing the softening length by a factor of 5, changing $N$ by a factor of 10, or changing the tree code angle $\theta$ by a factor of 3. 
In App.~\ref{a:checks} we show that a full N-body simulation with $N=20$K agrees, within the numerical noise, with  GONBY tree code results.

One difference between our setup and that of~\cite{Inoue:2009wd} is the initialization of the halo. We construct initial data using the Eddington formalism, so the halo is stationary from the get-go. \cite{Inoue:2009wd} initialized the halo with a Maxwellian velocity distribution, then let it relax for 2~Gyr. The slightly different phase space distribution functions produced by these two methods may be at the root of the difference between our results. 
Either way, the physical scenario that interests us most in this paper involves a multi-GC system, where GC-GC interactions perturb the dynamics and modify the naive DF effect in the inner halo region (see also~\cite{Inoue:2009wd}). Thus we consider the level of agreement demonstrated in this section to be sufficient for our purpose.

Next, we consider the semianalytic calculation, shown by the solid black curve in Fig.~\ref{fig:Inoue}. In order to roughly implement core stalling in the semianalytic calculation, we adjust the DF time in the Chandrasekhar formula via
\be\label{eq:tDFsemian}\tau_{\rm DF}&\to&\frac{\tau_{\rm DF}}{\left(1-e^{-\left(\frac{r}{r_{\rm cs}}\right)^4}\right)^4},\ee
where $r_{\rm cs}\left(M_{\rm GC}\right)$ is the core stalling radius predicted by \cite{Kaur2018}, defined from the equation
\be r_{\rm cs}&=&\left[\frac{M_h(r_{\rm cs})}{M_{\rm GC}}+1\right]^{\frac{1}{3}}r_p\left(M_{\rm GC}\right),\ee
with $r_p$ denoting the radius at which the halo mass becomes equal to the mass of the perturber,
\be\label{eq:rp} M_h\left(r_p\left(M_{\rm GC}\right)\right)=M_{\rm GC}.\ee

The {right panel} of Fig.~\ref{fig:Inoue} shows GONBY and semianalytic simulations for a more massive GC, $M_{\rm GC}=1.6\times10^6~M_\odot$, orbiting the same halo. The early stage of the orbit is very well captured by the semianalytic calculation, but the core stalling prescription predicts stalling somewhat too early. 
Despite this slight mismatch, we chose not to fine-tune the semianalytic calibration further. 
Eq.~(\ref{eq:tDFsemian}) is simply an ad-hoc prescription to quench DF in semianalytic calculations in the rough vicinity of the core stalling region. Apart from slightly over estimating the core stalling radius for the heavy GC in the {right panel}, this prescription also misses the ``super-Chandrasekhar" DF~\citep[][]{Petts2015} that GONBY simulations (albeit not the simulation of~\cite{Inoue:2009wd}) predict just prior to the onset of core stalling for the lighter GC on the {left}. We chose not to go to excess lengths to optimize Eq.~(\ref{eq:tDFsemian}) because, again, GC-GC interactions (that are fully captured in the semi-analytic code) will modify the dynamics anyway, and become dominant once GCs arrive at the inner halo where the core stalling kicks-in. 
%

\subsection{Single GC tests for realistic UDG halo models}\label{ss:calibudg}
\begin{figure}[hbp!]
\centering
                  \includegraphics[scale=0.4]{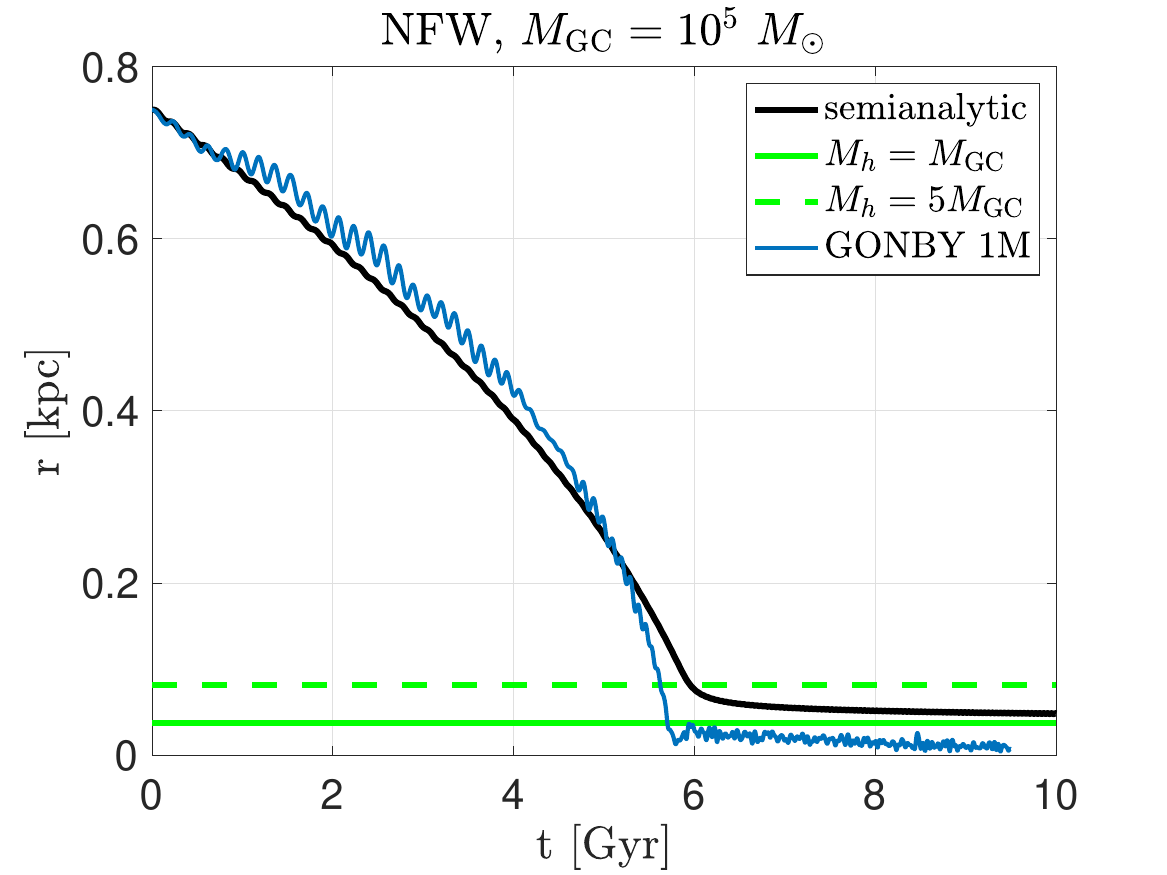}
            \includegraphics[scale=0.4]{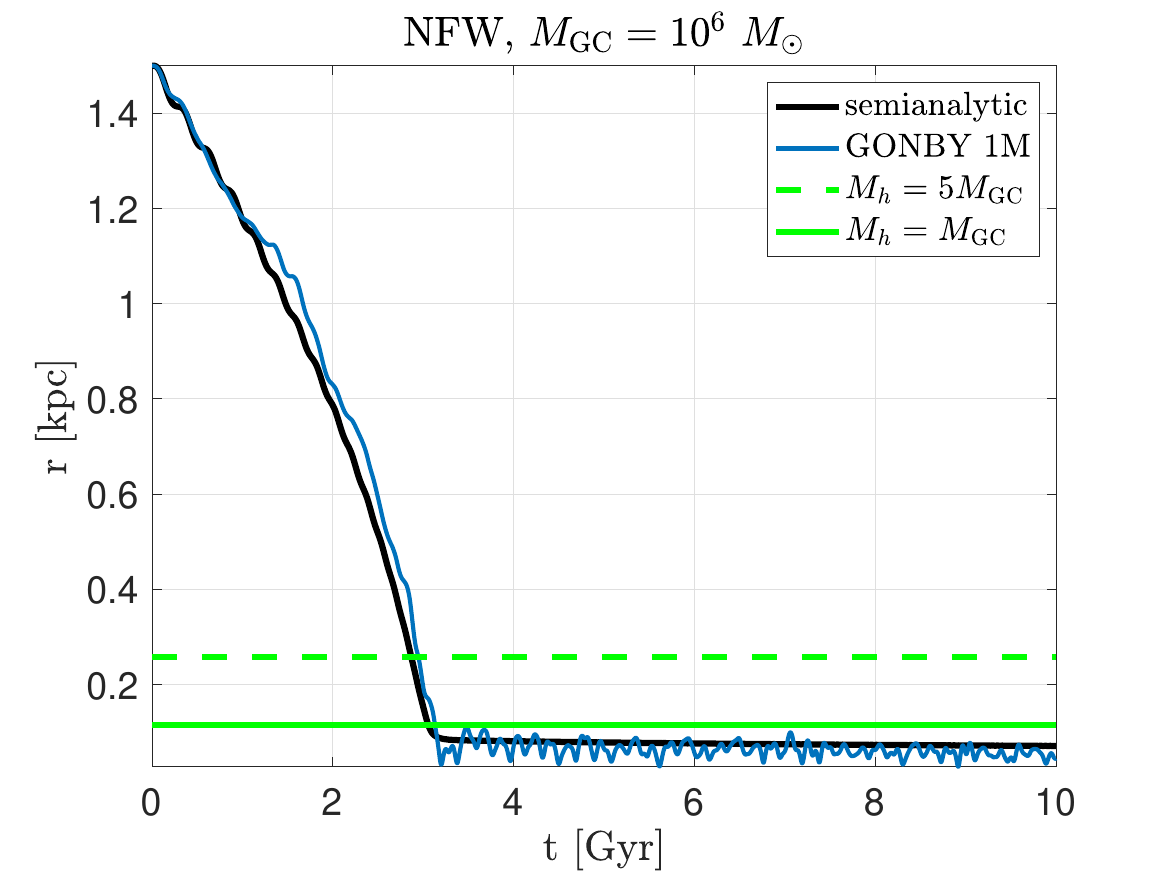}\\
                  \includegraphics[scale=0.4]{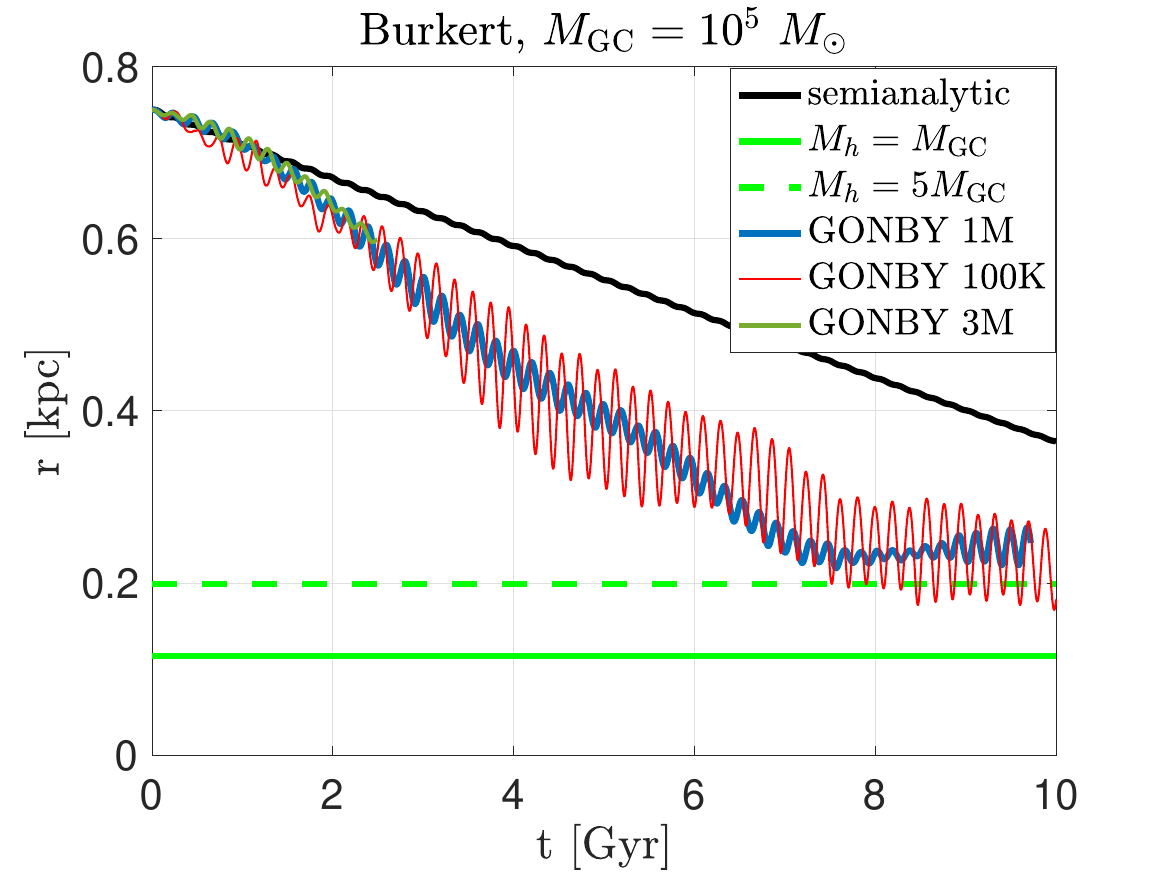}
            \includegraphics[scale=0.4]{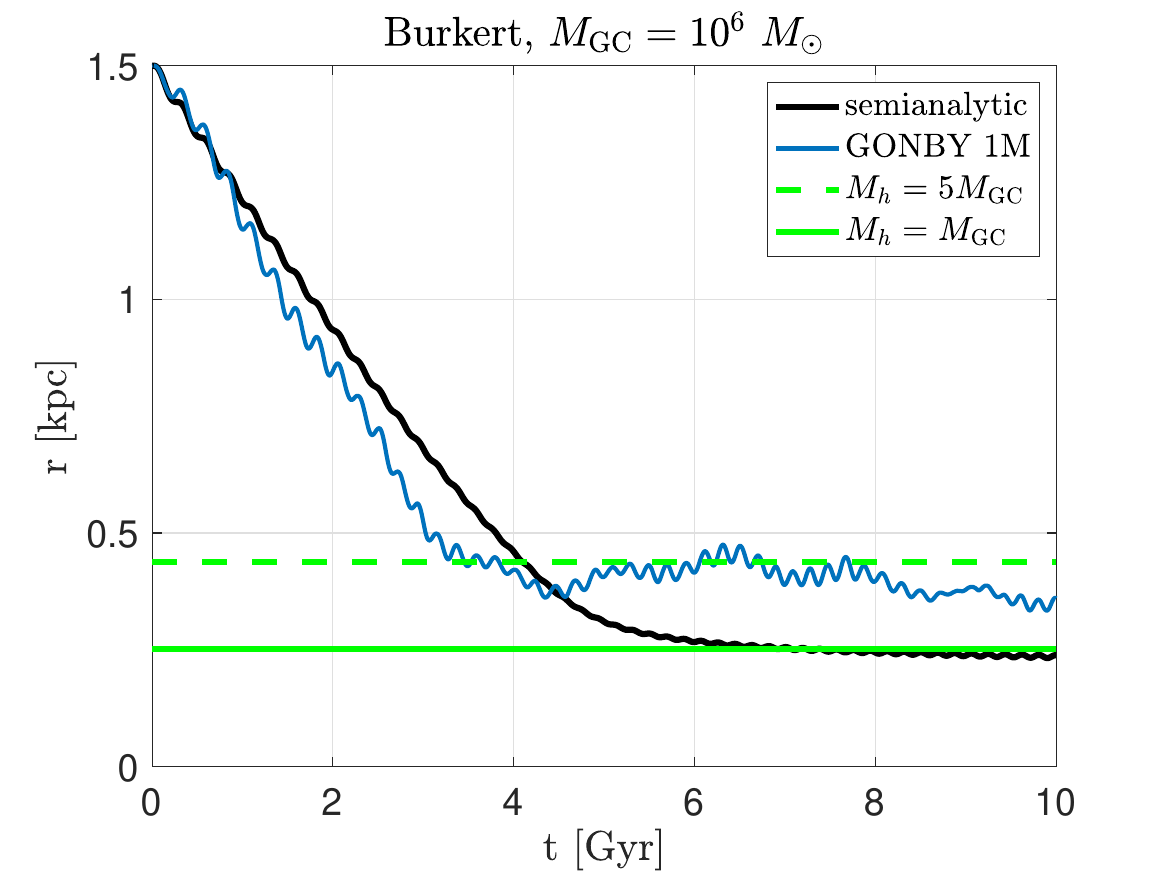}\\
                  \includegraphics[scale=0.4]{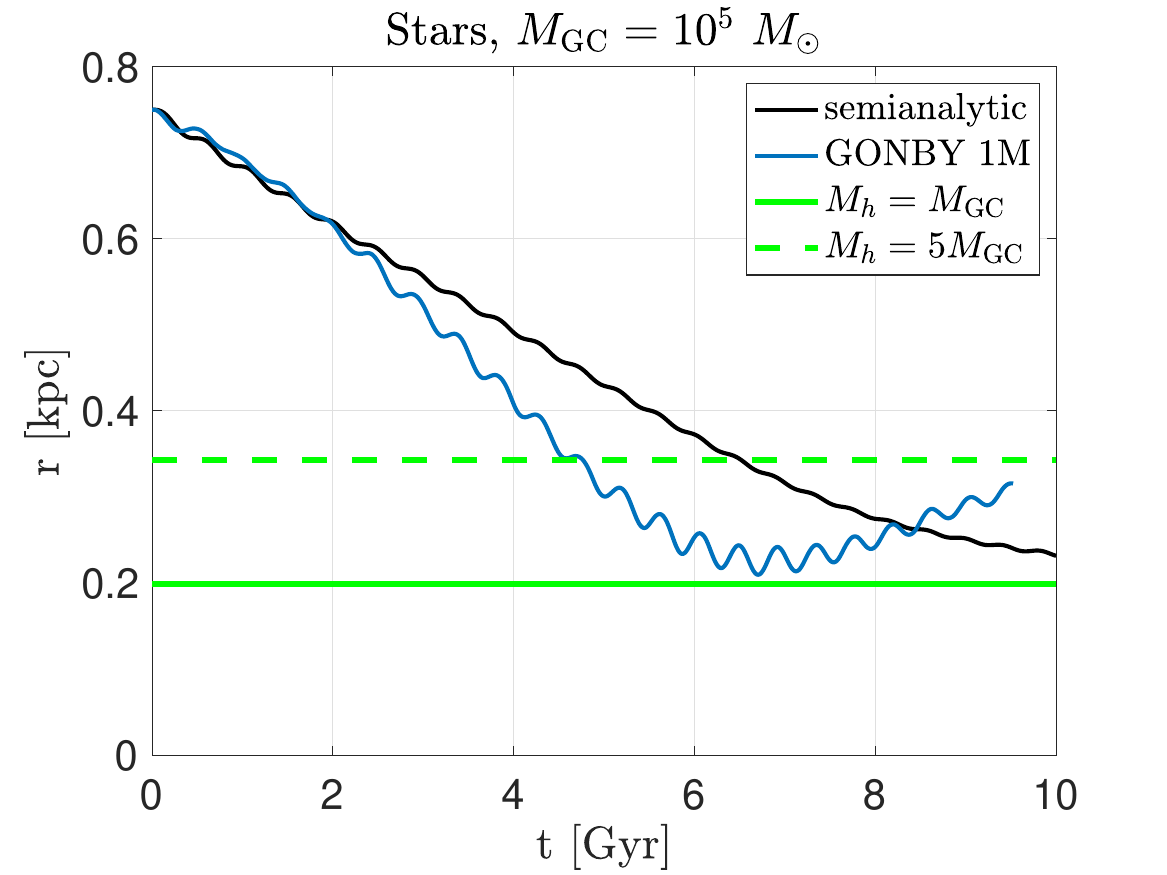}
            \includegraphics[scale=0.4]{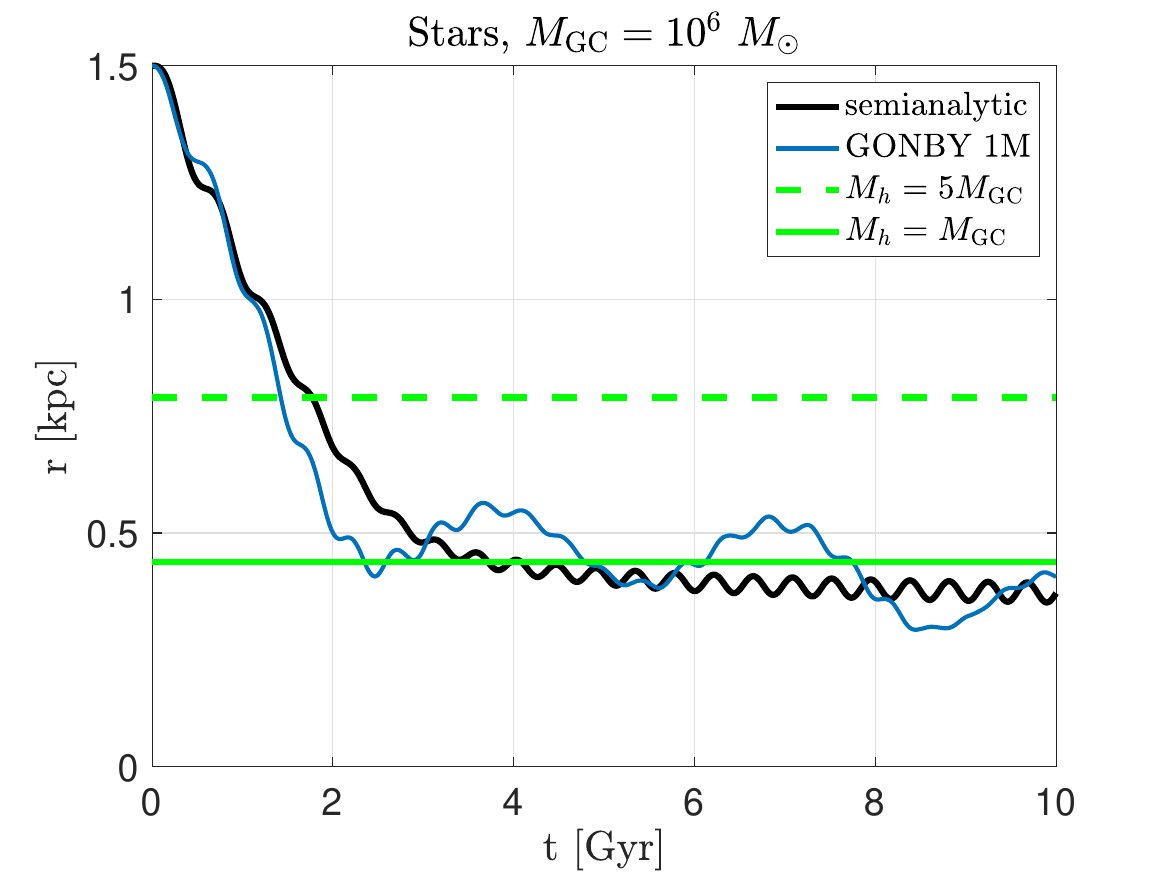}            
 \caption{NFW (top), Burkert (middle), and stars-only (bottom) halo, GONBY simulations (blue, red, and green lines) vs. semianalytical calculation (black line). Horizontal green dashed (solid) lines mark the radius where the enclosed halo mass becomes equal to 5 (1) times the GC mass. Left: $M_{\rm GC}=10^5~M_\odot$. Right: $M_{\rm GC}=10^6~M_\odot$.
 }
 \label{fig:CS}
\end{figure}

We continue to halo models designed to mimic UDG1. 
The first model is a cusp NFW halo (Eq.~(\ref{eq:nfw})) 
with $\rho_s=2\times10^6~M_\odot/{\rm kpc^3}$ and $R_s=6$~kpc. Results are shown in the top panels of Fig.~\ref{fig:CS}. Although the NFW halo has no core, we apply the same ``core stalling" prescription of Eq.~(\ref{eq:tDFsemian}). Roughly, in the cusp NFW case, this prescription halts DF when the GC mass becomes comparable to the halo mass contained in the GC orbit. In this situation we expect that central halo particles participate in effective two-body dynamics with the incoming GC. This will tidally disrupt the cusp in a region with scale radius of order $r\sim r_p(M_{\rm GC})$. Fig.~\ref{fig:CS} shows that the semianalytic calculation reproduces the N-body result down to $r\sim r_p$.

The second model is a core Burkert profile (Eq.~(\ref{eq:burk})) with $R_0=2$~kpc and $\rho_0=1.7\times10^7~M_\odot/{\rm kpc}^3$. Results are shown in the middle panels of Fig.~\ref{fig:CS}. The simulations are in rough agreement with the semianalytic prediction with some departure in the core stalling region. Focusing on $M_{\rm GC}=10^5~M_\odot$ (left panel), we made a series of convergence tests to the N-body simulation, including variation of $N$; softening radius; $R_{\rm max}$; tree code opening angle $\theta$; and integration time step, by factors of a few in each case. All tests show convergence, and we include some examples in the plot. 

The third model mimics a case in which UDG1 has only stars and no DM. We model the stellar density with a Burkert profile\footnote{Here we used Burkert rather than S\'ersic profile to model the stars; in the main text, the stellar body is modeled by S\'ersic as explained in Sec.~\ref{ss:dmhalo}.} with $R_0=2$~kpc (same as for the DM Burkert model). A mass-to-light ratio of $M/L=2$ in solar units for field stars gives $\rho_0\approx0.34\times10^7~M_\odot/{\rm kpc^3}$, a factor of 5 lower than the Burkert DM model halo mass density. We show the results for this model in the bottom panels of Fig.~\ref{fig:CS}. 

In each panel of Fig.~\ref{fig:CS}, solid (dashed) horizontal green line marks the radius at which the contained halo mass is equal to the GC mass (five times the GC mass). Another scale to keep in mind is the stellar S\'ersic radius of UDG1, $R_{\rm e}\approx2$~kpc. In all of the cases we explore, the semianalytic implementation of DF provides a good approximation to the N-body result at $r\gtrsim0.5R_{\rm e}$. This point guides us when we define observables to constrain DM models of the halo.

\subsection{Multiple GC tests for realistic UDG halo models}\label{ss:calibudg2}
\begin{figure*}
\centering
      \includegraphics[ scale= 0.38]{UDG1_core_fDM0_fMmx1_fReGC1_LCDFnorm_cx}
      \includegraphics[ scale= 0.44]{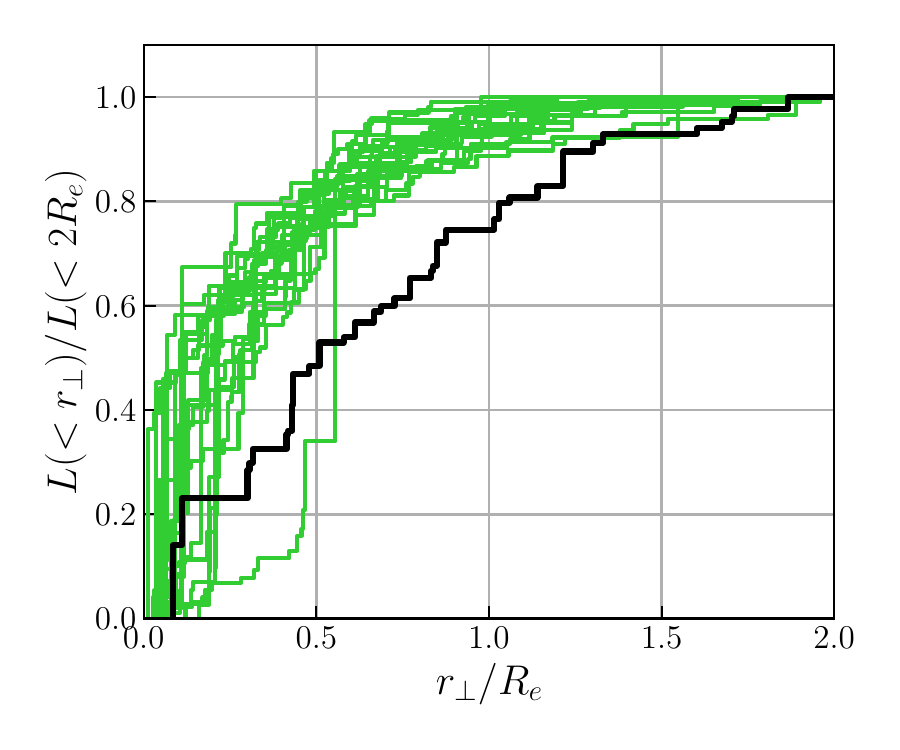} \\             
            \includegraphics[ scale= 0.38]{UDG1_core_fDM2_fMmx1_fReGC1_LCDFnorm_cx}
      \includegraphics[ scale= 0.44]{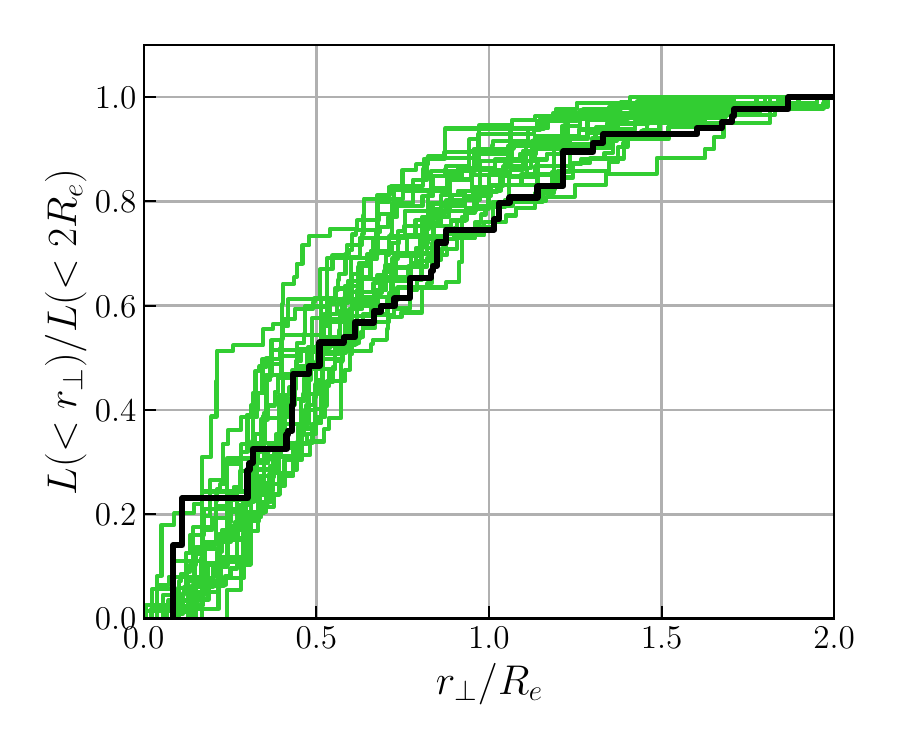}
 \caption{Comparison of results from semianalytic GC orbit integration and N-body tree code simulations. {\bf Top} panels: DM-free model fDM0, semianalytic ({\bf left} panel, same as top-left panel of Fig.~\ref{fig:Burk_LCDFs}) vs. GONBY ({\bf right} panel). {\bf Bottom} panels: fDM2 model, semianalytic ({\bf left} panel, same as top-third from left panel of Fig.~\ref{fig:Burk_LCDFs}) vs. GONBY ({\bf right} panel).
 }
 \label{fig:GONBYtest}
\end{figure*}

In Fig.~\ref{fig:GONBYtest} we compare tree code simulations ({\bf right} panel) to the semianalytic computation ({\bf left} panel). The {\bf top} panels show the DM-free scenario fDM0. The GONBY run used 100K star particles and $R_{\rm sim}=2$~kpc, with mass-per-particle $m\approx300$~M$_\odot$. We find good agreement between the semianalytic and tree code calculations in the region $r_\perp\geq0.5R_{\rm e}$, and both methods would assign ${\bf \color{red}X}$ at $r_\perp=0.5R_{\rm e}$ and at $r_\perp=1R_{\rm e}$. One out of 20 GONBY runs produced a luminosity CDF that falls below UDG1 data at $r_\perp=0.5R_{\rm e}$, and all 20 GONBY runs live above UDG1 data at $r_\perp=1R_{\rm e}$. (We can trace the low-CDF run to a random projection effect that brought a single, exceptionally massive, large-$r$ GC to sit at small $r_\perp$. Such a situation could equally well happen in the semianalytic code.) The {\bf bottom} panels of Fig.~\ref{fig:GONBYtest} show the core DM model fDM2. This model shows much better consistency with the data, and good agreement between the semianalytic and tree calculations for $r_\perp\lesssim1R_{\rm e}$. Around $r_\perp\approx R_{\rm e}$ the tree code CDF has lower spread than the semianalytic result, and tends to overshoot the data, which would earn it an ${\bf \color{red}X}$ at that point, compared with the semianalytic result which is a ${\bf \color{blue}\surd}$. Since the tree code overshoots the observed CDF, DF is slightly {\it too strong} in the model. This means that the model needs some {\it more} DM when evaluated in the N-body simulation than it does with the semianalytic approximation, so results based on the semianalytic code are slightly conservative. As seen in the previous paragraph of this appendix, the physical reason for the tree code vs. semianalytic difference is super-Chandrasekhar DF in the former (see Fig.~\ref{fig:CS}, middle-left panel), that occurs prior to the onset of core stalling. We conclude, from Fig.~\ref{fig:GONBYtest}, that detailed features of the luminosity CDF require careful simulations and could be sensitive to details of the halo model. However, up to minor adjustments, the broad trends are seen correctly in the semianalytic calculations, and we focus on these in the rest of the main text. 
%

\section{Comment on the Chandrasekhar coefficient}\label{a:chandra}

In our semianalytic analysis, we calculate $C_{\rm DF}$ directly from the distribution function using Eq.~(\ref{eq:ChandraC}). For reference, for a Maxwellian distribution $f_v(v)=\frac{1}{\left(2\pi\sigma^2\right)^{\frac{3}{2}}}e^{-\frac{v^2}{2\sigma^2}}$ one finds
\be C_{\rm DF}&=&{\rm erf}(X)-\frac{2X}{\sqrt{\pi}}e^{-X^2},\;\;\;\;\;\;X=\frac{V}{\sqrt{2}\sigma},\;\;\;\;\;\;\left({\rm Maxwellian\;\;gas}\right).\ee
When the probe's velocity $V$ is much larger than the 3D velocity dispersion of the gas, $V\gg\sigma$, we have $C_{\rm DF}\to1$. When the probe is slow, $V\ll\sigma$, we have $C_{\rm DF}\to\frac{\sqrt{2}}{3\sqrt{\pi}}\frac{V^3}{\sigma^3}$. In the latter limit we have
\be{\tau}_{\rm DF}&\to&\frac{3\sigma^3}{\sqrt{32\pi} G^2M_{\rm GC}\rho \,\ln\Lambda},\;\;\;\;\;\;\left(V\ll\sigma,\;\;\;{\rm Maxwellian\;\;gas}\right).\ee

\begin{figure}[hbp!]
\centering
       \includegraphics[scale=0.4]{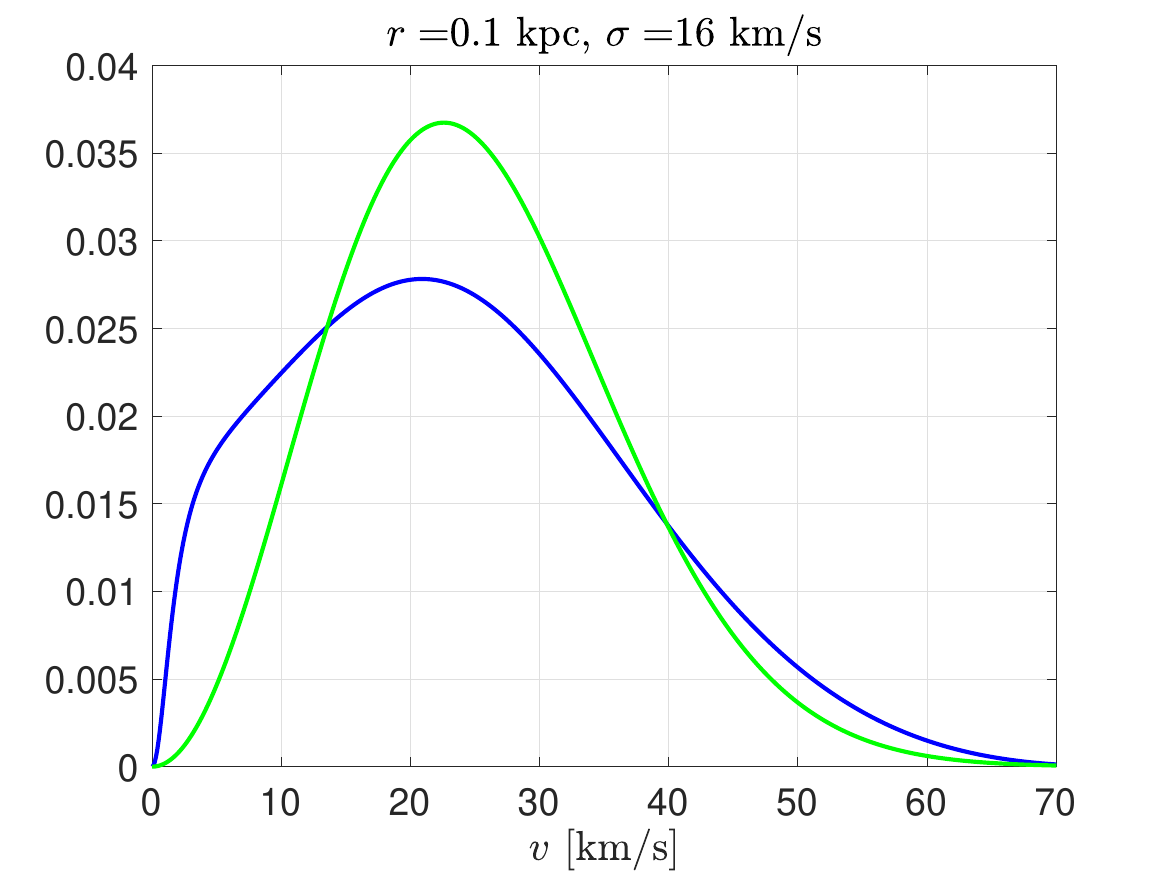} 
       \includegraphics[scale=0.4]{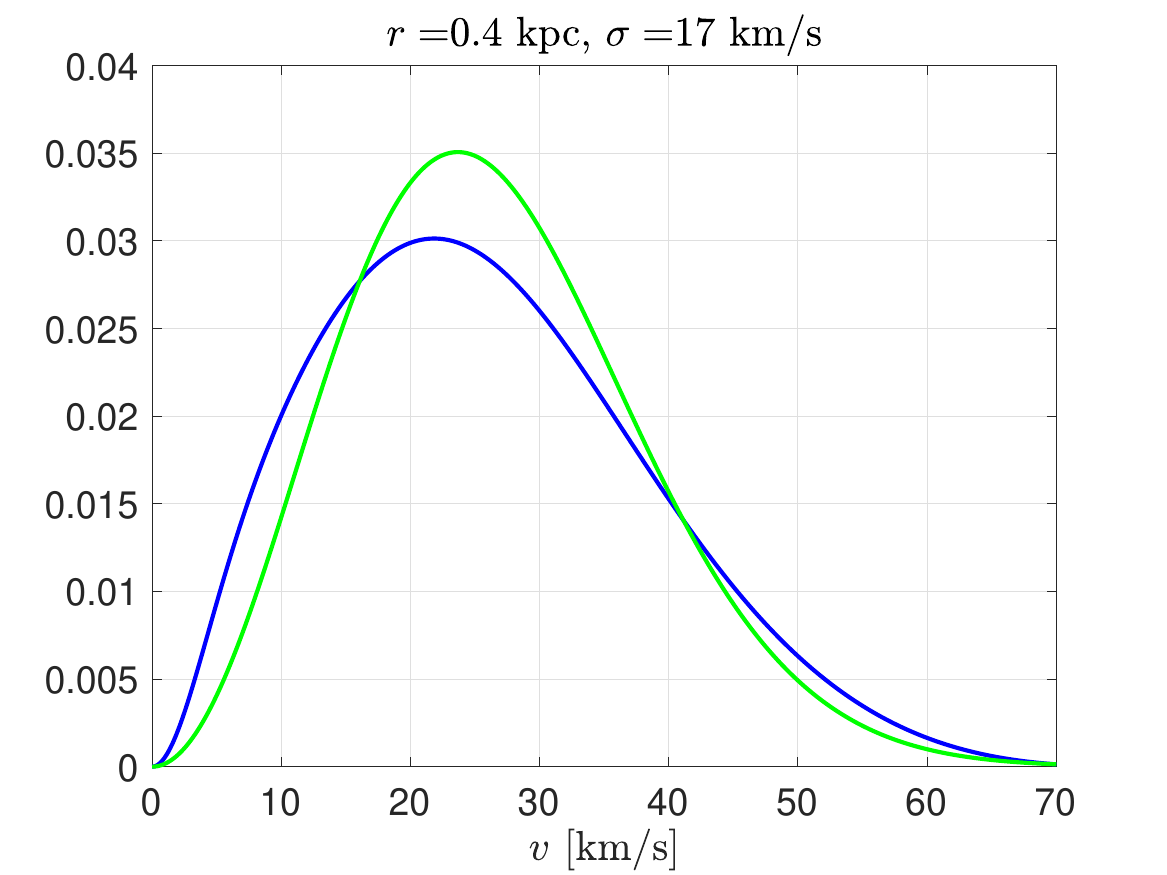} \\
                     \includegraphics[scale=0.4]{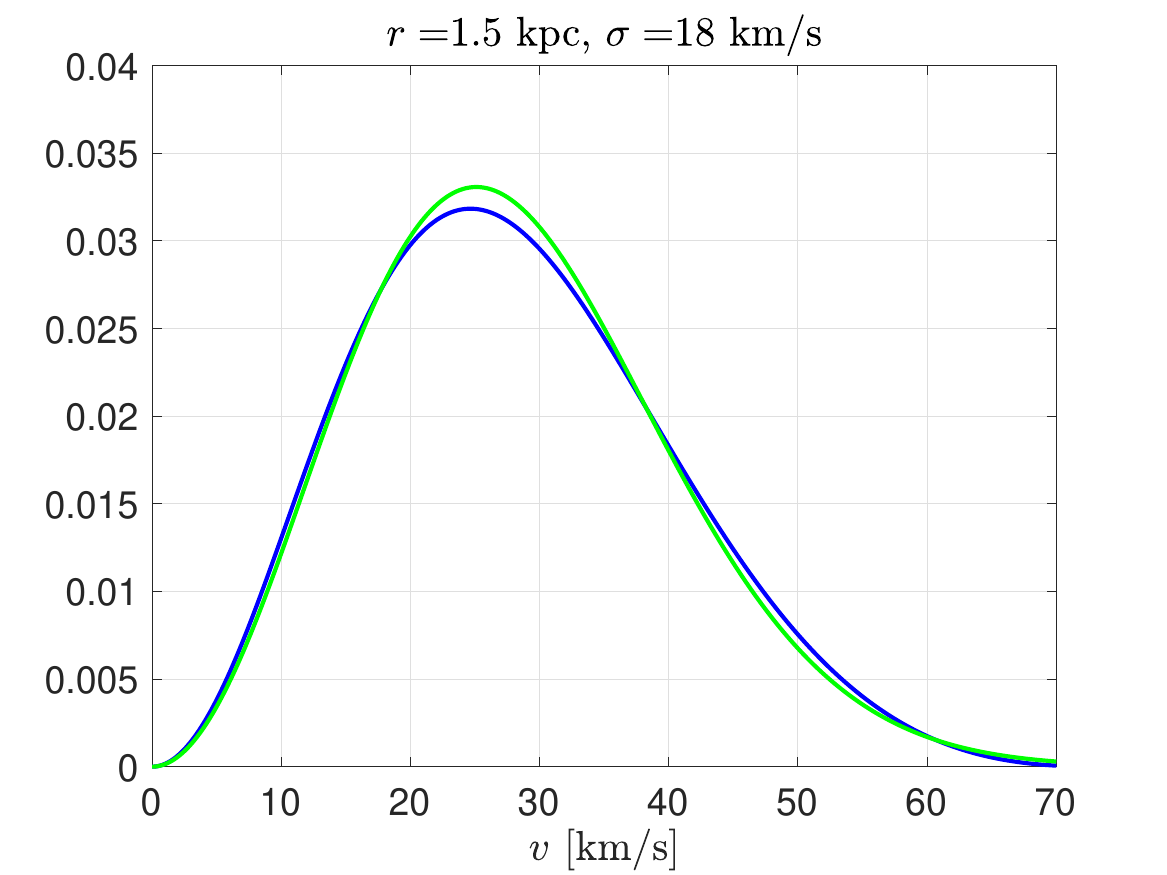} 
                                   \includegraphics[scale=0.4]{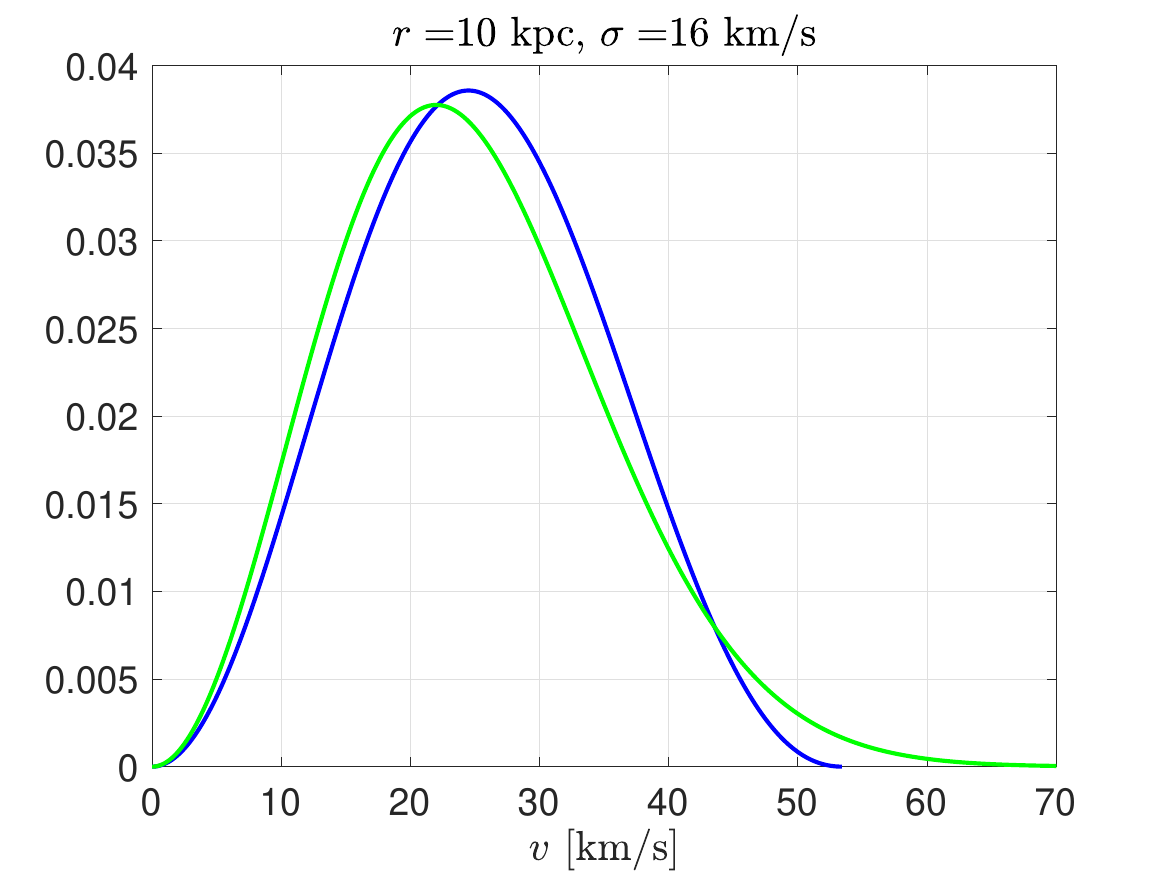} \\
                     \includegraphics[scale=0.4]{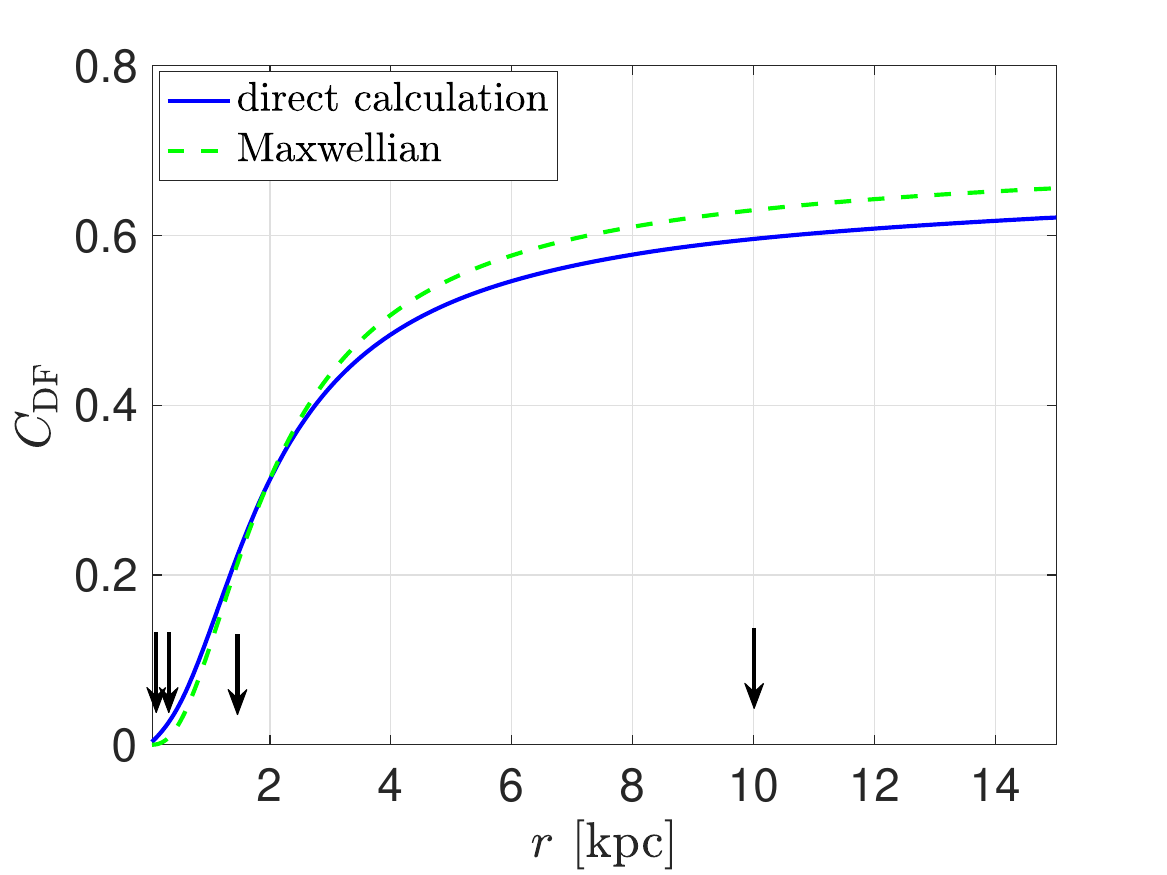} 
                     \includegraphics[scale=0.4]{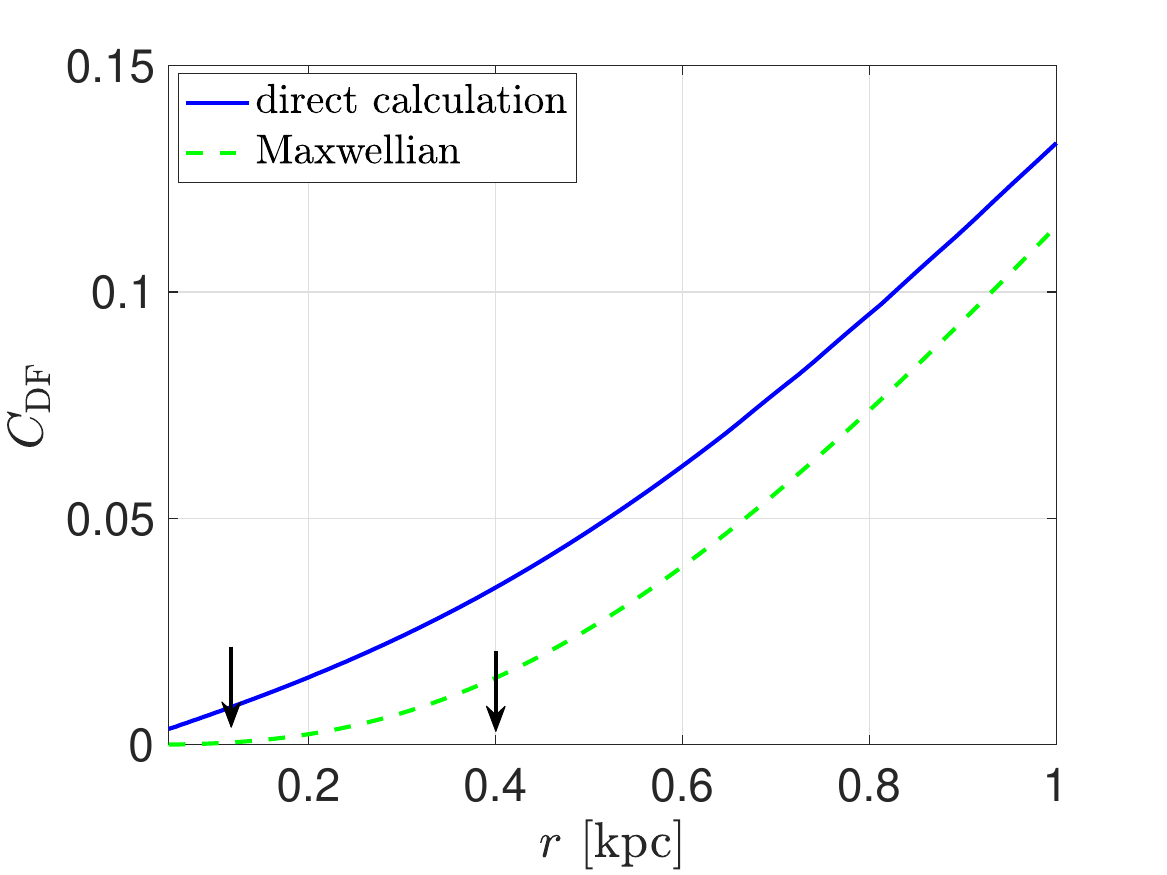} 
                      \caption{Comparison of the Maxwellian approximation to a direct numerical calculation of the velocity distribution function, done for the Burkert halo model. {Top four panels:} velocity distributions obtained at different points in the halo. {Bottom panels:}  $C_{\rm DF}$; the right panel zooms in on small-$r$. Arrows show approximate positions of the top panels.
 }
 \label{fig:phasespace}
\end{figure}

It is interesting to compare the Maxwellian approximation to a numerical calculation of the stationary distribution function and resulting $C_{\rm DF}$. We show an example in Fig.~\ref{fig:phasespace}, using the Burkert model (Eq.~(\ref{eq:burk})) with $\rho_0=1.7\times10^{-2}~M_\odot/{\rm pc^3}$ and $R_0=2$~kpc. The Maxwellian distribution is computed using the equivalent local value of $\sigma$, obtained at each point from the velocity distribution via $3\sigma^2=4\pi\int_0^\infty dvv^4f(\varepsilon(r,v))$. At small $r$, the numerical velocity distribution function develops a ``bump" of slow-moving particles, absent in the Maxwellian. This excess of slow particles causes DF to be more efficient than would be expected if one replaced the integral in $C_{\rm DF}$ by the analytic Maxwellian result. This is seen in the bottom panels, with the full $C_{\rm DF}$ exceeding the Maxwellian prediction at small $r$. The bottom-right panel is a zoom-in version of the bottom-left. For this plot we define $C_{\rm DF}$ by equating the GC velocity to the circular velocity (the same procedure as in \cite{Inoue:2009wd}). Note that in semianalytic calculations in the main body of the paper, we keep the actual GC velocity $V$ in the definition of $C_{\rm DF}$, without assuming a circular orbit.

\cite{Inoue:2009wd} used the Maxwell distribution to initialize the N-body halo, rather than solving Eq.~(\ref{eq:f}). Fig.~\ref{fig:phasespace} suggests that the details of the phase space distribution can make a significant impact on DF, particularly in the inner halo as relevant for core stalling. We expect that this issue is at the basis of the numerical difference between our results and those of~\cite{Inoue:2009wd}, discussed in Sec~\ref{ss:Inoue}.

\section{Additional checks}\label{a:checks}

\subsection{Single GC and multiple GC tests: comparison of full N-body simulations and semianalytic results}
\begin{figure}[hbp!]
\centering
      \includegraphics[scale=0.3]{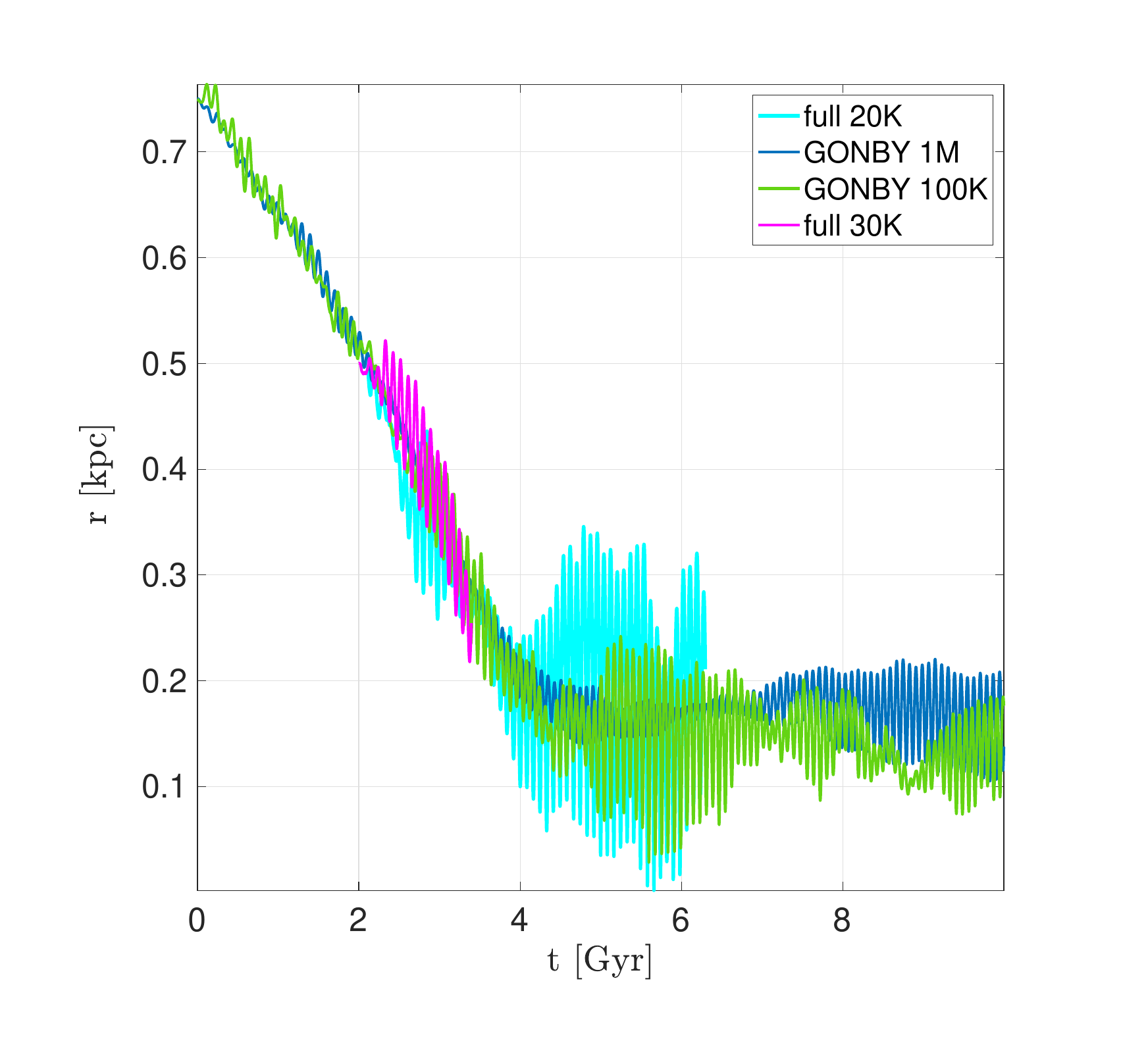}
 \caption{Comparison between full N-body (''full") and tree code (''GONBY") results. The halo set-up is as in Fig.~\ref{fig:Inoue}.
 }
 \label{fig:InoueFullTree}
\end{figure}

\begin{figure}[hbp!]
\centering
      \includegraphics[scale=0.45]{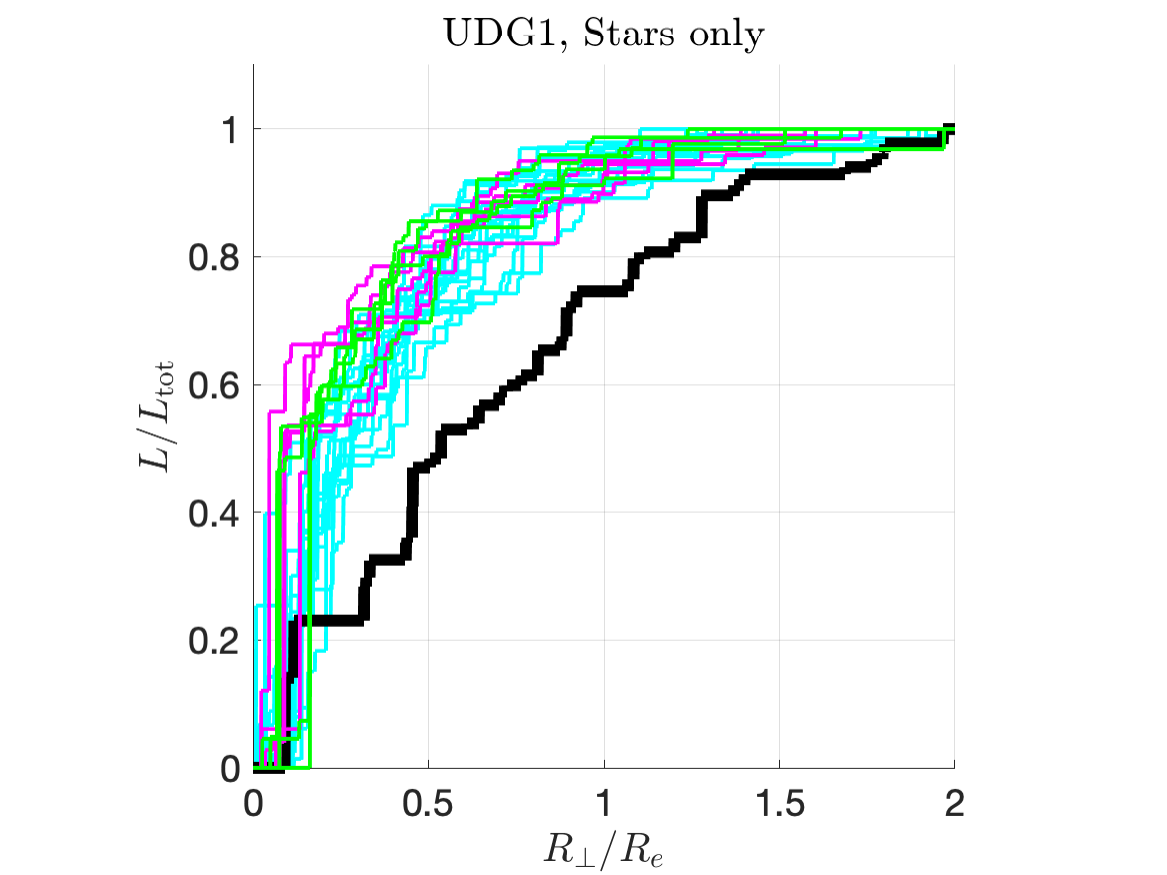} 
       \includegraphics[scale=0.45]{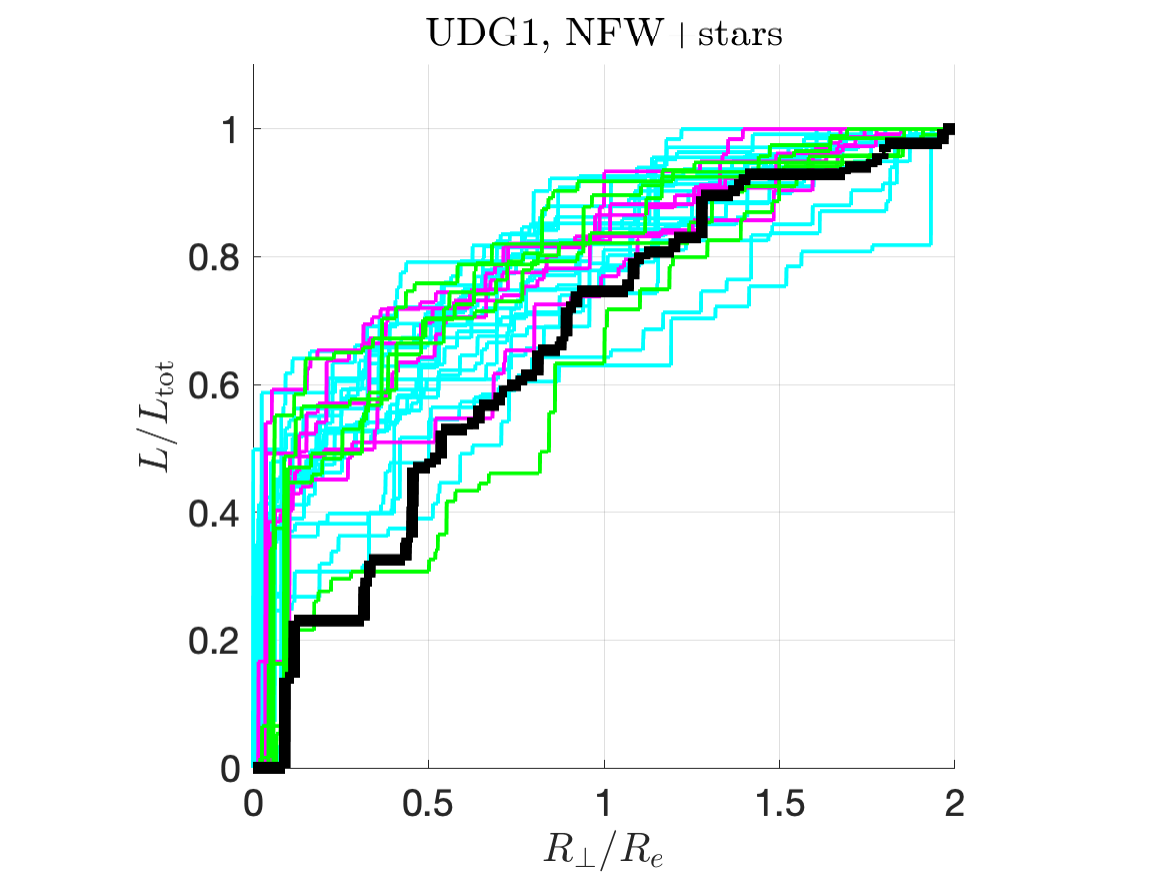} 
        \includegraphics[scale=0.45]{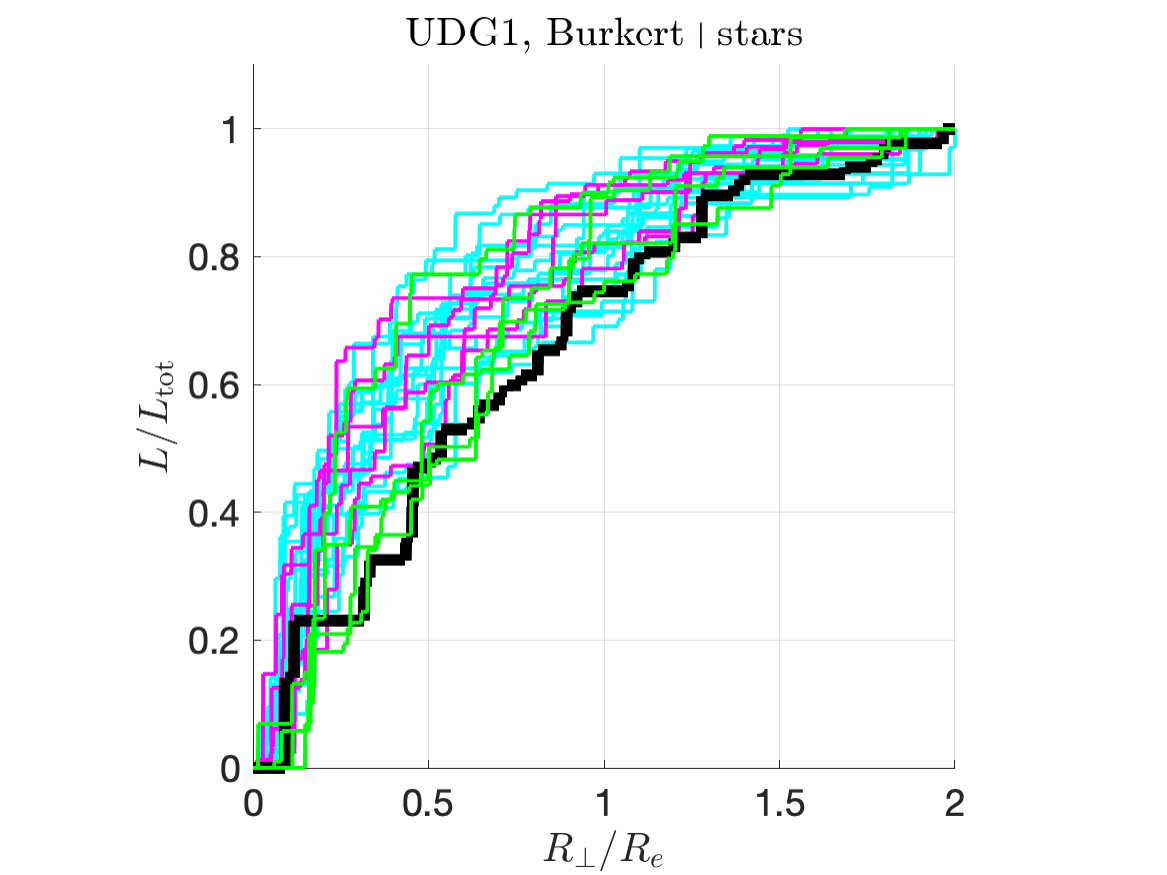}
 \caption{Luminosity CDF of GCs. Thick black: observed in UDG1. Cyan: semianalytic calculation. Purple (green): full N-body simulations with $R_{\rm sim}=1$~kpc and 5K particles ($R_{\rm sim}=2$~kpc and 10K particles). Evolution duration: 10~Gyr. The GCIMF in the simulations is identical to the currently observed GCMF.  
 }
 \label{fig:UDG1_Lgc}
\end{figure}

Comparison of full N-body and tree code simulations: Fig.~\ref{fig:InoueFullTree} repeats the same setup as in the {left panel} of Fig.~\ref{fig:Inoue}, comparing full N-body to tree code results in the orbital evolution of a single GC in a UDG1-like halo.
Next, we compare full N-body results to semianalytic calculations for a scenario with multiple GCs and a UDG1-like halo. For the purpose of this exercise we disable GC mass loss, and set the GCIMF to match the currently observed GCMF. 
Fig.~\ref{fig:UDG1_Lgc} shows GC luminosity CDF. Cyan, purple, and green curves show results of the semianalytic method, full N-body with 5K halo particles, and full N-body with 10K halo particles, respectively. Thick black curve shows the observed distribution. The three panels correspond to the NFW DM halo model, the Burkert DM halo model, and a model containing only the stellar body\footnote{The halo models used in this exercise are slightly different from the models used in the main paper; roughly, they correspond to fDM0 (stars only), cusp fDM5 (NFW+stars), and core fDM1 (Burkert+stars).}. We find good agreement between the (relatively low resolution)  N-body calculations and the semianalytic calculation. Some discrepancy can be seen in the star-only halo: here, the N-body simulations predict a luminosity CDF that is more contracted at small $r_\perp$ compared to the semianalytic prediction. This suggests that the semianalytic constraints on the DM content of the galaxy are slightly conservative.  
%

\subsection{Sensitivity of the results to GC merger prescription}\label{a:nomerge}
\begin{figure}[hbp!]
\centering
      \includegraphics[ scale= 0.4]{UDG1_cusp_fDM10_fMmx1_fReGC1_LCDFnorm_cx} 
       \includegraphics[ scale= 0.4]{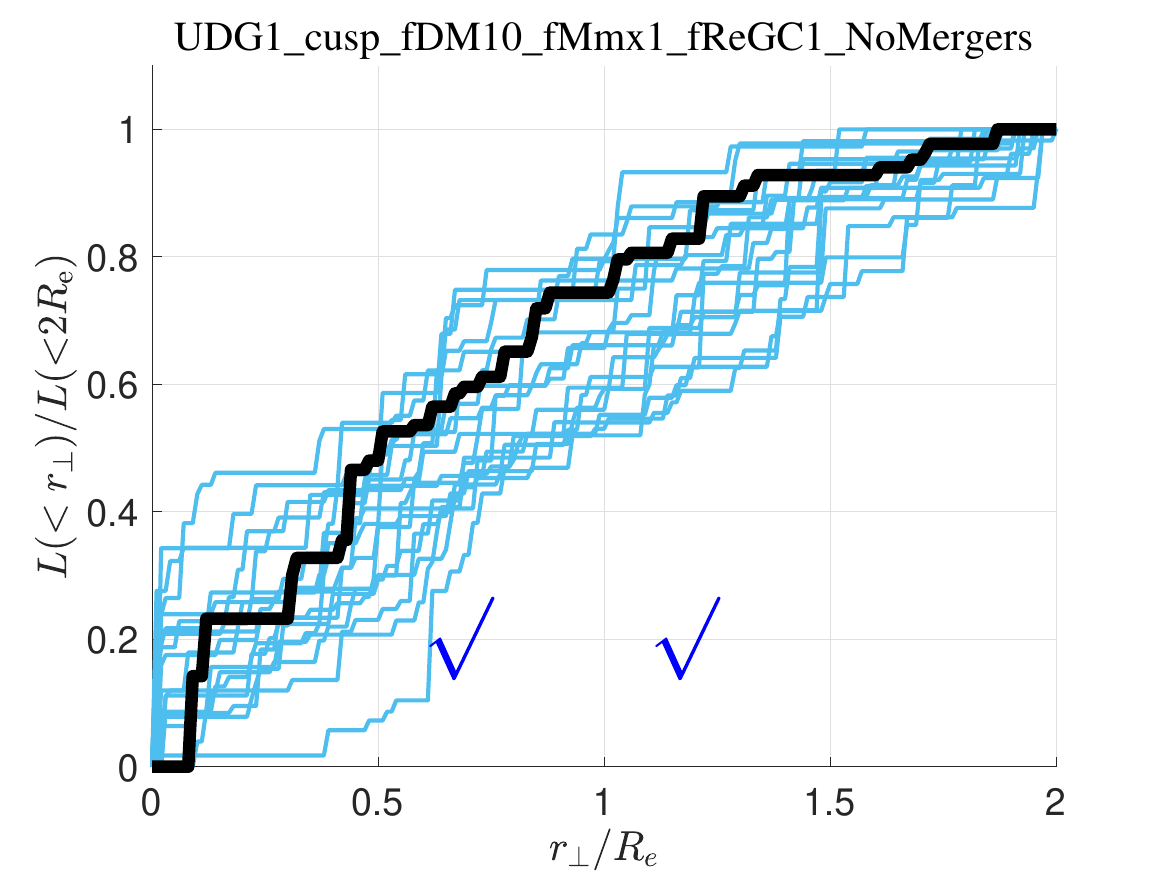}    \\
             \includegraphics[ scale= 0.4]{UDG1_cusp_fDM2_fMmx1_fReGC1_LCDFnorm_cx} 
       \includegraphics[ scale= 0.4]{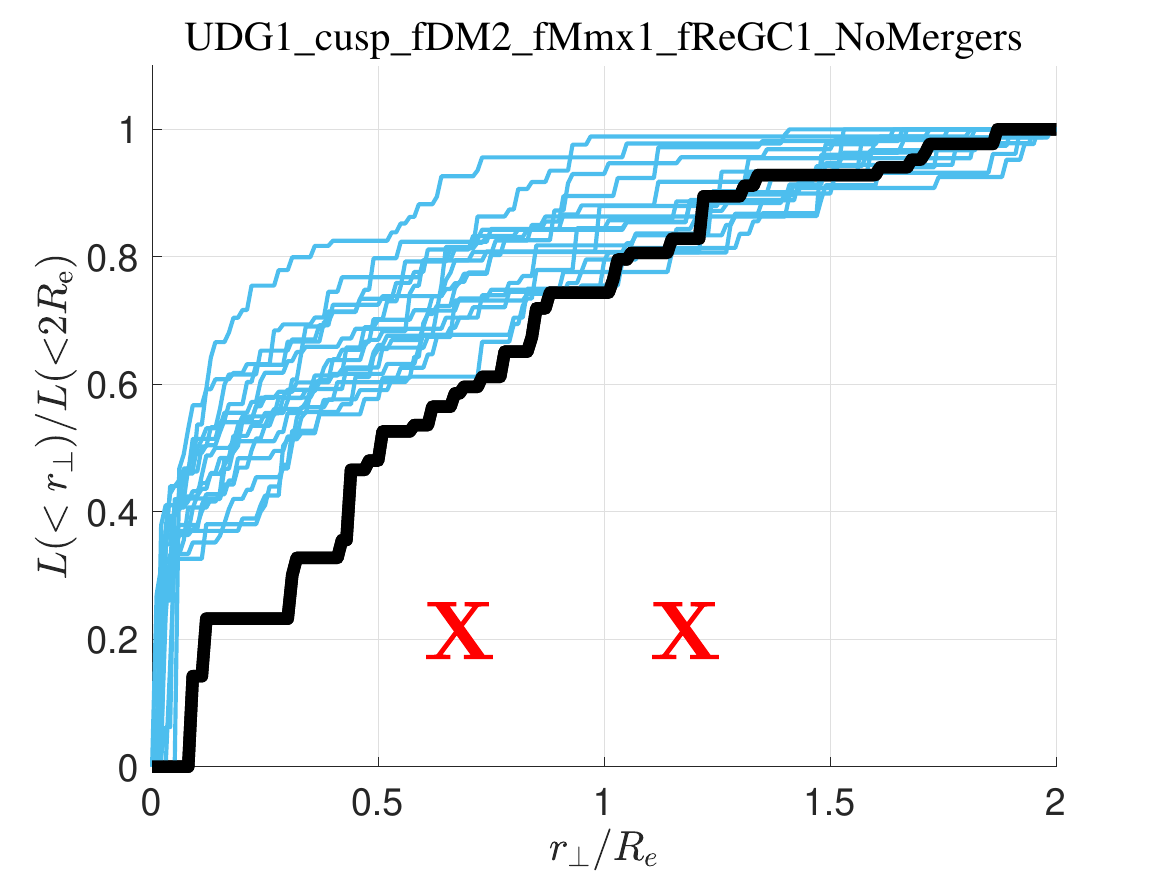} 
\caption{Sensitivity to GC merger prescription. Left panels: baseline prescription used in the body of the paper. Right panels: GC mergers disabled.
 }
 \label{fig:mergers}
\end{figure}

Our GC merger prescription is a rough approximation of a complex process. As a check of the sensitivity of our results to the implementation of mergers, Fig.~\ref{fig:mergers} compares simulations using the baseline prescriptions to simulations in which GC mergers were disabled altogether. We focus on NFW cusp models of UDG1, because these models maximize the merger probability. As can be expected, disabling mergers changes the luminosity CDF near $r_\perp\approx0$ and, in particular, removes the NSCs that are otherwise a generic prediction of the NFW model. However, the CDF at $r_\perp\gtrsim0.5R_{\rm e}$ is affected only marginally, and our consistency conditions at $r_\perp=0.5R_{\rm e}$ and at $r_\perp=R_{\rm e}$ yield the same result with or without mergers.
%

\subsection{Sensitivity of the results to GC mass loss prescription}\label{a:massloss}
\begin{figure}[hbp!]
\centering
      \includegraphics[ scale= 0.24]{UDG1_core_fDM0_fMmx1_fReGC1_LCDFnorm_cx} 
      \includegraphics[ scale= 0.24]{UDG1_core_fDM1_fMmx1_fReGC1_LCDFnorm_cx}       
      \includegraphics[ scale= 0.24]{UDG1_core_fDM2_fMmx1_fReGC1_LCDFnorm_cx}       
      \includegraphics[ scale= 0.24]{UDG1_core_fDM10_fMmx1_fReGC1_LCDFnorm_cx}     \\  
      \includegraphics[ scale= 0.24]{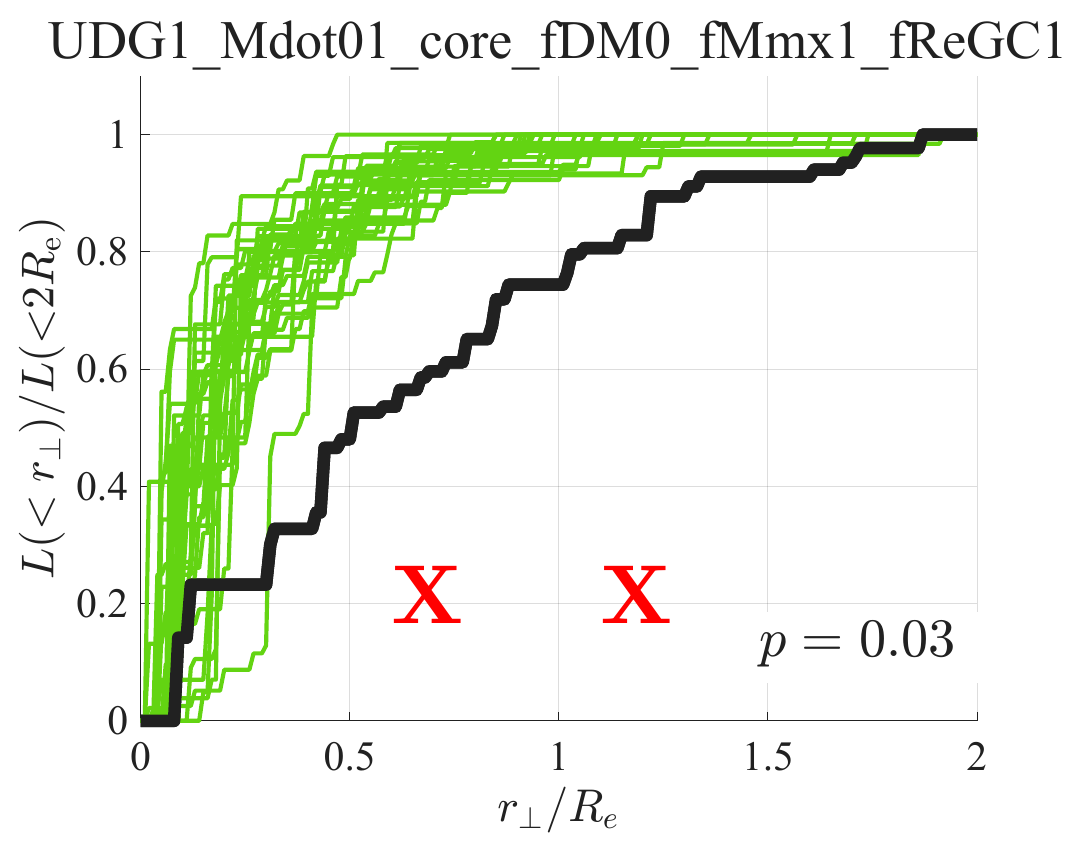} 
      \includegraphics[ scale= 0.24]{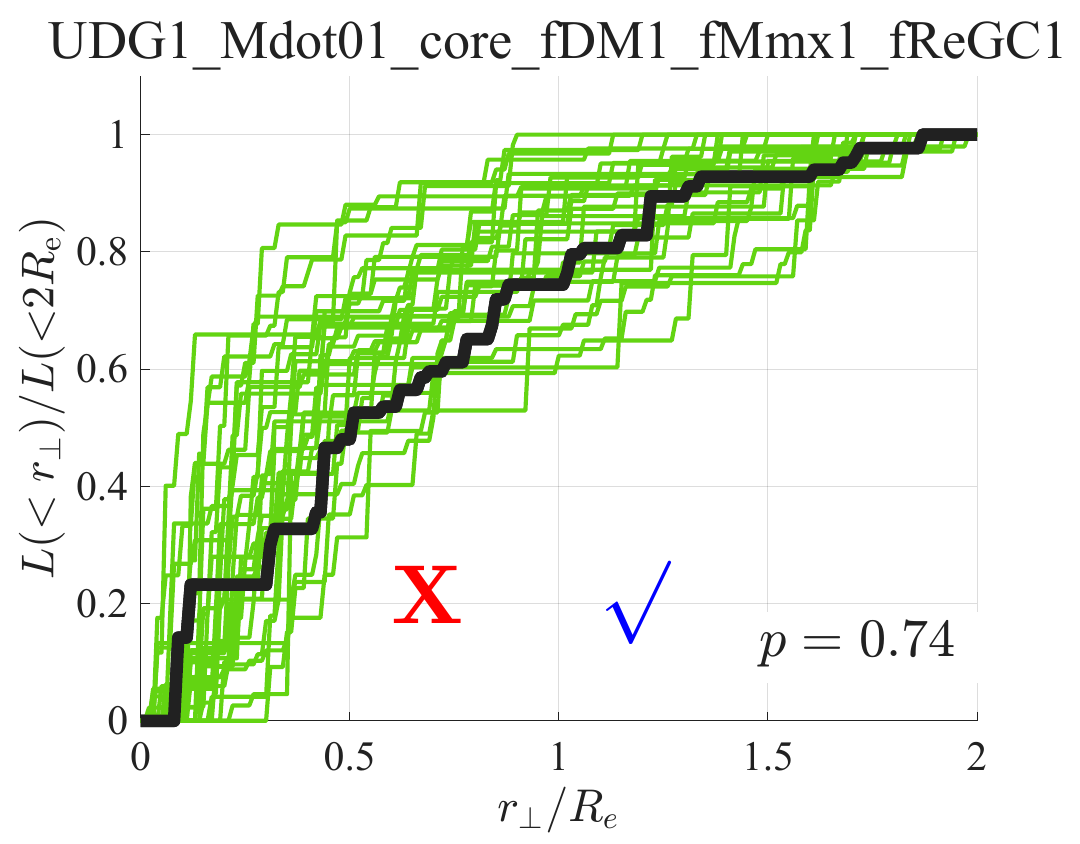}       
      \includegraphics[ scale= 0.24]{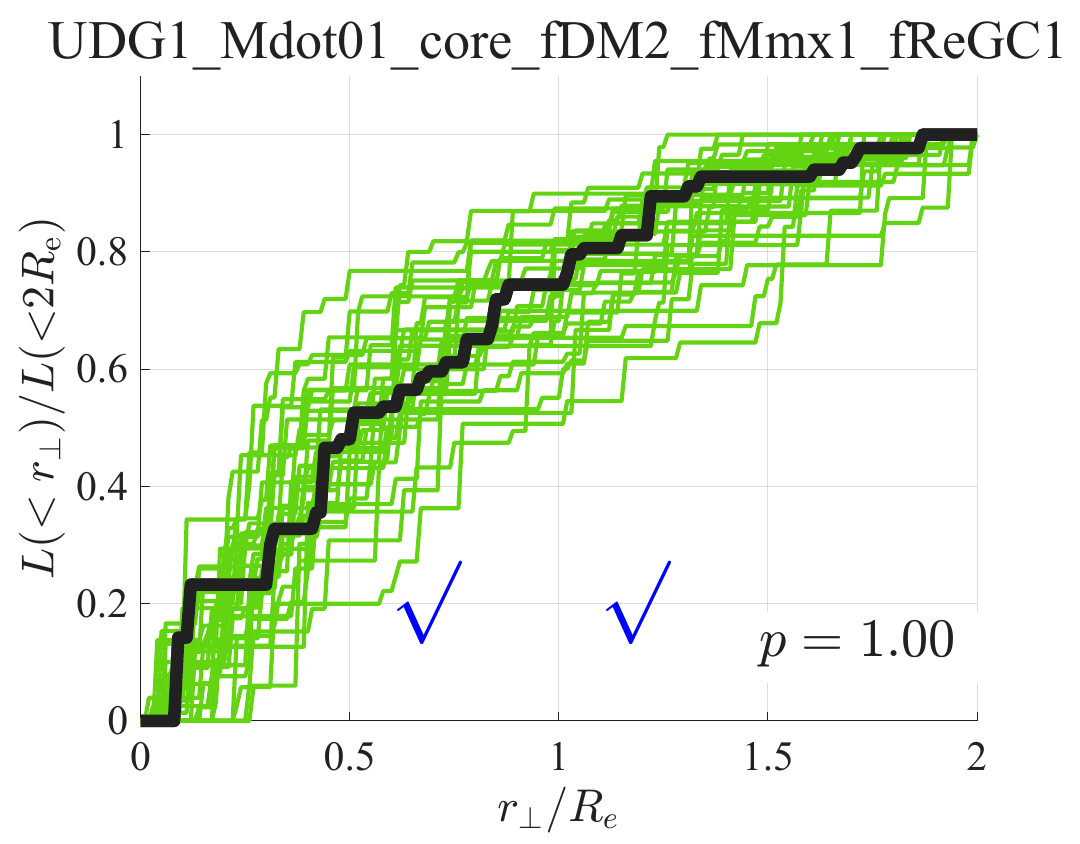}       
      \includegraphics[ scale= 0.24]{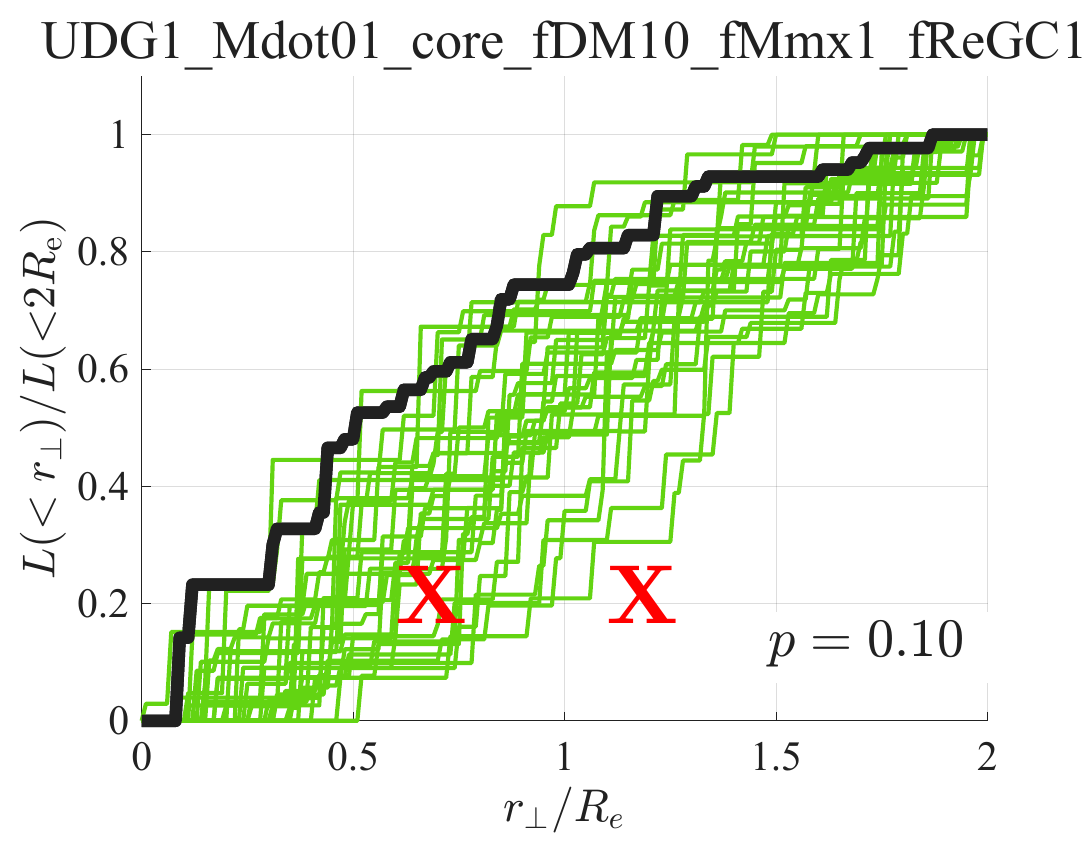}             
 \caption{Luminosity CDF for core models: sensitivity to GC mass loss prescription. GC initial radial distribution same as stars. Top panels: fiducial mass loss rate. Bottom panels: same models, with mass loss rate in isolation lower by a factor of ten. 
 }
 \label{fig:Burk_LCDFs_Mdot01}
\end{figure}

\begin{figure}[hbp!]
\centering
      \includegraphics[ scale= 0.24]{UDG1_cusp_fDM1_fMmx1_fReGC1_LCDFnorm_cx} 
      \includegraphics[ scale= 0.24]{UDG1_cusp_fDM2_fMmx1_fReGC1_LCDFnorm_cx}       
      \includegraphics[ scale= 0.24]{UDG1_cusp_fDM5_fMmx1_fReGC1_LCDFnorm_cx}       
      \includegraphics[ scale= 0.24]{UDG1_cusp_fDM10_fMmx1_fReGC1_LCDFnorm_cx}     \\  
      \includegraphics[ scale= 0.24]{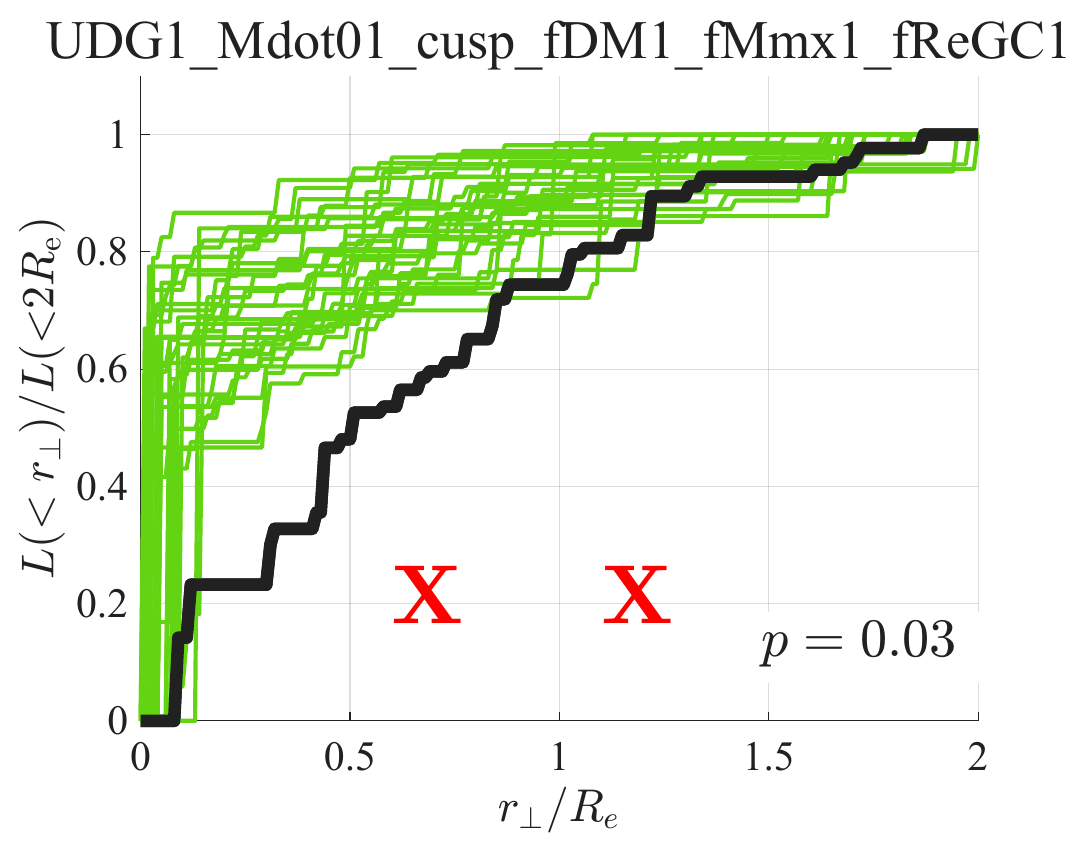} 
      \includegraphics[ scale= 0.24]{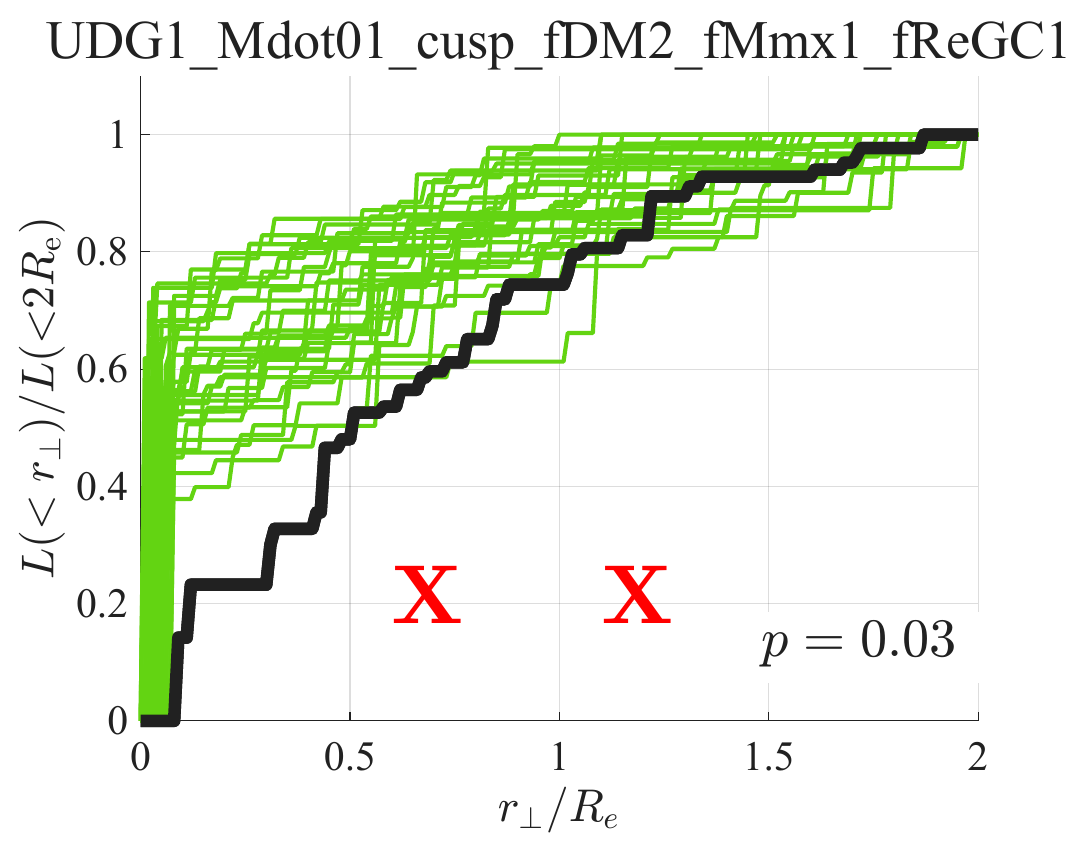}       
      \includegraphics[ scale= 0.24]{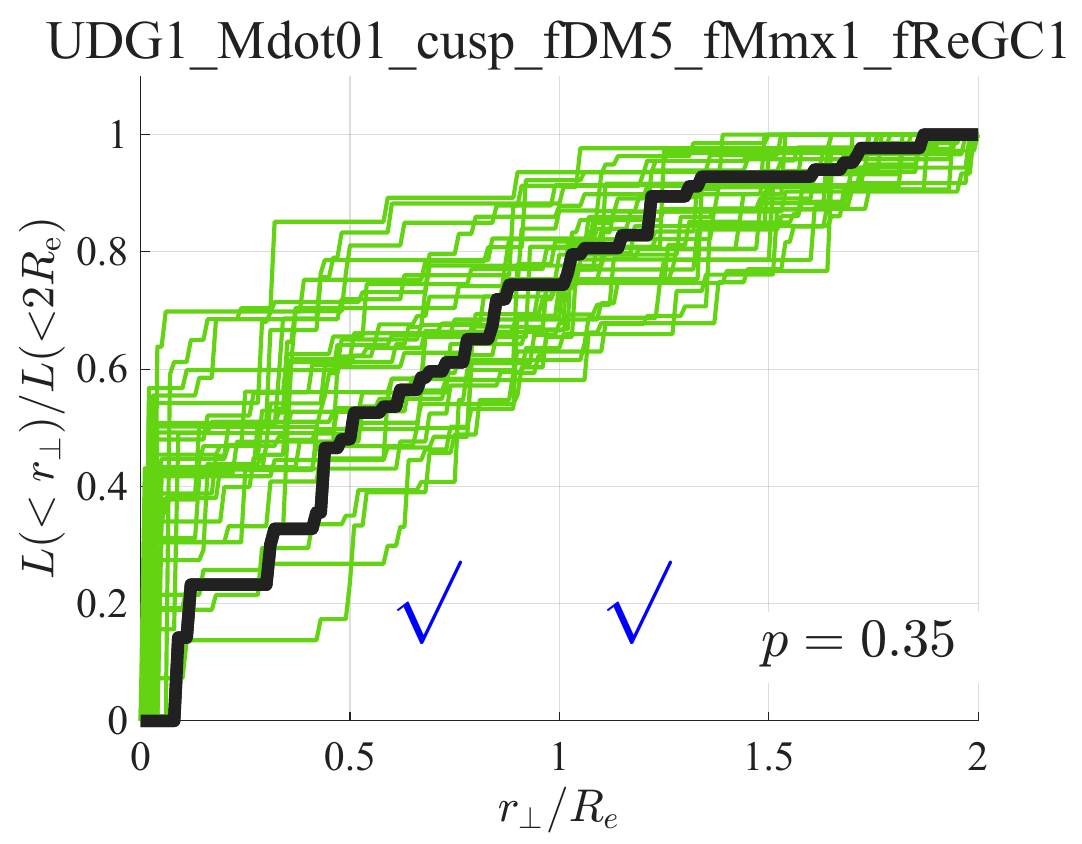}       
      \includegraphics[ scale= 0.24]{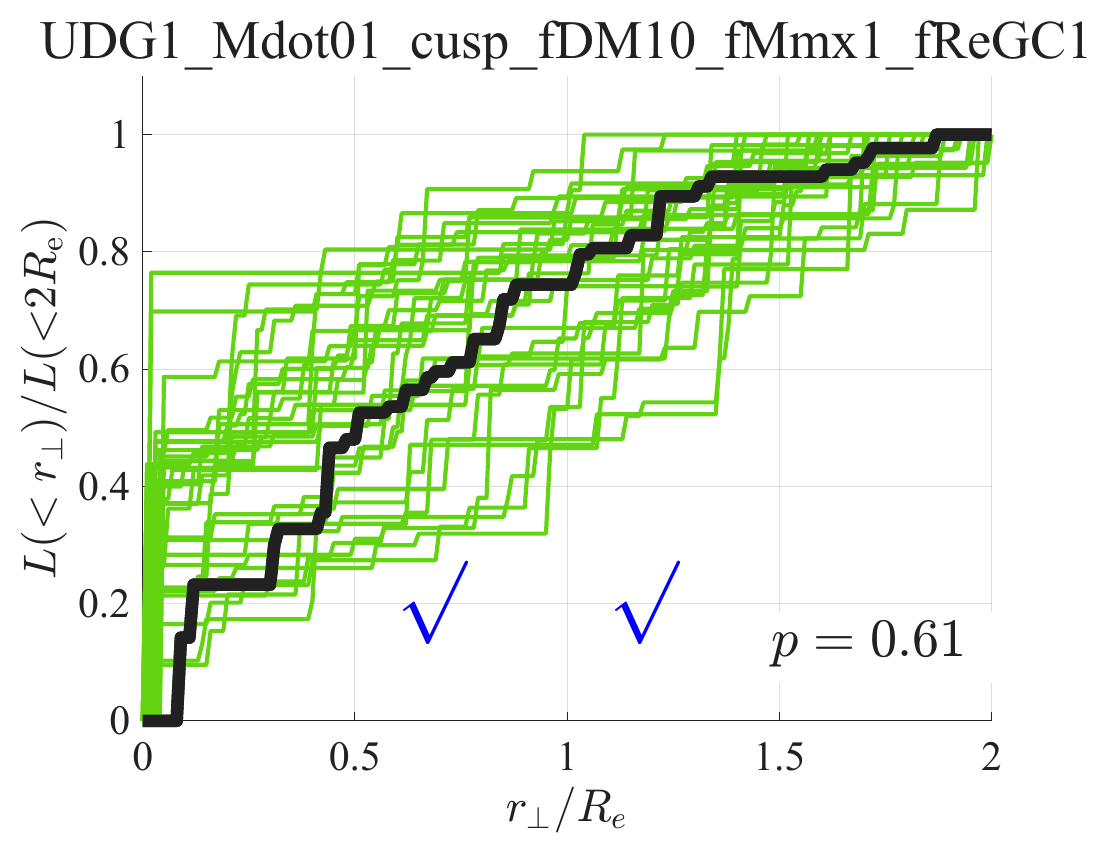}             
 \caption{Luminosity CDF for cusp models: sensitivity to GC mass loss prescription. GC initial radial distribution same as stars. Top panels: fiducial mass loss rate. Bottom panels: same models, with mass loss rate in isolation lower by a factor of ten. 
 }
 \label{fig:NFW_LCDFs_Mdot01}
\end{figure}

\begin{figure}[hbp!]
\centering
      \includegraphics[ scale= 0.24]{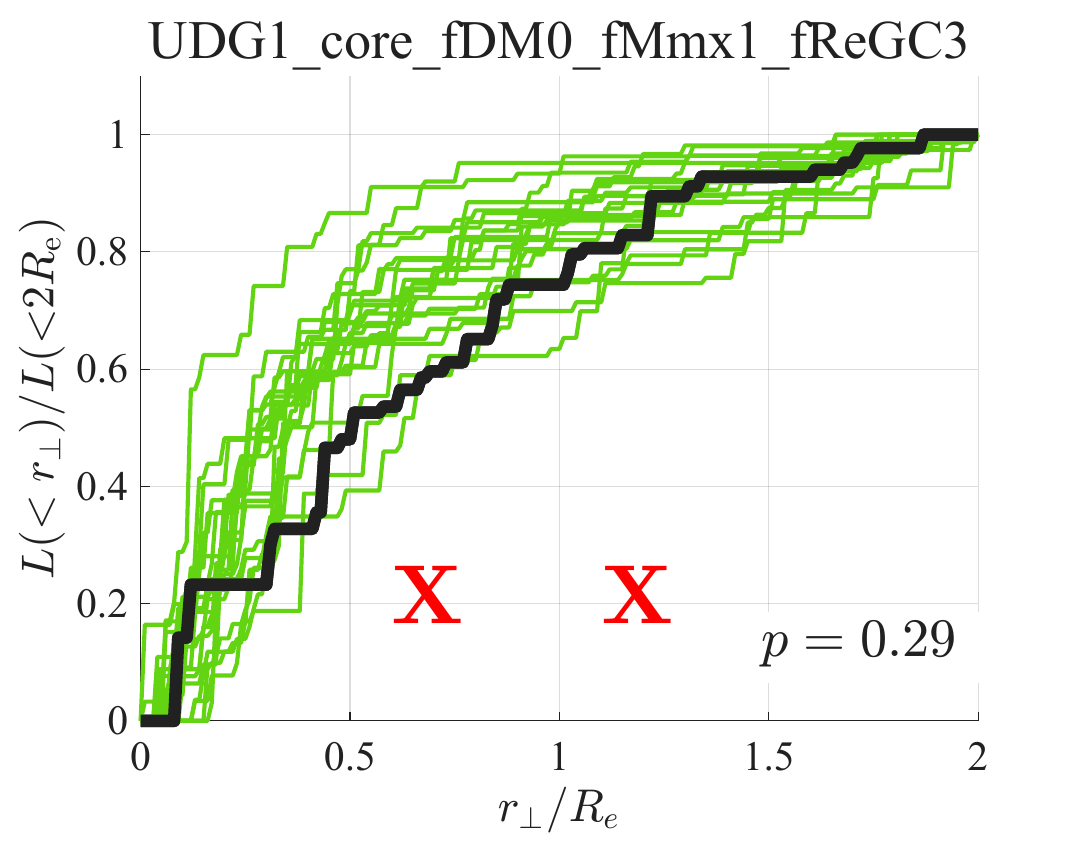} 
      \includegraphics[ scale= 0.24]{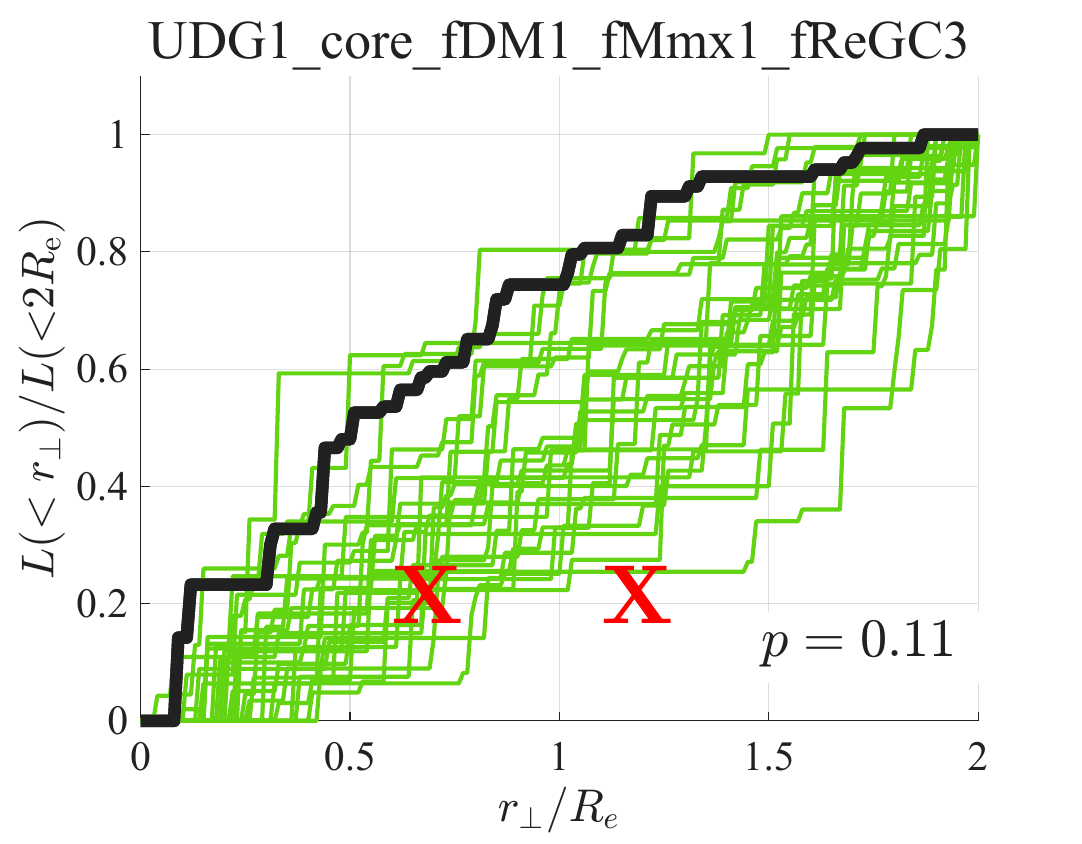}       
      \includegraphics[ scale= 0.24]{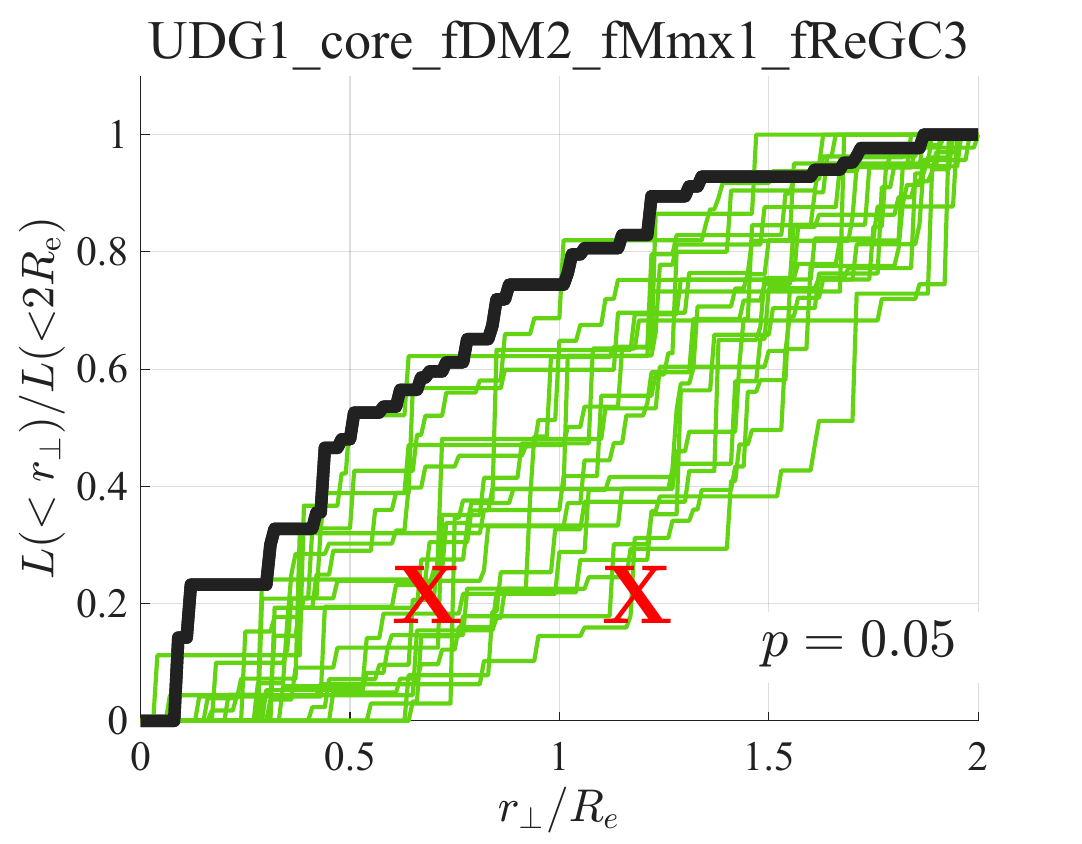}       
      \includegraphics[ scale= 0.24]{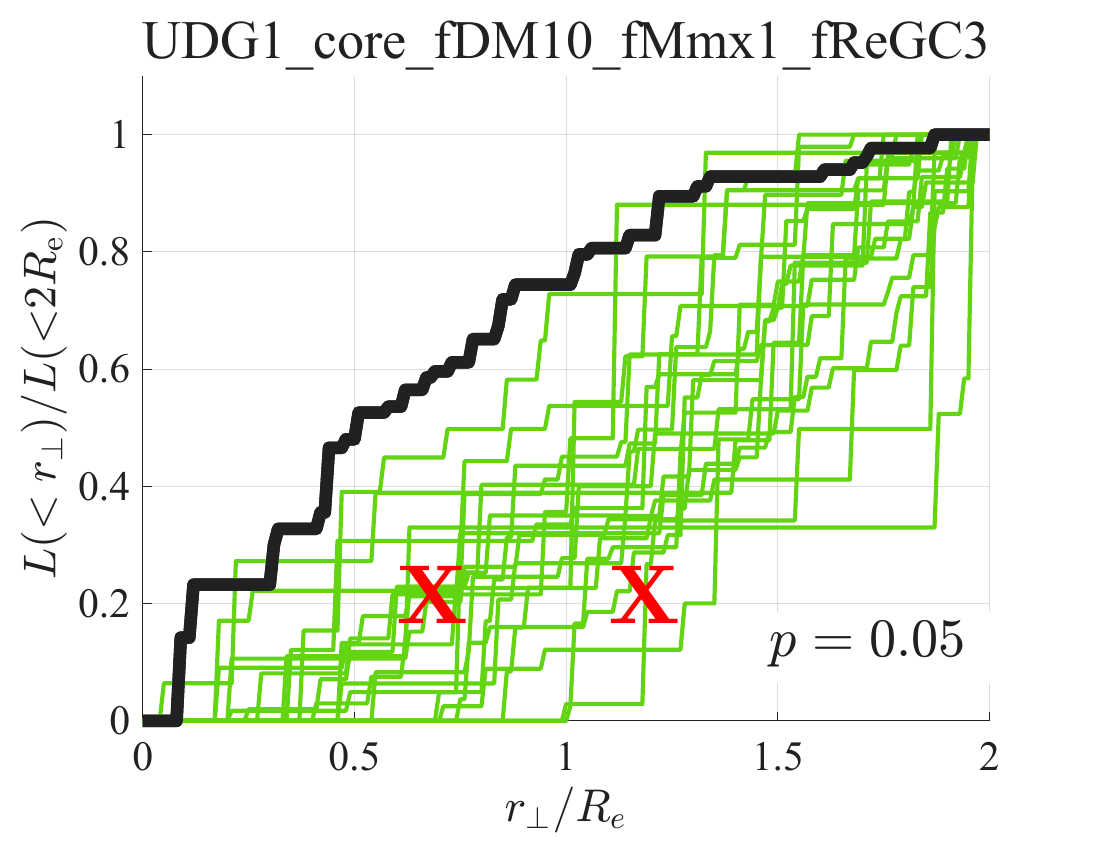}     \\  
      \includegraphics[ scale= 0.24]{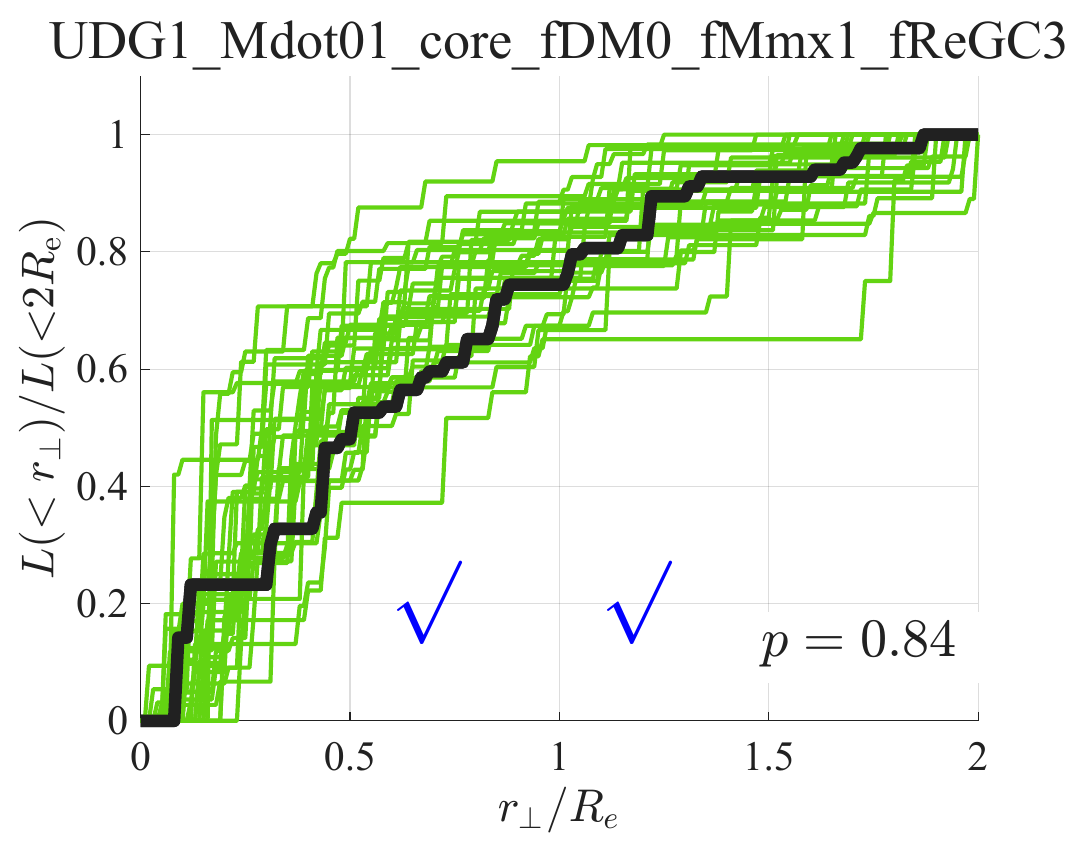} 
      \includegraphics[ scale= 0.24]{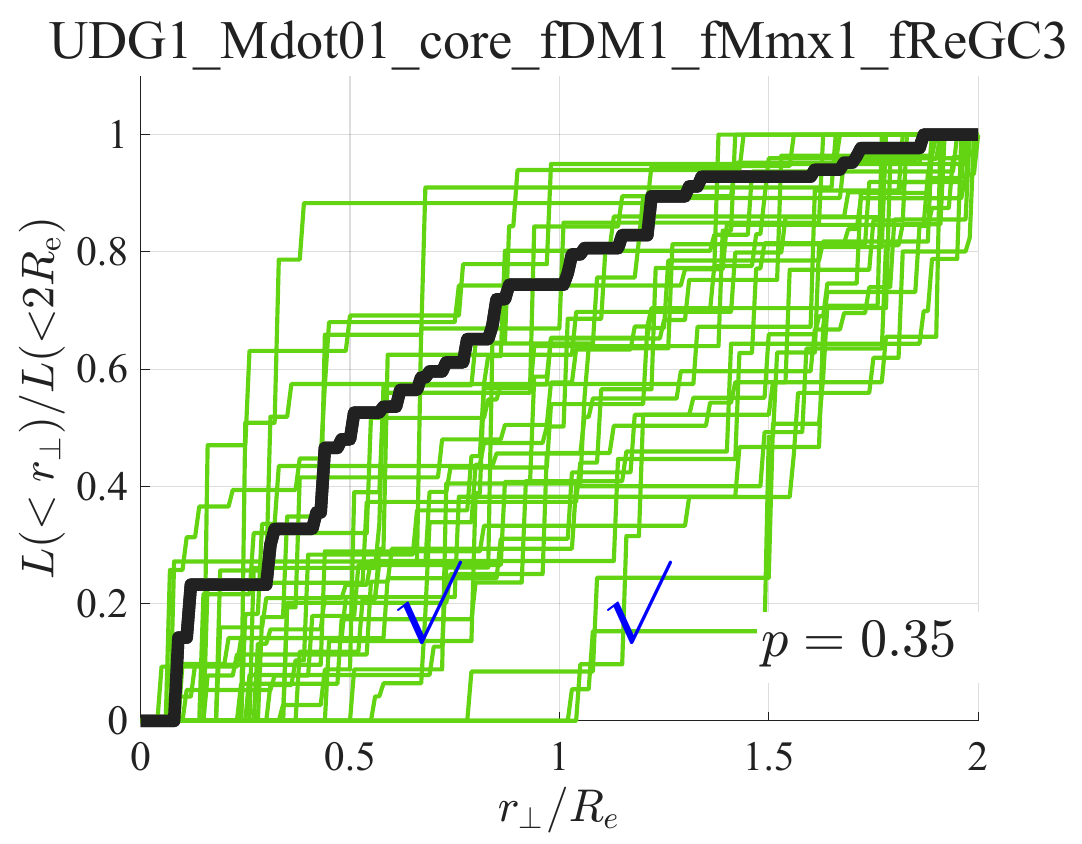}       
      \includegraphics[ scale= 0.24]{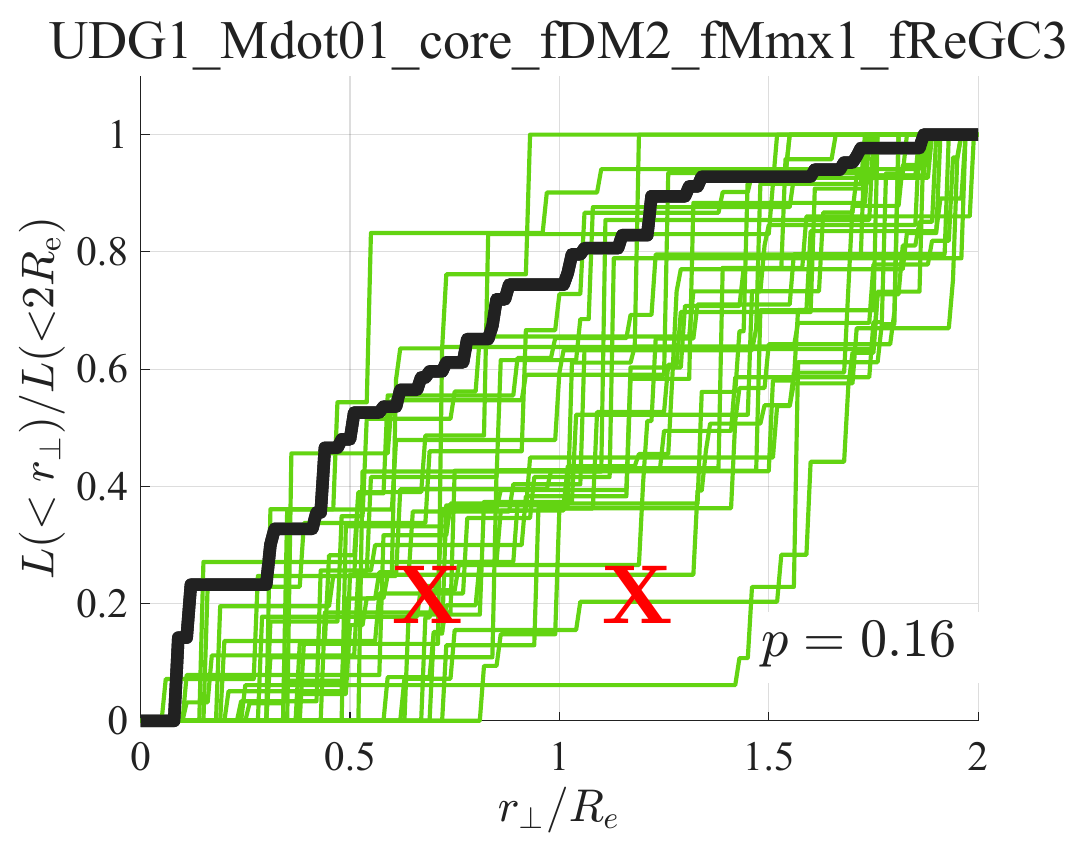}       
      \includegraphics[ scale= 0.24]{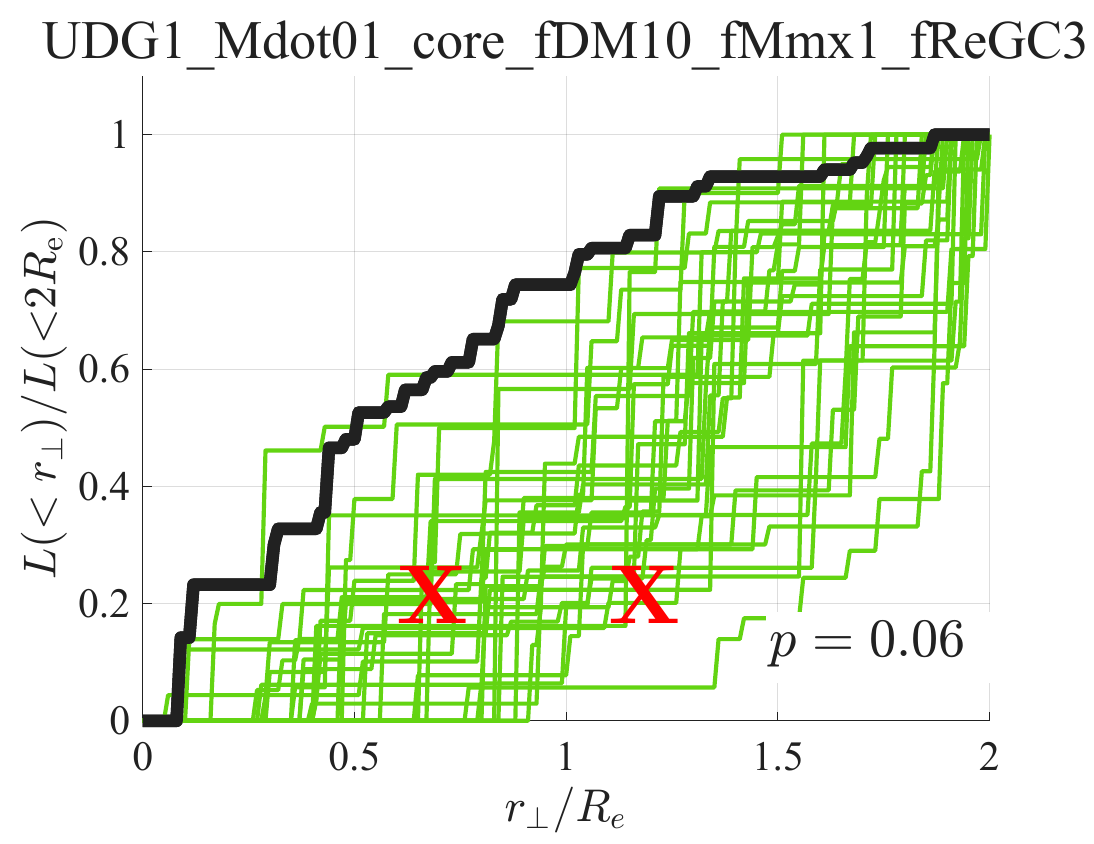}             
 \caption{Luminosity CDF for core models: sensitivity to GC mass loss prescription. GC initial radial distribution stretched by a factor of 3 w.r.t. the stars. Top panels: fiducial mass loss rate. Bottom panels: same models, with mass loss rate in isolation lower by a factor of ten. 
 }
 \label{fig:Burk_LCDFs_Mdot01_strtch3}
\end{figure}

\begin{figure}[hbp!]
\centering
      \includegraphics[ scale= 0.24]{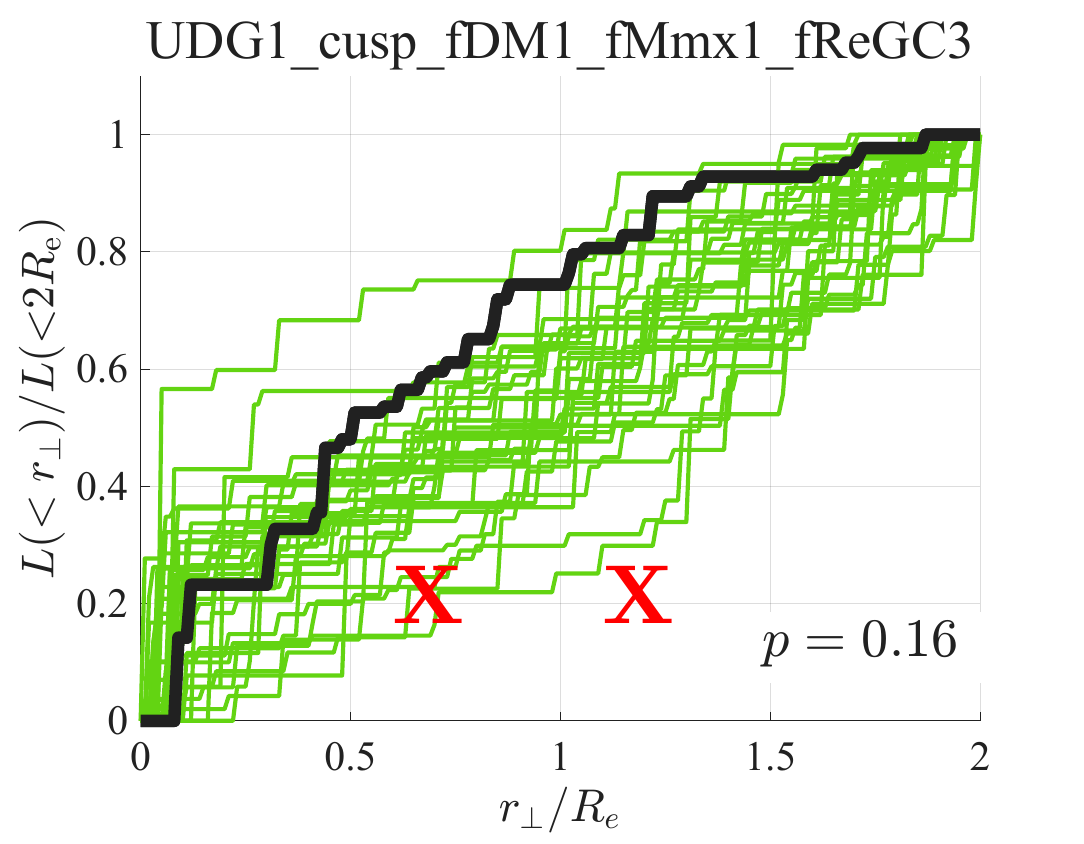} 
      \includegraphics[ scale= 0.24]{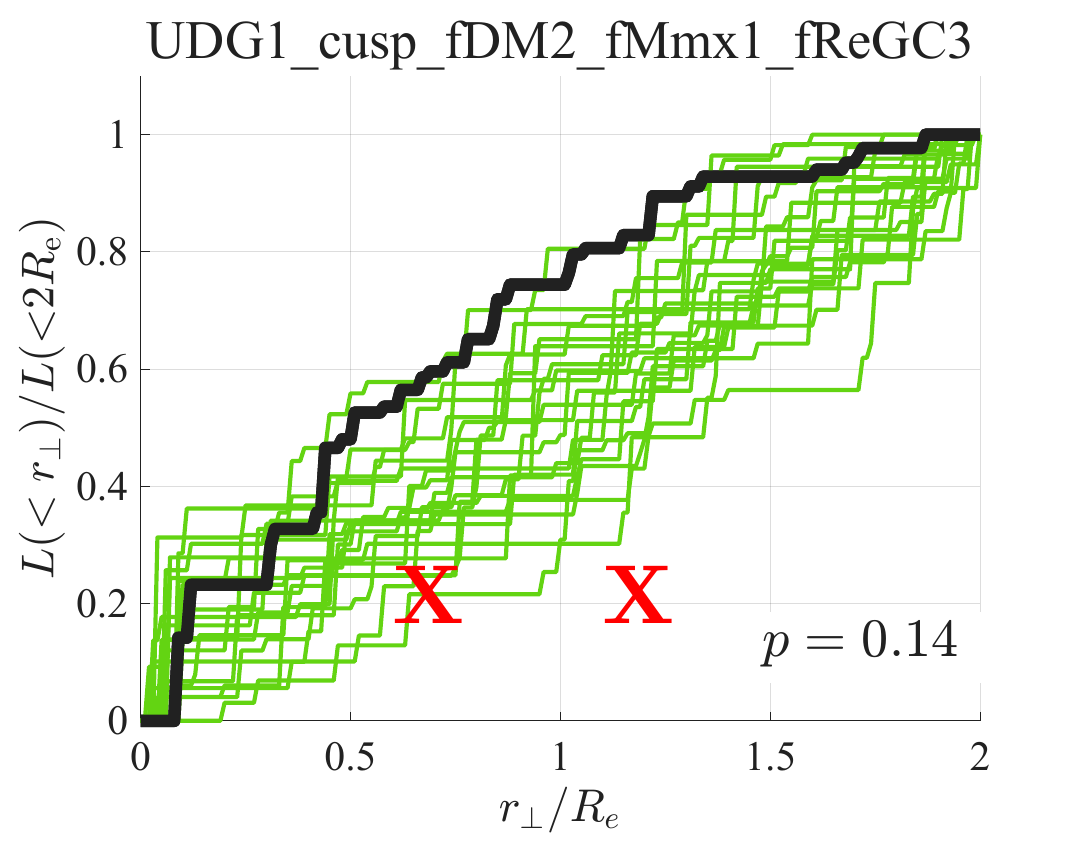}       
      \includegraphics[ scale= 0.24]{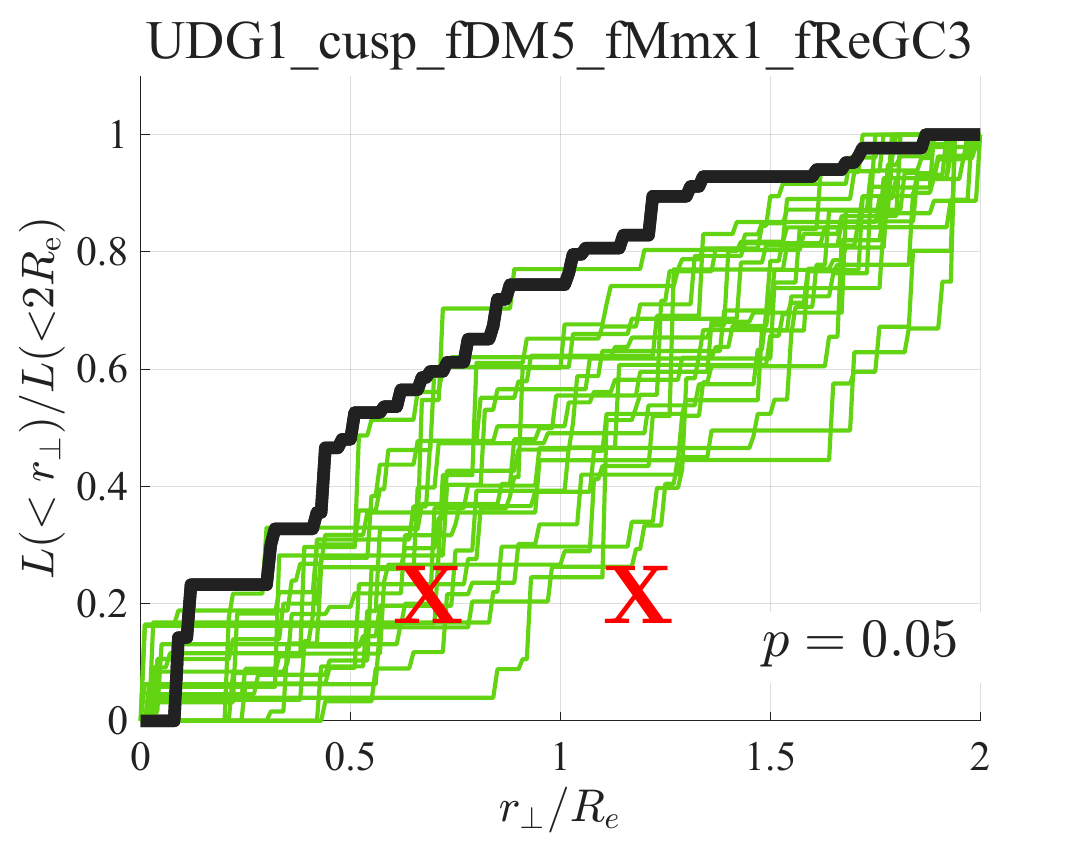}       
      \includegraphics[ scale= 0.24]{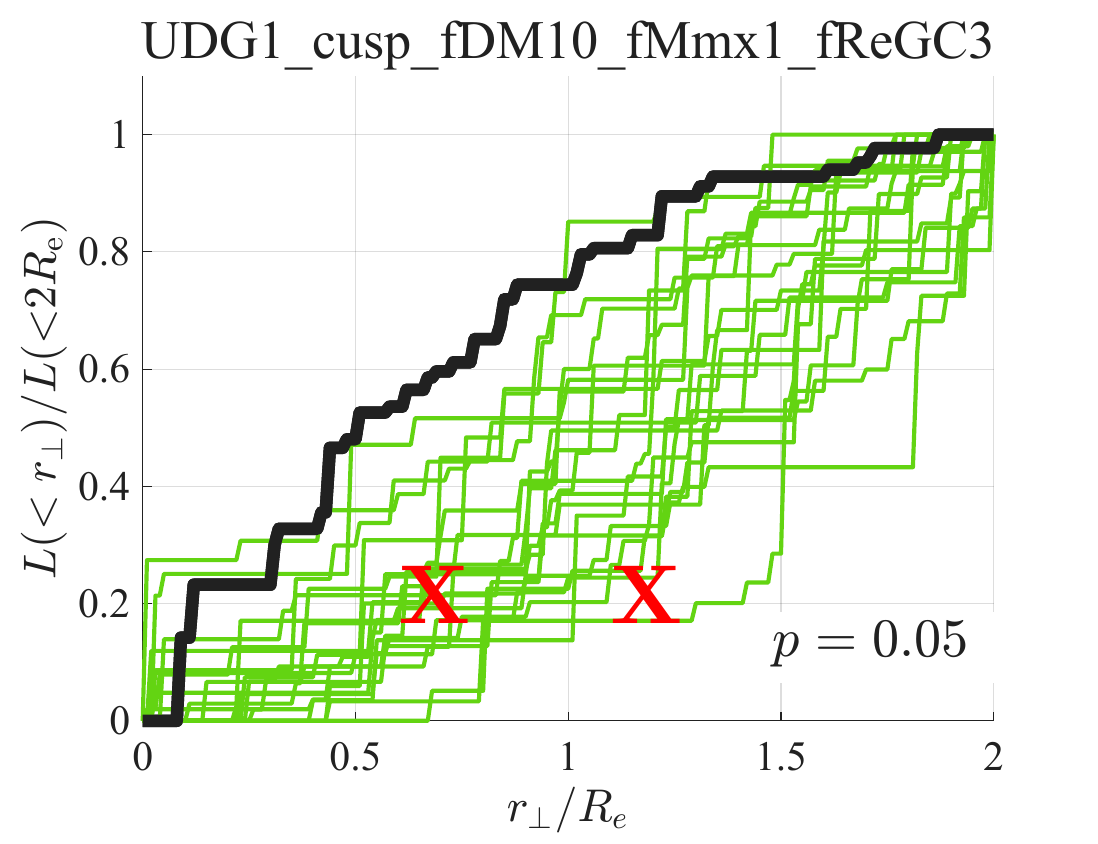}     \\  
      \includegraphics[ scale= 0.24]{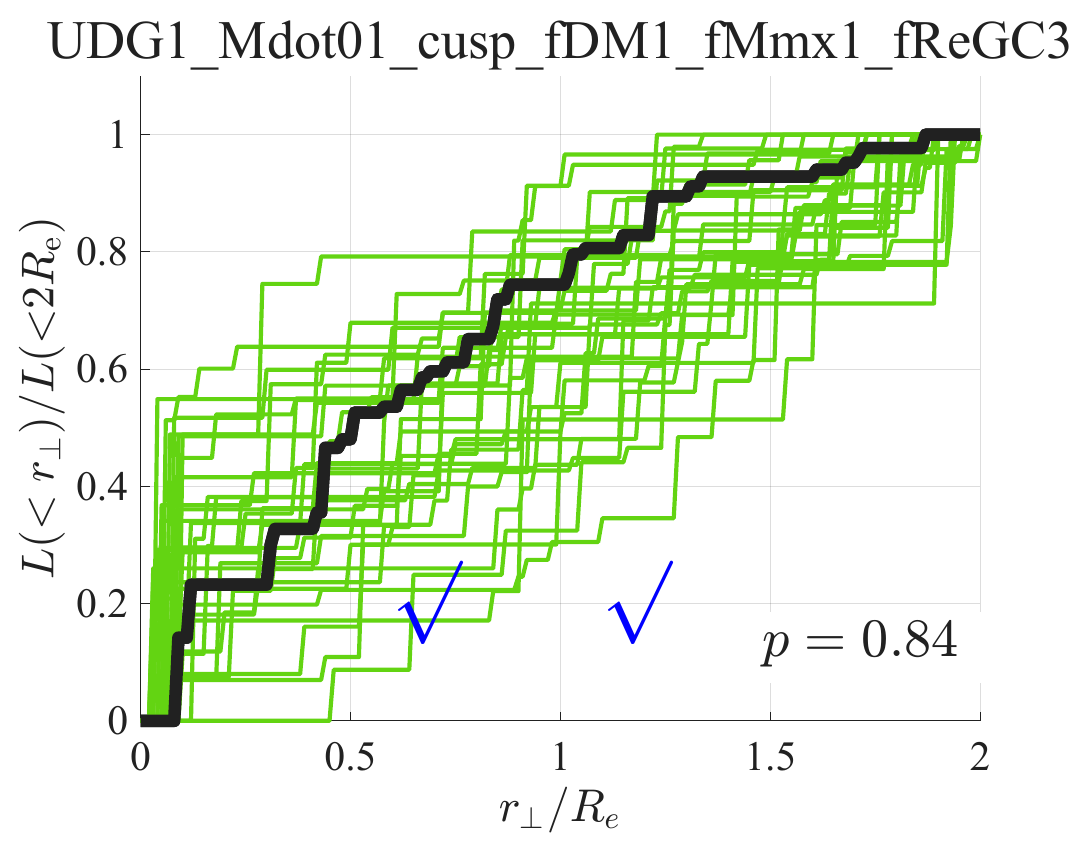} 
      \includegraphics[ scale= 0.24]{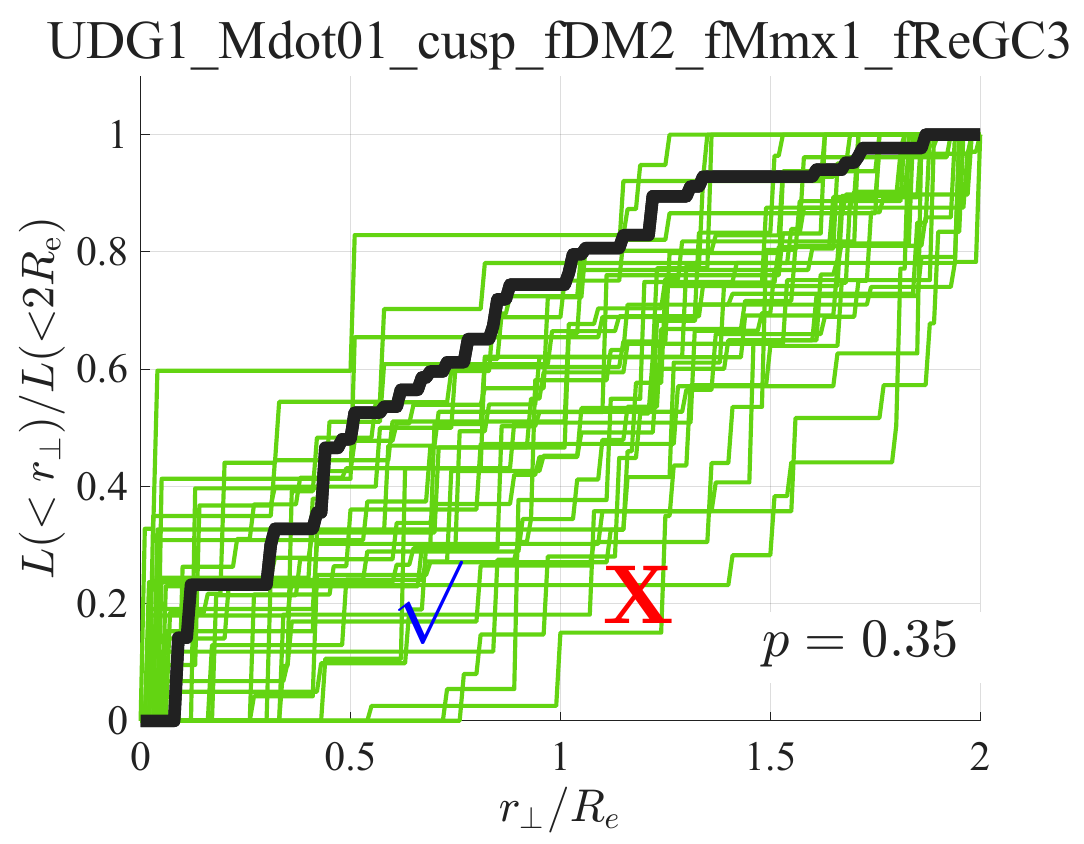}       
      \includegraphics[ scale= 0.24]{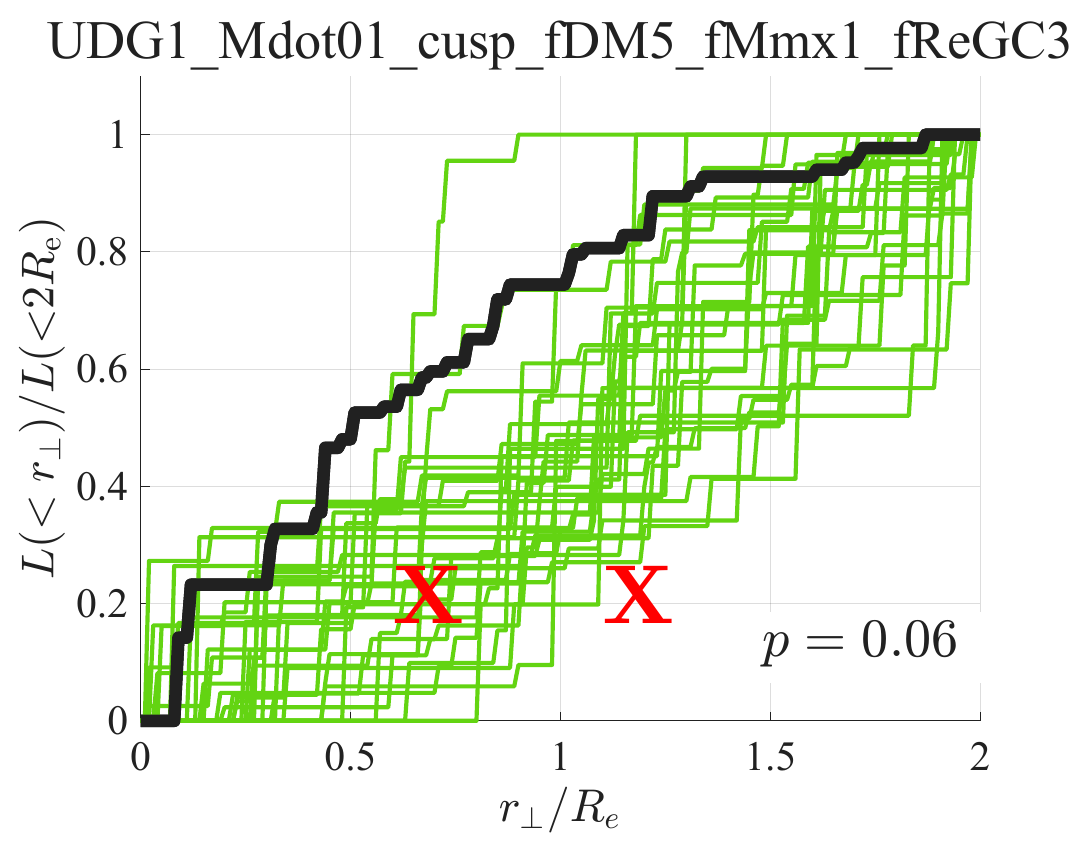}       
      \includegraphics[ scale= 0.24]{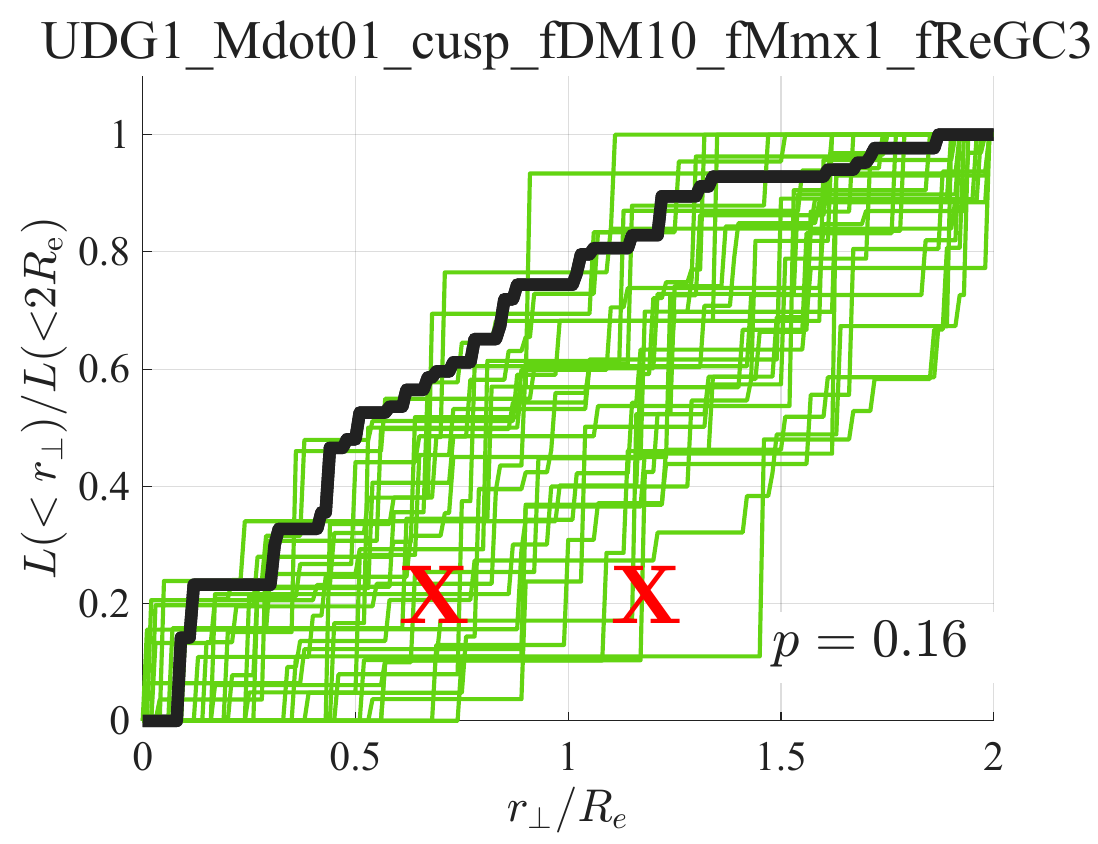}             
 \caption{Luminosity CDF for cusp models: sensitivity to GC mass loss prescription. GC initial radial distribution stretched by a factor of 3 w.r.t. the stars. Top panels: fiducial mass loss rate. Bottom panels: same models, with mass loss rate in isolation lower by a factor of ten. 
 }
 \label{fig:NFW_LCDFs_Mdot01_strtch3}
\end{figure}

Here we compare results of our fiducial GC mass loss prescription (in isolation; Eq.~(\ref{eq:Mdotiso}) to results with mass loss rate lower by a factor of ten. In Figs.~\ref{fig:Burk_LCDFs_Mdot01} and ~\ref{fig:NFW_LCDFs_Mdot01}, the top (bottom) panels refer to the fiducial (low) mass loss rate, for core and cusp halo models, respectively. In these plots the GC initial radial distribution matches that of the stellar body. 
Figs.~\ref{fig:Burk_LCDFs_Mdot01_strtch3} and ~\ref{fig:NFW_LCDFs_Mdot01_strtch3} repeat the analysis with an initial GC radial distribution stretch by a factor of 3.

\end{appendix}

\end{document}